%% file: main.tex
\documentclass{article}

\usepackage{amsmath}

\usepackage[preprint]{neurips_2026}
\AtBeginDocument{\newgeometry{left=1.2in, right=1.2in, top=1in, bottom=1in}}
\usepackage{titletoc}

\newtheorem{remark}{Remark}
\usepackage[utf8]{inputenc} 
\usepackage[T1]{fontenc}    
\usepackage{hyperref}       
\usepackage{url}            
\usepackage{wrapfig}
\usepackage{booktabs}       
\usepackage{subcaption}
\usepackage{amsfonts}       
\usepackage{adjustbox}
\usepackage{siunitx}
\usepackage[table]{xcolor}
\newcommand{\best}[1]{\textbf{#1}}
\newcommand{\bad}[1]{\cellcolor{red!7}\textcolor{red!70!black}{#1}}
\newcommand{\NA}{\multicolumn{1}{c}{--}}
\usepackage{nicefrac}       
\usepackage{microtype}      
\usepackage{xcolor}         
\usepackage{amssymb} 
\usepackage{multirow}
\usepackage{float}
\usepackage{tabularx}
\usepackage{makecell}
\usepackage{graphicx}

\usepackage[scaled=0.95]{helvet}    
\usepackage{pifont}                  
\usepackage{soul}                    
\usepackage{fontawesome5}            
\newcommand{\E}{\mathbb{E}}          

\newcommand{\acroletter}[1]{\textbf{\textsc{#1}}}

\definecolor{BestGreen}{HTML}{DDF3E4}
\definecolor{SecondGreen}{HTML}{EEF8F1}
\definecolor{BadRedOne}{HTML}{FFF0F0}
\definecolor{BadRedTwo}{HTML}{FCE2E2}
\definecolor{BadRedThree}{HTML}{F8D1D1}
\definecolor{BadRedFour}{HTML}{F4C0C0}
\definecolor{BadRedFive}{HTML}{EFAFAF}
\definecolor{BadRedSix}{HTML}{E99E9E}

\newcommand{\bestcell}[1]{\cellcolor{BestGreen}\bfseries #1}
\newcommand{\secondcell}[1]{\cellcolor{SecondGreen}#1}
\newcommand{\badone}[1]{\cellcolor{BadRedOne}#1}

\newcommand{\badthree}[1]{\cellcolor{BadRedThree}#1}
\newcommand{\badfour}[1]{\cellcolor{BadRedFour}#1}
\newcommand{\badfive}[1]{\cellcolor{BadRedFive}#1}
\newcommand{\badsix}[1]{\cellcolor{BadRedSix}#1}

\usepackage[most]{tcolorbox}
\usepackage{xcolor}

\definecolor{anchorholdbg}{HTML}{DCE9DE}   
\definecolor{midbg}{HTML}{E7EBF0}          
\definecolor{anchorconcedebg}{HTML}{F2E2D5}
\definecolor{accepterbg}{HTML}{E9DDE5}     
\definecolor{refuserbg}{HTML}{E5E5E5}      

\definecolor{anchorholdframe}{HTML}{97B49D}
\definecolor{midframe}{HTML}{AAB3BE}
\definecolor{anchorconcedeframe}{HTML}{C9A98E}
\definecolor{accepterframe}{HTML}{B9A3B0}
\definecolor{refuserframe}{HTML}{A8A8A8}

\newtcolorbox{typologybox}[3]{
  enhanced,
  breakable,
  colback=#1,
  colframe=#2,
  boxrule=0.8pt,
  arc=2mm,
  left=1.2mm,
  right=1.2mm,
  top=1mm,
  bottom=1mm,
  title={#3},
  fonttitle=\bfseries,
  coltitle=black
}

\newtcolorbox{quotebox}[1][]{
  enhanced,
  breakable,
  colback=white,
  colframe=black!15,
  boxrule=0.5pt,
  arc=1.5mm,
  left=1mm,
  right=1mm,
  top=0.8mm,
  bottom=0.8mm,
  sharp corners=southwest,
  before skip=2pt,
  after skip=4pt,
  #1
}

\usepackage{enumitem}

\definecolor{tbInk}{HTML}{111111}
\definecolor{tbCardinal}{HTML}{8C1515}
\definecolor{tbWarmGray}{HTML}{5F5D59}
\definecolor{tbMuteGray}{HTML}{77736F}
\definecolor{tbRule}{HTML}{D8D5CF}

\definecolor{haiMagenta}{HTML}{E50071}            
\definecolor{haiMagentaMuted}{HTML}{B85789}       
\definecolor{haiCyan}{HTML}{00A0E2}
\definecolor{haiLavender}{HTML}{C9A8D9}     
\definecolor{haiLavenderDeep}{HTML}{8E6FA7} 
\definecolor{haiLilac}{HTML}{F2EDF8}    
\definecolor{haiBlush}{HTML}{FCEAEF}    
\definecolor{haiSky}{HTML}{E5F4FB}      

\newcommand{\heartfill}{\ding{170}}
\sodef\kicker{}{0.18em}{0.5em plus.05em}{1em plus.1em minus.1em}

\usepackage{titlesec}
\titleformat{\section}
  {\Large\bfseries\color{haiCyan}}
  {\thesection}{0.6em}{}
\titleformat{\subsection}
  {\large\bfseries\color{haiMagentaMuted}}
  {\thesubsection}{0.5em}{}
\titleformat{\subsubsection}
  {\normalsize\bfseries\color{haiLavenderDeep}}
  {\thesubsubsection}{0.5em}{}
\titlespacing*{\section}{0pt}{1.0\baselineskip}{0.35\baselineskip}
\titlespacing*{\subsection}{0pt}{0.7\baselineskip}{0.25\baselineskip}
\titlespacing*{\subsubsection}{0pt}{0.55\baselineskip}{0.2\baselineskip}

\hypersetup{colorlinks=true, allcolors=haiCyan, pdfborder={0 0 0}}

\renewcommand{\maketitle}{}
\makeatletter\renewcommand{\@maketitle}{}\makeatother

\newcommand{\hairline}[2]{\textcolor{#1}{\rule[0.55ex]{#2}{0.45pt}}}

\newtcolorbox{featurecard}[1][haiLilac]{
  enhanced, breakable=false,
  colback=#1, colframe=#1,
  boxrule=0pt, arc=2.2mm,
  left=2.5mm, right=2.5mm, top=2.5mm, bottom=2.5mm,
}

\newtcolorbox{absbox}{
  enhanced, breakable,
  colback=white, colframe=haiLavender,
  boxrule=0.7pt, arc=2.5mm,
  left=5mm, right=5mm, top=4mm, bottom=4mm,
}

\newtcolorbox{resourcebox}{
  enhanced, breakable=false,
  colback=white, colframe=tbRule,
  boxrule=0.5pt, arc=2.5mm,
  left=4mm, right=4mm, top=2.5mm, bottom=2.5mm,
}

\newcommand{\termshero}{%
\vspace*{-0.7in}
\noindent
\begin{minipage}[c]{0.45\linewidth}
\includegraphics[height=36pt]{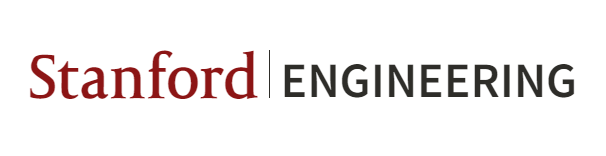}
\vspace{-0.8em}
\end{minipage}\hfill
\begin{minipage}[c]{0.45\linewidth}
\raggedleft
\includegraphics[height=23pt]{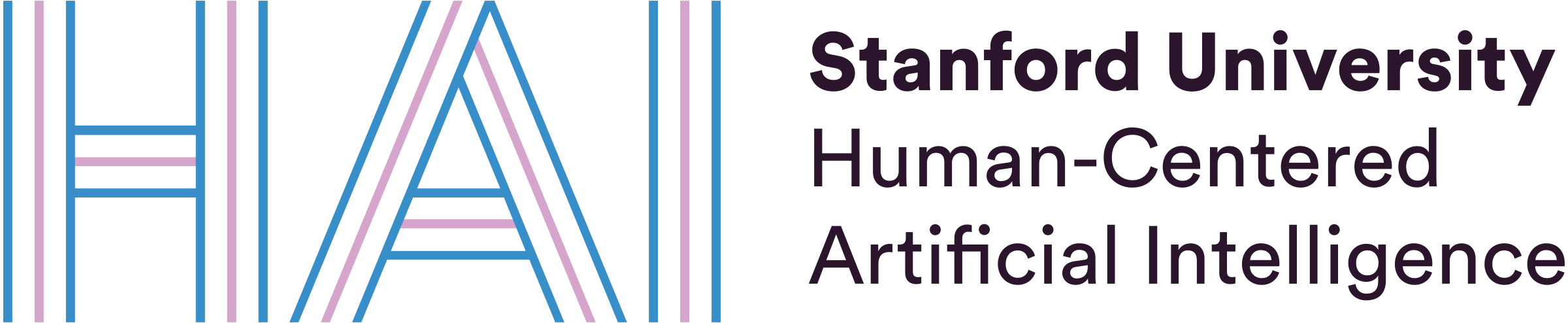}
\vspace{-1em}
\end{minipage}\par
\vspace{4pt}
\noindent\hairline{tbRule}{\linewidth}\par
\vspace{8pt}
\begin{center}
\includegraphics[height=45pt]{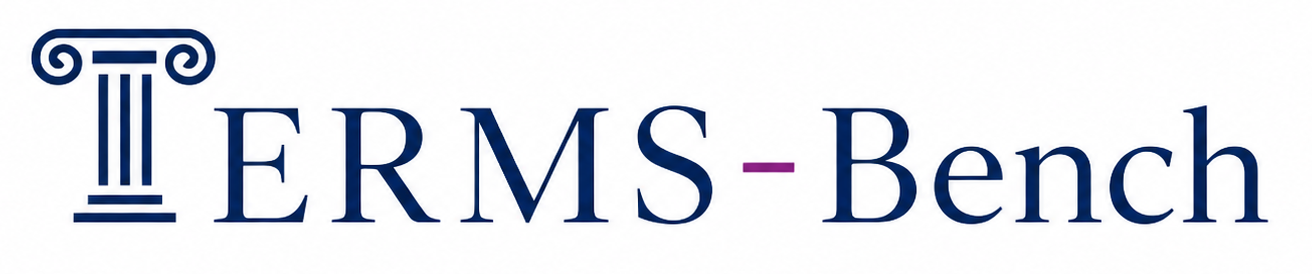}\par
\vspace{-0.2em}
{\fontsize{16}{20}\selectfont\sffamily\bfseries\color{tbInk}%
Diagnosing LLM Negotiation Agents Beyond Deal Rate\par}
\vspace{3pt}
\hairline{haiCyan}{36pt}\enspace\hairline{haiLavender}{36pt}\enspace\hairline{haiMagenta}{36pt}\par
\vspace{8pt}
{\color{tbInk}%
Erica Zhang\textsuperscript{\textcolor{haiCyan}{$\spadesuit$}},\,
Fangzhao Zhang\textsuperscript{\textcolor{haiCyan}{$\spadesuit$}},\,
Aneesh Pappu\textsuperscript{\textcolor{haiCyan}{$\spadesuit$}},\,
Batu El\textsuperscript{\textcolor{haiCyan}{$\spadesuit$}},\,
Jose Blanchet\textsuperscript{\textcolor{haiCyan}{$\spadesuit$}},\\[4pt]
Susan Athey\textsuperscript{\textcolor{haiMagenta}{$\clubsuit$}\textcolor{tbInk}{,}\textcolor{haiMagenta}{\heartfill}},\,
Jiashuo Liu\textsuperscript{\textcolor{haiCyan}{\S}\textcolor{tbInk}{,}\textcolor{haiMagenta}{\dag}},\,
James Zou\textsuperscript{\textcolor{haiCyan}{$\spadesuit$}\textcolor{tbInk}{,}\textcolor{haiMagenta}{\dag}}%
}\par\vspace{6pt}
\noindent{\small\color{tbWarmGray}%
\textsuperscript{\textcolor{haiCyan}{$\spadesuit$}}\,Stanford School of Engineering \quad
\textsuperscript{\textcolor{haiMagenta}{$\clubsuit$}}\,Stanford Department of Economics \quad
\textsuperscript{\textcolor{haiMagenta}{\heartfill}}\,Stanford Graduate School of Business\\[2pt]
\textsuperscript{\textcolor{haiCyan}{\S}}\,Independent \quad
\textsuperscript{\textcolor{haiMagenta}{\dag}}\,Equal Advising%
}\par
\end{center}
\vspace{5pt}
\begin{absbox}
\begin{center}\vspace{-2pt}
\textcolor{haiCyan}{$\bullet$}\,\hairline{haiCyan}{18pt}\quad
{\small\bfseries\color{haiCyan}\kicker{ABSTRACT}}\quad
\hairline{haiMagenta}{18pt}\,\textcolor{haiMagenta}{$\bullet$}\par
\vspace{4pt}
\end{center}
\begingroup
\small
\setlength{\parindent}{0pt}
\setlength{\parskip}{0pt}
Negotiation is a central mechanism of economic exchange, shaping markets,
procurement, labor agreements, and resource allocation. It is also a canonical
testbed for agentic language models, requiring multi-turn interaction under
hidden preferences, strategic communication, and binding constraints. These
properties make negotiation hard to evaluate: unlike math or code, it has no
intrinsic verifier. Existing LLM negotiation evaluations rely on LLM-vs.-LLM
interaction or aggregate outcomes such as deal rate, leaving failures opaque.
We introduce \textsc{Terms-Bench} (\acroletter{T}estbed for
\acroletter{E}conomic \acroletter{R}easoning in \acroletter{M}ulti-turn
\acroletter{S}trategy), a Bayesian-game framework that makes the environment
itself the verifier by specifying the counterpart's latent type, policy, and
payoff structure. We instantiate it in bilateral price negotiation, where the
counterpart's private state and simulator policy are hidden from the agent but
observable to the evaluator. This turns the counterpart from a black-box
opponent into a diagnostic instrument, enabling agent-attributable failure
analysis and oracle-reference optimality gaps.
Evaluating 13 LLM agents spanning frontier systems from major providers,
\textsc{Terms-Bench} turns negotiation evaluation from aggregate ranking into
actionable diagnosis: \emph{where} agents fail, \emph{why} they fail, and
\emph{what} to strengthen. Empirically, frontier models saturate deal rate yet
diverge in surplus extraction, cue use, belief calibration, and compliance,
revealing agent-specific bargaining bottlenecks masked by prior benchmarks.
\par
\endgroup
\end{absbox}
\vspace{6pt}
\begin{tcbraster}[raster columns=3, raster column skip=3mm, raster equal height=rows,
                  raster left skip=0pt, raster right skip=0pt]
\begin{featurecard}[haiSky]
{\small\color{haiCyan}\faIcon{shield-alt}}\enspace{\footnotesize\bfseries\sffamily\hyphenpenalty=10000\exhyphenpenalty=10000\color{haiCyan}Environment-as-Verifier}\par
\vspace{2pt}
{\footnotesize Specify the counterpart's latent type, policy, and payoff structure to make the environment verifiable.}
\end{featurecard}
\begin{featurecard}[haiLilac]
{\small\color{haiMagenta!75!violet}\faIcon{balance-scale}}\enspace{\footnotesize\bfseries\sffamily\hyphenpenalty=10000\exhyphenpenalty=10000\color{haiMagenta!75!violet}Bayesian Bargaining}\par
\vspace{2pt}
{\footnotesize A Bayesian-game framework for bilateral price negotiation under hidden information.}
\end{featurecard}
\begin{featurecard}[haiBlush]
{\small\color{haiMagenta}\faIcon{bullseye}}\enspace{\footnotesize\bfseries\sffamily\color{haiMagenta}Attributable Diagnostics}\par
\vspace{2pt}
{\footnotesize Pinpoint where agents fail, why they fail, and what to strengthen with oracle benchmarks.}
\end{featurecard}
\end{tcbraster}
\vspace{6pt}
\begin{resourcebox}
\noindent
\begin{minipage}[t]{0.32\linewidth}
{\color{haiCyan}\faIcon{trophy}}\quad{\bfseries\sffamily\color{haiCyan}Leaderboard}\par
{\footnotesize\href{https://terms-bench.github.io}{https://terms-bench.github.io}}
\end{minipage}\hfill
\begin{minipage}[t]{0.32\linewidth}
{\color{haiMagenta!75!violet}\faIcon{code}}\quad{\bfseries\sffamily\color{haiMagenta!75!violet}Project Hub}\par
{\footnotesize\href{https://github.com/Terms-bench}{https://github.com/Terms-bench}}
\end{minipage}\hfill
\begin{minipage}[t]{0.32\linewidth}
{\color{haiMagenta}\faIcon{envelope}}\quad{\bfseries\sffamily\color{haiMagenta}Correspondence}\par
{\footnotesize\texttt{\{yz4232,jamesz\}@stanford.edu}}
\end{minipage}
\end{resourcebox}
\vspace{8pt}
}

\title{TERMS-Bench: Diagnosing LLM Negotiation Agents Beyond Deal Rate}

\begin{document}

\termshero

\begin{figure}
    \centering
    \includegraphics[width=1\linewidth]{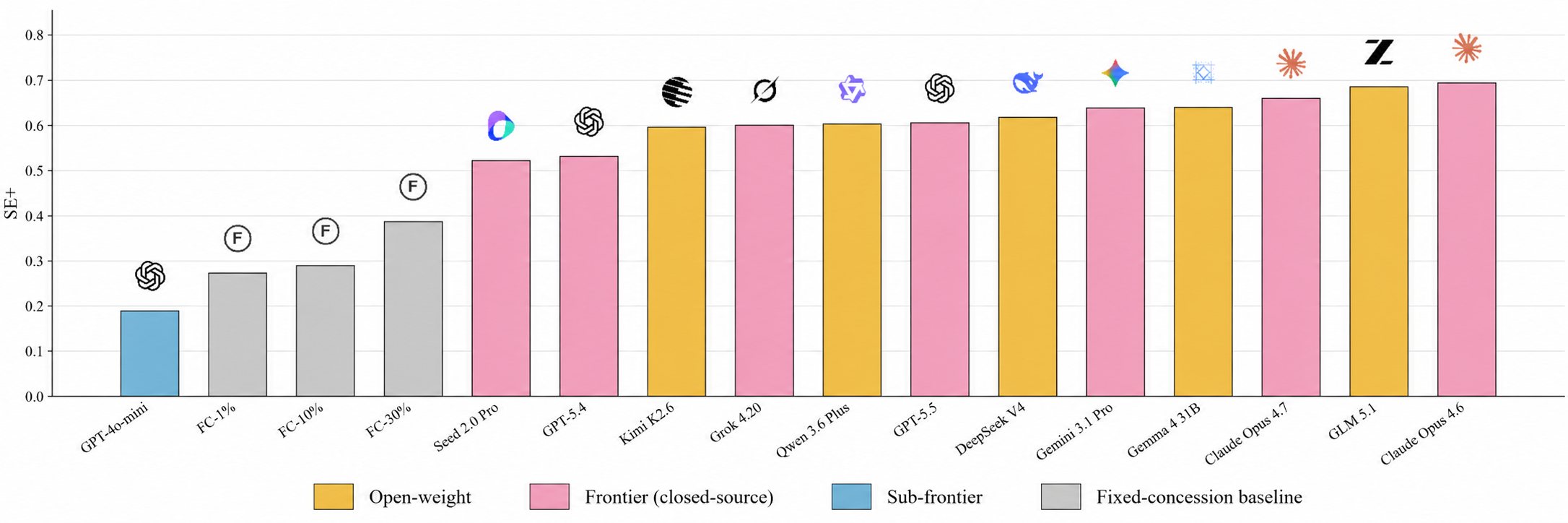}
    \caption{\small \textbf{Teaser terminal performance on the synthetic bilateral negotiation benchmark.} Surplus efficiency on feasible episodes ($SE_\pi^+$, normalized by ZOPA width) for 13 LLM agents and three fixed-concession baselines (FC-1\%,10\%,30\%), each evaluated on the same 1{,}800 seeded episodes against the \textsc{Terms-Bench} simulator. Bars are colored by tier. $SE_\pi^+$ is one of six diagnostic metrics spanning terminal value, agreement calibration, opponent modeling, and protocol compliance reported in \S\ref{sec:results}; product-grounded and other instantiations are deferred to \S\ref{sec:sim},\S\ref{sec:experiments}.}
    \label{fig:placeholder}
\end{figure}

\section{Introduction}
\input{sections/intro}

\vspace{-0.5em}
\section{Negotiation as a Diagnostic Game}\label{sec:methods}
\input{sections/method}
\vspace{-0.5em}
\section{Simulator Design}\label{sec:sim}
\input{sections/sim}
\vspace{-0.5em}
\section{Experiments}\label{sec:experiments}
\input{sections/exp}
\vspace{-0.2cm}
\section{Conclusion}\label{sec::conclusion}
\vspace{-0.2cm}
\input{sections/conclusion}
\section*{Acknowledgement}\label{sec:ack}
\input{acknowledgement}
\bibliographystyle{plainnat}
\bibliography{main}

\clearpage
\appendix

\thispagestyle{plain}
\addcontentsline{toc}{section}{Appendix}
\begin{center}
\vspace*{-1em}
{\scriptsize\color{tbCardinal}\kicker{APPENDIX}}\par
\vspace{6pt}
{\fontsize{26}{30}\selectfont\bfseries\color{tbCardinal}Contents\par}
\vspace{8pt}
\hairline{haiCyan}{36pt}\enspace\hairline{haiLavender}{36pt}\enspace\hairline{haiMagenta}{36pt}\par
\end{center}
\vspace{1.2em}

{
\hypersetup{hidelinks}
\small
\setlength{\parskip}{3pt}

\titlecontents{section}
  [3em]
  {\vspace{0.25em}\bfseries\color{haiCyan}}
  {\contentslabel{2.2em}}
  {}
  {\textcolor{tbRule}{\titlerule*[2pc]{.}}\;\textcolor{tbWarmGray}{\contentspage}}

\titlecontents{subsection}
  [4.8em]
  {\footnotesize\color{haiMagentaMuted}}
  {\contentslabel{2.8em}}
  {}
  {\textcolor{tbRule}{\titlerule*[1pc]{.}}\;\textcolor{tbMuteGray}{\contentspage}}

\startcontents[appendix]
\printcontents[appendix]{}{1}{\setcounter{tocdepth}{2}}
}

\newpage

\input{appendix/validity}
\input{appendix/formal_specification}
\input{appendix/counterpart_modeling}

\input{appendix/reference_oracle}
\input{appendix/belief}
\input{appendix/metrics}
\input{appendix/grader}
\input{appendix/implementation}
\input{appendix/ablation}

\input{appendix/leaderboard}
\input{appendix/prompts}

\end{document}

%% file: sections/intro.tex
Negotiation lies at the intersection of reasoning, communication, and social cognition, and serves as a canonical setting for agentic intelligence. It pervades commercial workflows such as procurement, contracting, pricing, logistics, where decisions are made under incomplete information, asymmetric incentives, and operational constraints \citep{raiffa1982art, bazerman1992negotiating}. Evaluation of LLM agents sits along a verifiability spectrum. At one end, math \citep{hendrycks2021measuring} and code \citep{chen2021evaluating, jimenez2024swebenchlanguagemodelsresolve} have intrinsic verifiers that yield automatic correctness signals and enable reinforcement learning from verifiable rewards \citep{lambert2024tulu3, deepseekai2025r1}. At the other end, open-ended generation has no reference solution and relies on human or LLM-as-a-judge proxies \citep{zheng2023judging, jia2025writingzero}. Negotiation occupies a \emph{semi-verifiable} middle ground: terminal outcomes such as agreement are objective, but the multi-turn reasoning and language strategy that produce them are not. This makes negotiation both a stress test for agentic systems and a methodological challenge for evaluation: the verifier must be constructed, and how it is constructed determines what can be diagnosed.

The design of LLM negotiation agents has evolved from end-to-end neural systems \citep{lewis2017deal} to modular architectures that decouple strategic reasoning from language realization \citep{he2018decoupling, yarats2018hierarchical}, strengthened through domain adaptation, structured reasoning, game-theoretic objectives, and behavioral conditioning \citep{chatterjee2024agreemate, yao2023react, hwang2023promptable, hua2024gametheoretic, cohen2025exploring}. This progress in agent design has been accompanied by parallel advances in evaluation. Building on early dialogue corpora \citep{lewis2017deal, he2018decoupling, chawla2021casino}, recent LLM negotiation benchmarks have advanced along several fronts: multi-turn, multi-agent, and multi-product market simulation with private reservation values and welfare-based scoring \citep{bianchi2024negotiationarena, liu2026agenticpaymultiagentllmnegotiation}; multi-party stakeholder games with cooperative, greedy, and adversarial incentives that surface manipulation and coalition dynamics \citep{abdelnabi2024cooperation}; and human-preference-validated utility metrics that move beyond profit-only evaluation \citep{oh2026meritfeedbackelicitsbetter}. In parallel, game-theoretic benchmarks such as \texttt{GTBench} \citep{duan2024gtbenchuncoveringstrategicreasoning} and \texttt{Alympics} \citep{mao2024alympicsllmagentsmeet} probe abstract strategic competence in canonical games, largely via LLM-vs-LLM play. Despite their differences, these frameworks share a common evaluation paradigm: the evaluated agent negotiates against another LLM, and competence is summarized by terminal outcomes such as deal rates or surplus shares. This paradigm has two structural limits. First, because the counterpart is itself a black-box policy with unspecified latent preferences, outcomes confound agent competence with counterpart variability and
preclude attribution to specific reasoning failures. Second, even
multi-component or human-preference-validated metrics ultimately collapse
performance into scalar rankings, obscuring where in the negotiation workflow an agent succeeds or fails.

Together, these limits point to a broader gap: existing evaluation measures \emph{what} outcome is reached, but not \emph{how} it arises from the agent's reasoning process. The LLM-vs.-LLM paradigm constructs verifiability by using another language model as a proxy judge \citep{zheng2023judging}, preserving non-verifiability inside the counterpart rather than resolving it. For deployable agents operating under uncertainty and operational constraints, this opacity matters: failures that \emph{would} be diagnosable under a transparent evaluator stay hidden. Empirical work shows that failures often arise within the negotiation process itself, through unstable reasoning, prompt sensitivity, and latent behavioral variability \citep{kwon-etal-2024-llms, schneider2024negotiating, huang2024personality}. Meaningful evaluation therefore requires an evaluator-transparent counterpart: one against which outcomes can be attributed to specific agent behaviors and systematic failures surfaced before deployment.

We introduce \textsc{Terms-Bench}, a Bayesian-game framework for negotiation evaluation that constructs the environment itself as the verifier. This is a third approach to evaluating semi-verifiable agentic tasks: where LLM-as-judge methods \citep{zheng2023judging, dubois2024alpacafarm} construct verifiability through a proxy model and outcome-rule benchmarks \citep{zhou2024webarena, liu2023agentbench} through hand-specified criteria, \textsc{Terms-Bench} constructs it through the environment, so agent behavior can be attributed rather than merely scored. Benchmark instantiations are specified by varying the action space, observation modality, reward structure, and counterpart policy. In this paper, we instantiate \textsc{Terms-Bench} in bilateral price negotiation under incomplete information. By fully specifying the negotiation kernel, the benchmark makes the counterpart a diagnostic instrument: the evaluator observes the latent state and policy needed to explain agent success or failure, enabling simulator-defined reference policies and optimality gaps.  Our contributions are:

\begin{itemize}[leftmargin=*, nosep]
\item \textbf{A Bayesian-game framework with attributable, reference-based evaluation.}
We introduce \textsc{Terms-Bench}, a framework that specifies the counterpart as
a parameterized kernel with latent type and language-mediated cues, supporting
benchmark instantiations along action-space, modality, and reward axes. We
instantiate the framework in bilateral price negotiation, where a fixed
counterpart policy enables agent-attributable failure analysis as well as simulator-defined oracle 
reference policies and optimality gaps.

\item \textbf{Capability-isolating counterpart families.}
We introduce six counterpart families that vary economic reactivity, cue
reliability, noise, and pressure. Cross-family degradation isolates failures in
cue use, latent-type inference, calibration under noisy evidence, and robustness
to adversarial pressure.

\item \textbf{Diagnostics and interventions beyond deal rate.}
We evaluate feasible and infeasible episodes separately across four axes:
terminal value, agreement calibration, opponent modeling, and protocol
compliance. Oracle-posterior and revealed-type interventions further decompose
performance gaps into inference, uncertainty, and control failures.
\end{itemize}

Together, these contributions turn negotiation evaluation into a diagnostic tool: not a leaderboard, but a controlled foundation for failure attribution that shows practitioners \emph{where} agents break down and \emph{what} to strengthen.

%% file: sections/method.tex
In this section, we formalize \textsc{Terms-Bench}, the Bayesian-game framework
that makes the environment itself the verifier, and instantiate it in bilateral
price negotiation under incomplete information. The framework and instantiation
are grounded in economic bargaining and negotiation-analysis primitives; see
Appendix~\ref{appdx:validity}.

\vspace{-0.5em}
\subsection{\textsc{Terms-Bench}: A Bayesian-Game Framework}
\label{sec:framework_general}

We model negotiation as an \emph{extensive-form, incomplete-information} game,
following classical Bayesian bargaining formulations
\citep{myerson1984twoperson, ausubel2002bargaining}. \textsc{Terms-Bench}
separates the \emph{bargaining environment and protocol} from the
\emph{agent policy} that acts within it. This fixed environment enables
diagnosis: performance differences can be attributed to agent policies rather
than counterpart variability.

\textbf{Bargaining environment and protocol.} A bargaining environment
is
\begin{equation}
\Gamma = (F, T_A, T_B, u_A, u_B, \mu),
\end{equation}
where $F$ is the set of feasible outcomes (including disagreement
$\bot$), $T_i$ is player $i$'s private type space, $u_i$ is the utility
function, and $\mu \in \Delta(T_A \times T_B)$ is the prior over latent
types.\footnote{We present the two-player form for clarity; the
formulation extends naturally to multi-party and multi-product settings
by replacing $(T_A, T_B)$ with $\{T_i\}_{i \in N}$ over a player set
$N$, with $\mu \in \Delta(\prod_{i \in N} T_i)$.} Each type
$t_i = (r_i, e_i, \kappa_i, \eta_i)$ encodes a reservation value $r_i$,
an outside-option payoff $e_i$, urgency $\kappa_i$, and a behavioral
parameter $\eta_i$ such as strategic stance. A protocol $\mathcal{P}$ is common knowledge and
fixes (i) the move ordering, (ii) admissible action space
$\mathcal{A} = \mathcal{D} \times \mathcal{L}$, (economic acts paired with natural-language realizations), (iii) information revelation rules, and
(iv) termination conditions. 

\textbf{Agent policy.}
Given $(\Gamma, \mathcal{P})$, an agent's behavior is characterized by a policy
$\pi: \mathcal{S} \to \mathcal{A}$ mapping information states to actions. At round $k$, the agent's information state
$
s_k = (h_k, x_A, b_k) \in \mathcal{S},
$
contains the public interaction history $h_k$, private side information $x_A$
(e.g., the agent's reservation value and market context), and the agent's
belief $b_k \in \Delta(T_B)$ over the counterpart's type.\footnote{Throughout, subscript $A$ denotes the evaluated agent and $B$ denotes the counterpart.} The policy therefore
captures both counterpart inference from observed signals and strategic action
selection. Together with the fixed counterpart policy $\pi_B$ and prior $\mu$,
$\pi$ induces terminal outcomes, allowing evaluation to compare policies under
the same environment. We defer the full information-state notation,
belief-update mechanics, and policy class to Appendix~\ref{appdx:formal_spec}.

\textsc{Terms-Bench} is intentionally general: instantiations may vary the
action space, observation modality, reward structure, and counterpart policy.
In this paper, we study a bilateral price-negotiation instantiation in which
every component of $(\Gamma, \mathcal{P}, \pi_B)$ is available to the evaluator.
\vspace{-0.5em}
\subsection{Bilateral Price-Negotiation Instantiation}
\label{sec:agora_base_setup}

In this instantiation, an evaluated buyer or seller negotiates a single scalar
price with an environment-simulated counterpart over up to $K$
alternating-offer rounds. The counterpart has a fully specified latent type that parameterizes its behavior and that the agent must infer from observed prices and messages. The agent, by contrast, is assigned only a reservation value $r_A$ as a hard individual-rationality constraint; its broader negotiation behavior is left to emerge through interaction.

\begin{figure}[t]
    \centering
    \includegraphics[width=1.0\linewidth]{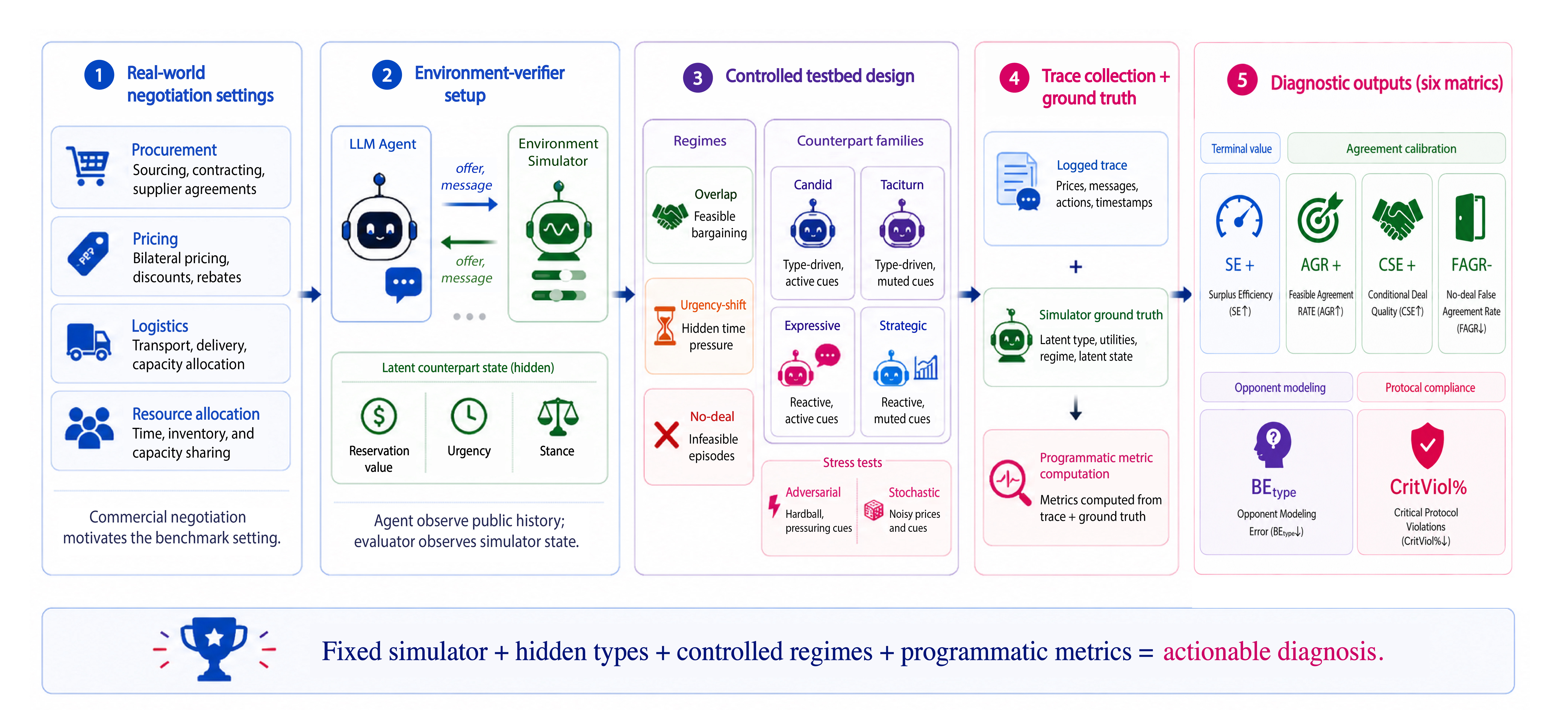}
    \vspace{-1em}
    \caption{\small \textbf{Overview of \textsc{Terms-Bench}.}
Commercial negotiation settings motivate an environment-verifier evaluation
pipeline: an LLM agent negotiates with a fixed simulator whose latent type,
policy, and payoff structure are hidden from the agent but observable to the
evaluator. Controlled regimes test feasible bargaining, urgency shifts, and
no-deal cases, while simulated counterpart families vary whether behavior is
driven mainly by hidden private constraints or recent agent behavior, and
whether language cues are informative or muted. Additional adversarial and
stochastic families provide stress tests. The evaluator combines logged traces
with simulator ground truth to compute six diagnostic metrics for surplus,
agreement, no-deal recognition, opponent modeling, and protocol compliance
(\S\ref{sec:evaluation}, \S\ref{sec:counterpart_model}).}
    \label{fig:overview}
\end{figure}
\vspace{-0.5em}
\subsubsection{Environment Specification}

\textbf{Counterpart type and regimes.}
For each episode, the counterpart draws a private type
$t_B = (r_B, \kappa_B, \eta_B) \sim \mu$, where $r_B \in \mathbb{R}_{+}$
is its reservation value (a buyer counterpart will not pay above $r_B$; a
seller counterpart will not accept below $r_B$), $\kappa_B \in [0,1]$ is its
urgency, and $\eta_B \in \{\texttt{conciliatory}, \texttt{neutral},
\texttt{aggressive}\}$ is its strategic stance.\footnote{We use stance rather
than personality as the primary behavioral driver; see
Appendix~\ref{app:stance_personality} for literature grounding
\citep{ma2005bigfive_conflict_negotiation, huang2024personality,
cohen2025exploring}.}
We fix the outside option $e=0$ for both parties. The type $t_B$ fully
parameterizes the counterpart kernel $\pi_B$
(\S\ref{sec:counterpart_model}) and must be inferred by the agent from observed
prices and messages.

To cover diverse bargaining scenarios, we define the environment prior as
$\mu = \sum_{m=1}^M w_m \, \mu_m$, with $w_m \geq 0 \ \text{and} \ \sum_m w_m = 1$,
where each component $\mu_m$ represents a structurally distinct regime.
We use three canonical regimes targeting different
competencies: (i) \emph{Overlap}, feasible bargaining and surplus extraction;
(ii) \emph{Urgency-shift}, time-pressure adaptation; and (iii) \emph{No-deal},
exit discipline under infeasibility. The regime draw specifies the joint
reservation geometry and hence both $r_B$ and the agent's reservation value
$r_A$. Full sampling distributions are given in
\S\ref{sec:regime_instantiation}.

\textbf{Protocol.} At round $k$, the agent selects
$a_k = (d_k, p_k, l_k)$, where $d_k \in \{\texttt{Offer}, \texttt{Accept},
\texttt{Reject}\}$ is the economic decision, $p_k$ is the proposed price (when
$d_k = \texttt{Offer}$), and $l_k$ is a natural-language message. An opener
variable $\chi \in \{\texttt{AgentOpens}, \texttt{CounterpartOpens}\}$
determines the first move. The agent observes the counterpart's price
$p_k^B$ and message $l_k^B$, but not its latent type, nor the structured
cues that parameterize $l_k^B$ (see \S\ref{sec:counterpart_model}). Offers
must respect price bounds and monotonic concession. \texttt{Accept} is
unavailable until a counterpart offer has been observed. Full constraints,
termination cases, and observation-space structure are given in
Appendix~\ref{appdx:formal_spec}.

\textbf{Outcomes and utility.} The outcome space is
$F = [p_{\min}, p_{\max}] \cup \{\bot\}$, with public price bounds and
disagreement $\bot$. Buyer and seller utilities are
$u_{\text{buyer}}(p) = r_{\text{buyer}} - p$ and
$u_{\text{seller}}(p) = p - r_{\text{seller}}$, both zero on $\bot$.
Agreement is individually rational iff $p \in [r_{\text{seller}},
r_{\text{buyer}}]$, yielding a non-empty zone of possible agreement (ZOPA)
when $r_{\text{buyer}} \geq r_{\text{seller}}$.

\subsubsection{The Counterpart as a Diagnostic Instrument}
\label{sec:counterpart_model}

Given a counterpart family $\mathcal{F}$ and
sampled private type $t_B = (r_B, \kappa_B, \eta_B)$, the counterpart is
governed by a fixed stochastic kernel
\begin{equation}
(d_k^B, p_k^B, \tilde{s}_k, \tilde{c}_k)
\sim \pi_B^{\mathcal{F}}(\cdot \mid t_B, h_{k-1}),
\label{eq:counterpart_kernel}
\end{equation}
where $(\tilde{s}_k, \tilde{c}_k)$ are latent sentiment and strategic-posture
cues that parameterize the counterpart's natural-language message but never
alter the committed economic action. Fixing $\pi_B$ and $\mu$ across evaluated agents attributes performance differences to $\pi$. Full specification admits a model-based optimal reference policy, yielding a simulator-defined optimality gap (Appendix~\ref{app:optimal_policy}).

\textbf{Capability-isolating counterpart families.}
We instantiate the counterpart kernel $\pi_B$ from
\S\ref{sec:counterpart_model} as six parameter presets spanning two diagnostic
axes (Figure~\ref{fig:overview}, right): \emph{economic reactivity}, how much
economic responses depend on recent agent behavior rather than latent type, and
\emph{cue reliability}, whether language cues reflect latent stance. The
$2\times2$ diagnostic core consists of \textsc{Candid},
\textsc{Taciturn}, \textsc{Expressive}, and \textsc{Strategic}; two stress
conditions add \textsc{Stochastic}, with noisy prices and weak cues, and
\textsc{Adversarial}, with hardball behavior, pressuring cues, and an
aggressive-skewed stance prior. Family identity is hidden from the agent, so cross-family degradation isolates
failures in cue use, latent-type inference, calibration under noise, and
robustness to adversarial pressure. Full presets and kernel mechanics are in
Appendix~\ref{sec:counterpart} (Table~\ref{tab:counterpart_families}).

\subsection{Objective, Diagnostics, and Attribution}
\label{sec:evaluation}
The agent's normative objective is to maximize expected utility against
$(\mu,\pi_B)$. Since disagreement has utility zero,
\begin{equation}
\mathbb{E}_{\mu,\pi,\pi_B}[u_A(f)]
= \mathbb{P}(f \neq \bot) \cdot \mathbb{E}[u_A(f) \mid f \neq \bot],
\label{eq:obj_decomp}
\end{equation}
decomposes performance into \emph{deal attainment} and \emph{value extraction
conditional on agreement}. In feasible regimes both are desirable; in no-deal
regimes agreement signals failed exit discipline. We therefore report agreement
calibration separately for feasible and infeasible episodes.

We report four diagnostic axes (Table~\ref{tab:metrics}):
(i) \emph{terminal value}, surplus extracted;
(ii) \emph{agreement calibration}, agreeing when feasible and exiting when not;
(iii) \emph{opponent modeling}, latent-type inference; and
(iv) \emph{protocol compliance}, constraint adherence.

\begin{table}[h]
\centering
\footnotesize
\renewcommand{\arraystretch}{0.95}
\begin{tabular}{llp{3.2cm}p{4.4cm}c}
\toprule
\textbf{Axis} & \textbf{Metric} & \textbf{Equation} & \textbf{What it captures} & \textbf{Dir.} \\
\midrule
Terminal value
& $SE_\pi^+$
& $\frac{1}{|\mathcal I^+|}\sum_{i\in\mathcal I^+} \frac{u_A(f_i)}{\Delta_i}$
& Surplus extracted, normalized by ZOPA width
& $\uparrow$ \\
\midrule
\multirow{3}{*}{Agreement calibration}
& $\mathrm{AGR}_\pi^+$
& $\frac{1}{|\mathcal I^+|}\sum_{i\in\mathcal I^+} \mathbf{1}[f_i \neq \bot]$
& Closes deals when feasible
& $\uparrow$ \\
& $CSE_\pi^+$
& $\frac{1}{|\mathcal I_{\mathrm{agr}}^+|}\sum_{i\in\mathcal I_{\mathrm{agr}}^+} \frac{u_A(f_i)}{\Delta_i}$
& Surplus given agreement
& $\uparrow$ \\
& $\mathrm{FAGR}_\pi^-$
& $\frac{1}{|\mathcal I^-|}\sum_{i\in\mathcal I^-} \mathbf{1}[f_i \neq \bot]$
& Agrees when infeasible
& $\downarrow$ \\
\midrule
Opponent modeling
& $BE_{\mathrm{type}}$
& $\tfrac{1}{3}(BE_r + BE_\kappa + \mathrm{Brier}_\eta)$
& Reservation, urgency, stance inference error
& $\downarrow$ \\
\midrule
Protocol compliance
& $\mathrm{CritViol}\%$
& $\frac{1}{N}\sum_i \mathbf{1}[V_i^{\mathrm{crit}} > 0]$
& Price-bound, IR, and invalid-action violations
& $\downarrow$ \\
\bottomrule
\end{tabular}
\vspace{0.2cm}
\caption{\small Headline diagnostic metrics, grouped by axis.
$f_i$ is episode $i$'s terminal outcome;
$\Delta_i := r_{\mathrm{buyer}}^{(i)} - r_{\mathrm{seller}}^{(i)}$;
$\mathcal{I}^+ := \{i : \Delta_i > 0\}$,
$\mathcal{I}^- := \{i : \Delta_i < 0\}$, and
$\mathcal{I}_{\mathrm{agr}}^+ := \{i \in \mathcal{I}^+ : f_i \neq \bot\}$.
Surplus efficiency decomposes as
$SE_\pi^+ = \mathrm{AGR}_\pi^+ \cdot CSE_\pi^+$, separating agreement rate
from conditional value extraction. Empirically, we compute belief error from the agent's reported belief. Secondary diagnostics are deferred to
Appendix~\ref{appdx:eval}.}
\label{tab:metrics}
\end{table}

\textbf{Attribution via oracle interventions.}
Outcome gaps do not reveal whether underperformance comes from poor
latent-type inference, residual uncertainty, or failure to act on correct
information. Since $(\Gamma,\mathcal P,\mu,\pi_B)$ is fully specified, we define
a dynamic-programming reference policy $\pi^\star$ (Appendix~\ref{app:optimal_policy}).
This oracle is simulator-relative: it is Bayes-optimal for the specified
prior, counterpart kernel, protocol, horizon, and utility function, not a claim
of human-optimal negotiation. We use it only as a benchmark-internal reference
point.

We isolate bottlenecks by varying the agent's information while holding the
episode and counterpart fixed. We compare \emph{base}, which infers $t_B$ from
prices and messages; \emph{oracle-posterior}, which receives the exact Bayesian
posterior $b_k$; and \emph{revealed-type}, which receives the realized $t_B$.
Letting $\bar{U}(\cdot)$ denote mean utility,
\vspace{0.5em}
\begin{equation}
\bar{U}(\pi^\star) - \bar{U}(\pi^{\mathrm{base}})
=
\underbrace{\bar{U}(\pi^{\mathrm{post}}) - \bar{U}(\pi^{\mathrm{base}})}_
{\Delta_{\mathrm{inf}}:\text{ inference}}
+
\underbrace{\bar{U}(\pi^{\mathrm{reveal}}) - \bar{U}(\pi^{\mathrm{post}})}_
{\Delta_{\mathrm{unc}}:\text{ uncertainty}}
+
\underbrace{\bar{U}(\pi^\star) - \bar{U}(\pi^{\mathrm{reveal}})}_
{\Delta_{\mathrm{ctrl}}:\text{ control}} .
\label{eq:gap_decomp}
\end{equation}
\vspace{0.5em}
Thus $\Delta_{\mathrm{inf}}$ measures the value of replacing the agent's own
inference with the simulator posterior, $\Delta_{\mathrm{unc}}$ the value of
resolving remaining type uncertainty, and $\Delta_{\mathrm{ctrl}}$ the residual
gap to the simulator oracle after type information is no longer hidden.  A negative
$\Delta_{\mathrm{inf}}$ means that the posterior intervention reduced realized
utility relative to the base prompt, while a negative $\Delta_{\mathrm{ctrl}}$
means that the full-reveal LLM condition exceeded the discretized oracle
reference on those episodes. Full intervention details are in
Appendix~\ref{sec:latent-type-inference}. 


%% file: sections/sim.tex
\vspace{-0.5em}
We factor the simulator into two fixed layers. First, the
\emph{economic regime generator} samples episode geometry and hidden state,
including reservation values, feasibility, and urgency. Second, the
\emph{environment-simulated counterpart policy} governs behavior within the
sampled scenario: openings, acceptance, walk-away, counter-offers, and
language-facing cues. Because both layers are fixed across evaluated agents,
performance differences can be cleanly attributed to the agent policy.

{\small \textbf{Notation.}
Let $R := p_{\max}-p_{\min}$ be the price range and define
$\Delta := r_{\mathrm{buyer}}-r_{\mathrm{seller}}$. Then $\Delta>0$ denotes a
feasible episode with ZOPA width $\Delta$, while $\Delta<0$ denotes a no-deal
episode with infeasibility gap $-\Delta$. We use $z>0$ for sampled feasible
widths and $q>0$ for sampled no-deal gaps.}
\vspace{-0.5em}
\subsection{Economic Regime Generator}
\label{sec:regime_instantiation}

The regime generator instantiates three canonical regimes by specifying
reservation geometry and urgency. Counterpart urgency is drawn from a baseline
law $D_\kappa=\mathrm{Beta}(\alpha_\kappa,\beta_\kappa)$ rescaled to $[0,1]$;
counterpart stance $\eta_B$ is drawn from a family-specific prior
(\S\ref{sec:counter_design}, Appendix~\ref{sec:counterpart}). 

{\small
\begin{itemize}[leftmargin=1.2em, itemsep=0.15em, topsep=0.2em, parsep=0em]
\item \textbf{Overlap} \emph{(surplus extraction; feasible-agreement calibration).}
Sample $z \sim \mathcal{U}[\Delta_{\min},\Delta_{\max}]$ and midpoint $m$
within price bounds; set
$r_{\mathrm{buyer}}=m+z/2$ and $r_{\mathrm{seller}}=m-z/2$. Difficulty varies
with $z$.

\item \textbf{Urgency-shift} \emph{(time-pressure adaptation).}
Use Overlap reservations, but draw
$\kappa_B \sim D_\kappa^{(s)}$ with mean shift
$s:=\mathbb{E}_{D_\kappa^{(s)}}[\kappa]-\mathbb{E}_{D_\kappa}[\kappa]$. Since urgency is hidden and payoff-irrelevant, this isolates adaptation to
pressure.

\item \textbf{No-deal} \emph{(exit discipline under infeasibility).}
Sample $q \sim \mathcal{U}[q_{\min},q_{\max}]$ and midpoint $m$; set
$r_{\mathrm{buyer}}=m-q/2$ and $r_{\mathrm{seller}}=m+q/2$. Smaller $q$ makes
infeasibility harder to distinguish from a narrow ZOPA.
\end{itemize}
}

The generator also supports a \emph{data-grounded} variant that swaps synthetic price geometry for empirical statistics from a real catalog; we describe it in \S\ref{sec:data_grounded} after the counterpart policy.
\vspace{-0.5em}
\subsection{Environment-Simulated Counterpart Policy}
\label{sec:counter_design}
Conditional on the sampled scenario and counterpart behavior family
$\mathcal{F}$ (\S\ref{sec:counterpart_model}), the counterpart kernel
$\pi_B$ comprises four components: (i) an opener-role protocol, (ii) an
acceptance and walk-away response model, (iii) a counter-offer rule, and (iv) a
cue-generation interface. Family-specific presets, history-feature
definitions, and default hyperparameters are deferred to
Appendix~\ref{sec:counterpart}.

\textbf{Opening role.}
The opener $\chi \in \{\texttt{AgentOpens}, \texttt{CounterpartOpens}\}$ is an
episode-level protocol attribute balanced across regime--family cells
(\S\ref{sec:exp_setup}). If the counterpart opens, its first price is sampled
from the randomized opening-offer model in
Appendix~\ref{appdx:opening-offer}. If the agent opens, the counterpart first
applies the response model below. Absent acceptance or walk-away, it samples its
first price from the same opening model, since no prior counterpart offer exists. 

\textbf{Acceptance and walk-away.} Given an agent offer $p_k^A$,
define the role-normalized favorability
\[
\bar{\Delta}_k :=
\begin{cases}
(p_k^A - r_B)/R, & \text{counterpart is seller},\\
(r_B - p_k^A)/R, & \text{counterpart is buyer},
\end{cases}
\]
so that $\bar{\Delta}_k \geq 0$ iff the offer is individually rational
for the counterpart. Using the concave deadline clock
$\widetilde{D}_k := \sqrt{k/K}$ and
$\widetilde{\bar{D}}_k := 1 - \widetilde{D}_k$, the acceptance
probability is
\begin{equation}
\label{eq:accept_probabilities}
a_k = \mathbf{1}\{\bar{\Delta}_k \geq 0\}\,
\sigma\!\left(g_\theta(p_k^A, t_B, k, h_{k-1})\right),
\end{equation}
\begin{equation}
\label{eq:accept_score}
g_\theta = \alpha \bar{\Delta}_k + \beta \kappa_B - \gamma \widetilde{\bar{D}}_k
+ \rho_{\mathcal{F}}(\eta_B)\,\mathrm{ConcedeSpeed}_k
+ \xi_{\mathcal{F}}(\eta_B)\,\mathrm{Rigidity}_k.
\end{equation}
Conditional on non-acceptance, the counterpart may walk away
(recorded $d_k^B = \texttt{Reject}$) with hazard
\begin{equation}
\label{eq:walkaway_probability}
\omega_k = \mathbf{1}\{k \geq k_{\mathrm{walk}}\}\,
\mathbf{1}\{\bar{\Delta}_k < 0\}\,
\sigma\!\left(\phi_0 + \phi_\Delta[-\bar{\Delta}_k]_+ + \phi_T \tau_k^W\right),
\end{equation}
where $\tau_k^W \in [0,1]$ is the walk-away clock active after a grace
period. Walk-away is therefore enabled only after $k_{\mathrm{walk}}$
rounds and only when the current offer violates the counterpart's
reservation constraint. The resulting response distribution is
$(a_k,\ (1-a_k)\omega_k,\ \mathbf{1}\{k<K\}(1-a_k)(1-\omega_k))$ over
$(\texttt{Accept}, \texttt{Reject}, \texttt{Offer})$, with remaining
mass at $k = K$ resulting in round-limit disagreement.

\textbf{Counter-offer generation.} When $d_k^B = \texttt{Offer}$
and the counterpart has made a prior offer, it forms a type-dependent
concession score
\begin{align}
\label{eq:counterpart_concession_score}
\tilde{\lambda}_B(h_{k-1})
&= \lambda_0 + \lambda_1 \kappa_B
- \lambda_{2,\mathcal{F}}(\eta_B)\,\mathrm{ConcedeMagnitude}_k \nonumber \\
&\quad - \lambda_3 \mathbf{1}\{\eta_B = \texttt{aggressive}\}
+ \lambda_4 \mathbf{1}\{\eta_B = \texttt{conciliatory}\},
\end{align}
clipped to $\lambda_B = \min\{1, \max\{0, \tilde{\lambda}_B\}\}$. The
noisy concession candidate $\tilde{p}_k^B = p_{k-1}^B - \lambda_B
(p_{k-1}^B - r_B) + \varepsilon_k$, $\varepsilon_k \sim
\mathcal{N}(0, \sigma_p^2)$, is projected onto the role-dependent
monotone feasible interval $\mathcal{M}_B(k)$ to enforce individual
rationality and monotonic concession:
\begin{equation}\label{eq:counter_project}
p_k^B = \Pi_{\mathcal{M}_B(k)}(\tilde{p}_k^B),
\quad
\mathcal{M}_B(k) =
\begin{cases}
[r_B, p_{k-1}^B], & \text{seller},\\
[p_{k-1}^B, r_B], & \text{buyer}.
\end{cases}
\end{equation}
If no prior counterpart offer exists, $p_k^B$ is drawn from the
opening-offer model.

\textbf{Cue generation and language realization.}
After committing the economic action, the simulator samples latent cues
$(\tilde{s}_k,\tilde{c}_k)$ from a family-specific cue model
(Appendix~\ref{appdx:cue_generation}), with
$\tilde{s}_k \in \{\texttt{positive},\texttt{neutral},\texttt{negative}\}$ and
$\tilde{c}_k \in \{\texttt{Concede},\texttt{Hold},\texttt{Pressure}\}$.
These cues condition the counterpart's message but never alter the committed
action or price.

\vspace{-0.5em}
\subsection{Data-Grounded Extension}\label{sec:data_grounded}
\textsc{Terms-Bench} admits a \emph{data-grounded} extension that lets
practitioners evaluate agents on their own market data while preserving the
diagnostic guarantees of the synthetic suite. Only the regime generator and
the observable product context are replaced; the counterpart kernel, oracle
policy, information-intervention decomposition, and metric definitions are
unchanged, so any data-grounded instantiation reuses the same diagnostic axes
as the synthetic main experiment and supports direct per-model comparison
across instantiations.

\textbf{Interface.}
A data-grounded instantiation is specified by three practitioner-supplied
inputs. (i)~A \emph{product catalog} of items
$j \in \mathcal{J}$, each with summary price statistics
$(\hat p_{\mathrm{ref}}^{(j)}, \hat p_{\mathrm{lo}}^{(j)}, \hat p_{\mathrm{hi}}^{(j)})$
(reference price plus historical low and high), used to calibrate the public
reference price and the latent reservation wedges around it.
(ii)~\emph{Category-level price bounds} $[p_{\min}^{(c)}, p_{\max}^{(c)}]$
that set the public action range. (iii)~\emph{Observable product context}
(item name, category, salient attributes, observable market range), which is
appended to the agent's prompt; private reservations, counterpart urgency,
and stance remain hidden. Given these inputs, the regime generator samples a category and product, applies the same overlap, urgency-shift, and no-deal geometries to the latent wedges, and passes episodes to the unchanged counterpart and evaluation pipeline. Thus, practitioners can plug in any catalog by supplying these statistics, without modifying the benchmark machinery.

\textbf{Instantiation in this paper.}
For our experiments we instantiate the extension on
\textit{AmazonHistoryPrice}~\citep{xia2024measuringbargainingabilitiesllms},
a CamelCamelCamel-derived catalog of 831 products across 14 categories with
per-product historical statistics. We use it as both an
external-validity check on the synthetic findings and a difficulty stress
test, since real catalogs typically present a much wider, long-tailed public
action range that makes the product reference price a more decisive anchor
than under the synthetic geometry. Per-model results, the geometric
comparison to the synthetic suite, and the full construction are deferred to
\S\ref{sec:results_dg} and Appendix~\ref{appdx:data-grounded}.

%% file: sections/exp.tex
\vspace{-0.5em}
\begin{table*}[h]
\centering
\small
\setlength{\tabcolsep}{4pt}
\renewcommand{\arraystretch}{1}
\begin{tabular}{@{}lccccccc@{}}
\toprule
\textbf{Agent}
& {$SE_\pi^+ \uparrow$}
& {$\mathrm{AGR}_\pi^+ \uparrow$}
& {$CSE_\pi^+ \uparrow$}
& {$\mathrm{FAGR}_\pi^- \downarrow$}
& {$BE_{\mathrm{type}} \downarrow$}
& {$\mathrm{CritViol} \downarrow$}
& {\%Oracle $\uparrow$} \\
\midrule
Claude Opus 4.6
 & \bestcell{0.694$_{\pm0.014}$} & 99.3$_{\pm0.5}$ & \secondcell{0.699$_{\pm0.013}$} & 0.00$_{\pm0.00}$ & 0.222$_{\pm0.013}$ & 0.00$_{\pm0.00}$ & \secondcell{76.3$_{\pm1.5}$} \\
GLM-5.1
 & \secondcell{0.686$_{\pm0.016}$} & 95.1$_{\pm1.2}$ & \bestcell{0.721$_{\pm0.014}$} & 0.00$_{\pm0.00}$ & \secondcell{0.218$_{\pm0.008}$} & \badfour{1.33$_{\pm0.53}$} & \bestcell{76.8$_{\pm1.8}$} \\
Claude Opus 4.7
 & 0.660$_{\pm0.014}$ & 98.2$_{\pm0.8}$ & 0.672$_{\pm0.014}$ & 0.00$_{\pm0.00}$ & 0.229$_{\pm0.006}$ & 0.00$_{\pm0.00}$ & 72.9$_{\pm1.6}$ \\
Gemma-4-31B-IT
 & 0.640$_{\pm0.014}$ & \secondcell{99.8$_{\pm0.2}$} & 0.641$_{\pm0.014}$ & 0.00$_{\pm0.00}$ & 0.260$_{\pm0.009}$ & \badone{0.06$_{\pm0.11}$} & 69.9$_{\pm1.5}$ \\
Gemini-3.1-Pro
 & 0.639$_{\pm0.013}$ & 99.7$_{\pm0.3}$ & 0.641$_{\pm0.013}$ & 0.00$_{\pm0.00}$ & 0.271$_{\pm0.012}$ & 0.00$_{\pm0.00}$ & 69.5$_{\pm1.4}$ \\
DeepSeek-V4-Pro
 & 0.618$_{\pm0.016}$ & 97.5$_{\pm0.9}$ & 0.633$_{\pm0.016}$ & 0.00$_{\pm0.00}$ & 0.228$_{\pm0.005}$ & \badthree{0.61$_{\pm0.36}$} & 69.1$_{\pm1.8}$ \\
GPT-5.5
 & 0.606$_{\pm0.014}$ & 99.2$_{\pm0.5}$ & 0.611$_{\pm0.014}$ & 0.00$_{\pm0.00}$ & 0.228$_{\pm0.010}$ & 0.00$_{\pm0.00}$ & 66.6$_{\pm1.6}$ \\
Qwen3.6-Plus
 & 0.604$_{\pm0.016}$ & 98.2$_{\pm0.7}$ & 0.614$_{\pm0.015}$ & \badone{0.17$_{\pm0.33}$} & 0.237$_{\pm0.006}$ & \badsix{2.06$_{\pm0.66}$} & 67.7$_{\pm1.8}$ \\
Grok 4.20
 & 0.601$_{\pm0.015}$ & 99.1$_{\pm0.5}$ & 0.606$_{\pm0.014}$ & 0.00$_{\pm0.00}$ & \bestcell{0.212$_{\pm0.006}$} & \badfive{1.50$_{\pm0.56}$} & 65.9$_{\pm1.6}$ \\
Kimi-K2.6
 & 0.597$_{\pm0.016}$ & 97.1$_{\pm1.0}$ & 0.614$_{\pm0.015}$ & 0.00$_{\pm0.00}$ & 0.236$_{\pm0.009}$ & 0.00$_{\pm0.00}$ & 66.7$_{\pm1.8}$ \\
GPT-5.4
 & 0.531$_{\pm0.016}$ & 99.4$_{\pm0.4}$ & 0.535$_{\pm0.016}$ & 0.00$_{\pm0.00}$ & 0.242$_{\pm0.009}$ & 0.00$_{\pm0.00}$ & 59.4$_{\pm1.8}$ \\
Doubao-Seed-2.0-Pro
 & 0.522$_{\pm0.014}$ & \bestcell{99.9$_{\pm0.2}$} & 0.523$_{\pm0.014}$ & 0.00$_{\pm0.00}$ & 0.247$_{\pm0.009}$ & \badone{0.06$_{\pm0.11}$} & 56.6$_{\pm1.5}$ \\
GPT-4o-mini
 & 0.189$_{\pm0.013}$ & 52.2$_{\pm2.8}$ & 0.363$_{\pm0.016}$ & 0.00$_{\pm0.00}$ & 0.251$_{\pm0.005}$ & 0.00$_{\pm0.00}$ & 22.2$_{\pm1.6}$ \\
\midrule
Fixed 30\%
 & 0.387$_{\pm0.015}$ & \bestcell{99.9$_{\pm0.2}$} & 0.387$_{\pm0.015}$ & 0.00$_{\pm0.00}$ & \NA & 0.00$_{\pm0.00}$ & 42.7$_{\pm1.6}$ \\
Fixed 10\%
 & 0.290$_{\pm0.013}$ & 94.5$_{\pm1.3}$ & 0.307$_{\pm0.013}$ & 0.00$_{\pm0.00}$ & \NA & 0.00$_{\pm0.00}$ & 33.3$_{\pm1.5}$ \\
Fixed 1\%
 & 0.273$_{\pm0.012}$ & 92.2$_{\pm1.5}$ & 0.296$_{\pm0.013}$ & 0.00$_{\pm0.00}$ & \NA & 0.00$_{\pm0.00}$ & 31.3$_{\pm1.4}$ \\
\bottomrule
\end{tabular}
\caption{\small Aggregate agent performance in the bilateral instantiation of
\textsc{Terms-Bench} with 95\% CIs. Green: top-two per metric; red: violations/false agreements (darker = worse). Arrows indicate
preferred direction.
The roster spans frontier systems from major providers at evaluation time,
plus \texttt{GPT-4o-mini} as a sub-frontier reference.}
\label{tab:main_results}
\vspace{-0.5em}
\end{table*}
\subsection{Experimental Setup}\label{sec:exp_setup}

\textbf{Benchmark suite.}
We evaluate agents across three regimes
(\S\ref{sec:regime_instantiation}) and six counterpart behavior families
(\S\ref{sec:counterpart_model}). For each regime--family pair, we block over
agent role $\{\texttt{Buyer},\texttt{Seller}\}$ and opener assignment
$\{\texttt{AgentOpens},\texttt{CounterpartOpens}\}$, with 25 episodes per
role--opener cell, yielding 100 episodes per regime--family\footnote{We focus on the counterpart-more-urgent direction because it directly probes
exploitation of counterpart time pressure; Appendix~\ref{appdx:urgency-down}
reports the reverse direction as a directionality check.
} pair and
1{,}800 episodes per agent.  Across agents, we reuse the same seeded
specifications, fixing reservation geometry, latent counterpart type, urgency
draws, opening harshness, opener assignment, and simulator random streams.

\textbf{Agents and inference.}
Agents negotiate against a fixed environment-simulated counterpart: a
parameterized stochastic kernel with a language-realization layer
(\S\ref{sec:counter_design}). We compare three fixed-concession baselines
(conceding 1\%, 10\%, and 30\% of the remaining distance to reservation per
\texttt{Offer}) to proprietary and open-weight LLM agents
(Table~\ref{tab:models}) under a shared wrapper.
LLMs are called via
\texttt{OpenRouter} and use identical role-conditioned prompts, a structured
per-round \texttt{JSON} interface, deterministic decoding (temperature $=0$),
and one rollout per seeded episode. When supported, reasoning effort is set to
\texttt{xhigh} to give each model its strongest available configuration. The
counterpart voice model is fixed to \texttt{GPT-5.2} and only renders committed
kernel actions in language; it does not affect prices, acceptances, walk-aways,
or outcomes.

\textbf{Reported metrics.}
Agents output a message, economic action, and belief estimate over the
counterpart's latent type. We report six primary metrics across four diagnostic
axes (\S\ref{sec:evaluation}), and additionally $\%\mathrm{Oracle}$: surplus
efficiency relative to a dynamic-programming oracle
(Appendix~\ref{app:optimal_policy}). While perfectly correlated with
$CSE_\pi^+$, it makes the oracle gap explicit. Details are deferred to
Appendix~\ref{sec:implementation_details}.

\textbf{Evaluation surfaces.}
Beyond the synthetic main suite, we evaluate two extensions that reuse the
counterpart kernel, oracle, and metrics verbatim. The \emph{data-grounded
variant} (\S\ref{sec:data_grounded}, \S\ref{sec:results_dg}) replaces synthetic
price geometry and abstract product references with Amazon-catalog statistics
(831 products, 14 categories) and exposes product context to the agent.
The \emph{commercial extension} (\S\ref{sec:results_commerce}) wraps each
episode in unit economics (\textsc{commerce mode}) and chains episodes into
multi-period sessions with cash and belief carryover (\textsc{bankroll mode});
construction details are in Appendix~\ref{app:leaderboard}. Both extensions
reuse $SE_\pi^+$, $\mathrm{AGR}_\pi^+$, and the latent-type belief
diagnostics, and add surface-specific metrics (regret rate for
commerce; terminal balance, survival rate, and an optional memory
premium for bankroll).
\vspace{-0.5em}
\subsection{Synthetic Main Experiment}\label{sec:results}
Table~\ref{tab:main_results} reports aggregate performance for 13 LLM agents
(Table~\ref{tab:models}) and 3 fixed baselines. Per-family and per-regime
breakdowns appear in Table~\ref{tab:family_regime_breakdown}; full metrics in
Tables~\ref{tab:aggregate_terminal_full}--\ref{tab:aggregate_protocol_full}.
Surplus efficiency varies 3.7-fold across LLMs, from \texttt{GPT-4o-mini} to
\texttt{Claude Opus 4.6}; the simplest fixed-concession baseline outperforms
\texttt{GPT-4o-mini}, so LLMs do not uniformly clear hand-coded heuristics. We unpack these gaps in five findings: aggregate structure (F1), cue and information failures (F2–F3), per-agent bottleneck attribution (F4), and stable behavioral fingerprints (F5).

\begin{figure}[t]
    \centering
    \includegraphics[width=1.0\linewidth]{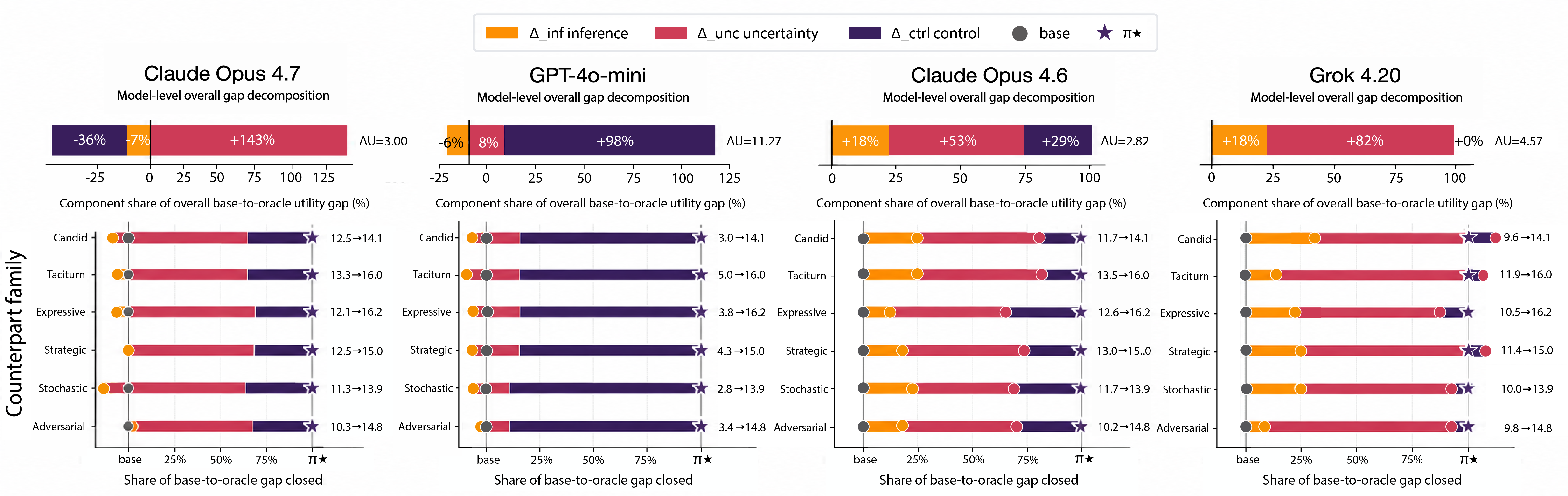}
    \caption{\small \textbf{Oracle gap decomposition.} Rows are normalized to each family's base-to-oracle
utility gap, so the signed inference, uncertainty, and control components sum
to $100\%$. Negative
$\Delta_{\mathrm{inf}}$: posterior injection hurt utility. Negative  $\Delta_{\mathrm{ctrl}}$: full-reveal LLM beat the discretized oracle (see \S\ref{sec:evaluation} for details).}
\label{fig:oracle_decomposition}
\vspace{-1em}
\end{figure}
\textbf{Finding 1: High deal rate hides what matters.}
Feasible agreement is nearly saturated: excluding
\texttt{GPT-4o-mini}, all evaluated LLMs achieve
$\mathrm{AGR}_\pi^+ \in [93.4\%,99.9\%]$
(Table~\ref{tab:main_results}). Yet agreement does not imply value. Among
agents with $\mathrm{AGR}_\pi^+ \ge 97\%$, $SE_\pi^+$ ranges from $0.522$
(\texttt{Doubao-Seed-2.0-Pro}) to $0.694$ (\texttt{Claude Opus 4.6});
\texttt{Doubao} attains the highest agreement rate ($99.9\%$) but much lower
conditional surplus than \texttt{Claude} ($CSE_\pi^+=0.523$ vs.\ $0.699$).
Other axes
reveal orthogonal failures: \texttt{GLM-5.1} has the best conditional surplus
and oracle attainment but critical
violations, while \texttt{Grok~4.20} has the lowest type-belief error
($0.212$) but only mid-tier surplus. 
Thus our metrics capture
distinct capabilities that deal-rate alone would mask.

\textbf{Finding 2: Decoupling strategy from voice reveals a cue-use failure.}
The counterpart families decouple language-facing cues from the economic kernel:
\begin{wrapfigure}[20]{r}{0.6\linewidth}
\centering
\includegraphics[width=\linewidth]{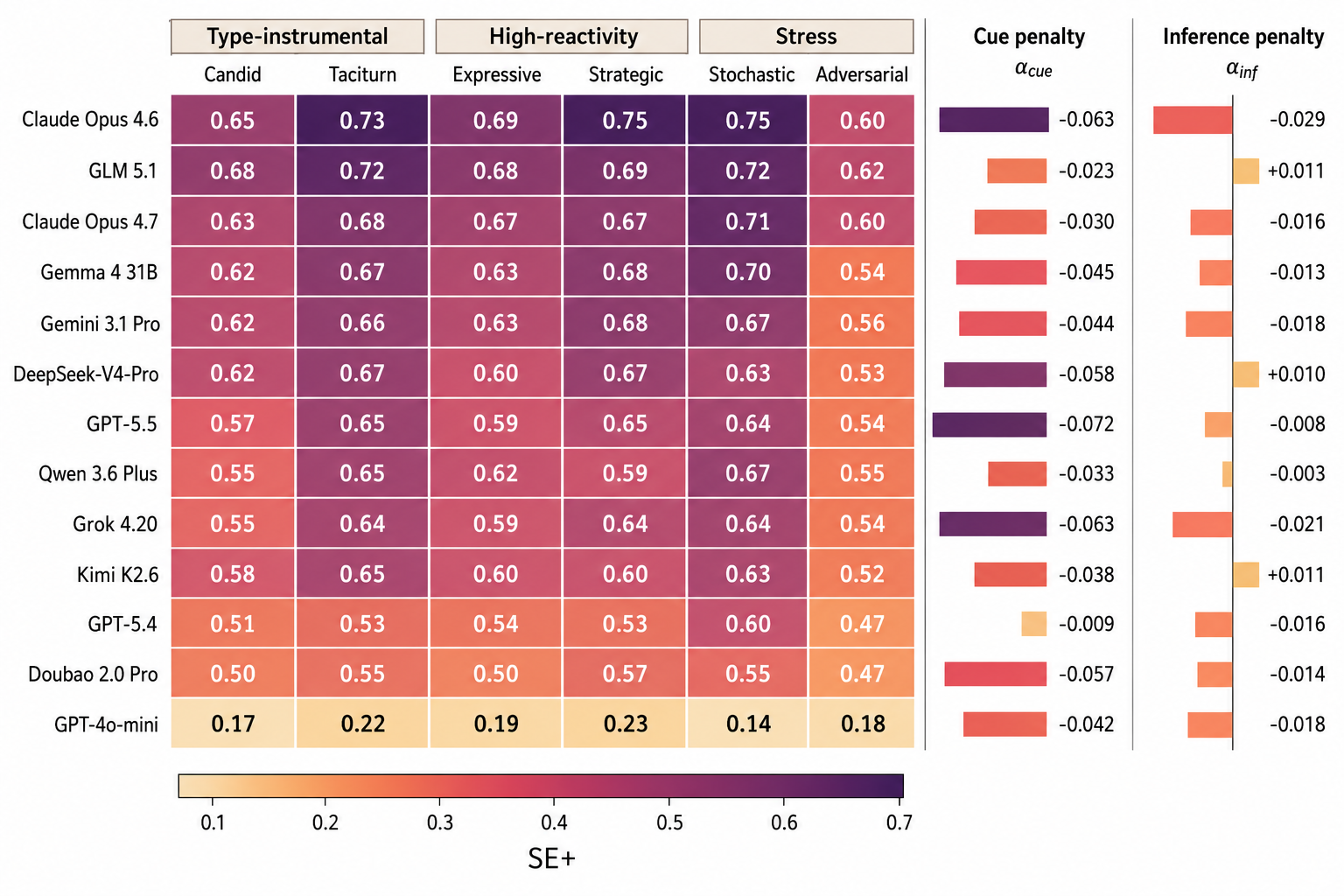}
\vspace{-2em}
\caption{\small Per-family surplus efficiency, cue and inference penalty.}
\label{fig:family_heatmap}
\end{wrapfigure}
\textsc{Candid}/\textsc{Expressive} expose informative cues; paired \textsc{Taciturn}/\textsc{Strategic} mute them. We define the cue penalty $\alpha_{\mathrm{cue}} = \overline{SE_\pi^+}(\text{cue-revealing}) - \overline{SE_\pi^+}(\text{cue-muted})$. For every LLM, $\alpha_{\mathrm{cue}}<0$ (Fig.~\ref{fig:family_heatmap}): agents extract \emph{less} surplus when informative cues are present. Penalties range from $-0.009$ (\texttt{GPT-5.4}) to $-0.063$ (\texttt{Claude Opus 4.6}, \texttt{Grok~4.20}), with 9 of 13 per-model 95\% bootstrap CIs excluding zero and an across-model Wilcoxon test at $p<10^{-3}$ (Appendix~\ref{app:alpha_significance}). Stress families sharpen the picture: \textsc{Stochastic} remains competitive despite corrupted prices and cues, while \textsc{Adversarial} is the lowest-surplus condition for most agents. The failure is not environmental noise but cue sensitivity used against the agent: warm cues induce over-concession, pressure cues trigger brittle behavior.

\begin{figure}
    \centering  
    \includegraphics[width=1 \linewidth]{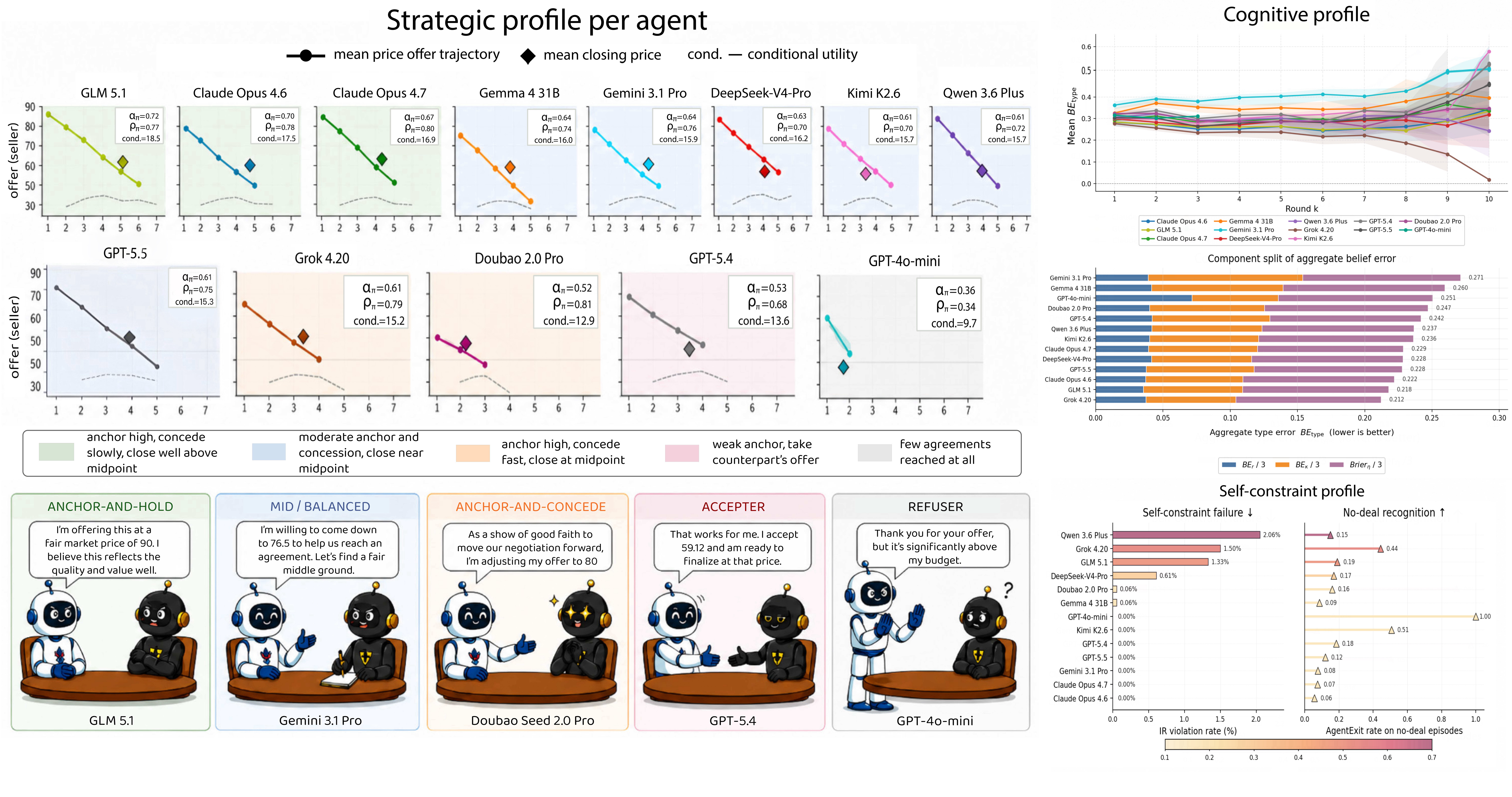}
    \vspace{-2.0em}
\caption{\small
Behavioral profiles across three axes. \textbf{Strategic profile} (left) shows
mean offer price trajectories in the (seller-opens, overlap) regime, where lines are clipped at mean closing round per-agent; diamonds mark mean
closing price, and panel annotations report trajectory coefficient $\alpha_\pi$,
closer rate $\rho_\pi$, and conditional utility (\texttt{cond.}). Background
tints define five bargaining typologies: \emph{anchor-and-hold},
\emph{mid/balanced}, \emph{anchor-and-concede}, \emph{accepter}, and
\emph{refuser}. \textbf{Cognitive profile} (top right) reports mean aggregate
belief error by round and its decomposition into reservation, urgency, and
stance components. \textbf{Self-constraint profile} (bottom right) shows
IR-violation rate (left) and $\mathrm{AgentExit}^{-}$ (\S \ref{appdx:eval}; the share of no-deal episodes where the agent
explicitly identifies infeasibility and rejects) on no-deal episodes (right). Lower IR-violation rate
and higher $\mathrm{AgentExit}^{-}$ indicate cleaner no-deal recognition.
}
\vspace{-1.5em}
    \label{fig:full}
\end{figure}

\textbf{Finding 3: Information does not reliably translate into surplus.}
F2 shows that frontier LLMs lose value from counterpart cues. An
information--action gap further appears across three diagnostics.
First, posterior injection has weak and inconsistent returns: the
\emph{oracle-posterior} intervention supplies a Bayesian posterior
$b_k\in\Delta(T_B)$ each round
(\S\ref{sec:evaluation}, \S\ref{sec:latent-type-inference}), but in the
decomposition it closes only $1\%$ of the base-to-oracle gap
for \texttt{Claude Opus 4.7} and is negative for \texttt{GPT-4o-mini}
($-6\%$; Fig.\ref{fig:oracle_decomposition}). Second, beliefs
do not improve online: for every frontier LLM, joint belief error is flat or
\emph{increases} over 10 rounds (Fig.\ref{fig:full}, top-right), with
$\mathrm{Brier}_\eta$ dominating $BE_r$ and $BE_\kappa$. Third, family-level
contrasts show the cost. Define
$
\alpha_{\mathrm{inf}}
=
\overline{SE_\pi^+}(\text{type-instrumental core})
-
\overline{SE_\pi^+}(\text{high-reactivity core}),
$
where type-instrumental families (\textsc{Candid}, \textsc{Taciturn}) reward
latent-type inference more than high-reactivity families
(\textsc{Expressive}, \textsc{Strategic}). A positive value indicates
benefit from latent-type structure; instead, $\alpha_{\mathrm{inf}}<0$ for
10 of 13 LLMs (Fig.\ref{fig:family_heatmap}), with largest penalties for
\texttt{Claude Opus 4.6} ($-0.029$) and \texttt{Grok~4.20} ($-0.021$). The inference penalty is rank-Wilcoxon-significant at the population
level ($p=0.007$) although the cruder sign test is only marginal
($p=0.073$) and no individual CI excludes zero
(Appendix~\ref{app:alpha_significance}). Thus agents fail to convert both surface cues
and payoff-relevant latent structure into better strategic action.

\textbf{Finding 4: Oracle interventions identify different agents fail for different reasons.}
The oracle decomposition (Eq.~\ref{eq:gap_decomp}) separates surplus gaps into inference, uncertainty, and control bottlenecks. \texttt{GPT-4o-mini} is control-limited: control accounts for $98\%$ of its much larger gap, while posterior and type revelation recover little surplus. \texttt{Claude Opus 4.7} and \texttt{Grok~4.20} are uncertainty-limited ($130\%$ and $82\%$), with small inference gains ($1\%$, $18\%$) and negligible residual control loss. \texttt{Claude Opus 4.6} is mixed across inference ($18\%$), uncertainty ($53\%$), and control ($29\%$). Oracle interventions thus point to distinct bottlenecks: smaller agents struggle to act on the revealed type, while stronger agents are bottlenecked by decisions under unresolved uncertainty.

\textbf{Finding 5: Stable bargaining fingerprints decouple concession from self-constraint.}
Trace-level profiles decompose behavior into \emph{concession dynamics} in feasible deals and \emph{boundary monitoring} under infeasibility. In overlap episodes, trajectory shape, agent-closer rate $\rho_\pi$, and conditional utility separate five styles: anchor-and-hold, mid/balanced, anchor-and-concede, accepter, and refuser(Fig.\ref{fig:full}, left). In no-deal episodes, IR-violation rate and $\mathrm{AgentExit}^-$
form two independent axes of self-constraint, with agents spread along both (Fig.\ref{fig:full}, bottom-right) These axes are independent: \texttt{GLM-5.1} anchors strongly in overlap yet breaches reservation in no-deal, while \texttt{Doubao-Seed-2.0-Pro} compromises quickly in overlap yet holds cleanly under infeasibility. Within-agent variation across counterpart families is much smaller than between-agent variation and no agent crosses a typology boundary (Appendix~\ref{app:strategic_profile}). Each model thus has a stable bargaining fingerprint, but strategic style and self-constraint are distinct capabilities that terminal metrics collapse.

\vspace{-0.5em}
\subsection{Data-Grounded Variant}\label{sec:results_dg}
\begin{figure}[t]
\centering
\includegraphics[width=\linewidth]{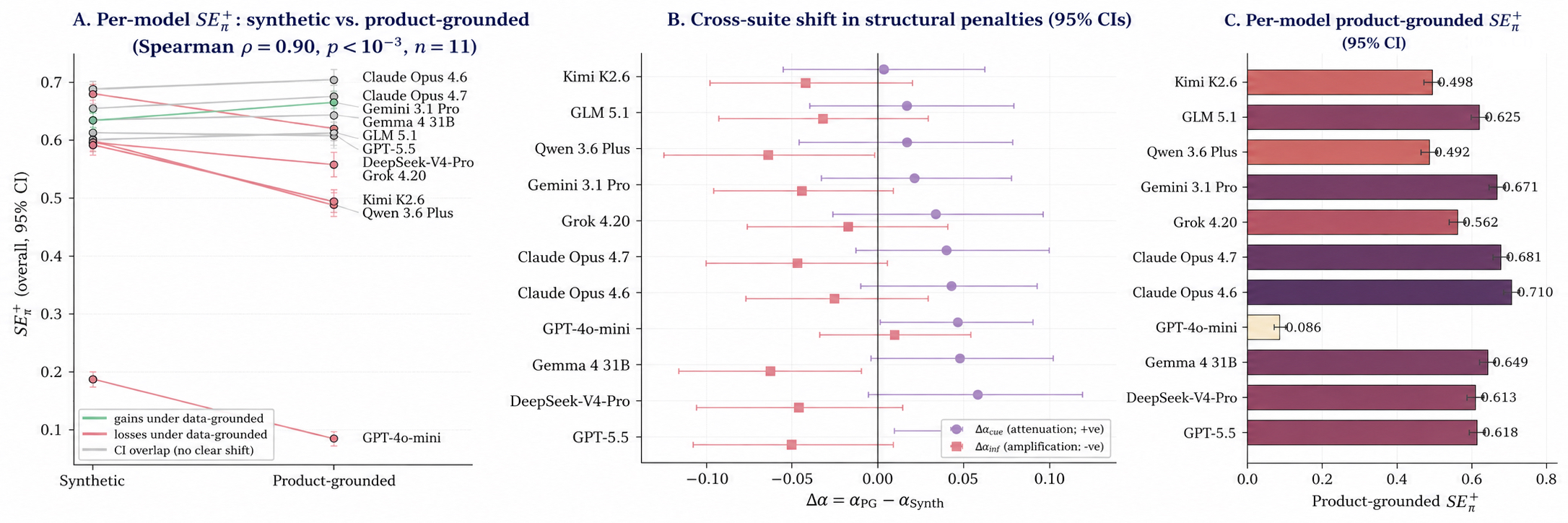}
\caption{\small \textbf{Data-grounded variant summary} (eleven paired models).
\textbf{A.}~Per-model $SE_\pi^+$ slopegraph from the synthetic suite (left)
to the data-grounded suite (right) with 95\% bootstrap CIs. Lines are
colored green where the data-grounded CI lies entirely above the synthetic
CI (gains under data-grounded), pink where it lies entirely below (losses
under data-grounded), and grey otherwise; rank
order is largely preserved.
\textbf{B.}~Cross-suite shift in the structural penalties
$\Delta\alpha = \alpha_{\mathrm{PG}}-\alpha_{\mathrm{Synth}}$ with 95\%
bootstrap CIs ($B{=}2000$), sorted by $\Delta\alpha_{\mathrm{cue}}$.
All eleven models attenuate the cue-use penalty
($\Delta\alpha_{\mathrm{cue}}>0$, paired Wilcoxon $p=0.001$); ten of the
eleven also amplify the inference penalty ($\Delta\alpha_{\mathrm{inf}}<0$,
$p=0.002$), with \texttt{GPT-4o-mini} the lone exception.
\textbf{C.}~Per-model product-grounded $SE_\pi^+$ with 95\% bootstrap CIs.}
\label{fig:pg_main_summary}
\end{figure}

We sweep a representative subset of the full model roster on the data-grounded instantiation
(\S\ref{appdx:data-grounded}), holding the counterpart kernel, oracle, and
metrics fixed and replacing only the price geometry and observable product
context with Amazon-catalog statistics (100 episodes per regime, identical
seeds across models). Geometry is non-trivially shifted: the public range
becomes long-tailed and the relative ZOPA collapses by roughly an order of
magnitude (\S\ref{appdx:data-grounded}), so agents must anchor on the public
product reference rather than search uniformly over $[p_{\min},p_{\max}]$.
Figure~\ref{fig:pg_main_summary} summarises the per-model cross-suite
shift in $SE_\pi^+$ and the two structural penalties from F2--F3; full
per-model tables, geometry comparisons, and bootstrap CIs are in
Appendix~\ref{appdx:product_grounded_results}.

\textbf{Finding 6: Diagnostic ordering and bottlenecks survive a market-realistic geometry.}
Across eleven paired models, the rank correlation between synthetic and
data-grounded $SE_\pi^+$ is Spearman $\rho{=}0.90$ ($p<10^{-3}$;
Fig.~\ref{fig:pg_main_summary}A,C): \texttt{Claude Opus 4.6} remains on top
and the bottom group is preserved, but the cross-suite shift is bimodal across
the leaderboard, with upper-half models tending to gain and lower-half models
tending to lose. Strong models exploit the public product anchor while weaker
models flounder in the wider absolute price range, so the data-grounded suite
acts as a difficulty multiplier that \emph{amplifies} capability differences.

The two structural penalties from F2--F3 also replicate
(Fig.~\ref{fig:pg_main_summary}B): $\alpha_{\mathrm{cue}}$ attenuates for all
eleven models (median paired shift $+0.040$, paired Wilcoxon $p=0.001$) as
the product anchor partly insulates agents from cue over-reaction, while
$\alpha_{\mathrm{inf}}$ amplifies for ten of eleven ($-0.045$, $p=0.002$) as
wider, skewed action ranges make latent-type inference more decisive; only
\texttt{GPT-4o-mini} bucks the latter shift. No-deal recognition is also
cleaner: $\mathrm{FAGR}_\pi^{-}{=}0.000$ for all eleven models. The
agent-specific bottlenecks are therefore properties of the agents, not of
the synthetic distribution; full per-model numbers, the geometry comparison,
and the underlying Wilcoxon tables are in
Appendix~\ref{appdx:product_grounded_results}.

\begin{figure}
    \centering
    \includegraphics[width=0.95\linewidth]{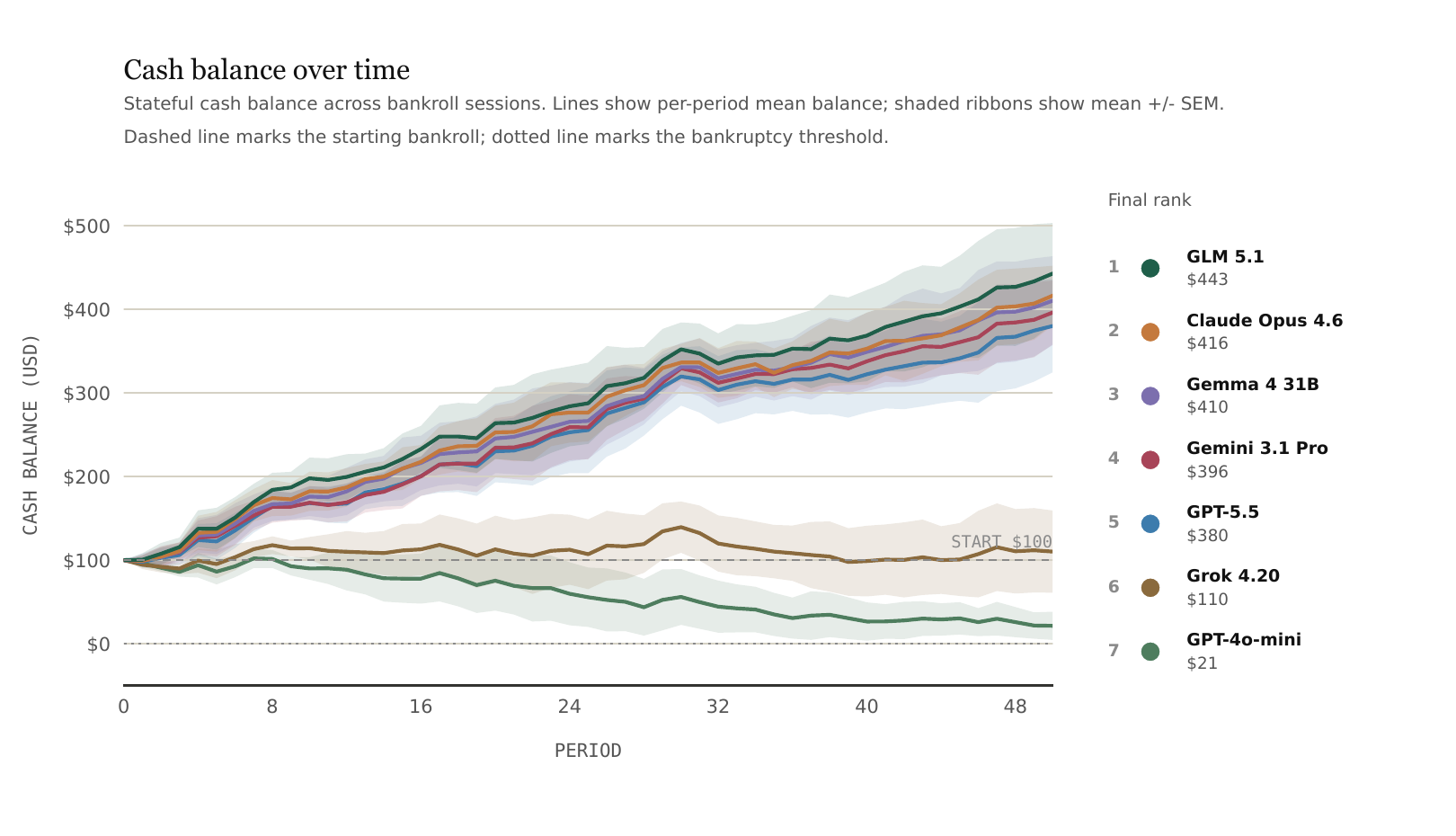}
    \vspace{-1.5em}
    \caption{\small \textbf{Cash-balance trajectories under the bankroll
    chain.} Per-period mean cash balance for seven LLM merchants across
    stateful sessions; shaded ribbons show $\pm 1$\,SEM and the right-edge
    ladder ranks agents by terminal balance. The dashed line marks the
    starting bankroll and the dotted line marks the bankruptcy threshold.
    Five LLMs compound to $\$380$--$\$443$ with full survival;
    \texttt{Grok~4.20} reaches $\$110$ ($75\%$ survival) and
    \texttt{GPT-4o-mini} ends near $\$21$ ($50\%$ survival). Full
    calibration in Appendix~\ref{app:leaderboard}.}
    \label{fig:bankroll_trajectories}
\end{figure}
\subsection{Commercial Extension: Commerce \& Bankroll}\label{sec:results_commerce}
The negotiation-only diagnostics describe \emph{how} an agent negotiates but
not what those choices cost in dollars. The commercial extension wraps each
episode in unit economics ($v$ resale value, $c$ fulfillment cost, $h$ fixed
overhead, lot size $n{\in}\{1,\dots,50\}$, regime-conditioned outside option
$o$). The merchant role earns $\Pi(p) = n(v - p - c) - h$ on agreement and
$o$ on walk-away; the vendor role is symmetric. The headline single-episode
metric is \emph{regret rate}, the scale-invariant gap to the per-scenario
best feasible profit summed over feasible scenarios:
$\mathrm{Regret}=1-\sum_i \Pi_i / \sum_i \Pi_i^*$.

\emph{Bankroll mode} chains $T$ commerce episodes into a single session,
threading a cash ledger and per-supplier beliefs across periods. The
cash balance evolves as
\begin{equation}
C_t = C_{t-1} + \Pi_t - (b + r \cdot R_t),
\qquad
C_0 = \text{starting capital},
\label{eq:bankroll_ledger}
\end{equation}
where $\Pi_t$ is period $t$'s commerce profit and $(b, r)$ are fixed
per-period and per-round operating costs over the $R_t$ rounds actually
played. The session terminates at \emph{hard ruin} the first time
$C_t < \tau$ (default $\tau=0$); remaining periods produce zero profit.
The headline session-level metrics are \emph{terminal balance} $C_T$
and \emph{survival rate} $\Pr[T_{\mathrm{ruin}} > T]$. Supplier-mode
variants, the optional memory-premium diagnostic, and full defaults
are in Appendix~\ref{app:leaderboard}.

\textbf{Finding 7: Diagnostic gaps translate into dollar-scale regret.}
On a 192-scenario commerce sweep (merchant role, default regime mixture;
Appendix~\ref{app:leaderboard}), the synthetic $SE_\pi^+$ ordering carries
over but its magnitude becomes operational. \texttt{Claude Opus 4.6} earns
\$68{,}592 against \$48{,}776 for the strongest fixed-concession baseline, a
$+\$19{,}816$ swing on identical scenarios at the same walk-away rate, and
leaves only $29\%$ of feasible profit on the table, against $50\%$ for the
baseline and $80\%$ for \texttt{GPT-4o-mini}. \texttt{GPT-4o-mini} is
simultaneously over-cautious ($66\%$ walk-away) and miscalibrated ($32.6\%$
negative-profit closes); regret rate exposes this where average margin
cannot, since margin is comparable across the bottom tier ($\approx 26\%$)
while regret rates diverge by tens of points. Profit-summed-across-all-
feasible-scenarios is therefore strictly more diagnostic than averaging
over closed deals.

\textbf{Finding 8: Bankroll surfaces solvency failures invisible to single-episode metrics.}
Chaining commerce episodes into a stateful session with per-period
operating drag amplifies the diagnostic separation
(Fig.~\ref{fig:bankroll_trajectories}). The five strongest LLMs
(\texttt{GLM 5.1}, \texttt{Claude Opus 4.6}, \texttt{Gemma 4 31B},
\texttt{Gemini 3.1 Pro}, \texttt{GPT-5.5}) survive every session and
compound to \$380--\$443 in terminal cash; \texttt{Grok 4.20} reaches
only \$110 with one of four sessions ruining; \texttt{GPT-4o-mini}
ends near \$21 with half its sessions bankrupt before the horizon.
The terminal-balance spread between the strongest and weakest LLM is
$\sim 21\times$, sharper than the $\sim 14\times$ spread in synthetic
$SE_\pi^+$, because small per-period concession or walk-away errors
compound across the chain into solvency failures that single-episode
metrics cannot see. Calibration, per-model numbers, and reproduction
scripts are in Appendix~\ref{app:leaderboard}.

\textbf{Additional Results.}
Appendix experiments further stress-test the main findings. Difficulty
stratification (Appendix~\ref{sec:difficulty}), prompt ablations
(Appendix~\ref{app:prompt_ablations}), and extreme-reveal ablations
(Appendix~\ref{app:language_reasoning_ablation}) show that high agreement is
not saturation: frontier agents close only a limited fraction of the oracle
gap even when hidden environment information is exposed. Voice, role, opener,
urgency, and runtime--cost analyses confirm that the observed failures are not
artifacts of any single configuration, and trace-level typology reveals stable
model-specific bargaining styles
(Appendices~\ref{appdx:deferred_exp},~\ref{app:strategic_profile}).

%% file: sections/conclusion.tex
\textsc{Terms-Bench} reframes negotiation evaluation away from black-box
LLM-vs.-LLM play toward agent-attributable diagnosis. By using the environment
itself as verifier, its bilateral price-negotiation instantiation exposes
structured failures hidden by outcome-only evaluation and grounds those
diagnoses in economic bargaining primitives. Its goal is diagnostic: environment-side verifiability
enables scalable, human-annotation-free capability isolation, while human
studies and richer multi-issue settings remain important next steps for
validating transfer beyond bilateral price negotiation.



%% file: acknowledgement.tex
We thank \textcolor{haiCyan}{Professor Ludwig Schmidt} for his insightful feedback during the early development of this benchmark. We also thank \textcolor{haiCyan}{Professor Alvin Roth} for his valuable perspectives on the economics of bargaining theory and its relevance to our work. His foundational work on \emph{Axiomatic Models of Bargaining} has been an important source of inspiration for the economic grounding of our framework.

\textcolor{haiLavender}{E.Z.} acknowledges support from the Stanford Graduate Fellowship (SGF) for Sciences and Engineering and a Jump Trading PhD Fellowship. \textcolor{haiLavender}{F.Z.} acknowledges support from SGF. \textcolor{haiLavender}{A.P.} and \textcolor{haiLavender}{B.E.} acknowledge support from the Knight-Hennessy scholarship. \textcolor{haiLavender}{J.B.} gratefully acknowledges support from DoD through the grants Air Force Office of Scientific Research under award number FA9550-20-1-0397 and ONR 1398311, also support from NSF via grants 2229012, 2312204, 2403007 is gratefully acknowledged.

%% file: appendix/validity.tex
\section{Economic Basis of \textsc{Terms-Bench}}\label{appdx:validity}

\textsc{Terms-Bench} is a general framework for constructing verifiable
negotiation environments for the purpose of evaluating LLM agents. In this paper, we instantiate it in a stylized
bilateral price-negotiation setting. The goal is not absolute realism, since
real-world negotiations vary widely across contexts, but diagnostic control: to
distill core bargaining primitives into a minimal environment where agent
behavior can be attributed to specific capabilities and failures.

Although controlled, the instantiation is not arbitrary. It combines a canonical
Bayesian bargaining backbone with practitioner-facing features such as bounded
prices, explicit no-deal cases, urgency, language-mediated signaling, and
data-grounded public context. We organize the literature grounding around five
simulator components: (i) the Bayesian bargaining environment; (ii) the
interaction protocol; (iii) the regime design; (iv) the counterpart-family
mechanism; and (v) the evaluation metrics. This preserves the economic structure
of bargaining while surfacing concrete decision problems for deployed agents,
and leaves room for customization toward procurement, pricing, and marketplace
applications.

\paragraph{(i) Bayesian bargaining environment setup.}
At the environment level, the bilateral instantiation follows the classical
bargaining representation of a feasible set together with a disagreement
outcome, as in Nash's bargaining problem and Roth's axiomatic treatment of
bargaining \citep{nash1950bargaining,roth1979axiomatic,thomson2009bargaining}.
It then introduces incomplete information over payoff-relevant parameters, most
centrally reservation values, consistent with the standard economics literature
on bargaining under incomplete information
\citep{chatterjee1983bargaining,ausubel2002bargaining}. Hart's discussion of
the axiomatic tradition is especially relevant here: when the bargaining
problem is specified without an extensive form, procedural assumptions are
``necessarily \ldots\ ad hoc,'' so solutions should rest on ``general
principles.''\footnote{Hart argues that when the extensive form of bargaining is
not specified, such procedural assumptions are ``necessarily \ldots\ ad hoc,''
so solutions should be based on ``general principles''
\citep[p.~317]{hart1992axiomatic}.} In this sense, the bilateral instantiation
should be read as a controlled protocol layered on top of a standard bargaining
scaffold, rather than as an attempt to recover a \emph{uniquely correct}
real-world bargaining procedure \citep{hart1992axiomatic,thomson2022axiomatic}.

\paragraph{(ii) Interaction protocol design.}
Private reservation values are a canonical abstraction in bargaining theory
\citep{chatterjee1983bargaining,ausubel2002bargaining}, and they remain common
in empirical pricing and customized-negotiation models, where transaction
outcomes are explained using latent reserve prices, willingness-to-pay, and
bargaining power \citep{grennan2013price,kadiyala2023predicting}. Likewise,
sequential offer exchange is a standard bargaining protocol in both economic
theory and automated-negotiation research
\citep{ausubel2002bargaining,faratin1998negotiation,baarslag2016learning}, and
it has also been documented in naturally occurring field settings
\citep{dindaroglu2024empirical}. These ingredients make the interaction
protocol more than a generic alternating-offer game: it captures the recurring
setting in which an agent must act under private constraints, interpret
counterpart behavior from sequential signals, and decide whether to continue,
agree, or walk away.

\paragraph{(iii) Regime design.}
The regime design is likewise grounded in core distinctions from bargaining and
negotiation research. The overlap and no-deal regimes formalize whether the
bargaining problem admits a mutually acceptable agreement at all, which
corresponds to the basic distinction between feasible agreement and disagreement
as the individually rational fallback outcome in classical bargaining theory
\citep{nash1950bargaining,roth1977individual}. Modeling this distinction
explicitly is important because negotiation research has long emphasized that
impasse is not merely a failure to optimize price; it is often a substantively
different outcome that must be analyzed in its own right
\citep{schweinsberg2022impasses}. The urgency-shift regime isolates a second
economically meaningful axis: time pressure. A broad literature shows that
deadlines and time pressure shape concession behavior, agreement likelihood, and
the incidence or timing of disagreement
\citep{stuhlmacher1998timepressure,karagozoglu2019bargaining}. By holding
reservation geometry fixed while shifting urgency, the benchmark separates
adaptation to counterpart time pressure from simple variation in feasible
surplus. This is precisely the kind of controlled design needed to diagnose
whether an agent is responding to latent temporal pressure rather than to a
trivially easier or harder price geometry.

\paragraph{(iv) Counterpart-family mechanism and design.}
The counterpart policy should be understood as a literature-grounded benchmark
kernel, \emph{not} as a structurally estimated or purely hand-engineered model
of human behavior. Its components---opening behavior, concession dynamics,
acceptance conditions, walk-away under time pressure, and latent-type inference
from offers and messages---mirror dimensions that are widely studied in
economics, organizational behavior, and automated negotiation
\citep{faratin1998negotiation,baarslag2014acceptance,baarslag2016learning}.
Prior work shows that first offers materially affect negotiated outcomes
\citep{petrowsky2025power}, concession patterns shape beliefs about reservation
values and later counteroffers \citep{tey2021impact}, and deadlines or time
pressure influence bargaining dynamics and disagreement patterns
\citep{stuhlmacher1998timepressure,karagozoglu2019bargaining}. The role of the
counterpart families is therefore not to claim an exhaustive taxonomy of real
negotiators, but to induce interpretable stress conditions over these
well-studied dimensions. This makes it possible to test whether an agent is
brittle to sparse evidence, over-reacts to cue style, mishandles time pressure,
or fails under harder bargaining postures.

\paragraph{(v) Evaluation metrics.}
The metric design is motivated by the same principle. The disagreement outcome
and individual rationality are constitutive of the bargaining problem itself
\citep{nash1950bargaining,roth1977individual}, and under incomplete
information, individual rationality binds any feasible bargaining mechanism
\citep{myerson1983efficient}, so a negotiation benchmark should not collapse all
performance into a single agreement-rate number. The negotiation-analysis and
automated-negotiation literatures accordingly evaluate systems along multiple
dimensions rather than a single scalar score
\citep{raiffa2002negotiation,baarslag2012measuring,baarslag2013practical,
baarslag2016learning}. Accordingly, the headline metrics are grouped around
terminal value, agreement calibration, opponent modeling, and protocol
compliance. Feasible terminal value captures surplus realization relative to the
\citet{myerson1983efficient} theoretical ceiling; agreement calibration reflects
the long-standing finding that systematic miscalibration produces impasse even
when a feasible agreement exists \citep{babcock1997explaining}; false agreement
in no-deal regimes operationalizes individual rationality as a first-class
violation; belief accuracy reflects opponent modeling as a prerequisite for
efficient negotiation \citep{baarslag2016learning} and a distinct axis in
recent LLM negotiation evaluation \citep{kwon-etal-2024-llms}; and protocol
compliance reflects the \citet{schelling1960strategy} view that constraint
adherence is itself strategic behavior, carried into modern LLM-agent
evaluation \citep{abdelnabi2024cooperation}.

Accordingly, the bilateral instantiation is intended to measure whether an
agent can satisfy prerequisite competencies that recur in bilateral
buyer--seller settings: preserving individual rationality, recognizing no-deal
cases, adapting to urgency and concession behavior, responding to opening
anchors, and using language cues without allowing language to override
structured economic actions. This focus is consistent with empirical work
showing that negotiated outcomes in business markets depend on latent
willingness-to-pay, bargaining ability, and sequential interaction
\citep{grennan2014ability,dindaroglu2024empirical}, that real B2B negotiations
exhibit systematic buyer--seller differences in tactics and information use
\citep{sigurdardottir2019buyer}, and that AI negotiation style can affect
discounts, speed, trust, and willingness for future interaction in
buyer--supplier negotiations \citep{herold2025brave}. Taken together, these
findings support the practitioner-facing relevance of this instantiation: while
it is not a perfect simulator of deployed negotiations, it evaluates
capabilities that are closely tied to real negotiated pricing and procurement
workflows.

Because \textsc{Terms-Bench} fixes the counterpart kernel while hiding latent
state from the evaluated agent, it supports failure attribution at a level that
outcome-only negotiation arenas do not: poor performance can be traced to
inference errors, concession-control failures, brittle cue use, or explicit
policy violations rather than being absorbed into a single deal-rate or surplus
number. In this sense, the bilateral price-negotiation setting serves as a
controlled diagnostic instantiation of \textsc{Terms-Bench}.

%% file: appendix/formal_specification.tex
\section{Specification of the Bilateral Price-Negotiation Instantiation}
\label{appdx:formal_spec}

This section consolidates the formal specification deferred from
Section \ref{sec:methods}. It complements \S\ref{sec:methods} by enumerating agent information state, the action and observation spaces, termination
cases, and hard constraints in detail.
\vspace{-0.5em}
\subsection{Information State and Belief Dynamics}
\label{app:info_state}

We expand on the information-state space and belief update deferred from
\S\ref{sec:framework_general}. The information state $s_k = (h_k, x_A, b_k)$
lives in $\mathcal{S} := \mathcal{H} \times \mathcal{X}_A \times \mathcal{B}$,
with $\mathcal{H}$ the space of public interaction histories, $\mathcal{X}_A$
the space of private side information available to the agent (e.g., market
statistics or profile-level context), and $\mathcal{B} \subseteq \Delta(T_B)$
the space of internal belief states over opponent types. The belief is updated
as $b_k = \Psi(h_k)$, where the update mechanism $\Psi$ is part of the agent
policy and is not specified by $\mu$. The true counterpart type $t_B$ is not
directly observable; the agent must infer it from the sequence of prices and
language realizations.

The policy $\pi: \mathcal{S} \to \mathcal{A}$ thus jointly determines
(i) how the agent performs latent-state inference over opponent types from
observed signals, and (ii) how it selects strategic actions to optimize
outcomes under the fixed bargaining environment and protocol. We do not
restrict the implementation of $\pi$: it may be a prompted LLM, a rule-based
agent, a learned policy, or a model-based planner.
\vspace{-0.5em}
\subsection{Action and Observation Spaces}
\label{app:action_obs_space}

\paragraph{Action.}
At each round $k$ the agent selects $a_k = (d_k, p_k, l_k)$, with
$d_k \in \{\texttt{Offer}, \texttt{Accept}, \texttt{Reject}\}$,
$p_k \in [p_{\min}, p_{\max}]$ active only when $d_k = \texttt{Offer}$, and
$l_k \in \mathcal{L}$ a free-form natural-language message. \texttt{Accept}
binds at the counterpart's last offered price; \texttt{Reject} terminates
with disagreement $\bot$. The message $l_k$ lets agents justify offers,
request information, signal constraints, or employ other communicative
strategies. Messages may influence counterpart behavior through the cue
layer of $\pi_B$ (Appendix~\ref{sec:counterpart}), but only the economic
decision $d_k$ and price $p_k$ directly determine negotiation outcomes
and constraint compliance.

\paragraph{Observation.}
At each round $k$, the evaluated agent observes
\begin{equation}
o_k = (p_k^B, l_k^B) \in \mathcal{O},
\end{equation}
where $p_k^B \in [p_{\min}, p_{\max}]$ is the counterpart's price offer and
$l_k^B \in \mathcal{L}$ is the counterpart's natural-language message. The
simulator-internal cue variables
$(\tilde{s}_k, \tilde{c}_k) \in
\{\texttt{positive}, \texttt{neutral}, \texttt{negative}\}
\times \{\texttt{Concede}, \texttt{Hold}, \texttt{Pressure}\}$ are not part
of $\mathcal{O}$ (see \S\ref{sec:agora_base_setup}); they parameterize the
realization of $l_k^B$ but are never themselves observed. The public
interaction history up to round $k$ is
$h_k = (o_1, a_1, \ldots, o_k)$, accumulating all observations and the
agent's past actions. The agent's private context $x_A$ includes its
assigned role, its own reservation price $r_A$, and any additional side
information.
\vspace{-0.5em}
\subsection{Termination and Constraints}
\label{app:termination_constraints}

\paragraph{Termination.} A negotiation episode ends in one of five
scenarios:
\begin{enumerate}
    \item The agent chooses \texttt{Accept}; the outcome is the
    counterpart's last offered price.
    \item The agent chooses \texttt{Reject}; the outcome is
    disagreement $\bot$.
    \item The counterpart accepts the agent's offer; the outcome is
    the agent's proposed price.
    \item The counterpart terminally rejects (walk-away); the outcome
    is disagreement $\bot$.
    \item The round limit $K$ is reached without agreement; the
    outcome is disagreement $\bot$.
\end{enumerate}

\paragraph{Constraints.} All \texttt{Offer} actions must satisfy:
(i) price bounds $p_{\min} \leq p_k \leq p_{\max}$;
(ii) monotonic concession: for buyer agents, $p_k \geq p_{k-1}$ where
$k, k-1$ index the agent's own offers; for seller agents,
$p_k \leq p_{k-1}$;
(iii) turn budget $k \leq K$.
Violations of (i) or of individual rationality (accepting or offering
a price strictly worse than $r_A$) are recorded as critical violations
in $\mathrm{CritViol}\%$. Monotonicity and turn-budget violations are
recorded as secondary procedural diagnostics
(Appendix~\ref{appdx:eval}).
\vspace{-0.5em}
\subsection{Type Parameterization: Stance vs.\ Personality}
\label{app:stance_personality}

Our type parameterization $t_i = (r_i, \kappa_i, \eta_i)$ excludes
personality traits, despite their documented correlations with
negotiation outcomes in recent LLM studies
\citep{cohen2025exploring}. Following
\citet{ma2005bigfive_conflict_negotiation} and
\citet{huang2024personality}, we instead model strategic stance
$\eta_i \in \{\texttt{conciliatory}, \texttt{neutral},
\texttt{aggressive}\}$ as the primary behavioral driver, since
conflict-management styles have been shown to mediate the relationship
between personality and bargaining behavior. Incorporating personality
as an orthogonal communication-level dimension (e.g., as a separate
language-realization parameter that does not affect the economic
kernel) remains a direction for future work.

%% file: appendix/counterpart_modeling.tex
\section{Counterpart Behavior Families}
\label{sec:counterpart}

Counterpart behavior in \textsc{Terms-Bench}'s bilateral price-negotiation is governed by the fixed policy in
Section~\ref{sec:counter_design} and the cue generation mechanism in
Appendix~\ref{appdx:cue_generation}. Behavior families correspond to presets of
this common simulator. They do not introduce separate acceptance, walk-away, or
offer-generation formulas.

We use six main families. Four form a diagnostic core over two axes:
\begin{enumerate}[label=(\roman*)]
  \item \textbf{Economic reactivity.} How strongly does the counterpart's
  economic behavior depend on the agent's recent offer trajectory, relative to
  the counterpart's latent type $(r_B,\kappa_B,\eta_B)$?
  \item \textbf{Cue reliability.} Do the language-facing sentiment and posture
  cues reliably reflect the counterpart's latent stance $\eta_B$, or are they
  collapsed to noncommittal states?
\end{enumerate}
Two additional families,
\texttt{Stochastic} and \texttt{Adversarial}, are stress conditions: 
\texttt{Stochastic} provides a noise-floor condition under noisy price and cue
channels, while \texttt{Adversarial} provides a hardball condition with an
aggressive-skewed stance prior and pressuring cues.

\begin{enumerate}

\item \emph{Candid counterparts}
\textup{[type-instrumental economics, accurate cues]}.
This family replaces the earlier \texttt{Truthful} family. The counterpart's
economic behavior is low-noise and strongly type-conditioned: reservation value
sets the feasible boundary, urgency changes acceptance and concession timing,
and stance changes the payoff consequences of rigidity and concession.
Sentiment and strategic cues are generated from the base cue model, so language
is informative about latent stance. This family is designed so that correct
inference of $(r_B,\kappa_B,\eta_B)$ is instrumentally valuable for surplus
extraction.

\item \emph{Taciturn counterparts}
\textup{[type-instrumental economics, uninformative cues]}.
Economic behavior follows the same type-instrumental preset as
\texttt{Candid}, but the cue channel is collapsed to neutral,
noncommittal states. This isolates inference from economic behavior alone: an
agent that degrades relative to \texttt{Candid} is relying heavily
on linguistic or stylistic cues.

\item \emph{Expressive counterparts}
\textup{[high-reactivity economics, accurate cues]}.
The cue channel remains accurate, but economic behavior is more strongly
history-reactive. Counter-offers and acceptance probabilities respond more to
the agent's recent concession pattern and rigidity. This family tests whether
agents can use reliable cues while avoiding confusion between latent type and
state-dependent reactions to their own behavior.

\item \emph{Strategic counterparts}
\textup{[high-reactivity economics, uninformative cues]}.
Economic behavior is strongly history-reactive, and the cue channel is
uninformative. The counterpart is linguistically guarded while adapting
tactically through price and acceptance behavior. This is the hardest core
family for opponent modeling because both the economic and language channels
provide imperfect evidence.

\item \emph{Adversarial counterparts}
\textup{[hardball economics, pressuring cues]}.
This family is an explicit stress test rather than part of the core factorial.
The stance prior is skewed toward aggressive counterparts, economic reactivity
is high, concessionary behavior is strongly exploited, and rigidity is punished
for aggressive types. The cue channel is biased toward negative sentiment and
\texttt{Pressure}. This family tests whether agents can recognize and adapt to
hardball negotiation without either over-conceding to pressure or holding firm
until walk-away.

\item \emph{Stochastic counterparts}
\textup{[moderate-reactivity economics, noisy/weak cues]}.
This family degrades both the price and cue channels through noise rather than
through deliberate strategic concealment. Economic behavior uses a moderate
reactivity preset, but price noise is high, so offer trajectories are less
diagnostic of the underlying concession rule. The cue channel remains weakly
coupled to latent stance but is made noisy through elevated sentiment variance
and strategic-cue temperature, rather than being collapsed to neutral states.
This family provides a noise-floor test for belief calibration and surplus
extraction: a robust agent should avoid over-interpreting noisy price movements
or unstable linguistic cues.

\end{enumerate}

The family identity and parameter configuration are hidden from the negotiating
agent and logged only for evaluator analysis. We report performance by family
to distinguish failures of cue use, economic-channel inference, type-vs-state
decomposition, and robustness to hardball pressure. 

\begin{table}[t]
\centering
\scriptsize
\setlength{\tabcolsep}{4pt}
\renewcommand{\arraystretch}{0.95}
\begin{tabular}{p{2.0cm} p{2.0cm} p{1.8cm} p{2.2cm} p{2.0cm} p{1.0cm} p{1.2cm}}
\toprule
\textbf{Family} & \textbf{Economic preset} & \textbf{Cue channel}
& $\lambda_2(\texttt{C},\texttt{N},\texttt{A})$
& $\eta_B$ prior & $\bar\sigma_p$ & \textbf{Role} \\
\midrule
Candid
& type-instrumental
& accurate
& $(0.30,0.50,1.00)$
& uniform
& low
& core \\

Taciturn
& type-instrumental
& uninformative
& $(0.30,0.50,1.00)$
& uniform
& low
& core \\

Expressive
& high-reactivity
& accurate
& $(0.45,0.90,1.80)$
& uniform
& mod.
& core \\

Strategic
& high-reactivity
& uninformative
& $(0.45,0.90,1.80)$
& uniform
& mod.
& core \\

Stochastic
& moderate-reactivity
& noisy/weak
& $(0.35,0.70,1.40)$
& uniform
& high
& noise floor \\

Adversarial
& hardball
& pressuring
& $(0.60,1.40,2.60)$
& aggressive-skewed
& low
& stress \\
\bottomrule
\end{tabular}
\vspace{0.5em}
\caption{\scriptsize
Summary of counterpart behavior families. The first four families form the
diagnostic core, crossing economic reactivity with cue reliability. The tuple
$\lambda_2(\texttt{C},\texttt{N},\texttt{A})$ gives the concession-reciprocity
coefficient for conciliatory, neutral, and aggressive stance types. The
\texttt{Stochastic} family is a noise-floor condition with high price noise and
weak/noisy cues, while \texttt{Adversarial} is a hardball stress condition with
an aggressive-skewed stance prior and pressuring cues.
}
\label{tab:counterpart_families}
\vspace{-2em}
\end{table}
\paragraph{Implementation under the canonical simulator.}
Table~\ref{tab:counterpart_families} describes the six main counterpart
behavior families. All six families are instantiated by varying only parameter
presets already present in the fixed counterpart policy of
Section~\ref{sec:counter_design} and the cue-generation mechanism of
Appendix~\ref{appdx:cue_generation}. In particular, all families share the same
opening-role protocol, randomized opening-offer model, acceptance model,
walk-away hazard, counter-offer rule, and cue-generation interface.

Family differences arise through two sets of presets. First, the economic
preset controls stance-dependent response parameters, including
$\rho_{\mathcal F}(\eta_B)$, $\xi_{\mathcal F}(\eta_B)$, and
$\lambda_{2,\mathcal F}(\eta_B)$, together with the price-noise scale and the
stance prior over $\eta_B$. Second, the cue preset controls whether the
language-facing sentiment and strategic-posture cues are informative,
uninformative, noisy, or pressuring. The \texttt{Candid} and
\texttt{Taciturn} families share the same type-instrumental economic preset and
differ only in cue reliability; the \texttt{Expressive} and \texttt{Strategic}
families share the same high-reactivity economic preset and differ only in cue
reliability. The \texttt{Stochastic} family provides a noise-floor condition by
using moderate economic reactivity with high price noise and weak/noisy cues,
while the \texttt{Adversarial} family uses a hardball economic preset with an
aggressive-skewed stance prior and pressuring cues.

\subsection{Family-Specific Economic Presets}
\label{appdx:economic_presets}

The acceptance and counter-offer equations in Section~\ref{sec:counter_design}
use stance-dependent coefficients
$\rho_{\mathcal F}(\eta_B)$, $\xi_{\mathcal F}(\eta_B)$, and
$\lambda_{2,\mathcal F}(\eta_B)$. We order stance types as
\[
(\texttt{C},\texttt{N},\texttt{A})
:=
(\texttt{conciliatory},\texttt{neutral},\texttt{aggressive}).
\]
Table~\ref{tab:economic_presets} gives the economic preset used by each
counterpart family. These presets instantiate the same acceptance and
counter-offer formulas for all families.

\begin{table}[H]
\centering
\scriptsize
\setlength{\tabcolsep}{4.5pt}
\renewcommand{\arraystretch}{1.0}
\begin{tabular}{p{2.0cm} p{2.8cm} p{2.2cm} p{2.2cm} p{3.2cm}}
\toprule
\textbf{Preset} & $\rho(\texttt{C},\texttt{N},\texttt{A})$
& $\xi(\texttt{C},\texttt{N},\texttt{A})$
& $\lambda_2(\texttt{C},\texttt{N},\texttt{A})$
& \textbf{Used by} \\
\midrule

Type-instrumental
& $(0,\,-0.25,\,-0.75)$
& $(+0.40,\,0,\,-0.50)$
& $(0.30,\,0.50,\,1.00)$
& \texttt{Candid}, \texttt{Taciturn} \\

High-reactivity
& $(0,\,-0.75,\,-1.50)$
& $(+0.40,\,0,\,-0.75)$
& $(0.45,\,0.90,\,1.80)$
& \texttt{Expressive}, \texttt{Strategic} \\

Moderate-stochastic
& $(0,\,-0.50,\,-1.10)$
& $(+0.35,\,0,\,-0.60)$
& $(0.35,\,0.70,\,1.40)$
& \texttt{Stochastic} \\

Hardball
& $(-0.25,\,-1.25,\,-2.25)$
& $(0,\,-0.50,\,-1.20)$
& $(0.60,\,1.40,\,2.60)$
& \texttt{Adversarial} \\

\bottomrule
\end{tabular}
\vspace{0.5em}
\caption{\scriptsize
Family-specific economic presets. Coefficients are ordered by stance type:
conciliatory, neutral, aggressive. Conciliatory counterparts are most receptive to firm but feasible bargaining:
they reward rigidity in the acceptance model, but do not directly reward fast
concession. Aggressive counterparts, by contrast, penalize pure rigidity and
exploit rapid concession, making measured movement preferable. Thus, the
agent's optimal concession posture depends on correctly inferring latent
stance: firm
bargaining is distinctly more effective against conciliatory counterparts,
while aggressive counterparts punish pure rigidity and exploit rapid concession.
}
\label{tab:economic_presets}
\end{table}
\vspace{-2em}
The type-instrumental preset is designed so that latent stance is economically
meaningful without making recent agent history dominate the latent-type signal.
In this preset, conciliatory counterparts are receptive to firm but feasible
offers, neutral counterparts behave as the middle case, and aggressive
counterparts penalize pure rigidity while exploiting rapid concession. This
makes stance inference useful: agents should \emph{not} use the same
concession posture against all counterpart types.

On the other hand, the high-reactivity preset increases the influence of the agent's recent offer
trajectory while preserving the same stance semantics. It is harder than the
type-instrumental preset because observed price dynamics are more strongly
confounded by the agent's own behavior. The stochastic preset uses moderate
economic reactivity but is paired with elevated price and cue noise. The
hardball preset combines strong exploitation of fast concession with penalties
for rigidity, especially under aggressive stance. The adversarial family
additionally uses an aggressive-skewed stance prior:
\[
\Pr(\eta_B=\texttt{conciliatory})=0.05, \ \
\Pr(\eta_B=\texttt{neutral})=0.15, \ \
\Pr(\eta_B=\texttt{aggressive})=0.80.
\]

\subsection{Economic Response Details}
\label{appdx:economic_response_details}

This section provides additional details for the economic response model in
Section~\ref{sec:counter_design}. We use the same notations as in Section~\ref{sec:counter_design}. The response model separates three
counterpart outcomes after an agent offer: acceptance of the current offer,
terminal walk-away, and continued bargaining through a counter-offer. All
counterpart behavior families use the same response equations; families differ
only through parameter presets specified in Appendix~\ref{appdx:economic_presets}.

\subsubsection{Acceptance}
Recall that $\bar{\Delta}_k\ge 0$ means that the offer is individually rational for
the counterpart. The acceptance gate
$\mathbf{1}\{\bar{\Delta}_k\ge 0\}$ enforces this individual-rationality
constraint: seller counterparts never accept prices below $r_B$, and buyer
counterparts never accept prices above $r_B$.

We use the concave deadline clock
\[
D_k:=\frac{k}{K},
\qquad
\widetilde D_k:=\sqrt{D_k},
\qquad
\widetilde{\bar D}_k:=1-\widetilde D_k.
\]
The square-root transformation spreads deadline pressure over the later part of
the negotiation instead of concentrating it almost entirely in the final round.

Conditional on individual rationality, acceptance is stochastic:
\[
a_k
:=
\pi_B(d_k^B=\texttt{Accept}\mid p_k^A,t_B,h_{k-1})
=
\mathbf{1}\{\bar{\Delta}_k\ge 0\}
\sigma\!\left(g_\theta(p_k^A,t_B,k,h_{k-1})\right),
\]
where
\[
g_\theta(p_k^A,t_B,k,h_{k-1})
=
\alpha\bar{\Delta}_k
+\beta\kappa_B
-\gamma\widetilde{\bar D}_k
+\rho_{\mathcal F}(\eta_B)\,\mathrm{ConcedeSpeed}_k
+\xi_{\mathcal F}(\eta_B)\,\mathrm{Rigidity}_k .
\]
Here $\mathcal F$ denotes the counterpart behavior family. The terms
$\mathrm{ConcedeSpeed}_k$ and $\mathrm{Rigidity}_k$ are deterministic history
features defined in Appendix~\ref{sec:history_features}. The coefficients
$\rho_{\mathcal F}(\eta_B)$ and $\xi_{\mathcal F}(\eta_B)$ are
family- and stance-dependent. Negative values of
$\rho_{\mathcal F}(\eta_B)$ mean that fast recent agent concession reduces the
counterpart's willingness to accept, while positive values make concession more
reciprocated. Positive values of $\xi_{\mathcal F}(\eta_B)$ make rigidity more
effective; negative values make rigidity costly. In the type-instrumental
families, this makes stance inference directly payoff-relevant: firmness,
concession, and balanced movement have different consequences for
conciliatory, neutral, and aggressive counterparts.

\subsubsection{Terminal Walk-Away}
The walk-away branch is a reduced-form participation constraint. It prevents
continuation from being costless when the agent repeatedly proposes prices that
are outside the counterpart's individually rational region. The default hazard
is deliberately sparse:
\[
\omega_k
=
\mathbf{1}\{k\ge k_{\mathrm{walk}}\}
\mathbf{1}\{\bar{\Delta}_k<0\}
\sigma\!\left(
\phi_0+\phi_\Delta[-\bar{\Delta}_k]_+ + \phi_T\tau_k^W
\right),
\]
where
\[
\tau_k^W
=
\min\left\{
1,
\max\left\{
0,
\frac{k-k_{\mathrm{walk}}}{K-k_{\mathrm{walk}}}
\right\}
\right\}.
\]
Here $[-\bar{\Delta}_k]_+$ is the normalized reservation shortfall of the
agent's current offer from the counterpart's perspective, and $\tau_k^W$ is a
walk-away clock that equals zero when terminal exit first becomes available and
approaches one near the round limit. In the default configuration we use
\[
k_{\mathrm{walk}}=\left\lceil\frac{K}{2}\right\rceil,
\qquad
\phi_0=-4.5,
\qquad
\phi_\Delta=30.0,
\qquad
\phi_T=1.5.
\]

The hard gate $\mathbf{1}\{\bar{\Delta}_k<0\}$ implies that walk-away does not
alter ordinary feasible bargaining when the agent remains inside the
counterpart's individually rational region. It is active primarily in no-deal
episodes and in feasible episodes where the agent opens or persists outside the
counterpart's reservation constraint. This makes the mechanism diagnostic
rather than broadly punitive: a competent agent can still bargain firmly within
the feasible region, but persistent non-individually-rational offers become
increasingly risky late in the interaction.

\paragraph{Why the default hazard omits urgency and stance.}
We intentionally omit direct urgency and stance terms from the default
walk-away hazard. Urgency already affects acceptance through $\beta\kappa_B$
and concession dynamics through $\lambda_1\kappa_B$. Stance already affects
acceptance through $\rho_{\mathcal F}(\eta_B)$ and
$\xi_{\mathcal F}(\eta_B)$, counter-offer dynamics through
$\lambda_{2,\mathcal F}(\eta_B)$ and $\lambda_3,\lambda_4$, and cue generation
through the family-specific cue model. Adding direct urgency or stance effects
to walk-away would give the same latent variables additional behavioral
channels and risk making them overly dominant. Keeping terminal exit
type-neutral preserves the walk-away branch as a narrow participation-constraint
mechanism.

\subsubsection{Termination Accounting}\label{sec:termination_accounting}
For $k<K$, the counterpart response probabilities are
\[
\pi_B(\texttt{Accept})=a_k,\qquad
\pi_B(\texttt{Reject})=(1-a_k)\omega_k,\qquad
\pi_B(\texttt{Offer})=(1-a_k)(1-\omega_k).
\]
At the terminal round, the remaining non-acceptance and non-walk-away mass is
assigned to round-limit disagreement. In evaluation, we separately log the
termination source:
\[
\begin{aligned}
\tau_{\mathrm{term}} \in \{&
\texttt{AgentAccept},
\texttt{CounterpartAccept},
\texttt{AgentReject},\\
&
\texttt{CounterpartWalkAway},
\texttt{Timeout},
\texttt{AgreementViolation}
\}.
\end{aligned}
\]
This allows no-deal performance to be distinguished between disciplined agent
exit, counterpart walk-away, timeout, and individually irrational agreement.

\subsection{History Feature Definitions}\label{sec:history_features}

$\mathrm{ConcedeMagnitude}_k$, $\mathrm{ConcedeSpeed}_k$, and
$\mathrm{Rigidity}_k$ are fixed scalar summaries of the agent's recent offer
dynamics, computed deterministically from $h_{k-1}$. To make these features
comparable across cheap and expensive products and across negotiation regimes,
we normalize all offer changes by the public price range $R$, which is fixed within each episode and strictly positive by construction.

Let
\[
s_A :=
\begin{cases}
+1, & \text{if the agent role is buyer},\\
-1, & \text{if the agent role is seller},
\end{cases}
\]
so that $s_A(p_j^A-p_{j-1}^A)>0$ always corresponds to a
\emph{concessionary} move by the agent, while
$s_A(p_j^A-p_{j-1}^A)<0$ corresponds to hardening or reversal. Let
\[
\mathcal{J}_k := \{j : \max(2,k-3) \le j \le k-1\}
\]
index the most recent rounds for which two consecutive agent offers are
available.

\begin{itemize}

\item \emph{Concession magnitude.}
We define the normalized recent concession magnitude by
\begin{equation}
\label{eq:concede_magnitude}
\mathrm{ConcedeMagnitude}_k =
\begin{cases}
\displaystyle
\frac{1}{|\mathcal{J}_k|}
\sum_{j\in\mathcal{J}_k}
\frac{\max\{0,\; s_A(p_j^A-p_{j-1}^A)\}}{R},
& \text{if } |\mathcal{J}_k| \ge 1,\\[2.0ex]
0, & \text{otherwise.}
\end{cases}
\end{equation}
This feature measures how much the agent has recently conceded, on average,
expressed as a fraction of the episode's public price range. Non-concessionary
moves, including hardening or reversal, contribute zero rather than being
counted as eagerness.

\item \emph{Concession speed.}
We define the normalized directional concession speed by
\begin{equation}
\label{eq:concede_speed}
\mathrm{ConcedeSpeed}_k =
\begin{cases}
\displaystyle
\frac{1}{|\mathcal{J}_k|}
\sum_{j\in\mathcal{J}_k}
\frac{s_A(p_j^A-p_{j-1}^A)}{R},
& \text{if } |\mathcal{J}_k| \ge 1,\\[2.0ex]
0, & \text{otherwise.}
\end{cases}
\end{equation}
By construction, larger values indicate more concessionary recent behavior,
values near zero indicate relatively stationary or mixed behavior, and negative
values indicate recent hardening or reversal. This sign convention is
role-invariant: positive $\mathrm{ConcedeSpeed}_k$ has the same behavioral
meaning for buyers and sellers.

\item \emph{Rigidity.}
We define the rigidity indicator by
\begin{equation}
\label{eq:rigidity}
\mathrm{Rigidity}_k =
\begin{cases}
1, & \text{if } \dfrac{\max\{0,\; s_A(p_{k-1}^A-p_{k-2}^A)\}}{R}
< \tau_{\mathrm{rigid}},\\[2.0ex]
0, & \text{otherwise,}
\end{cases}
\qquad \tau_{\mathrm{rigid}} \in (0,1),
\end{equation}
whenever the last two agent offers are available; otherwise
$\mathrm{Rigidity}_k := 0$. Thus, $\mathrm{Rigidity}_k=1$ indicates that the
agent made only a minimal recent concession relative to the episode's price
scale. In particular, holding firm or moving in a hardening direction both
count as rigid behavior under this definition.

\end{itemize}

Unless otherwise stated, we set $\tau_{\mathrm{rigid}}=0.1$.

\paragraph{Boundary conditions with mixed opener roles.}
History features are computed from realized public histories and are well-defined
under both opener roles. If fewer than two agent offers have been observed, then
\[
\mathrm{ConcedeMagnitude}_k
=
\mathrm{ConcedeSpeed}_k
=
\mathrm{Rigidity}_k
=
0.
\]
Thus, when the agent opens, the counterpart's first response to the agent's
opening offer is evaluated without artificial history-reactivity. Once the
agent has made at least two offers, the usual role-normalized definitions apply.

Similarly, counterpart-side concession features used in cue generation are set
to zero until two counterpart offers have been observed. This covers both cases:
the ordinary counterpart-opens case, where the first counterpart offer is the
episode anchor, and the agent-opens case, where the first counterpart offer may
occur only after the counterpart declines to accept the agent's opening offer.

\subsection{Opening Role and Opening-Offer Generation}
\label{appdx:opening-offer}

At the start of each episode, the protocol specifies an opener role
\[
\chi\in\{\texttt{AgentOpens},\texttt{CounterpartOpens}\}.
\]
The opener role is an episode attribute of the interaction protocol, not a
separate counterpart behavior family. In the main experimental suite, $\chi$ is
assigned by blocked allocation so that agent-opens and counterpart-opens
episodes are exactly balanced within each regime--family cell
(Section~\ref{sec:exp_setup}). More general stochastic generators may instead
draw $\chi$ from a user-specified distribution.

If $\chi=\texttt{CounterpartOpens}$, the counterpart produces the first price
proposal. If $\chi=\texttt{AgentOpens}$, the evaluated agent produces the first
price proposal, and $\texttt{Accept}$ is unavailable because no counterpart
offer has yet been observed. The counterpart then responds using the economic
response model in Section~\ref{sec:counter_design}. If it neither accepts nor
walks away, its first price proposal is generated by the opening-offer model
below, since no previous counterpart offer exists.

\paragraph{Randomized counterpart opening harshness.}
To avoid making the counterpart reservation value nearly invertible from its
first price, we randomize the harshness of the counterpart's first offer. The
episode-level harshness variable is
\[
d_{0,e}\sim \mathcal D_{\mathrm{open}},
\qquad
\mathcal D_{\mathrm{open}}
=
\mathrm{Uniform}(d_{\min},d_{\max}),
\]
with default
$
d_{\min}=0.20,
\qquad
d_{\max}=0.80.
$
For controlled ablations, $\mathcal D_{\mathrm{open}}$ may be replaced by
stratum-specific intervals corresponding to soft, medium, or harsh counterpart
anchors. The realized $d_{0,e}$ is hidden from the agent and logged for
analysis. It is used whenever the counterpart generates its first offer, whether
the counterpart opens the episode or makes its first offer after declining an
agent opening. If no counterpart offer is ever generated, the variable is
unused for that trajectory.

In difficulty grading, counterpart opening harshness is treated as an
environment-side opening difficulty variable only when
$\chi=\texttt{CounterpartOpens}$. When $\chi=\texttt{AgentOpens}$, the
episode's first price is chosen by the evaluated agent and is therefore a
policy decision rather than an instance property. If the counterpart later
produces its first offer in an agent-opens episode, the realized anchor is
logged as a response diagnostic but is not included in the
environment difficulty score.

\paragraph{Counterpart first-offer model.}
When the counterpart must produce its first offer, we generate an initial
counterpart price relative to its reservation value. Define the counterpart's
directional slack
\[
S_B^{\mathrm{open}}
:=
\begin{cases}
p_{\max}-r_B, & \text{if the counterpart is a seller},\\[0.8ex]
r_B-p_{\min}, & \text{if the counterpart is a buyer},
\end{cases}
\]
and direction
\[
s_B :=
\begin{cases}
+1, & \text{if the counterpart is a seller},\\
-1, & \text{if the counterpart is a buyer}.
\end{cases}
\]
Thus $s_B(p-r_B)>0$ means that price $p$ lies in the counterpart's favorable
direction. The role-dependent feasible interval for the counterpart's first
offer is
\[
\mathcal B_B
:=
\begin{cases}
[r_B,p_{\max}], & \text{if the counterpart is a seller},\\
[p_{\min},r_B], & \text{if the counterpart is a buyer}.
\end{cases}
\]

The counterpart's first offer is
\begin{equation}
\label{eq:opening_offer}
p_{\mathrm{init}}^B
=
\Pi_{\mathcal B_B}
\left(
r_B
+
s_B d_{0,e}\phi(\kappa_B,\eta_B)S_B^{\mathrm{open}}
+
\varepsilon_0
\right),
\qquad
\varepsilon_0\sim\mathcal N(0,\sigma_0^2).
\end{equation}
Projection onto $\mathcal B_B$ ensures that the first counterpart offer respects
both public price bounds and the counterpart's own reservation value.

The modulation factor is
\begin{equation}
\label{eq:opening_modulation}
\phi(\kappa_B,\eta_B)
=
\operatorname{clip}
\left(
1-\omega_\kappa\kappa_B
+\omega_\eta\mathbf{1}\{\eta_B=\texttt{aggressive}\}
-\omega_\eta'\mathbf{1}\{\eta_B=\texttt{conciliatory}\},
\phi_{\min},\phi_{\max}
\right).
\end{equation}
More urgent counterparts open less aggressively, aggressive counterparts claim
more favorable slack, and conciliatory counterparts open closer to reservation.
Because $d_{0,e}$ is randomized and hidden, the counterpart's first offer remains
informative but is not a point-identifying signal for $r_B$.

\subsection{Cue Generation and Language Realization}
\label{appdx:cue_generation}

We give the complete specification of the hidden cue variables
\[
(\tilde s_k,\tilde c_k)
\in
\{\texttt{positive},\texttt{neutral},\texttt{negative}\}
\times
\{\texttt{Concede},\texttt{Hold},\texttt{Pressure}\}
\]
used by the environment-simulated counterpart. These cues are not revealed
directly to the evaluated agent; instead, they parameterize the counterpart's
natural-language message $m_k^B$ conditional on the already-committed economic
action $(d_k^B,p_k^B)$.

\paragraph{Counterpart concession magnitude.}
Let $p_{k-1}^B$ and $p_k^B$ denote the counterpart's two most recent offers,
when both exist. We define the normalized concession magnitude
\[
C_k^B :=
\begin{cases}
\min\!\left\{1,\,
\dfrac{|p_k^B-p_{k-1}^B|}
{|p_{k-1}^B-r_B|+\varepsilon_c}
\right\},
& \text{if } d_k^B=\texttt{Offer} \text{ and a previous counterpart offer exists},\\[1.2ex]
0, & \text{otherwise},
\end{cases}
\]
where $\varepsilon_c>0$ is a small constant used only for numerical stability.
This definition is role-agnostic: for a seller, concessions correspond to
lowering price; for a buyer, concessions correspond to increasing price.

\paragraph{Deadline clock.}
We define normalized round progress by
\[
D_k:=\frac{k}{K},
\]
and use the concave deadline clock
\[
\widetilde D_k:=\sqrt{D_k}.
\]
For equations written in terms of remaining time, we use
\[
\widetilde{\bar D}_k:=1-\widetilde D_k.
\]
The square-root clock spreads deadline pressure over the later part of the
interaction instead of concentrating it only at the final round. In the
acceptance model, $\widetilde{\bar D}_k$ replaces the linear remaining-time
term; in the strategic cue model, $\widetilde D_k$ replaces the linear
deadline-proximity term.

\paragraph{Boundary conditions under mixed opener roles.}
Because either party may open the episode, boundary values are defined by the
available public history rather than by a fixed round-one pattern. If the
counterpart has made fewer than two offers, then no previous counterpart offer
exists for computing $C_k^B$, and we set
\[
C_k^B=0.
\]
If the agent has made fewer than two offers, then the agent-history features
used by the acceptance and counter-offer models are set to their boundary
values:
\[
\mathrm{ConcedeMagnitude}_k=0,
\qquad
\mathrm{ConcedeSpeed}_k=0,
\qquad
\mathrm{Rigidity}_k=0.
\]
Thus, when $\chi=\texttt{CounterpartOpens}$, the counterpart's opening message
uses $C_k^B=0$ because no previous counterpart offer exists. When
$\chi=\texttt{AgentOpens}$, the counterpart's first response to the agent's
opening offer is evaluated with zero counterpart-concession history and zero
agent-history reactivity unless the relevant prior offers exist. These boundary
values ensure that cue likelihoods, acceptance probabilities, and belief
updates are well-defined under both opener roles without special-casing the
protocol.

\subsubsection{Strategic Cue Generation}

The coarse strategic cue $\tilde c_k$ captures the counterpart's current
bargaining posture while remaining partially informative about its latent
stance type $\eta_B$. We generate $\tilde c_k$ from a hybrid model that combines
(i) a latent-type prior over postures with (ii) adjustments driven by realized
economic action, current concession magnitude, and transformed deadline
proximity.

For terminal-style actions, we use deterministic mappings:
\[
\tilde c_k^{\mathrm{base}}
=
\begin{cases}
\texttt{Concede}, & \text{if } d_k^B=\texttt{Accept},\\
\texttt{Pressure}, & \text{if } d_k^B=\texttt{Reject}.
\end{cases}
\]
Here $d_k^B=\texttt{Reject}$ denotes terminal counterpart walk-away.

For offer actions, define a latent-type bias vector over
$\{\texttt{Concede},\texttt{Hold},\texttt{Pressure}\}$:
\[
b(\eta_B)=
\begin{cases}
(b_C,\,0,\,-b_C), & \eta_B=\texttt{conciliatory},\\
(0,\,b_H,\,0), & \eta_B=\texttt{neutral},\\
(-b_P,\,0,\,b_P), & \eta_B=\texttt{aggressive},
\end{cases}
\]
where the coordinates are ordered as
$(\texttt{Concede},\texttt{Hold},\texttt{Pressure})$. Given an offer action
$d_k^B=\texttt{Offer}$, we define posture logits
\[
\ell_k(\texttt{Concede})
=
b_{\eta_B}^{(\texttt{Concede})}
+\alpha_C(C_k^B-\tau_{\mathrm{conc}}),
\]
\[
\ell_k(\texttt{Hold})
=
b_{\eta_B}^{(\texttt{Hold})},
\]
\[
\ell_k(\texttt{Pressure})
=
b_{\eta_B}^{(\texttt{Pressure})}
+\alpha_P(\widetilde D_k-\tau_{\mathrm{dead}})
-\beta_C C_k^B.
\]
The base strategic cue is then sampled as
\[
\tilde c_k^{\mathrm{base}}
\mid
(d_k^B=\texttt{Offer},\eta_B,C_k^B,\widetilde D_k)
\sim
\mathrm{Categorical}
\left(
\operatorname{softmax}(\ell_k)
\right).
\]
Thus, conciliatory types are biased toward \texttt{Concede}, aggressive types
toward \texttt{Pressure}, and neutral types toward \texttt{Hold}, while realized
concession and deadline proximity modulate these latent tendencies.

\subsubsection{Sentiment Cue Generation}

The affective sentiment cue $\tilde s_k$ is sampled from a three-level latent
score model whose mean depends on the counterpart's stance type
$\eta_B\in\{\texttt{conciliatory},\texttt{neutral},\texttt{aggressive}\}$.
Let
\[
\mu(\eta_B)=
\begin{cases}
+\mu_s, & \eta_B=\texttt{conciliatory},\\
0, & \eta_B=\texttt{neutral},\\
-\mu_s, & \eta_B=\texttt{aggressive},
\end{cases}
\qquad
z_k=\mu(\eta_B)+\epsilon_k,
\qquad
\epsilon_k\sim\mathcal N(0,\sigma_s^2).
\]
We then define
\[
\tilde s_k^{\mathrm{base}}
=
\begin{cases}
\texttt{positive}, & z_k>\tau_s,\\
\texttt{neutral}, & |z_k|\le \tau_s,\\
\texttt{negative}, & z_k<-\tau_s,
\end{cases}
\]
where $\tau_s>0$ is a fixed threshold.

Equivalently, conditional on $\eta_B$, the categorical probabilities are
\[
\Pr(\tilde s_k^{\mathrm{base}}=\texttt{positive}\mid \eta_B)
=
1-\Phi\!\left(\frac{\tau_s-\mu(\eta_B)}{\sigma_s}\right),
\]
\[
\Pr(\tilde s_k^{\mathrm{base}}=\texttt{neutral}\mid \eta_B)
=
\Phi\!\left(\frac{\tau_s-\mu(\eta_B)}{\sigma_s}\right)
-
\Phi\!\left(\frac{-\tau_s-\mu(\eta_B)}{\sigma_s}\right),
\]
\[
\Pr(\tilde s_k^{\mathrm{base}}=\texttt{negative}\mid \eta_B)
=
\Phi\!\left(\frac{-\tau_s-\mu(\eta_B)}{\sigma_s}\right),
\]
where $\Phi$ is the standard normal CDF. Larger $\sigma_s^2$ weakens the
coupling between sentiment and latent stance; smaller $\sigma_s^2$ makes tone
more tightly tied to $\eta_B$.

\subsubsection{Family-Specific Cue Parameterization}
\label{appdx:family_cues}

The constructions above define base sentiment and strategic cues. The six
counterpart behavior families are instantiated through simple family-specific
cue settings applied to these base cues.

For \texttt{Candid} and \texttt{Expressive}, we use the base cue
model directly:
\[
\tilde s_k := \tilde s_k^{\mathrm{base}},
\qquad
\tilde c_k := \tilde c_k^{\mathrm{base}}.
\]

For \texttt{Taciturn} and \texttt{Strategic}, the cue channel is made
uninformative by collapsing to noncommittal middle states:
\[
\tilde s_k := \texttt{neutral},
\qquad
\tilde c_k := \texttt{Hold}.
\]
This removes direct linguistic evidence of latent stance while leaving the
economic channel unchanged.

For \texttt{Stochastic}, cues are weakly informative but noisy. We sample
sentiment from the same latent-score model with elevated noise
$\sigma_{s,\mathrm{stoch}}$ and sample strategic posture using a softened
distribution:
\[
\tilde c_k
\sim
\mathrm{Categorical}
\left(
\operatorname{softmax}\!\left(\frac{\ell_k}{T_{\mathrm{stoch}}}\right)
\right),
\qquad
T_{\mathrm{stoch}}>1.
\]
Unless otherwise stated, we use
\[
\sigma_{s,\mathrm{stoch}}=2.0,
\qquad
T_{\mathrm{stoch}}=2.5.
\]
This preserves a noisy relationship between cues and latent stance, unlike the
collapsed cue channel used by \texttt{Taciturn} and \texttt{Strategic}.

For \texttt{Adversarial}, the cue channel is pressuring:
\[
\tilde s_k := \texttt{negative},
\qquad
\tilde c_k := \texttt{Pressure}.
\]
This produces a systematically hardball linguistic surface. Because the
adversarial family also uses an aggressive-skewed stance prior and hardball
economic preset, these cues are often directionally aligned with pressure, but
they should not be interpreted as calibrated posterior probabilities over
stance.

\subsubsection{Language Realization}

Natural-language realizations are generated only after the economic kernel has
already committed to $(d_k^B,p_k^B,\tilde s_k,\tilde c_k)$. The voice layer
therefore cannot alter economic outcomes. The language model receives a
rendering prompt containing:
\begin{itemize}
    \item the counterpart role, buyer or seller;
    \item the fixed economic action $d_k^B$;
    \item the fixed price $p_k^B$ if $d_k^B=\texttt{Offer}$;
    \item the sentiment cue $\tilde s_k$;
    \item the strategic cue $\tilde c_k$; and
    \item a short summary of the public history $h_{k-1}$.
\end{itemize}
It is instructed to produce a brief message that is semantically consistent
with the fixed action and price, while expressing the specified tone and
posture. In particular:
\begin{itemize}
    \item \texttt{Concede} encourages compromise-oriented phrasing;
    \item \texttt{Hold} encourages firm but non-escalatory phrasing;
    \item \texttt{Pressure} encourages urgency- or deadline-oriented phrasing.
\end{itemize}
The sentiment cue $\tilde s_k$ controls whether that phrasing is
polite/constructive (\texttt{positive}), matter-of-fact (\texttt{neutral}), or
tense (\texttt{negative}).

\paragraph{Default simulator hyperparameters and calibration.}
Unless otherwise stated, we use
\[
\tau_{\mathrm{conc}}=0.10,\qquad
\tau_{\mathrm{dead}}=0.80,\qquad
\mu_s=1.0,\qquad
\tau_s=0.5,\qquad
\sigma_s=0.75,
\]
together with strategic-cue parameters
\[
b_C=1.0,\qquad b_H=0.5,\qquad b_P=1.0,
\]
\[
\alpha_C=2.0,\qquad
\alpha_P=2.0,\qquad
\beta_C=1.0.
\]
These values are chosen so that latent stance and interaction state exert
comparable influence on emitted posture: under the base cue model,
conciliatory types making visibly concessionary offers are more likely to emit
\texttt{Concede}, aggressive types late in the negotiation with low concession
are more likely to emit \texttt{Pressure}, and neutral types under moderate
conditions tend to emit \texttt{Hold}.

\subsection{Default Parameter Summary}\label{appdx:default_param_counter}
For readability, we compile the default hyperparameters 
for counterpart simulator in Table \ref{tab:counterpart_economic_params}-\ref{tab:counterpart_opening_cue_params} and note that regime-specific task generation parameters used in experiments are provided in Table~\ref{tab:regime_params} in Section \ref{sec:implementation_details}. 

We note that the shared counter-offer defaults are deliberately conservative.
Under baseline urgency $\kappa_B=0.5$, no recent agent concession, and before
family-specific reciprocity effects, the shared baseline yields
$\lambda_0+\lambda_1\kappa_B=0.12+0.28(0.5)=0.26$. Thus, absent noise and
additional stance-dependent effects, the counterpart retains approximately
$(1-0.26)^2=0.55$ of its initial distance to reservation after two
counter-offers and $(1-0.26)^3=0.41$ after three. This avoids near-saturation
from overly fast concession while still allowing urgency and family-specific
stance presets to modulate behavior. In particular, stance affects not only
concession speed but also whether agent rigidity and recent concession are
rewarded or exploited.

\begin{table*}[ht!]
\centering
\scriptsize
\setlength{\tabcolsep}{5pt}
\renewcommand{\arraystretch}{1}
\begin{tabular}{p{3.0cm} p{3.0cm} p{2.2cm} p{5.6cm}}
\toprule
\textbf{Component} & \textbf{Parameter} & \textbf{Default} & \textbf{Interpretation} \\
\midrule

Acceptance model
& $\alpha$ & $6.0$
& Sensitivity of acceptance to normalized offer favorability $\bar{\Delta}_k$; larger values make acceptance more responsive to how favorable the current offer is for the counterpart. \\

Acceptance model
& $\beta$ & $1.0$
& Sensitivity of acceptance to counterpart urgency $\kappa_B$; larger values make urgent counterparts more willing to accept. \\

Acceptance model
& $\gamma$ & $2.0$
& Sensitivity to transformed remaining time $\widetilde{\bar D}_k=1-\sqrt{k/K}$; larger values make early-round acceptance less likely while using a concave deadline clock to avoid concentrating deadline pressure only in the final round. \\

Acceptance model
& $\rho_{\mathcal F}(\eta_B)$ & family-specific
& Stance-dependent sensitivity to $\mathrm{ConcedeSpeed}_k$; values are specified by the family economic preset in Table~\ref{tab:economic_presets}. \\

Acceptance model
& $\xi_{\mathcal F}(\eta_B)$ & family-specific
& Stance-dependent sensitivity to $\mathrm{Rigidity}_k$; values are specified by the family economic preset in Table~\ref{tab:economic_presets}. \\

\midrule

Walk-away model
& $k_{\mathrm{walk}}$ & $\lceil K/2\rceil$
& First agent-response round in which terminal counterpart exit is enabled. This gives the agent a grace period to infer infeasibility and reject before the counterpart can preemptively exit. \\

Walk-away model
& $\phi_0$ & $-4.5$
& Intercept of the walk-away hazard. The negative value makes walk-away rare unless the current offer is outside the counterpart's reservation constraint. \\

Walk-away model
& $\phi_\Delta$ & $30.0$
& Sensitivity to normalized reservation shortfall $[-\bar{\Delta}_k]_+$. Larger values make strongly non-individually-rational offers more likely to trigger terminal exit. \\

Walk-away model
& $\phi_T$ & $1.5$
& Sensitivity to the walk-away clock $\tau^W_k$. Larger values make persistent infeasible offers more likely to trigger terminal exit as the round limit approaches. \\

\midrule

Counter-offer model
& $\lambda_0$ & $0.12$
& Baseline latent concession tendency in the unconstrained concession score $\tilde\lambda_B(h_{k-1})$. This lower default avoids overly rapid convergence to the counterpart's reservation value. \\

Counter-offer model
& $\lambda_1$ & $0.28$
& Urgency sensitivity in the latent concession score. At baseline urgency $\kappa_B=0.5$, a neutral counterpart with no agent concession has $\tilde\lambda_B\approx 0.26$. \\

Counter-offer model
& $\lambda_{2,\mathcal F}(\eta_B)$ & family-specific
& Stance-dependent reciprocity sensitivity to the agent's recent role-normalized concession magnitude. Larger values make the counterpart hold firmer against fast-conceding agents. Values are specified by the family economic preset in Table~\ref{tab:economic_presets}. \\

Counter-offer model
& $\lambda_3,\,\lambda_4$ & $0.10,\,0.10$
& Stance adjustments in the concession score. $\lambda_3$ slows aggressive counterparts; $\lambda_4$ accelerates conciliatory counterparts. \\

Counter-offer model
& $\bar{\sigma}_p^{\mathrm{low}},\,\bar{\sigma}_p^{\mathrm{mod}},\,\bar{\sigma}_p^{\mathrm{high}}$ & $0.01,\,0.03,\,0.08$
& Normalized additive price-noise scale, where $\bar{\sigma}_p := \sigma_p/(p_{\max}-p_{\min})$. Actual offer noise is $\sigma_p=\bar{\sigma}_p(p_{\max}-p_{\min})$. \\

Counter-offer model
& clipping of $\lambda_B$ & $[0,1]$
& Effective concession rate is clipped to $[0,1]$ so that the deterministic price update moves weakly toward reservation and never anti-concedes. \\

Counter-offer model
& offer projection & $\mathcal{M}_B(k)$
& Subsequent counter-offers are projected onto the dynamic monotone feasible interval
$\mathcal{M}_B(k)=[r_B,p_{k-1}^B]$ for seller counterparts and
$\mathcal{M}_B(k)=[p_{k-1}^B,r_B]$ for buyer counterparts. This enforces both
counterpart individual rationality and weak monotonicity; if noise would reverse
the concession direction, the counterpart holds at its previous offer. \\

\midrule

History features
& $\tau_{\mathrm{rigid}}$ & $0.10$
& Threshold for the rigidity indicator. If the agent's most recent role-normalized concession is below this level, the agent is treated as rigid. \\

\bottomrule
\end{tabular}
\caption{\footnotesize
Default economic-response and price-dynamics hyperparameters used in
\textsc{Terms-Bench}'s bilateral price-negotiation instantiation. Parameters are specified in normalized state space whenever
possible. The revised counter-offer defaults slow baseline concession relative
to earlier settings, while the walk-away hazard introduces terminal counterpart
exit only for offers outside the counterpart's individually rational region.
}
\label{tab:counterpart_economic_params}
\end{table*}

\begin{table*}[t]
\centering
\scriptsize
\setlength{\tabcolsep}{5pt}
\renewcommand{\arraystretch}{1}
\begin{tabular}{p{3.0cm} p{3.0cm} p{2.2cm} p{5.6cm}}
\toprule
\textbf{Component} & \textbf{Parameter} & \textbf{Default} & \textbf{Interpretation} \\
\midrule

Opening-offer model
& $d_{0,e}$ & $\mathrm{Unif}(0.20,0.80)$
& Episode-level opening harshness. Randomizing $d_{0,e}$ prevents the counterpart's first offer from being nearly invertible into its reservation value. Controlled ablations may use narrower stratum-specific intervals. \\

Opening-offer model
& $\omega_\kappa$ & $0.30$
& Urgency-discount coefficient in the opening-offer modulation factor $\phi(\kappa_B,\eta_B)$; larger values make more urgent counterparts open less aggressively. \\

Opening-offer model
& $\omega_\eta,\,\omega_\eta'$ & $0.15,\,0.15$
& Stance modulation coefficients in $\phi(\kappa_B,\eta_B)$. $\omega_\eta$ makes aggressive counterparts claim more favorable slack; $\omega_\eta'$ makes conciliatory counterparts open closer to reservation. \\

Opening-offer model
& $\phi_{\min},\,\phi_{\max}$ & $0.5,\,1.5$
& Clipping range for the opening-offer modulation factor. Ensures the type/urgency adjustment remains positive and bounded. \\

Opening-offer model
& $\bar{\sigma}_0$ & $0.02$
& Normalized opening-offer noise scale, where $\sigma_0=\bar{\sigma}_0(p_{\max}-p_{\min})$. \\

\midrule

Strategic cue model
& $\tau_{\mathrm{dead}}$ & $0.80$
& Threshold applied to transformed deadline proximity $\widetilde D_k=\sqrt{k/K}$ in the strategic-cue logits. \\

Strategic cue model
& $\alpha_P$ & $2.0$
& Sensitivity of the \texttt{Pressure} logit to transformed deadline proximity $\widetilde D_k$. \\

Strategic cue model
& $b_C,\,b_H,\,b_P$ & $1.0,\,0.5,\,1.0$
& Baseline latent-type bias toward \texttt{Concede}, \texttt{Hold}, and \texttt{Pressure}, respectively. These determine how strongly $\eta_B$ influences the strategic cue absent strong state signals. \\

Strategic cue model
& $\alpha_C,\,\alpha_P,\,\beta_C$ & $2.0,\,2.0,\,1.0$
& State-sensitivity coefficients. $\alpha_C$ increases the \texttt{Concede} logit after realized counterpart concession; $\alpha_P$ increases \texttt{Pressure} near the deadline; $\beta_C$ suppresses \texttt{Pressure} after visible concession. \\

\midrule

Sentiment cue model
& $\mu_s,\,\tau_s,\,\sigma_s$ & $1.0,\,0.5,\,0.75$
& Sentiment-generation parameters. $\mu_s$ controls separation between conciliatory and aggressive sentiment means, $\tau_s$ discretizes latent sentiment, and $\sigma_s$ controls baseline sentiment noise. \\

\midrule

Family-specific cue settings
& Candid / Expressive
& base cue model
& Accurate cue channel: sentiment and strategic posture are sampled from the base cue model and remain informative about latent stance. \\

Family-specific cue settings
& Taciturn / Strategic
& $\tilde s_k=\texttt{neutral},\ \tilde c_k=\texttt{Hold}$
& Uninformative cue channel: sentiment and strategic posture are collapsed to noncommittal middle states. \\

Family-specific cue settings
& $\sigma_{s,\mathrm{stoch}},\,T_{\mathrm{stoch}}$
& $2.0,\,2.5$
& Noisy cue channel for the \texttt{Stochastic} family: elevated sentiment noise and strategic-cue temperature produce weak but non-collapsed cues. \\

Family-specific cue settings
& Adversarial
& $\tilde s_k=\texttt{negative},\ \tilde c_k=\texttt{Pressure}$
& Pressuring cue channel: adversarial counterparts use systematically aggressive sentiment and pressure-oriented posture. \\
\bottomrule
\end{tabular}
\caption{\footnotesize
Default opening-offer and cue-generation hyperparameters used in
\textsc{Terms-Bench}'s bilateral price-negotiation. Opening-offer parameters determine the counterpart's initial
anchor, while cue-generation parameters control the informativeness and tone of
the language-facing signal channel.
}
\label{tab:counterpart_opening_cue_params}
\end{table*}

%% file: appendix/reference_oracle.tex
\section{Oracle-Cue Bayes-Optimal Reference Policy}\label{app:optimal_policy}

Because the counterpart policy $\pi_B$ and environment prior $\mu$ are fully specified and fixed, \textsc{Terms-Bench} admits computation of a Bayes-optimal benchmark policy $\pi^*$ via backward induction. The negotiation reduces to a partially observable Markov decision process in which the hidden state comprises the counterpart type $t_B = (r_B, \kappa_B, \eta_B)$ together with an episode-level opening-harshness nuisance variable $d_{0,e}$ (Appendix~\ref{app:opening_likelihood}). We solve this decision process on a discretized belief space to obtain the reference policy and the associated optimality gap.

In the benchmark's observation model, the agent observes $(p^B_k, m^B_k)$: the counterpart's price and a natural-language message. The sentiment cue $\tilde{s}_k$ and strategic cue $\tilde{c}_k$ are \emph{latent} variables embedded in $m^B_k$ and must be inferred. To define a computable upper bound, we derive the reference policy under an oracle-cue assumption: the agent directly observes the counterpart's economic action $d^B_k \in \{\texttt{Offer}, \texttt{Accept}, \texttt{Reject}\}$, the sentiment cue $\tilde{s}_k$, and the strategic cue $\tilde{c}_k$, together with a counter-offer price $p^B_k$ when $d^B_k = \texttt{Offer}$. The three-way economic action set reflects the three-branch counterpart response model of Section~\ref{sec:counterpart_model}, in which the counterpart may terminate negotiation through a walk-away action in addition to accepting or counter-offering. Any real LLM agent operates under the true observation model with imperfect cue extraction, so the oracle policy provides an upper bound on achievable performance. Throughout this appendix, we specialize to the linear-surplus case $g_b(x) = g_s(x) = x$, so that $u_{\mathrm{buyer}}(p) = r_A - p$ and $u_{\mathrm{seller}}(p) = p - r_A$.

The oracle additionally conditions on the opener role $\chi \in \{\texttt{AgentOpens}, \texttt{CounterpartOpens}\}$, which is sampled once at the start of each episode and is common knowledge. The backward induction is executed over both initial conditions: under $\chi = \texttt{CounterpartOpens}$, the round-1 state already contains a counterpart opening offer $p^B_1$ and the agent's action set at round~1 is $\{\texttt{Accept}, \texttt{Reject}\} \cup \{\texttt{Offer}(p) : p \in \mathcal{P}\}$; under $\chi = \texttt{AgentOpens}$, the round-1 action set is $\{\texttt{Offer}(p) : p \in \mathcal{P}\}$ alone because no counterpart offer has yet been observed.

\paragraph{Notational convention.}
Throughout this appendix, $a_k(p^A_k, t_B, \psi_k)$ (with arguments shown) denotes the counterpart's \emph{acceptance probability} at round~$k$ conditional on the agent's offer $p^A_k$, while the plain symbol $a_k$ (without arguments, only in contexts like the belief update below) denotes the agent's round-$k$ action. Where the two could be confused within a single equation, we write the acceptance probability with its full argument list.

\subsection{Belief Discretization}\label{app:belief_disc}

We discretize the type space $\mathcal{T}_B$ into a finite grid:
\begin{itemize}[nosep]
    \item Reservation: $\mathcal{R} = \{r_{\min},\, r_{\min} + \delta_r,\, \ldots,\, r_{\max}\}$ with spacing $\delta_r = 50$;
    \item Urgency: $\mathcal{K} = \{0.1,\, 0.3,\, 0.5,\, 0.7,\, 0.9\}$ (5 levels);
    \item Stance: $\mathcal{H} = \{\texttt{C}, \texttt{N}, \texttt{A}\}$ for conciliatory, neutral, and aggressive (3 categories).
\end{itemize}
This yields $N = |\mathcal{R}| \times 5 \times 3$ discrete types, where $|\mathcal{R}| = (r_{\max} - r_{\min})/\delta_r$. The belief state at round~$k$ is a probability vector $b_k \in \Delta(\mathcal{R} \times \mathcal{K} \times \mathcal{H})$, initialized from the environment prior~$\mu$.

\paragraph{Family-specific initial belief.}
For all families except \textsc{Adversarial}, the stance marginal of $\mu$ is uniform on $\mathcal{H}$ and the initial belief $b_0$ inherits this uniform marginal. For \textsc{Adversarial}, the stance marginal is skewed toward aggressive counterparts,
\begin{equation}\label{eq:adv_prior}
    \Pr(\eta_B = \texttt{C}) = 0.05,\quad \Pr(\eta_B = \texttt{N}) = 0.15,\quad \Pr(\eta_B = \texttt{A}) = 0.80,
\end{equation}
and $b_0$ is set accordingly; the reservation and urgency marginals remain as in other families.

\paragraph{Opening harshness as a nuisance variable.}
Opening harshness $d_{0,e} \sim \mathrm{Uniform}(d_{\min}, d_{\max})$ (defaults $d_{\min}=0.20$, $d_{\max}=0.80$) is realized once per episode and influences only the counterpart's opening offer under $\chi = \texttt{CounterpartOpens}$, or the counterpart's round-1 response under $\chi = \texttt{AgentOpens}$ when that response is $\texttt{Offer}$. Because $d_{0,e}$ does not enter any subsequent-round dynamics, we do not expand the belief over $\mathcal{T}_B$ to include it; instead, we marginalize $d_{0,e}$ analytically in the round-1 likelihood (Appendix~\ref{app:opening_likelihood}). An alternative implementation expands $\mathcal{T}_B$ by a coarse $d_0$-grid (for example, five levels), increasing $N$ by a factor of five; we adopt the marginalization approach because $d_{0,e}$ is information-free after round~1 and further posterior refinement over it has no downstream value.

\subsection{Augmented Information State}\label{app:aug_state}

The belief $b_k$ alone is not a sufficient statistic for the agent's decision problem. The counterpart model depends on additional known quantities:
\begin{enumerate}[nosep]
    \item \textbf{Opener role} $\chi \in \{\texttt{AgentOpens}, \texttt{CounterpartOpens}\}$, assigned at episode start and constant across rounds.
    \item \textbf{History-reactive features} $\phi_k := (\mathrm{ConcedeSpeed}_k,\, \mathrm{Rigidity}_k,\, \mathrm{ConcedeMagnitude}_k)$, which are deterministic functions of the agent's past offer sequence (Appendix~\ref{sec:history_features}) and parameterize the counterpart's acceptance probability, walk-away hazard, and concession rate.
    \item \textbf{Offer-history summary} $h^B_k := (p^B_k,\, p^B_{k-1})$, which records the counterpart's current and previous offers. These are needed to evaluate acceptance utility $u_A(p^B_k)$, to compute the price-likelihood mean~\eqref{eq:price_mean}, to derive the counterpart's concession magnitude $C^B_k$ used in the strategic-cue model, and to specify the role-dependent monotone feasible interval $\mathcal{M}_B(k)$ onto which counter-offers are projected.
\end{enumerate}
Additionally, the monotonic concession constraint and the update of $\phi_{k+1}$ require the agent's recent own-offer history. We store this as $\xi^A_k := (p^A_{k-3},\, p^A_{k-2},\, p^A_{k-1})$, with entries set to $\varnothing$ before they exist. All components besides $b_k$ are deterministic given the public history. We define the augmented information state
\begin{equation}\label{eq:aug_state}
    \psi_k \;:=\; (b_k,\, \phi_k,\, \xi^A_k,\, h^B_k,\, \chi).
\end{equation}
All value functions, $Q$-functions, likelihoods, and state transitions below are defined over $\psi_k$.

\paragraph{Boundary values under mixed opener roles.}
History features and the offer-history summary adopt boundary values whenever the requisite offers have not yet been realized. If the agent has made fewer than two own offers, then $\mathrm{ConcedeSpeed}_k = \mathrm{Rigidity}_k = \mathrm{ConcedeMagnitude}_k = 0$. If the counterpart has produced fewer than two offers, the entries of $h^B_k$ that do not yet exist are set to $\varnothing$ and $C^B_k = 0$. Under $\chi = \texttt{AgentOpens}$, the counterpart has made no offer prior to the agent's round-1 action, so these boundary values additionally extend into round~2 whenever the counterpart responded at round~1 with $\texttt{Offer}$ (yielding $p^B_1$) but no earlier counterpart offer exists. This is a natural extension of the round-1 boundary handling and does not require special-casing in the DP.

\subsection{Observation Likelihood}\label{app:obs_likelihood}

Under the oracle-cue assumption, the agent observes the counterpart's economic action $d^B_k \in \{\texttt{Offer}, \texttt{Accept}, \texttt{Reject}\}$, the sentiment cue $\tilde{s}_k$, and the strategic cue $\tilde{c}_k$, together with a counter-offer price $p^B_k$ when $d^B_k = \texttt{Offer}$. The observation likelihood decomposes according to the counterpart's sequential generative process: the economic action is drawn first; if $d^B_k = \texttt{Offer}$, the price is drawn next; then the sentiment cue is drawn independently from the stance type; and finally the strategic cue is drawn conditional on the realized economic action and price.

\subsubsection{Counterpart Action Distribution}\label{app:action_dist}

Given the agent's most recent offer $p_k^A$, the counterpart first decides whether to accept. The acceptance probability is (Eq.~\ref{eq:accept_probabilities}--\ref{eq:accept_score}):
\begin{equation}\label{eq:oracle_accept}
a_k(p_k^A, t_B, \psi_k)
\;:=\; \mathbf{1}\{\bar\Delta_k \geq 0\}\,
\sigma\!\left(g_\theta(p_k^A, t_B, k, \phi_k)\right),
\end{equation}
where $\sigma(\cdot)$ is the logistic function and
\begin{equation}\label{eq:oracle_score}
g_\theta
\;=\; \alpha\,\bar\Delta_k
\;+\; \beta\,\kappa_B
\;-\; \gamma\,\bar{\tilde{D}}_k
\;+\; \rho_F(\eta_B)\,\mathrm{ConcedeSpeed}_k
\;+\; \xi_F(\eta_B)\,\mathrm{Rigidity}_k,
\end{equation}
with the normalized quantities
\[
\bar\Delta_k :=
\begin{cases}
\dfrac{p_k^A - r_B}{p_{\max}-p_{\min}}, & \text{seller counterpart},\\[1.0ex]
\dfrac{r_B - p_k^A}{p_{\max}-p_{\min}}, & \text{buyer counterpart},
\end{cases}
\qquad
\bar{\tilde{D}}_k \;:=\; 1 - \sqrt{k/K}.
\]
The concave remaining-time term $\bar{\tilde{D}}_k$ replaces the previous linear clock; the default deadline weight is correspondingly raised to $\gamma = 2.0$ to partially compensate for the concavity. The coefficients $\rho_F(\eta_B)$ and $\xi_F(\eta_B)$ are \emph{stance- and family-dependent} (Appendix~\ref{appdx:economic_presets}), so the payoff consequences of conceding quickly and of holding firm differ across the three stance types within a fixed family.

Conditional on non-acceptance, the counterpart either terminates through a walk-away or produces a counter-offer. The walk-away hazard is
\begin{equation}\label{eq:walk_hazard}
\omega_k(p_k^A, t_B, \psi_k)
\;:=\; \mathbf{1}\{k \geq k_{\mathrm{walk}}\}\,\mathbf{1}\{\bar\Delta_k < 0\}\,
\sigma\!\left(\phi_0 + \phi_\Delta\,[-\bar\Delta_k]_+ + \phi_T\,\tau^W_k\right),
\end{equation}
where $\tau^W_k = \min\{1,\, \max\{0,\, (k - k_{\mathrm{walk}})/(K - k_{\mathrm{walk}})\}\}$ and the defaults are $k_{\mathrm{walk}} = \lceil K/2 \rceil$, $\phi_0 = -4.5$, $\phi_\Delta = 30.0$, $\phi_T = 1.5$. The hazard is nonzero only after a grace period and only when the current offer is outside the counterpart's individually rational region.

For $k < K$, and writing $a_k \equiv a_k(p^A_k, t_B, \psi_k)$ and $\omega_k \equiv \omega_k(p^A_k, t_B, \psi_k)$ for brevity, the counterpart action probabilities are
\begin{align}
P(d^B_k = \texttt{Accept} \mid t_B, \psi_k, p^A_k) &= a_k, \label{eq:p_accept}\\
P(d^B_k = \texttt{Reject} \mid t_B, \psi_k, p^A_k) &= (1 - a_k)\,\omega_k, \label{eq:p_reject}\\
P(d^B_k = \texttt{Offer}  \mid t_B, \psi_k, p^A_k) &= (1 - a_k)(1 - \omega_k). \label{eq:p_offer_distro}
\end{align}
At $k = K$, any remaining non-acceptance and non-walk-away mass results in round-limit disagreement.

\paragraph{Hard support constraint from \texttt{Reject} observations.}
Because $\omega_k$ carries the indicator $\mathbf{1}\{\bar\Delta_k < 0\}$, observing $d^B_k = \texttt{Reject}$ at any round $k \geq k_{\mathrm{walk}}$ is a \emph{hard constraint} on $t_B$: it can only occur for types whose reservation value $r_B$ makes the agent's current offer non-individually-rational. The Bayesian update following a \texttt{Reject} observation therefore assigns posterior mass zero to every type with $\bar\Delta_k \geq 0$, eliminating a contiguous region of the belief support.

\subsubsection{Counter-Offer Price Likelihood}\label{app:price_likelihood}

If $d^B_k = \texttt{Offer}$, the counterpart's price is generated by Eq.~\eqref{eq:counterpart_concession_score}-\eqref{eq:counter_project}:
\begin{equation}\label{eq:counter_offer_gen}
p^B_k \;=\; \Pi_{\mathcal{M}_B(k)}\!\left[\, p^B_{k-1} - \lambda_B(\phi_k;\, t_B)\cdot(p^B_{k-1} - r_B) + \varepsilon_k \,\right],
\quad \varepsilon_k \sim \mathcal{N}(0, \sigma_p^2),
\end{equation}
where the effective concession rate is
\begin{equation}
\lambda_B(\phi_k;\, t_B) \;=\; \mathrm{clip}_{[0,1]}\!\Big[\lambda_0 + \lambda_1\,\kappa_B - \lambda_{2,F}(\eta_B)\,\mathrm{ConcedeMagnitude}_k - \lambda_3\,\mathbf{1}\{\eta_B = \texttt{A}\} + \lambda_4\,\mathbf{1}\{\eta_B = \texttt{C}\}\Big],
\end{equation}
with revised defaults $\lambda_0 = 0.12$, $\lambda_1 = 0.28$, and stance-dependent $\lambda_{2,F}(\eta_B)$ per the family preset. The projection is onto the role-dependent monotone feasible interval
\begin{equation}\label{eq:M_B}
\mathcal{M}_B(k) \;:=\; \begin{cases} [r_B,\, p^B_{k-1}], & \text{seller counterpart},\\[2pt] [p^B_{k-1},\, r_B], & \text{buyer counterpart}, \end{cases}
\end{equation}
rather than onto the public bounds $[p_{\min}, p_{\max}]$. Define the deterministic pre-projection mean
\begin{equation}\label{eq:price_mean}
\bar{p}^B_k(t_B, \psi_k) \;:=\; p^B_{k-1} - \lambda_B(\phi_k;\, t_B)\cdot(p^B_{k-1} - r_B),
\end{equation}
so that the unprojected offer is $\tilde{p}^B_k \sim \mathcal{N}\!\big(\bar{p}^B_k,\, \sigma_p^2\big)$ and the observed price $p^B_k = \Pi_{\mathcal{M}_B(k)}(\tilde{p}^B_k)$. Writing $\mathcal{M}_B(k) = [a,b]$ with
\[
a \;=\; \begin{cases} r_B, & \text{seller},\\ p^B_{k-1}, & \text{buyer}, \end{cases}
\qquad
b \;=\; \begin{cases} p^B_{k-1}, & \text{seller},\\ r_B, & \text{buyer}, \end{cases}
\]
the price likelihood is a mixed distribution with point masses at $a$ and $b$:
\begin{equation}\label{eq:price_likelihood}
    f_{\mathrm{price}}(p^B_k \mid t_B, \psi_k) \;=\;
    \begin{cases}
        \Phi\!\!\left(\dfrac{a - \bar{p}^B_k}{\sigma_p}\right), & p^B_k = a, \\[10pt]
        \dfrac{1}{\sigma_p}\,\phi\!\!\left(\dfrac{p^B_k - \bar{p}^B_k}{\sigma_p}\right), & a < p^B_k < b, \\[10pt]
        1 - \Phi\!\!\left(\dfrac{b - \bar{p}^B_k}{\sigma_p}\right), & p^B_k = b,
    \end{cases}
\end{equation}
where $\phi$ and $\Phi$ denote the standard normal pdf and cdf, respectively.

\paragraph{Informativeness of the reservation-endpoint point mass.}
The lower endpoint $a$ of $\mathcal{M}_B(k)$ is $r_B$ for a seller counterpart; symmetrically, the upper endpoint $b$ is $r_B$ for a buyer counterpart. Observing $p^B_k$ pinned at the reservation endpoint is therefore highly informative about the hidden reservation value: it indicates that the Gaussian draw was clamped against the counterpart's own reservation constraint, revealing that the counterpart has conceded to the boundary of its feasible region. The other endpoint of $\mathcal{M}_B(k)$ is the previous counterpart offer $p^B_{k-1}$, which is already known from $h^B_k$; a draw pinned there signals that noise attempted to reverse the concession direction and carries information about $\kappa_B$ and $\eta_B$ through $\lambda_B(\phi_k; t_B)$ but not directly about $r_B$. The oracle's posterior over $r_B$ should therefore tighten substantially whenever a counter-offer is pinned at the reservation endpoint, particularly in later rounds when multiple such pinnings accumulate.

\subsubsection{Sentiment Cue Likelihood}\label{app:sent_likelihood}

The sentiment cue $\tilde{s}_k$ depends only on the stance type $\eta_B$ and is conditionally independent of the realized price. From the latent-score model in Appendix~\ref{appdx:cue_generation}:
\begin{align}\label{eq:sent_likelihood}
    P(\tilde{s}_k = \texttt{pos} \mid \eta_B) &= 1 - \Phi\!\!\left(\frac{\tau_s - \mu(\eta_B)}{\sigma_s}\right), \notag\\
    P(\tilde{s}_k = \texttt{neu} \mid \eta_B) &= \Phi\!\!\left(\frac{\tau_s - \mu(\eta_B)}{\sigma_s}\right) - \Phi\!\!\left(\frac{-\tau_s - \mu(\eta_B)}{\sigma_s}\right), \\
    P(\tilde{s}_k = \texttt{neg} \mid \eta_B) &= \Phi\!\!\left(\frac{-\tau_s - \mu(\eta_B)}{\sigma_s}\right), \notag
\end{align}
where $\mu(\texttt{C}) = +\mu_s$, $\mu(\texttt{N}) = 0$, $\mu(\texttt{A}) = -\mu_s$.

\subsubsection{Strategic Cue Likelihood}\label{app:strat_likelihood}

For terminal counterpart actions, the mapping is deterministic:
\begin{equation}\label{eq:strat_terminal}
    \tilde{c}_k \;=\; \begin{cases} \texttt{Concede}, & d^B_k = \texttt{Accept}, \\ \texttt{Pressure}, & d^B_k = \texttt{Reject}. \end{cases}
\end{equation}
For offer actions, the cue is sampled from $\mathrm{Categorical}(\mathrm{softmax}(\ell_k))$ with the concave deadline clock $\tilde{D}_k := \sqrt{k/K}$:
\begin{align}\label{eq:strat_logits}
    \ell_k(\texttt{Concede})  &= b^{(\texttt{C})}_{\eta_B} + \alpha_C\,(C^B_k - \tau_{\mathrm{conc}}), \notag\\
    \ell_k(\texttt{Hold})     &= b^{(\texttt{H})}_{\eta_B}, \\
    \ell_k(\texttt{Pressure}) &= b^{(\texttt{P})}_{\eta_B} + \alpha_P\,(\tilde{D}_k - \tau_{\mathrm{dead}}) - \beta_C\,C^B_k, \notag
\end{align}
where $b(\eta_B)$ are the stance-dependent bias vectors from Appendix~\ref{appdx:cue_generation} and
\begin{equation}
    C^B_k \;=\; \min\!\left(1,\; \frac{|p^B_k - p^B_{k-1}|}{|p^B_{k-1} - r_B| + \epsilon_c}\right)
\end{equation}
is the realized concession magnitude, computed from the offer pair $(p^B_k, p^B_{k-1}) \in h^B_k$ when both offers exist; otherwise $C^B_k = 0$ by the boundary convention of Appendix~\ref{app:aug_state}. We define the strategic-cue likelihood as a function of the stance type, deadline clock, and concession magnitude:
\begin{equation}\label{eq:strat_likelihood}
    P_{\mathrm{strat}}(\tilde{c}_k = c \mid \eta_B, \tilde{D}_k, C^B_k) \;:=\; \frac{\exp\!\big(\ell_k(c;\, \eta_B, \tilde{D}_k, C^B_k)\big)}{\sum_{c'}\exp\!\big(\ell_k(c';\, \eta_B, \tilde{D}_k, C^B_k)\big)}.
\end{equation}
In the generic round-$k$ case (rounds $k \geq 2$ with two counterpart offers available), $C^B_k$ is computed from $(p^B_k, p^B_{k-1}) \in h^B_k$, so we also write $P_{\mathrm{strat}}(\tilde c_k \mid t_B, h^B_k) \equiv P_{\mathrm{strat}}(\tilde c_k \mid \eta_B, \tilde D_k, C^B_k(h^B_k))$ as convenient shorthand. Family-specific overrides apply: \textsc{Stochastic} replaces $\sigma_s$ by $\sigma_{s,\mathrm{stoch}} = 2.0$ in~\eqref{eq:sent_likelihood} and uses softmax temperature $T_{\mathrm{stoch}} = 2.5$ in~\eqref{eq:strat_likelihood}; \textsc{Taciturn} and \textsc{Strategic} collapse cues to $\tilde{s}_k = \texttt{neu}$, $\tilde{c}_k = \texttt{Hold}$; \textsc{Adversarial} collapses cues to $\tilde{s}_k = \texttt{neg}$, $\tilde{c}_k = \texttt{Pressure}$. The remaining families use the base cue model above.

\subsubsection{Full Observation Likelihood}\label{app:full_likelihood}

The generative process is sequential: draw the economic action, then (if \texttt{Offer}) draw the price, then draw sentiment (independently), then draw the strategic cue (deterministic for \texttt{Accept}/\texttt{Reject}; conditional on the realized price for \texttt{Offer}). Using the shorthand $a_k \equiv a_k(p^A_k, t_B, \psi_k)$ and $\omega_k \equiv \omega_k(p^A_k, t_B, \psi_k)$ for the counterpart acceptance probability and walk-away hazard, the full likelihood for each observation type is:

\paragraph{Counter-offer observation ($d^B_k = \texttt{Offer}$):}
\begin{equation}\label{eq:full_likelihood_offer}
    \boxed{\;P(o_k \mid t_B, \psi_k, p^A_k) \;=\; \underbrace{(1 - a_k)(1 - \omega_k)}_{\text{offer branch}} \;\cdot\; \underbrace{f_{\mathrm{price}}(p^B_k \mid t_B, \psi_k)}_{\text{price}} \;\cdot\; \underbrace{P(\tilde{s}_k \mid \eta_B)}_{\text{sentiment}} \;\cdot\; \underbrace{P_{\mathrm{strat}}(\tilde c_k \mid \eta_B, \tilde D_k, C^B_k)}_{\text{strategic cue}}\;.}
\end{equation}
\paragraph{Acceptance observation ($d^B_k = \texttt{Accept}$):}
\begin{equation}\label{eq:full_likelihood_accept}
    P(o_k \mid t_B, \psi_k, p^A_k) \;=\; a_k \;\cdot\; P(\tilde{s}_k \mid \eta_B) \;\cdot\; \mathbf{1}\{\tilde{c}_k = \texttt{Concede}\}.
\end{equation}
\paragraph{Walk-away observation ($d^B_k = \texttt{Reject}$):}
\begin{equation}\label{eq:full_likelihood_reject}
    P(o_k \mid t_B, \psi_k, p^A_k) \;=\; (1 - a_k)\,\omega_k \;\cdot\; P(\tilde{s}_k \mid \eta_B) \;\cdot\; \mathbf{1}\{\tilde{c}_k = \texttt{Pressure}\}.
\end{equation}
Because $\omega_k$ contains $\mathbf{1}\{k \geq k_{\mathrm{walk}}\}\mathbf{1}\{\bar\Delta_k < 0\}$, \eqref{eq:full_likelihood_reject} is zero for every type with $\bar\Delta_k \geq 0$, inducing the hard support constraint of Section~\ref{app:action_dist}.

\subsubsection{Opening-Round Likelihood}\label{app:opening_likelihood}

The round-1 observation depends on the opener role. In both cases the counterpart's emissions at round~1 include cues, so the round-1 likelihood carries the same sentiment and strategic-cue factors as in later rounds; only the \emph{price} factor is replaced by the opening-offer model when the counterpart produces its first offer, and the hidden opening harshness $d_{0,e}$ is analytically marginalized. Because the round-1 strategic cue uses the boundary value $C^B_1 = 0$ (no previous counterpart offer), it depends only on $\eta_B$ and the deadline clock $\tilde{D}_1 = \sqrt{1/K}$, and we write it in the explicit form $P_{\mathrm{strat}}(\tilde c_1 \mid \eta_B, \tilde D_1, C^B_1{=}0)$ to emphasize that $p^B_1$ itself does not enter the cue logit.

\paragraph{Opening-offer price model.}
When the counterpart produces its first offer $p^B_1$ (either as an opening under $\chi = \texttt{CounterpartOpens}$, or as a response to the agent's opening under $\chi = \texttt{AgentOpens}$), the price is generated by the opening-offer model of Appendix~\ref{appdx:opening-offer}:
\begin{equation}\label{eq:opening_model}
    p^B_1 \;=\; \Pi_{\mathcal{B}_B}\!\left[r_B + s_B\,d_{0,e}\,\varphi(\kappa_B, \eta_B)\,\Delta^{\mathrm{slack}}_B + \varepsilon_0\right],\quad \varepsilon_0 \sim \mathcal{N}(0,\,\sigma_0^2),
\end{equation}
where $s_B = +1$ for a seller counterpart, $s_B = -1$ for a buyer; $\mathcal{B}_B = [r_B, p_{\max}]$ for a seller and $[p_{\min}, r_B]$ for a buyer; $\Delta^{\mathrm{slack}}_B$ is the counterpart's directional slack; and $\varphi(\kappa_B, \eta_B)$ is the modulation factor. Conditional on $(t_B, d_{0,e})$, the unprojected opening is Gaussian with mean $\mu_0(t_B, d) := r_B + s_B\,d\,\varphi(\kappa_B, \eta_B)\,\Delta^{\mathrm{slack}}_B$ and variance $\sigma_0^2$. Projection onto $\mathcal{B}_B = [a_0, b_0]$ produces a mixed distribution with point masses at the two endpoints:
\begin{equation}\label{eq:f_open}
f_{\mathrm{open}}(p^B_1 \mid t_B, d_{0,e}) \;=\;
\begin{cases}
\Phi\!\left(\dfrac{a_0 - \mu_0(t_B, d_{0,e})}{\sigma_0}\right), & p^B_1 = a_0, \\[8pt]
\dfrac{1}{\sigma_0}\,\phi\!\left(\dfrac{p^B_1 - \mu_0(t_B, d_{0,e})}{\sigma_0}\right), & a_0 < p^B_1 < b_0, \\[8pt]
1 - \Phi\!\left(\dfrac{b_0 - \mu_0(t_B, d_{0,e})}{\sigma_0}\right), & p^B_1 = b_0,
\end{cases}
\end{equation}
with $(a_0, b_0) = (r_B, p_{\max})$ for a seller counterpart and $(p_{\min}, r_B)$ for a buyer. In both cases one of the point masses sits at $r_B$, so the opening likelihood is reservation-informative whenever the noise clamps the raw opening against the counterpart's reservation constraint. Because $d_{0,e}$ is hidden, the oracle uses the marginal likelihood
\begin{equation}\label{eq:open_marginalized}
P_{\mathrm{open}}(p^B_1 \mid t_B) \;:=\; \frac{1}{d_{\max} - d_{\min}}\int_{d_{\min}}^{d_{\max}} f_{\mathrm{open}}(p^B_1 \mid t_B, d)\, \mathrm{d}d,
\end{equation}
computed numerically on a fixed quadrature grid $\{d^{(1)},\ldots,d^{(L)}\}$ (we use $L = 9$ composite-Simpson nodes).

\paragraph{\texttt{CounterpartOpens}.}
The counterpart opens with $p^B_1$ together with cues $(\tilde s_1, \tilde c_1)$, before the agent has acted. Because the sentiment and strategic-cue likelihoods are conditionally independent of $d_{0,e}$ given $(t_B, p^B_1)$, the full round-1 likelihood factorizes as
\begin{equation}\label{eq:open_full_likelihood}
\boxed{\;P(o_1 \mid t_B,\,\chi{=}\texttt{CounterpartOpens}) \;=\; \underbrace{P_{\mathrm{open}}(p^B_1 \mid t_B)}_{\text{opening price, }d_{0,e}\text{-marg.}}
  \;\cdot\; \underbrace{P(\tilde s_1 \mid \eta_B)}_{\text{sentiment}}
  \;\cdot\; \underbrace{P_{\mathrm{strat}}(\tilde c_1 \mid \eta_B,\, \tilde D_1,\, C^B_1{=}0)}_{\text{strategic cue, boundary}}\;.}
\end{equation}
Family-specific cue overrides from Appendix~\ref{appdx:cue_generation} apply. Because $d_{0,e}$ does not enter any subsequent-round dynamics, no posterior over it need be propagated; the round-1 belief update~\eqref{eq:belief_update} uses~\eqref{eq:open_full_likelihood} directly.

\paragraph{\texttt{AgentOpens}.}
No counterpart observation precedes the agent's round-1 action, so $b_1 = b_0$ and the agent chooses $a_1$ under the prior. The counterpart's response is emitted \emph{within} round~1 (conditional on the agent's round-1 offer $p^A_1$, with history-reactive features at their zero boundary values) and is consumed by the $b_2$ update; thus $p^B_1$ under $\chi = \texttt{AgentOpens}$ denotes the counterpart's first and only round-1 offer, occurring after the agent's action rather than before it. The response follows the three-branch distribution~\eqref{eq:p_accept}--\eqref{eq:p_offer_distro}, and the corresponding observation likelihoods are~\eqref{eq:full_likelihood_accept} for \texttt{Accept}, \eqref{eq:full_likelihood_reject} for walk-away, and the following three-factor analogue of~\eqref{eq:full_likelihood_offer} for \texttt{Offer}, with the opening-offer price model replacing the counter-offer price model and $p^A_1$ conditioning the acceptance and walk-away probabilities:

\begin{equation}\label{eq:agent_opens_offer_likelihood}
\begin{split}
P(o_1 \mid t_B, \psi_1, p_1^A)
&=
\bigl[1-a_1(p_1^A,t_B,\psi_1)\bigr]
\bigl[1-\omega_1(p_1^A,t_B,\psi_1)\bigr] \\
&\quad \cdot P_{\mathrm{open}}(p_1^B \mid t_B)\,
P(\tilde{s}_1 \mid \eta_B)\,
P_{\mathrm{strat}}(\tilde{c}_1 \mid \eta_B, \tilde{D}_1, C_1^B = 0).
\end{split}
\tag{55}
\end{equation}
Here we condition on $p^A_1$ rather than on the full agent action $a_1$ to avoid the notational clash with the counterpart acceptance probability $a_1(\cdot)$; since $d_1 = \texttt{Offer}$ is forced under $\chi = \texttt{AgentOpens}$ at round~1, $p^A_1$ fully identifies the agent's action. Both $a_1(p^A_1, t_B, \psi_1)$ and $\omega_1(p^A_1, t_B, \psi_1)$ are evaluated with the zero-boundary agent history in $\psi_1$.

\subsection{Belief Update}\label{app:belief_update}

Given prior belief $b_k$ and observation $o_{k+1}$ after the agent takes action $a_k$ with offer price $p^A_k$ (when $d_k = \texttt{Offer}$), the posterior follows from Bayes' rule:
\begin{equation}\label{eq:belief_update}
    b_{k+1}(t_B) \;=\; \frac{b_k(t_B)\;\cdot\; P(o_{k+1} \mid t_B, \psi_k, p^A_k)}{\displaystyle\sum_{t'_B \in \mathcal{T}_B} b_k(t'_B)\;\cdot\; P(o_{k+1} \mid t'_B, \psi_k, p^A_k)},
\end{equation}
where the observation likelihood is~\eqref{eq:full_likelihood_offer}, \eqref{eq:full_likelihood_accept}, or~\eqref{eq:full_likelihood_reject} according to the counterpart's action in rounds $k \geq 2$. The round-1 case is handled separately:
\begin{itemize}[nosep]
    \item Under $\chi = \texttt{CounterpartOpens}$, $b_1 = \mathrm{BayesUpdate}(b_0, o_1)$ uses the full opening likelihood~\eqref{eq:open_full_likelihood} (no agent action is required since the counterpart emits first). The agent then chooses its round-1 action from $b_1$.
    \item Under $\chi = \texttt{AgentOpens}$, $b_1 = b_0$ (no observation precedes the agent's action), and the first nontrivial update is $b_2 = \mathrm{BayesUpdate}(b_1, o_1, p^A_1)$, using~\eqref{eq:full_likelihood_accept} if $d^B_1 = \texttt{Accept}$, \eqref{eq:full_likelihood_reject} if $d^B_1 = \texttt{Reject}$, or~\eqref{eq:agent_opens_offer_likelihood} if $d^B_1 = \texttt{Offer}$.
\end{itemize}
Conditioning on $p^A_k$ is necessary because the counterpart's acceptance probability, walk-away hazard, and response distribution all depend on the agent's offered price. The components $\phi_k$, $\xi^A_k$, $h^B_k$, and $\chi$ of $\psi_k$ are deterministic given the public history and do not require Bayesian updating.

\paragraph{Per-stance likelihood evaluation.}
Because $\rho_F(\eta_B)$, $\xi_F(\eta_B)$, and $\lambda_{2,F}(\eta_B)$ all depend on stance, the acceptance score $g_\theta$, the walk-away hazard $\omega_k$, and the concession mean $\bar{p}^B_k$ must be evaluated separately for each of $\eta_B \in \{\texttt{C}, \texttt{N}, \texttt{A}\}$ when computing the per-type likelihoods. The belief support does not gain any new dimensions, but the per-$(r_B, \kappa_B)$ evaluation cost scales by the three stance branches. Because stance-dependent coefficients produce distinct acceptance probabilities and distinct concession rates for the same $(r_B, \kappa_B)$, the likelihood ratios across stances are sharper under the new formulation than under uniform coefficients, which should accelerate posterior concentration on $\eta_B$ in cue-informative families.

\subsection{Value Function and Bellman Equation}\label{app:bellman}

All value functions are defined over the augmented state $\psi_k = (b_k, \phi_k, \xi^A_k, h^B_k, \chi)$.

\paragraph{Terminal condition.}
At round $K$, if no agreement has been reached, the outcome is disagreement with utility $0$:
\begin{equation}
    V^*_{K+1}(\psi) \;=\; 0 \qquad \forall\,\psi.
\end{equation}

\paragraph{Action set.}
The agent's action set depends on what it has observed:
\begin{equation}\label{eq:action_set}
\mathcal{A}_k(\psi_k) \;=\;
\begin{cases}
\{\texttt{Offer}(p) : p \in \mathcal{P}\}, & k = 1 \text{ and } \chi = \texttt{AgentOpens},\\[2pt]
\{\texttt{Accept}, \texttt{Reject}\} \cup \{\texttt{Offer}(p) : p \in \mathcal{P}\}, & \text{otherwise (a counterpart offer exists in } h^B_k\text{)},
\end{cases}
\end{equation}
where $\mathcal{P} \subseteq [p_{\min}, p_{\max}]$ is the discretized admissible price set. The value function satisfies
\begin{equation}\label{eq:bellman}
    V^*_k(\psi_k) \;=\; \max_{a \,\in\, \mathcal{A}_k(\psi_k)}\; Q_k(\psi_k, a),
\end{equation}
with $Q$-functions as follows.

\paragraph{Accept.}
The agent accepts the counterpart's current offer $p^B_k \in h^B_k$:
\begin{equation}\label{eq:q_accept}
    Q_k(\psi_k, \texttt{Accept}) \;=\; u_A(p^B_k).
\end{equation}
Since $p^B_k$ is in $h^B_k$, no expectation over $t_B$ is needed. Individual rationality requires $u_A(p^B_k) \geq 0$.

\paragraph{Reject.}
The agent exits the negotiation:
\begin{equation}\label{eq:q_reject}
    Q_k(\psi_k, \texttt{Reject}) \;=\; 0.
\end{equation}

\paragraph{Offer.}
The agent proposes price $p$. The counterpart accepts (agreement at $p$, utility $u_A(p)$), walks away (disagreement, utility $0$), or counter-offers (continuing to round $k+1$):
\begin{equation}\label{eq:q_offer}
\begin{aligned}
Q_k(\psi_k, \texttt{Offer}(p))
\;=\; \sum_{t_B \in \mathcal{T}_B} b_k(t_B)\Bigg[\;
&a_k(p, t_B, \psi_k)\, u_A(p) \\
\;+\; &\underbrace{(1 - a_k(p, t_B, \psi_k))\,\omega_k(p, t_B, \psi_k)\cdot 0}_{\text{walk-away, }u_A=0} \\
\;+\; &(1 - a_k(p, t_B, \psi_k))(1 - \omega_k(p, t_B, \psi_k))\,W_{k+1}(\psi_k, t_B, p) \Bigg].
\end{aligned}
\end{equation}
The walk-away term is retained explicitly to emphasize that it consumes probability mass without contributing to value. The continuation term marginalizes over counter-offer observations only,
\begin{equation}\label{eq:continuation}
    W_{k+1}(\psi_k, t_B, p) \;:=\; \sum_{o \,\in\, \mathcal{O}_{\mathrm{cont}}} P(o \mid t_B, \psi_k, p, d^B_k = \texttt{Offer})\;\cdot\; V^*_{k+1}\!\big(\psi_{k+1}(o, p)\big),
\end{equation}
where $P(o \mid t_B, \psi_k, p, d^B_k = \texttt{Offer})$ is the counterpart's response distribution \emph{conditional on producing a counter-offer} (so that the price, sentiment, and strategic-cue factors in~\eqref{eq:full_likelihood_offer} are normalized by the offer-branch probability). The next-round augmented state is
\begin{equation}\label{eq:state_transition}
    \psi_{k+1}(o, p) \;=\; \big(
    \underbrace{b_{k+1}}_{\mathrm{BayesUpdate}},\;
    \underbrace{\phi(\xi^A_k, p)}_{\phi_{k+1}},\;
    \underbrace{(p^A_{k-2}, p^A_{k-1}, p)}_{\xi^A_{k+1}},\;
    \underbrace{(p^B_{k+1}, p^B_k)}_{h^B_{k+1}},\;
    \underbrace{\chi}_{\text{inherited}}
    \big).
\end{equation}

\paragraph{Discretized observation space.}
The continuation sum ranges over counter-offer observations:
\begin{equation}\label{eq:obs_cont}
    \mathcal{O}_{\mathrm{cont}} \;=\; \{(p_j, s, c) \,:\, p_j \in \mathcal{P}_{\mathrm{bin}},\; s \in \mathcal{S},\; c \in \mathcal{C}\},
\end{equation}
with $|\mathcal{O}_{\mathrm{cont}}| = 9M$. Because the counter-offer distribution~\eqref{eq:price_likelihood} is supported on $\mathcal{M}_B(k)$ rather than on $[p_{\min}, p_{\max}]$, the price bins must be intersected with $\mathcal{M}_B(k)$ before integrating:
\begin{equation}\label{eq:bin_prob}
    P_{\mathrm{bin}}(j \mid t_B, \psi_k) \;=\; \int_{\mathrm{bin}\,j \,\cap\, \mathcal{M}_B(k)} f_{\mathrm{price}}(p \mid t_B, \psi_k)\, \mathrm{d}p.
\end{equation}
Bins entirely outside $\mathcal{M}_B(k)$ receive zero probability. In practice, we treat the two endpoints of $\mathcal{M}_B(k)$ as dedicated atoms (with masses from~\eqref{eq:price_likelihood}) and distribute the remaining interior mass across interior bins via midpoint integration. Acceptance and walk-away observations are \emph{not} included in $\mathcal{O}_{\mathrm{cont}}$: both terminate the episode, the accept branch is handled by the $a_k u_A(p)$ term in~\eqref{eq:q_offer}, and the walk-away branch contributes zero utility via the explicit middle term in~\eqref{eq:q_offer}.

\begin{remark}[Joint marginalization]\label{rem:joint}
    The continuation term in~\eqref{eq:q_offer} cannot be factored as $(1 - \bar{a}_k)(1 - \bar{\omega}_k)\,\bar{W}_{k+1}$ using belief-averaged response rates $\bar{a}_k$, $\bar{\omega}_k$. All three of $a_k$, $\omega_k$, and the counter-offer distribution are $t_B$-dependent and generally correlated: for example, types whose reservation values make the agent's current offer non-individually-rational have simultaneously $a_k = 0$ and nonzero walk-away mass, so treating acceptance and walk-away as independent across types would double-count hazard and under-count continuation. The expectation must be taken jointly over $(t_B, d^B_k, o)$.
\end{remark}

\subsection{Optimal Policy}\label{app:optimal_action}

The optimal action at round $k$ is
\begin{equation}\label{eq:optimal_policy}
    \pi^*_k(\psi_k) \;=\; \arg\max_{a \,\in\, \mathcal{A}_k(\psi_k)}\, Q_k(\psi_k, a),
\end{equation}
subject to the monotonic concession constraint on admissible prices (Section~\ref{app:termination_constraints}). Backward induction proceeds from round $K$ to round $1$. Because $\chi$ is part of $\psi_k$, the DP handles both opener roles uniformly; equivalently, one may run the recursion separately for each value of $\chi$. Under $\chi = \texttt{AgentOpens}$, the round-1 value is $V^*_1(\psi_1) = \max_{p \in \mathcal{P}} Q_1(\psi_1, \texttt{Offer}(p))$ with no \texttt{Accept} term, and the continuation $W_2$ accounts for the three-branch counterpart response to the agent's opening using~\eqref{eq:agent_opens_offer_likelihood} as the counter-offer-branch observation likelihood. Under $\chi = \texttt{CounterpartOpens}$, the round-1 belief $b_1$ is obtained by Bayesian update of $b_0$ against the full opening likelihood~\eqref{eq:open_full_likelihood}, and the usual action set~\eqref{eq:action_set} applies.

\subsection{Computational Complexity}\label{app:complexity}

With $N = |\mathcal{T}_B|$ discretized types, $M$ price levels, $H$ discretized history-feature configurations, $K$ rounds, and $L$ quadrature nodes for the $d_0$-marginalization:
\begin{itemize}[nosep]
    \item \textbf{Belief updates:} $O(N)$ per observation, with a constant-factor increase from evaluating three-branch action probabilities and per-stance likelihoods.
    \item \textbf{Opening-round update:} $O(L \cdot N)$ per unique value of $p^B_1$, required under both opener roles whenever $d^B_1 = \texttt{Offer}$.
    \item \textbf{$Q$-function for $\texttt{Offer}(p)$:} $O(N \cdot |\mathcal{O}_{\mathrm{cont}}|)$ per price level.
    \item \textbf{Value iteration per round:} $O(H \cdot M \cdot N^2 \cdot |\mathcal{O}_{\mathrm{cont}}|)$.
    \item \textbf{Total:} $O(K \cdot H \cdot M \cdot N^2 \cdot |\mathcal{O}_{\mathrm{cont}}|)$, evaluated for each opener role.
\end{itemize}
With $N \approx 300$, $M = 50$, $K = 10$, $|\mathcal{O}_{\mathrm{cont}}| = 9M = 450$, and $L = 9$, this remains tractable on a single machine. In practice, we cache acceptance probabilities and walk-away hazards, reuse stance-specific concession means across price bins, and prune belief states with negligible posterior mass to reduce wall-clock time.

\subsection{Optimality Gap}\label{app:opt_gap}

Given the oracle reference policy $\pi^*$ and an evaluated policy $\pi$, the optimality gap is
\begin{equation}\label{eq:opt_gap}
    \mathrm{OptGap} \;=\; \bar{U}_{\pi^*} - \bar{U}_{\pi},\qquad \bar{U}_{\pi} := \frac{1}{N}\sum_{i=1}^{N} u_A(f_i),
\end{equation}
where the terminal outcomes $(f_i)_{i=1}^N$ are aggregated across all five termination sources enumerated in Appendix~\ref{app:termination_constraints}: agent accept, agent reject, counterpart accept, counterpart walk-away, and round-limit timeout. Because $\pi^*$ and $\pi$ face the same counterpart model and environment prior, the gap isolates strategic-reasoning quality from intrinsic task difficulty. Since the oracle-cue policy observes at least as much as a real agent under the benchmark observation model, $\bar{U}_{\pi^*_{\mathrm{oracle}}} \geq \bar{U}_{\pi^*_{\mathrm{true}}}$; hence the oracle-based optimality gap is an \emph{upper bound} on the optimality gap under the true observation model. Agent rejection and counterpart walk-away both produce $u_A = 0$, so partitioning disagreement mass between them does not affect $\mathrm{OptGap}$; however, the two sources should be logged separately so that termination-distribution mismatches between $\pi$ and $\pi^*$ can be diagnosed post hoc.

\begin{remark}[Counterpart family specialization]\label{rem:family}
    Family-specific parameterizations enter the oracle at three distinct levels.
    \textbf{(i) Cue channel:} \textsc{Candid} and \textsc{Expressive} use the base sentiment and strategic-cue models; \textsc{Taciturn} and \textsc{Strategic} collapse cues to $\texttt{neutral}/\texttt{Hold}$, removing cue-channel information; \textsc{Stochastic} uses $\sigma_{s,\mathrm{stoch}} = 2.0$ and softmax temperature $T_{\mathrm{stoch}} = 2.5$; \textsc{Adversarial} collapses cues to $\texttt{neg}/\texttt{Pressure}$.
    \textbf{(ii) Economic preset:} each family supplies its own stance-dependent $\rho_F(\eta_B)$, $\xi_F(\eta_B)$, $\lambda_{2,F}(\eta_B)$, and price-noise scale $\sigma_p$, which replace the shared placeholders in the acceptance, walk-away, and counter-offer likelihoods.
    \textbf{(iii) Stance prior:} the initial belief $b_0$ inherits the stance marginal of $\mu$. This is uniform for all families except \textsc{Adversarial}, which uses the aggressive-skewed prior~\eqref{eq:adv_prior}. Consequently, the oracle's round-1 belief for \textsc{Adversarial} is already strongly concentrated on $\eta_B = \texttt{A}$ before any observation.
    The previous draft's characterization of \textsc{Adversarial} by a universal $\xi < 0$ sign flip is obsolete: under stance-dependent coefficients, every family has $\xi_F(\texttt{A}) < 0$ via its economic preset, and the \textsc{Adversarial} distinctiveness lies in the hardball economic preset, the skewed stance prior, and the pressuring cue channel taken together.
\end{remark}

%% file: appendix/belief.tex
\section{Oracle Intervention Analysis}
\label{sec:latent-type-inference}

This section gives the full protocol for the oracle interventions summarized in
Section~\ref{sec:evaluation}. The goal is not merely to test
whether an agent can infer the counterpart's latent type, but to attribute
performance gaps to distinct information and control bottlenecks. In particular,
we ask whether an agent underperforms because it forms inaccurate beliefs, because
the correct posterior remains uncertain, or because it fails to act effectively
even when the relevant latent information is supplied.

\paragraph{Latent-state control problem.}
In the bilateral price-negotiation instantiation of \textsc{Terms-Bench}, the counterpart has hidden type
\[
t_B = (r_B,\kappa_B,\eta_B),
\]
where \(r_B\) is reservation value, \(\kappa_B\) is urgency, and \(\eta_B\) is
strategic stance. The evaluated agent does not observe \(t_B\) directly. It acts
from the public interaction history, its own private context, and any additional
side information exposed by the benchmark interface. Thus each episode is a
sequential decision problem under latent state.

Let
$
s_k = (h_k, x_A, b_k)
$
denote an information state at round \(k\), where \(h_k\) is public history,
\(x_A\) is the agent's private context, and
$
b_k \in \Delta(T_B)
$
is a belief over counterpart types. The benchmark separately elicits belief
reports and evaluates them using \(BE_r\), \(BE_\kappa\), and stance Brier score
(Section~\ref{sec:evaluation}). However, belief accuracy is only one part of the
diagnosis: an accurate belief is useful only if it changes downstream bargaining
actions in utility-improving ways. The oracle interventions therefore evaluate
both belief access and policy response.

\subsection{Oracle Bayesian Posterior}

Because the simulator \((\Gamma,\pi_B)\) is fully specified, we can compute an
oracle Bayesian posterior over counterpart types from the realized public
history. Given history \(h_k\), the oracle filter computes
$
b_k^{\mathrm{orc}}(t_B)
=
P(t_B \mid h_k;\mu,\pi_B),
$
where the likelihood is induced by the benchmark prior \(\mu\), the counterpart
policy \(\pi_B\), and the observed sequence of prices, actions, and messages.
The likelihood includes all channels of the counterpart model: acceptance,
walk-away, counter-offer, opening-offer, sentiment, and strategic-cue generation.
The dynamic-programming reference policy \(\pi^\star\) in
Appendix~\ref{app:optimal_policy} acts directly from this oracle belief state and
the known simulator model.

\paragraph{Oracle-Posterior intervention.}

The oracle-posterior intervention gives the evaluated LLM access to the oracle
posterior while leaving the rest of the interaction unchanged. At each round, the
LLM receives the standard benchmark observation plus a serialized posterior
summary
\begin{align}
z_k^{\mathrm{post}}
=
\big(
&\mathbb{E}_{b_k^{\mathrm{orc}}}[r_B],
\mathrm{CI}_{0.90}^{b_k^{\mathrm{orc}}}(r_B),
\mathbb{E}_{b_k^{\mathrm{orc}}}[\kappa_B],
b_k^{\mathrm{orc}}(\kappa_B), \nonumber \\
&\mathbf{p}_{\eta,k}^{\mathrm{orc}},
H(b_k^{\mathrm{orc}})
\big),
\end{align}
where
\[
\mathbf{p}_{\eta,k}^{\mathrm{orc}}
:=
\big(
b_k^{\mathrm{orc}}(\eta_B=\texttt{conciliatory}),
b_k^{\mathrm{orc}}(\eta_B=\texttt{neutral}),
b_k^{\mathrm{orc}}(\eta_B=\texttt{aggressive})
\big).
\]
The LLM still chooses both the economic action and the natural-language message
through the same wrapper, prompt template, decoding settings, and protocol
constraints as in the base condition. Thus the intervention removes
posterior-formation error while preserving posterior uncertainty and preserving
the LLM's downstream strategic-control problem.

We reveal a low-dimensional posterior summary rather than the full discretized
belief vector for two reasons. First, the full vector depends on grid resolution
and ordering, making it less stable across simulator implementations. Second,
the summary exposes the quantities most directly relevant for decision-making:
the posterior mean and uncertainty for reservation value, the urgency marginal,
the stance marginal, and posterior entropy. These are also aligned with the
benchmark's reported belief metrics.

\subsection{Nested Information Conditions}

We compare four nested conditions. Each condition is evaluated on the same
episode distribution and with the same environment dynamics.

\begin{enumerate}
    \item \textbf{Base agent.}
    The evaluated LLM observes the standard benchmark interface and must infer
    \(t_B\) from prices, actions, and messages.

    \item \textbf{Oracle-posterior agent.}
    The evaluated LLM observes the standard interface plus
    \(z_k^{\mathrm{post}}\) at each round. This removes posterior-formation
    error while preserving uncertainty over \(t_B\).

    \item \textbf{Revealed-type agent.}
    The evaluated LLM is given the true latent type \(t_B\) directly. This
    removes both posterior-formation error and residual latent-state
    uncertainty.

    \item \textbf{Model-based oracle.}
    The dynamic-programming policy \(\pi^\star\) acts from the oracle belief
    state and the known simulator model. This removes LLM planning, execution,
    and prompt-following errors.
\end{enumerate}

These conditions form an intervention ladder. Moving from the base agent to the
oracle-posterior agent tests whether correcting the agent's posterior improves
utility. Moving from the oracle-posterior agent to the revealed-type agent tests
the residual value of eliminating uncertainty. Moving from the revealed-type
agent to the model-based oracle tests whether the LLM can convert perfect
latent-state information into effective bargaining decisions.

\subsection{Gap Decomposition}

Let \(\bar U(\cdot)\) denote mean utility under a fixed intervention condition.
We define
\[
\Delta_{\mathrm{inf}}
:=
\bar U(\pi^{\mathrm{post}})-\bar U(\pi^{\mathrm{base}}),
\qquad
\Delta_{\mathrm{unc}}
:=
\bar U(\pi^{\mathrm{reveal}})-\bar U(\pi^{\mathrm{post}}),
\qquad
\Delta_{\mathrm{ctrl}}
:=
\bar U(\pi^\star)-\bar U(\pi^{\mathrm{reveal}}).
\]
Equivalently,
\[
\bar{U}(\pi^\star) - \bar{U}(\pi^{\mathrm{base}})
=
\Delta_{\mathrm{inf}}
+
\Delta_{\mathrm{unc}}
+
\Delta_{\mathrm{ctrl}}.
\]

The three terms have the following interpretation.

\begin{itemize}
    \item \(\Delta_{\mathrm{inf}}\) measures the value of replacing the
    agent's internally formed beliefs with the oracle posterior. It captures
    posterior-formation error only insofar as correcting that error changes
    downstream utility.

    \item \(\Delta_{\mathrm{unc}}\) measures the value of collapsing posterior
    uncertainty to the realized latent type. This term is large when even the
    correct posterior leaves economically meaningful uncertainty about the
    counterpart.

    \item \(\Delta_{\mathrm{ctrl}}\) measures the remaining gap between an LLM
    with perfect latent-type information and the model-based oracle. This term
    isolates strategic-control failures: failures to choose offers,
    acceptances, rejections, or walk-away decisions that effectively exploit the
    available information.
\end{itemize}

These quantities are intervention effects, not guaranteed nonnegative error
components. For example, oracle-posterior access may have limited value if the
family is cue-muted, if behavior is highly history-reactive, or if the LLM does
not know how to translate the posterior summary into a better pricing policy. In
some cases, extra information may even perturb the prompt-level policy. We
therefore interpret the decomposition empirically and report it by behavior
family (see below).

\paragraph{Family-Dependent Value of Information}

The value of oracle information depends on the counterpart family. In
type-driven families, behavior is more tightly coupled to stable latent traits,
so improved posterior information should be more useful for agreement
calibration and surplus extraction. In cue-muted families, the public language
channel carries less information about stance, which can reduce the value of
language-based inference. In history-reactive families, observed behavior is
partly driven by the agent's own trajectory, so better estimates of fixed latent
type may not fully determine the best response. In adversarial or stochastic
families, informative structure may be weakened further by hardball behavior,
pressuring cues, or noisy price and cue channels.

This family dependence is central to the diagnostic role of the benchmark. A
small \(\Delta_{\mathrm{inf}}\) does not by itself imply that the base agent
already inferred the type well; it may instead mean that type information has
limited marginal decision value in that family. Conversely, a large
\(\Delta_{\mathrm{ctrl}}\) indicates that the bottleneck is not information
access but policy execution: even when the agent is told the relevant latent
state, it does not act like the model-based oracle.

\paragraph{Reported Quantities}

For each family and intervention condition, we report three groups of metrics.
First, we report belief quality through \(BE_r\), \(BE_\kappa\), and stance
Brier score. Second, we report downstream bargaining performance, including mean
utility \(\bar U_\pi\), surplus efficiency \(SE_\pi\), agreement rate
\(\mathrm{AGR}_\pi\), and conditional surplus. Third, we report the incremental
intervention effects
\[
\Delta_{\mathrm{inf}}, \qquad
\Delta_{\mathrm{unc}}, \qquad
\Delta_{\mathrm{ctrl}}.
\]
Together, these quantities distinguish failures of latent-state recovery from
failures of strategic use. This is the main purpose of the oracle intervention
analysis: to move beyond aggregate outcome gaps and identify which part of the
agentic negotiation pipeline should be strengthened.


%% file: appendix/metrics.tex
\section{Evaluation Metrics}
\label{appdx:eval}

This appendix gives the full implementation details for the metrics reported in
Section~\ref{sec:evaluation}. We also follow the notation convention in Section~\ref{sec:evaluation}. Recall that the main text reports a compact primary metric
set:
\[
SE_\pi^+,\qquad
\mathrm{AGR}_\pi^+,\qquad
CSE_\pi^+,\qquad
\mathrm{FAGR}_\pi^-,\qquad
BE_{\mathrm{type}},\qquad
\mathrm{CritViol\%}.
\]
Here we define these metrics precisely and provide secondary decompositions used
for diagnostic analysis. To start, we first recall the outcome and utility setup.

\paragraph{Terminal outcomes and utility.}
Let $f_i$ denote the terminal outcome of episode $i$. If agreement occurs at
price $p_i$, the evaluated agent's utility is
\[
u_A(f_i)
=
\begin{cases}
r_A^{(i)}-p_i, & \text{buyer agent and deal at price }p_i,\\
p_i-r_A^{(i)}, & \text{seller agent and deal at price }p_i.
\end{cases}
\]
If the episode terminates in disagreement, $f_i=\bot$, then
$
u_A(f_i)=0.
$
A deal outside the agent's own reservation constraint yields negative utility
and is counted as a critical reservation-price violation.

\subsection{Feasible Terminal Performance}

\paragraph{Feasible surplus efficiency.}
The primary feasible terminal-value metric is surplus efficiency:
\begin{equation}
\label{eq:appendix_se_plus}
SE_\pi^+
=
\frac{1}{|\mathcal I^+|}
\sum_{i\in\mathcal I^+}
\frac{u_A(f_i)}{\Delta_i}.
\end{equation}
Disagreement in feasible episodes contributes zero utility. Loss-making
agreements are not clipped; they contribute negative utility and are also
captured by the critical violation metric.

\paragraph{Feasible agreement rate.}
Feasible agreement rate measures whether the agent closes deals when a
mutually beneficial agreement exists:
\begin{equation}
\label{eq:appendix_agr_plus}
\mathrm{AGR}_\pi^+
=
\frac{1}{|\mathcal I^+|}
\sum_{i\in\mathcal I^+}
\mathbf{1}[f_i\neq\bot].
\end{equation}

\paragraph{Conditional feasible deal quality.}
Let
$
\mathcal A^+
:=
\{i\in\mathcal I^+: f_i\neq\bot\}
$
be the set of agreed feasible episodes. Conditional feasible deal quality is
\begin{equation}
\label{eq:appendix_cse_plus}
CSE_\pi^+
=
\frac{1}{|\mathcal A^+|}
\sum_{i\in\mathcal A^+}
\frac{u_A(f_i)}{\Delta_i},
\end{equation}
whenever $|\mathcal A^+|>0$. If an agent reaches no feasible agreements, we
report $CSE_\pi^+$ as undefined rather than imputing a value.

Together, these metrics decompose feasible terminal performance:
\begin{equation}
\label{eq:appendix_terminal_decomposition}
SE_\pi^+
=
\mathrm{AGR}_\pi^+
\cdot
CSE_\pi^+
\end{equation}
whenever $|\mathcal A^+|>0$. Thus, $SE_\pi^+$ gives the overall normalized
terminal value, while $\mathrm{AGR}_\pi^+$ and $CSE_\pi^+$ distinguish reliable
closers from agents that extract high surplus only on the subset of episodes
where agreement occurs.

\paragraph{Raw utility.}
We also report empirical mean utility as a secondary, scale-dependent metric:
$
\bar U_\pi
=
\frac{1}{N}
\sum_{i=1}^N u_A(f_i).
$
Because raw utility depends on the price scale of the instance, it is not used
as the primary cross-regime value metric.

\paragraph{Oracle gap.}
When a model-based reference policy $\pi^\star$ is available, we report the
optimality gap
$
\mathrm{Gap}_\pi
=
\bar U_{\pi^\star}
-
\bar U_\pi.
$
This gap is computed using the same regime weights as the evaluated benchmark
suite. If reference policies are available only for a subset of regimes,
scenario sources, or simulator variants, the gap is reported only on that
matched subset.

\subsection{No-Deal Calibration and Exit Behavior}

\paragraph{No-deal false agreement rate.}
In no-deal episodes, agreement is a failure because there is no price that is
individually rational for both parties. We therefore report
\begin{equation}
\label{eq:appendix_fagr_minus}
\mathrm{FAGR}_\pi^-
=
\frac{1}{|\mathcal I^-|}
\sum_{i\in\mathcal I^-}
\mathbf{1}[f_i\neq\bot],
\end{equation}
where lower is better. Equivalently, the no-deal safe termination rate is
$
\mathrm{SafeTerm}_\pi^-
=
1-\mathrm{FAGR}_\pi^-.
$

\paragraph{Termination-source diagnostics.}
For diagnostic analysis, we log the terminal source
$
\tau_{\mathrm{terminal}}
\in
\{
\texttt{AgentAccept},
\texttt{CounterpartAccept},
\texttt{AgentReject},
\texttt{CounterpartWalkAway},
\texttt{Timeout}
\}.
$
Agreement corresponds to
$
\tau_{\mathrm{terminal}}
\in
\{\texttt{AgentAccept},\texttt{CounterpartAccept}\}.
$
For no-deal episodes, we additionally report the agent-initiated exit rate:
\begin{equation}
\label{eq:appendix_agent_exit}
\mathrm{AgentExit}_\pi^-
=
\frac{1}{|\mathcal I^-|}
\sum_{i\in\mathcal I^-}
\mathbf{1}[
\tau_{\mathrm{terminal}}^{(i)}
=
\texttt{AgentReject}
].
\end{equation}
This distinguishes disciplined infeasibility detection from cases where the
agent is rescued by counterpart walk-away or timeout.

\subsection{Opponent-Modeling Metrics}

Opponent-modeling metrics are computed only for agents that expose explicit
belief estimates. If an agent does not expose beliefs, these metrics are left
undefined rather than imputed.

\paragraph{Reservation-value error.}
For counterpart reservation value, we report normalized mean absolute error:
\begin{equation}
\label{eq:appendix_be_r}
BE_r
=
\frac{1}{K_r}
\sum_{(i,k)\in\mathcal K_r}
\frac{
|\hat r_{B,i}^{\,k}-r_B^{(i)}|
}{
p_{\max}^{(i)}-p_{\min}^{(i)}
}.
\end{equation}
Here $\mathcal K_r$ is the set of episode-round pairs for which a valid
reservation estimate is available, and $K_r=|\mathcal K_r|$.

\paragraph{Urgency error.}
For counterpart urgency, we report
\begin{equation}
\label{eq:appendix_be_kappa}
BE_\kappa
=
\frac{1}{K_\kappa}
\sum_{(i,k)\in\mathcal K_\kappa}
|\hat\kappa_{B,i}^{\,k}-\kappa_B^{(i)}|.
\end{equation}

\paragraph{Stance Brier score.}
For stance, when the agent returns a probability vector
$
\hat p_i^k
=
(\hat p_{i,c}^{\,k})_{c\in\mathcal C}
$
over
$
\mathcal C
=
\{\texttt{conciliatory},\texttt{neutral},\texttt{aggressive}\},
$
we report the normalized multiclass Brier score
\begin{equation}
\label{eq:appendix_brier_eta}
\mathrm{Brier}_\eta
=
\frac{1}{K_B}
\sum_{(i,k)\in\mathcal K_B}
\frac{1}{2}
\sum_{c\in\mathcal C}
\left(
\hat p_{i,c}^{\,k}
-
\mathbf{1}[c=\eta_B^{(i)}]
\right)^2.
\end{equation}
The factor $1/2$ normalizes the three-class Brier score to lie in $[0,1]$.

\paragraph{Aggregate type belief error.}
The main text reports the aggregate type belief error
\begin{equation}
\label{eq:appendix_be_type}
BE_{\mathrm{type}}
=
\frac{
BE_r
+
BE_\kappa
+
\mathrm{Brier}_\eta
}{3},
\end{equation}
computed only when all three components are available. 

\paragraph{Stance hit rate.}
As a secondary, less calibration-sensitive summary, we define the point estimate
$
\hat\eta_{B,i}^{\,k}
=
\arg\max_{c\in\mathcal C}
\hat p_{i,c}^{\,k}
$
and report
$
\mathrm{StanceAcc}
=
\frac{1}{K_\eta}
\sum_{(i,k)\in\mathcal K_\eta}
\mathbf{1}[
\hat\eta_{B,i}^{\,k}
=
\eta_B^{(i)}
].
$
This metric is reported only as a supplement to the Brier score because it
discards probability calibration.

\subsection{Protocol Compliance and Violation Accounting}

We separate critical violations from secondary procedural diagnostics. The main
text reports critical violation rate; this appendix reports the full breakdown.

\paragraph{Critical violations.}
An episode has a critical violation if one or more of the following occur.

\emph{Price-bound violation.}
The agent proposes an offer outside the public price bounds:
$
p\notin[p_{\min},p_{\max}].
$

\emph{Reservation-price / individual-rationality violation.}
The agent accepts or offers a price that would give it negative utility if
executed. For a buyer agent, such prices satisfy
$
p>r_A;
$
for a seller agent, they satisfy
$
p<r_A.
$
This category includes accepting a counterpart offer outside the agent's
reservation constraint, proposing an own offer outside the agent's reservation
constraint, or reaching a terminal agreement with
$
u_A(f_i)<0.
$

\emph{Invalid-action violation.}
The agent returns an action that is invalid in the current information state.
Examples include choosing \texttt{Accept} when no counterpart offer has been
observed, returning a non-null price with \texttt{Accept} or \texttt{Reject},
returning a null price with \texttt{Offer}, or producing an action outside the
allowed action set.

Let $V_i^{\mathrm{crit}}$ denote the number of critical violations in episode
$i$. The primary compliance metric is
\begin{equation}
\label{eq:appendix_critviol}
\mathrm{CritViol\%}
=
\frac{1}{N}
\sum_{i=1}^N
\mathbf{1}[V_i^{\mathrm{crit}}>0].
\end{equation}

\paragraph{Critical violation breakdown.}
We also report component-wise critical violation rates:
$
\mathrm{BoundViol\%}
=
\frac{1}{N}
\sum_{i=1}^N
\mathbf{1}[V_i^{\mathrm{bound}}>0],
$
$
\mathrm{ResViol\%}
=
\frac{1}{N}
\sum_{i=1}^N
\mathbf{1}[V_i^{\mathrm{res}}>0],
$
$
\mathrm{InvalidAct\%}
=
\frac{1}{N}
\sum_{i=1}^N
\mathbf{1}[V_i^{\mathrm{act}}>0].
$

\paragraph{Secondary procedural diagnostics.}
The following diagnostics are logged but are not included in the headline
critical violation rate unless otherwise stated.

\emph{Monotonicity violation.}
Buyer offers must be weakly increasing over the agent's own offer sequence, and
seller offers must be weakly decreasing. Let $p_k^A$ be the agent's current
offer and $p_{\mathrm{prev}}^A$ its previous own offer. A monotonicity violation
occurs when
$
p_k^A < p_{\mathrm{prev}}^A
\quad
\text{for buyer agents},
$
or
$
p_k^A > p_{\mathrm{prev}}^A
\quad
\text{for seller agents}.
$
We report $\mathrm{MonoViol\%}$ separately because monotonicity failures
diagnose bargaining coherence but are less severe than price-bound or
reservation-price failures.

\emph{Turn-budget violation.}
An action after terminal negotiation or outside the round budget is counted as
a turn-budget violation. In most experiments, the harness prevents such actions,
so this metric is primarily an implementation sanity check.

\emph{Schema or parse violation.}
If an agent is required to output JSON and fails to return a parseable object
matching the required schema, the output is marked as a schema violation. If the
malformed output prevents recovery of a valid economic action, it is also
counted as an invalid-action violation.

\emph{Information-leakage flag.}
When enabled, we flag messages that explicitly reveal private information such
as the agent's reservation value. Because this check can depend on
string-matching heuristics, it is reported separately from the primary critical
violation rate.

\paragraph{Any-violation rate.}
For completeness, we define
$
\mathrm{AnyViol\%}
=
\frac{1}{N}
\sum_{i=1}^N
\mathbf{1}
[
V_i^{\mathrm{crit}}
+
V_i^{\mathrm{proc}}
>0
],
$
where $V_i^{\mathrm{proc}}$ counts secondary procedural diagnostics. This is an
audit statistic, not the headline compliance metric.




\paragraph{Undefined or small-denominator metrics.}
Some conditional metrics may be undefined for agents that never reach the
conditioning event. In particular, $CSE_\pi^+$ is undefined when
$|\mathcal A^+|=0$, and opponent-modeling metrics are undefined when no valid
belief estimates are available. We report such entries as undefined rather than
imputing zero. For conditional metrics, tables should include or make available
the relevant denominator to avoid over-interpreting small-sample estimates.

%% file: appendix/grader.tex
\section{Difficulty Grader}
\label{sec:difficulty}

A useful negotiation benchmark should not only rank agents on average, but also
expose how those rankings change as the bargaining problem becomes
structurally harder. \textsc{Terms-Bench} therefore characterizes each episode
by a pre-interaction difficulty score derived from the sampled environment,
allowing us to report performance across difficulty tiers rather than rely on
a single aggregate mixture. For feasible bargaining, difficulty rises when the
ZOPA is narrow, the evaluated agent faces greater relative time pressure, the
counterpart stance is more demanding, or the horizon is shorter; for no-deal
regimes, difficulty rises when infeasibility is harder to detect: the
reservation gap is near-feasible, cue evidence is weak or pressuring, and
surface behavior encourages continued bargaining.

The mixed-opener protocol requires one important distinction. Some difficulty
variables are properties of the sampled environment, such as ZOPA width,
urgency asymmetry, counterpart stance, cue reliability, and deadline. Other
quantities depend on the realized opening action. When the counterpart opens,
the realized counterpart anchor is an environment-side difficulty factor. When
the agent opens, however, the opening price is chosen by the evaluated policy
and is therefore a policy behavior rather than an instance property. We
therefore report:
$D^{\mathrm{env}}$,
a role-comparable environment difficulty score, together with opener-role
strata
\[
\chi\in\{\texttt{AgentOpens},\texttt{CounterpartOpens}\}.
\]
For counterpart-opens episodes, we additionally report a counterpart-opening
anchor score. For agent-opens episodes, we log the agent's opening
aggressiveness as a policy diagnostic rather than including it in
environment difficulty.

\subsection{Difficulty Dimensions}

We consider two broad regimes: (i) overlap regimes, in which agreement is
feasible, and (ii) no-deal regimes, in which rational behavior requires walking
away.

\paragraph{Overlap regime.}
Episodes with feasible bargaining space vary in difficulty along several
structural dimensions:

\begin{table}[H]
\centering
\scriptsize
\setlength{\tabcolsep}{5pt}
\renewcommand{\arraystretch}{2.0}
\begin{tabular}{p{2.8cm} p{5.3cm} p{2.2cm} p{2.5cm}}
\toprule
\textbf{Dimension} & \textbf{Score / variable} & \textbf{Hard direction} & \textbf{Use Case} \\
\midrule

ZOPA width
&
$d_{\mathrm{zopa}}=1-\dfrac{\Delta}{R}$, $\Delta=r_{\mathrm{buyer}}-r_{\mathrm{seller}}$
&
$d_{\mathrm{zopa}}\uparrow$
&
env. score \\

Urgency pressure
&
$d_{\mathrm{press}}
=
\left[
\dfrac{\kappa_A-\kappa_B}
{\kappa_A+\kappa_B+\varepsilon_\kappa}
\right]_+$
&
$d_{\mathrm{press}}\uparrow$
&
env. score \\

Counterpart stance
&
$d_{\mathrm{stance}}
=
\begin{cases}
0, & \eta_B=\texttt{conciliatory}\\
0.5, & \eta_B=\texttt{neutral}\\
1, & \eta_B=\texttt{aggressive}
\end{cases}$
&
$d_{\mathrm{stance}}\uparrow$
&
env. score \\

Deadline
&
$d_K
=
1-\dfrac{K-K_{\min}}{K_{\max}-K_{\min}}$
&
$d_K\uparrow$
&
env. score if $K$ varies \\

Counterpart opening
&
$d_{\mathrm{open}}^B
=
\min\!\left\{
1,\dfrac{2|p_{\mathrm{open}}^B-r_B|}
{\Delta+\varepsilon_d}
\right\}$
&
$d_{\mathrm{open}}^B\uparrow$
&
only if $\chi=\texttt{CounterpartOpens}$ \\

Opener role
&
$\chi\in\{\texttt{AgentOpens},\texttt{CounterpartOpens}\}$
&
stratify
&
not scalar difficulty \\

Agent opening
&
$a_{\mathrm{open}}^\pi
=
\min\!\left\{
1,\dfrac{2|p_{\mathrm{open}}^A-r_A|}
{\Delta+\varepsilon_d}
\right\}$
&
diagnostic
&
only if $\chi=\texttt{AgentOpens}$ \\

\bottomrule
\end{tabular}
\vspace{0.5em}
\caption{\scriptsize
Overlap-regime difficulty dimensions. The scalar environment score uses only
pre-interaction instance properties: ZOPA width, urgency pressure, stance, and
deadline when $K$ varies. Counterpart opening harshness is included only within
the counterpart-opens subset. Agent opening aggressiveness is logged as a policy
diagnostic because the opening price is chosen by the evaluated agent.
}
\label{tab:overlap_difficulty_dimensions}
\end{table}

Thus, harder overlap episodes have narrower feasible bargaining regions,
greater time pressure on the agent, harder counterpart stance, and shorter
horizons. Counterpart opening harshness is used only within the
counterpart-opens subset. In agent-opens episodes, the first price is chosen by
the evaluated agent and is analyzed separately as an opening-policy diagnostic.

\paragraph{No-deal regime.}
In episodes without feasible agreement, the challenge is not surplus extraction
but correct recognition of infeasibility. Difficulty arises from:

\begin{table}[H]
\centering
\scriptsize
\setlength{\tabcolsep}{5pt}
\renewcommand{\arraystretch}{1.35}
\begin{tabular}{p{3.0cm} p{5.4cm} p{2.2cm} p{2.0cm}}
\toprule
\textbf{Dimension} & \textbf{Score / variable} & \textbf{Hard direction} & \textbf{Use Case} \\
\midrule

Infeasibility gap
&
$d_{\mathrm{gap}}
=
\exp\!\left(
\dfrac{\Delta}
{\sigma_{\mathrm{scale}}+\varepsilon_\sigma}
\right)$
&
$d_{\mathrm{gap}}\uparrow$
&
env. score \\

Cue channel
&
$d_{\mathrm{cue}}
=
\begin{cases}
0, & \text{accurate}\\
0.5, & \text{uninformative}\\
0.75, & \text{weak/noisy}\\
1, & \text{pressuring}
\end{cases}$
&
$d_{\mathrm{cue}}\uparrow$
&
env. score \\

Surface behavior
&
$d_{\mathrm{surf}}
=
\begin{cases}
1, & \eta_B=\texttt{conciliatory}\\
0.5, & \eta_B=\texttt{neutral}\\
0, & \eta_B=\texttt{aggressive}
\end{cases}$
&
$d_{\mathrm{surf}}\uparrow$
&
env. score \\

Opener role
&
$\chi\in\{\texttt{AgentOpens},\texttt{CounterpartOpens}\}$
&
stratify
&
not scalar difficulty \\

Termination source
&
$\tau_{\mathrm{term}}$
&
diagnostic
&
post hoc only \\

\bottomrule
\end{tabular}
\vspace{0.5em}
\caption{\scriptsize
No-deal difficulty dimensions. In no-deal episodes,
$\Delta=r_{\mathrm{buyer}}-r_{\mathrm{seller}}<0$. Harder instances are
near-feasible, meaning $\Delta$ is close to zero from below, so
$d_{\mathrm{gap}}$ is large. Difficulty also increases when cues are weak,
uninformative, or pressuring, and when surface behavior sustains bargaining
despite infeasibility. Opener role is reported as a stratification variable.
Termination source, specified by $\{\texttt{AgentReject},\texttt{CounterpartWalkAway},
\texttt{Timeout},\texttt{Agreement}\}$, is not part of pre-interaction difficulty, but is logged to
distinguish disciplined agent exit from counterpart walk-away, timeout, or
irrational agreement.
}
\label{tab:nodeal_difficulty_dimensions}
\end{table}

Near-feasible gaps combined with weak, noisy, uninformative, or pressuring
signals create the hardest instances for detecting that no agreement should
occur. As in overlap regimes, opener role is treated as a stratification
variable rather than folded into the scalar environment-difficulty score.

\subsection{Formal Difficulty Scores}

\paragraph{Overlap environment difficulty.}
For overlap regimes, let
\[
\Delta := r_{\mathrm{buyer}}-r_{\mathrm{seller}}>0,
\qquad
R := p_{\max}-p_{\min}.
\]
Define normalized ZOPA difficulty
\[
d_{\mathrm{zopa}}
:=
1-\frac{\Delta}{R}.
\]
Define urgency-pressure difficulty from the evaluated agent's perspective as
\[
d_{\mathrm{press}}
:=
\max\!\left\{
0,\;
\frac{\kappa_{\mathrm{agent}}-\kappa_{\mathrm{cp}}}
{\kappa_{\mathrm{agent}}+\kappa_{\mathrm{cp}}+\varepsilon_\kappa}
\right\},
\]
so that difficulty increases when the agent is more time-pressured than the
counterpart. Define stance hardness as
\[
d_{\mathrm{stance}}
=
\begin{cases}
1.0, & \eta_B=\texttt{aggressive},\\
0.5, & \eta_B=\texttt{neutral},\\
0.0, & \eta_B=\texttt{conciliatory}.
\end{cases}
\]
If multiple horizons are used, define
\[
d_{\mathrm{deadline}}
=
1-\frac{K-K_{\min}}{K_{\max}-K_{\min}}
\in[0,1],
\]
so that shorter horizons are harder.

The role-comparable overlap difficulty score excludes realized opening
actions:
\begin{equation}
\label{eq:overlap_env_difficulty}
D_{\mathrm{overlap}}^{\mathrm{env}}
=
\frac{
w_z d_{\mathrm{zopa}}
+
w_p d_{\mathrm{press}}
+
w_s d_{\mathrm{stance}}
+
\mathbf{1}\{K_{\max}>K_{\min}\}w_k d_{\mathrm{deadline}}
}{
w_z+w_p+w_s+\mathbf{1}\{K_{\max}>K_{\min}\}w_k
}.
\end{equation}
Unless otherwise stated, we use
\[
(w_z,w_p,w_s,w_k)=(0.45,\,0.25,\,0.20,\,0.10).
\]
When $K$ is fixed across all episodes, the deadline term is omitted and the
remaining weights are renormalized by the denominator in
\eqref{eq:overlap_env_difficulty}.

\paragraph{Counterpart-opening anchor difficulty.}
When $\chi=\texttt{CounterpartOpens}$, the counterpart's realized opening
anchor is an additional environment-side difficulty factor. Let
$p_{\mathrm{open}}^B$ denote the counterpart's first offer. We define
\[
d_{\mathrm{open}}^B
:=
\min\!\left\{
1,\;
\frac{2|p_{\mathrm{open}}^B-r_B|}
{\Delta+\varepsilon_d}
\right\}.
\]
This score is large when the counterpart opens far from its reservation value
relative to the width of the feasible bargaining region. Because the opening
distance parameter is now randomized at the episode level, this realized score
captures the combined effect of $d_{0,e}$, urgency, stance, directional slack,
and opening noise.

For analyses restricted to counterpart-opens episodes, we optionally define an
anchor-augmented difficulty score
\begin{equation}
\label{eq:overlap_anchor_difficulty}
D_{\mathrm{overlap}}^{\mathrm{cp\mbox{-}open}}
=
(1-\omega_o)D_{\mathrm{overlap}}^{\mathrm{env}}
+
\omega_o d_{\mathrm{open}}^B,
\qquad
\omega_o=0.20.
\end{equation}
This score should not be used to compare counterpart-opens and agent-opens
episodes directly; cross-opener comparisons use
$D_{\mathrm{overlap}}^{\mathrm{env}}$ and stratify by $\chi$.

\paragraph{Agent-opening policy diagnostic.}
When $\chi=\texttt{AgentOpens}$, the first price is chosen by the evaluated
agent and is not part of environment difficulty. For diagnostic purposes, we
log the agent's opening aggressiveness in feasible episodes as
\[
a_{\mathrm{open}}^\pi
:=
\min\!\left\{
1,\;
\frac{2|p_{\mathrm{open}}^A-r_A|}
{\Delta+\varepsilon_d}
\right\},
\]
where $p_{\mathrm{open}}^A$ is the agent's first offer and $r_A$ is the agent's
reservation value. This quantity measures how far the agent anchors from its
own reservation relative to the feasible bargaining width. It is reported as a
policy-behavior statistic, not as a pre-interaction difficulty score.

\paragraph{No-deal environment difficulty.}
For no-deal regimes, define
\[
\Delta := r_{\mathrm{buyer}}-r_{\mathrm{seller}} < 0.
\]
Let $\sigma_{\mathrm{scale}}$ denote the relevant price-variation scale:
$\sigma_{\mathrm{scale}}=\sigma_{\mathrm{mkt}}$ in data-grounded episodes and
$\sigma_{\mathrm{scale}}=p_{\max}-p_{\min}$ in purely synthetic episodes unless
a regime-specific scale is specified. We define
\[
d_{\mathrm{gap}}
:=
\exp\!\left(
\frac{\Delta}{\sigma_{\mathrm{scale}}+\varepsilon_\sigma}
\right).
\]
This term is close to one for near-feasible no-deal instances
($\Delta\approx 0^{-}$) and decreases toward zero as the infeasibility gap
widens.

The no-deal environment difficulty score is
\begin{equation}
\label{eq:nodeal_env_difficulty}
D_{\mathrm{nodeal}}^{\mathrm{env}}
=
v_\Delta d_{\mathrm{gap}}
+
v_c d_{\mathrm{cue}}
+
v_s d_{\mathrm{surf}},
\qquad
v_\Delta+v_c+v_s=1.
\end{equation}
Unless otherwise stated, we use
\[
(v_\Delta,v_c,v_s)=(0.60,\,0.25,\,0.15).
\]
As in overlap regimes, opener role $\chi$ is reported as a separate
stratification variable rather than being folded into
$D_{\mathrm{nodeal}}^{\mathrm{env}}$.

\subsection{Difficulty-Stratified Evaluation}

Using these scores, episodes are grouped into difficulty bins. For overlap and
urgency-shift regimes, the primary binning score is
$D_{\mathrm{overlap}}^{\mathrm{env}}$. For no-deal regimes, the primary binning
score is $D_{\mathrm{nodeal}}^{\mathrm{env}}$. In all regimes, we additionally
report metrics stratified by opener role:
\[
\chi=\texttt{AgentOpens}
\qquad\text{versus}\qquad
\chi=\texttt{CounterpartOpens}.
\]
Within the counterpart-opens subset, we may further stratify by
$d_{\mathrm{open}}^B$ or by the anchor-augmented score
$D_{\mathrm{overlap}}^{\mathrm{cp\mbox{-}open}}$. Within the agent-opens
subset, we report the policy diagnostic $a_{\mathrm{open}}^\pi$ to measure how
agents choose anchors when no counterpart price has yet been observed.

This decomposition separates three effects that would otherwise be confounded:
(i) structural environment difficulty, (ii) whether the agent acts as opener
or responder, and (iii) the quality of the agent's own opening policy.

\begin{figure}[t]
\centering
\includegraphics[width=0.82\linewidth]{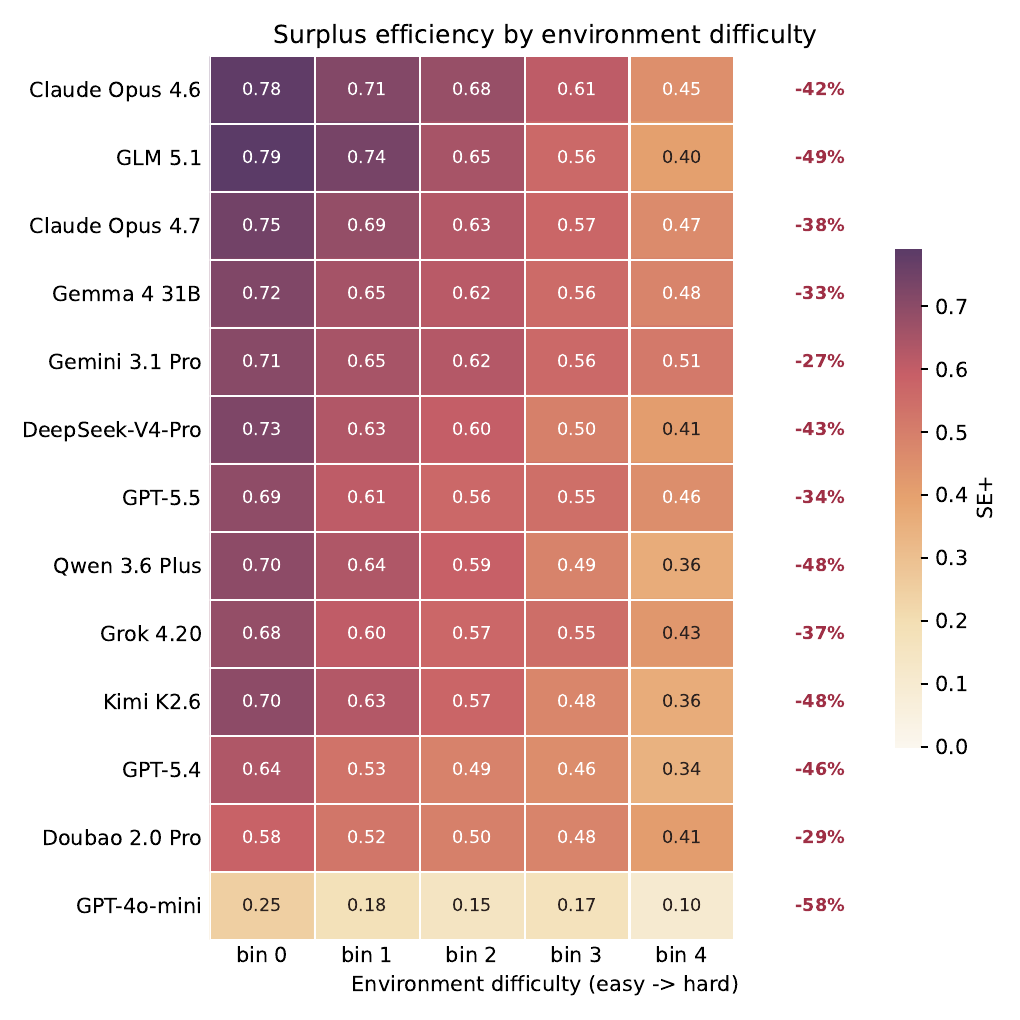}
\caption{\footnotesize Surplus efficiency across structural environment-difficulty bins.
Bins progress from easy to hard. Darker cells indicate higher $SE^+$. The
right-hand column reports the percentage drop from the easiest to the hardest
bin.}
\label{fig:difficulty-bins}
\end{figure}

\paragraph{Empirical bin-by-bin performance.}
Figure~\ref{fig:difficulty-bins} reports surplus efficiency across structural
difficulty bins for the full model sweep. Performance declines materially from
the easiest to the hardest bin for every evaluated agent: drops range from
$-27\%$ (\texttt{Gemini 3.1 Pro}) to $-58\%$ (\texttt{GPT-4o-mini}), with a
median drop of $-42\%$ across the thirteen LLMs. \texttt{Claude Opus 4.6}
collapses from $SE_\pi^+{=}0.78$ on the easiest bin to $0.45$ on the hardest
($-42\%$); \texttt{GLM 5.1} from $0.79$ to $0.40$ ($-49\%$). Even
high-performing agents that look strong in aggregate degrade substantially on
harder structural instances, confirming that \textsc{Terms-Bench} generates
graded evaluation tiers rather than a single undifferentiated distribution.

\begin{figure}[t]
\centering
\begin{subfigure}[t]{0.5\linewidth}
\centering
\includegraphics[width=\linewidth]{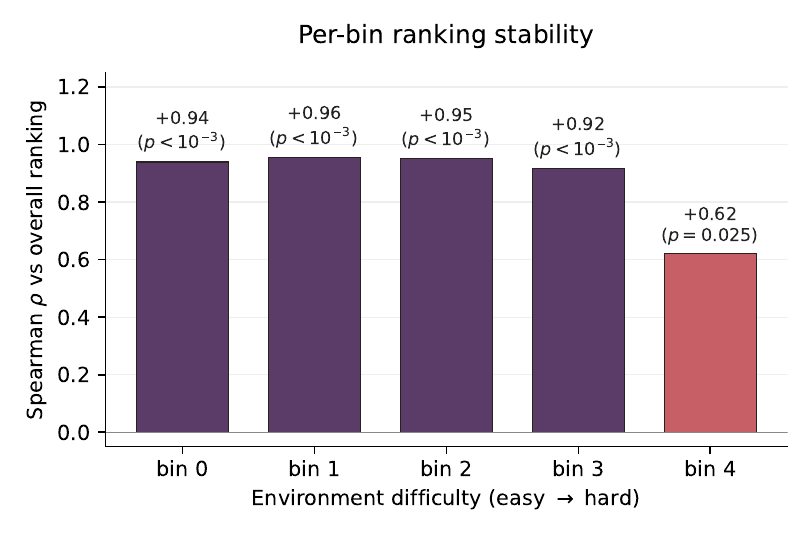}
\caption{\footnotesize Per-bin Spearman $\rho$ vs the aggregate ranking.
Bins~0--3 sit at $\rho\in[0.92,0.96]$; bin~4 (red) drops to
$\rho=0.62$.}
\label{fig:difficulty-rank-stability-bars}
\end{subfigure}\hfill
\begin{subfigure}[t]{0.5\linewidth}
\centering
\includegraphics[width=\linewidth]{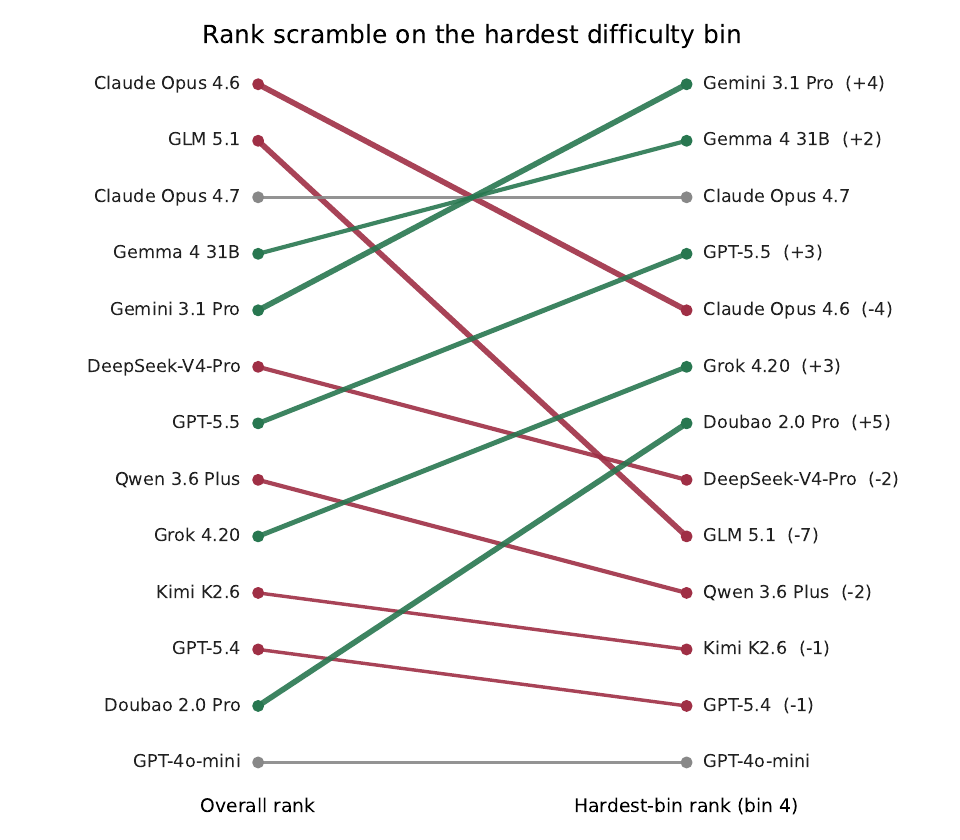}
\caption{\footnotesize Per-agent rank movement from the aggregate ordering
(left) to the hardest-bin ordering (right). Green = climbed (annotated
$+\Delta$), red = dropped, gray = unchanged.}
\label{fig:difficulty-rank-scramble}
\end{subfigure}
\caption{\footnotesize Rank stability across difficulty bins. Panel~A
quantifies the cliff between bins~0--3 and the hardest bin; Panel~B shows
where the rank churn actually occurs, with \texttt{GLM 5.1} ($-7$) and
\texttt{Doubao 2.0 Pro} ($+5$) the largest moves and
\texttt{Gemini 3.1 Pro} ($+4$) and \texttt{Gemma 4 31B} ($+2$) taking the
top two hardest-bin positions.}
\label{fig:difficulty-rank-stability}
\end{figure}

\paragraph{Rank stability across bins.}
The aggregate leaderboard is largely preserved across difficulty tiers but
partially scrambles on the hardest bin
(Fig.~\ref{fig:difficulty-rank-stability}). Spearman rank correlations between
each per-bin ranking and the aggregate ranking are
$\rho\in[0.92,0.96]$ for bins~0--3 ($p<10^{-3}$ throughout) and drop sharply
to $\rho=0.62$ on the hardest bin ($p=0.025$;
Fig.~\ref{fig:difficulty-rank-stability}A). The hardest-bin ordering reveals
where the scramble actually occurs (Fig.~\ref{fig:difficulty-rank-stability}B):
\texttt{Gemini 3.1 Pro} and \texttt{Gemma 4 31B} rise from overall ranks 5 and
4 to the top two positions ($SE_\pi^+$ of $0.51$ and $0.48$), \texttt{Doubao
2.0 Pro} climbs five ranks (12$\to$7), \texttt{GPT-5.5} climbs three ranks
(7$\to$4), while \texttt{GLM 5.1} drops seven ranks (2$\to$9) and
\texttt{Claude Opus 4.6} drops from rank 1 to rank 5. Robustness to high
structural difficulty is therefore a partially independent capability that
aggregate ranking does not fully capture.

%% file: appendix/implementation.tex
\section{Experiment Details}\label{sec:implementation_details}
\vspace{-0.5em}
\subsection{Implementation Details}
\label{app:exp_setup}

We provide the implementation details needed to reproduce the main
\textsc{Terms-Bench} evaluation, including the evaluated models, episode
construction, seeding, agent interface, inference settings, and logging
protocol.

\subsubsection{Model Details}
\label{appdx:model_details}

Table~\ref{tab:models} lists the 13 LLM agents evaluated in the bilateral
price-negotiation instantiation of \textsc{Terms-Bench}. All models are queried
through \texttt{OpenRouter} via API calls. Reasoning-capable models use the
same maximum reasoning-effort setting unless otherwise noted. We additionally
compare against fixed-concession baselines with concession rates
$\{0.01,0.10,0.30\}$.

\begin{table}[H]
\centering
\footnotesize
\begin{tabular}{llcl}
\toprule
Model & Provider & Reasoning & Reference \\
\midrule
Claude Opus 4.7              & Anthropic        & \texttt{xhigh} & \cite{anthropic_claude_opus_47_2026} \\
Claude Opus 4.6              & Anthropic        & \texttt{xhigh} & \cite{anthropic_claude_opus_46_2026} \\
Google Gemma 4 31B IT        & Google DeepMind  & \texttt{xhigh} & \cite{google_gemma4_31b_it_2026} \\
Gemini 3.1 Pro Preview       & Google DeepMind  & \texttt{xhigh} & \cite{google_gemini_31_pro_preview_2026} \\
DeepSeek V4 Pro              & DeepSeek         & \texttt{xhigh} & \cite{deepseek_v4_pro_2026} \\
Qwen3.6-Plus                 & Alibaba Qwen     & \texttt{xhigh} & \cite{qwen36_plus_2026} \\
Kimi K2.6                    & Moonshot AI      & \texttt{xhigh} & \cite{moonshot_kimi_k26_2026} \\
GPT-5.4                      & OpenAI           & \texttt{xhigh} & \cite{openai_gpt54_2026} \\
GPT-5.5                      & OpenAI           & \texttt{xhigh} & \cite{openai2026gpt55} \\
GPT-4o-mini                  & OpenAI           & ---             & \cite{openai_gpt4o_mini_2024} \\
Doubao-Seed-2.0-Pro          & ByteDance        & \texttt{xhigh} & \cite{bytedance_doubao_seed20_pro_2026} \\
Zhipu GLM-5.1                & Zhipu            & \texttt{xhigh} & \cite{zai_glm51_2026} \\
Grok 4.2                     & xAI              & \texttt{xhigh} & \cite{xai_grok42_2026} \\
\bottomrule
\end{tabular}
\vspace{0.5em}
\caption{\footnotesize Language models evaluated in the bilateral
price-negotiation instantiation of \textsc{Terms-Bench}. The \emph{Reasoning}
column reports the reasoning-effort setting used at evaluation:
\texttt{xhigh} denotes maximum reasoning allocation; ``---'' denotes models for
which reasoning is not available.}
\label{tab:models}
\end{table}

\subsubsection{Episode Construction and Regime Parameters}
\label{sec:param_setting}

The main evaluation suite spans three scenario regimes
(\textsc{Overlap}, \textsc{Urgency-Shift}, and \textsc{No-Deal}) and six
counterpart behavior families. For each $(\text{regime},\text{family})$ pair,
we construct 100 episodes using a balanced $2\times2$ allocation over agent
role and opener role:
\[
\scriptsize
(\texttt{Buyer},\texttt{CounterpartOpens}), \;
(\texttt{Buyer},\texttt{AgentOpens}), \;
(\texttt{Seller},\texttt{CounterpartOpens}), \;
(\texttt{Seller},\texttt{AgentOpens}),
\]
with 25 episodes in each sub-cell. Thus each evaluated agent is run on
\[
6 \times 3 \times 100 = 1{,}800
\]
main-suite episodes. The opener role is balanced at the episode level: when the
counterpart opens, $d_{0,e}$ governs the counterpart's first offer; when the
agent opens, the agent's first price is logged as an opening-policy diagnostic
rather than treated as an environment-generation parameter. Main paper
urgency-shift results use the counterpart-more-urgent direction; the reverse
direction is reported in Appendix~\ref{appdx:urgency-down}.
Table~\ref{tab:regime_params} summarizes the regime-specific task-generation
parameters. Shared simulator hyperparameters for the counterpart policy are
reported in Appendix~\ref{appdx:default_param_counter}.

\begin{table}[h]
\centering
\scriptsize
\setlength{\tabcolsep}{4pt}
\renewcommand{\arraystretch}{1.08}
\caption{Regime-specific task-generation parameters used in experiments.}
\vspace{0.5em}
\label{tab:regime_params}
\begin{tabular}{@{}l l p{5.5cm} p{5.6cm}@{}}
\toprule
\textbf{Regime} & \textbf{Parameter} & \textbf{Values} & \textbf{Notes} \\
\midrule

All regimes
& $d_{0,e}$
& $\mathrm{Unif}(0.20,\,0.80)$
& hidden episode-level opening harshness; used only when the counterpart makes its first offer \\

All regimes
& $\mathcal F$
& \begin{tabular}[t]{@{}l@{}}
$\{\texttt{Inference\mbox{-}Critical},\ \texttt{Taciturn},\ \texttt{Expressive},$ \\
$\texttt{Strategic},\ \texttt{Stochastic},\ \texttt{Adversarial}\}$
\end{tabular}
& counterpart behavior family; balanced by construction \\

All regimes
& $K$
& fixed benchmark horizon
& maximum number of negotiation rounds \\

\midrule

Overlap
& $\Delta$
& $[\Delta_{\min},\,\Delta_{\max}]$, $\Delta>0$
& ZOPA width, where $\Delta=r_{\mathrm{buyer}}-r_{\mathrm{seller}}$ \\

Overlap
& $D_\kappa$
& $\mathrm{Beta}(\alpha_\kappa,\beta_\kappa)$ on $[0,1]$
& baseline urgency law \\

Overlap
& $d_{\mathrm{open}}^B$
& logged post hoc
& realized counterpart-anchor difficulty, only if $\chi=\texttt{CounterpartOpens}$ \\

\midrule

Urgency shift
& $\Delta$
& $[\Delta_{\min},\,\Delta_{\max}]$, $\Delta>0$
& same feasible price geometry as overlap \\

Urgency shift
& $D_\kappa^{(s)}$
& $\mathrm{Beta}(\alpha_{\mathrm{shifted}},\,\beta_{\mathrm{shifted}})$ on $[0,1]$
& counterpart urgency drawn from a single shifted Beta law \\

Urgency shift
& $s$
& $s = \mathbb E[D_\kappa^{(s)}] - \mathbb E[D_\kappa]$
& realized mean shift; logged post hoc, not swept in the main suite \\

Urgency shift
& $d_{\mathrm{open}}^B$
& logged post hoc
& realized counterpart-anchor difficulty, only if $\chi=\texttt{CounterpartOpens}$ \\

\midrule

No-deal
& $\Delta$
& $[-g_{\max},\,-g_{\min}]$, $\Delta<0$
& infeasible gap, with $g=-\Delta=r_{\mathrm{seller}}-r_{\mathrm{buyer}}>0$ \\

No-deal
& $D_\kappa$
& $\mathrm{Beta}(\alpha_\kappa,\beta_\kappa)$ on $[0,1]$
& baseline urgency law unless otherwise stated \\

No-deal
& $\tau_{\mathrm{term}}$
& logged post hoc
& termination source:
$\texttt{AgentReject}$, $\texttt{CounterpartWalkAway}$,
$\texttt{Timeout}$, or $\texttt{Agreement}$ \\

\bottomrule
\end{tabular}
\end{table}

Numerical defaults for
$(\alpha_\kappa,\beta_\kappa,\alpha_{\mathrm{shifted}},\beta_{\mathrm{shifted}},
\Delta_{\min},\Delta_{\max},g_{\min},g_{\max})$
are listed in Appendix~\ref{appdx:default_param_counter}.

\subsubsection{Scenario Sampling and Seeding}
\label{appdx:scenario_sampling}

All evaluated agents are run on the same indexed scenario set. To support both
cross-model comparisons and within-cell regime contrasts, scenario latents are
drawn once per
\[
\texttt{(family, agent\_role, opener\_role, episode\_index)}
\]
cell and reused across regimes. Given a fixed \texttt{base\_seed}, we compute
\[
   \mathrm{cell}(b,f,r,o,e)
   =
   b \cdot 10^{7}
   + f_{\mathrm{idx}} \cdot 10^{5}
   + r_{\mathrm{idx}} \cdot 10^{4}
   + o_{\mathrm{idx}} \cdot 10^{3}
   + e \cdot 10,
\]
and use disjoint seed streams $\sigma_i=\mathrm{cell}(\cdot)+i$ for
$i\in\{1,\dots,5\}$ to draw
\[
(\eta_B,\kappa_A,\kappa_B^{(0)},\kappa_B^{(s)},d_{0,e}),
\]
corresponding to counterpart stance, agent urgency, counterpart baseline
urgency, counterpart shifted urgency, and opening harshness. A separate shared
stream draws a geometry percentile $u_e\sim\mathrm{Unif}(0,1)$ and maps it to
\[
   z_e = \Delta_{\min} + u_e(\Delta_{\max}-\Delta_{\min}),
   \qquad
   q_e = g_{\min} + u_e(g_{\max}-g_{\min}).
\]
The overlap and urgency-shift regimes use $z_e$ as the ZOPA width, while the
no-deal regime uses $q_e$ as the infeasibility gap. Thus overlap and no-deal
siblings in the same cell differ only in the sign of $\Delta$ and, for
urgency-shift, the realized counterpart urgency.

Because all latent streams are independent of the evaluated model, every model
sees the same 1,800 scenarios in the same order. Since the counterpart policy
is history-dependent, matched seeds do not force identical trajectories:
different agents may induce different acceptance, walk-away, counter-offer,
and cue realizations through their own actions. For any scalar metric $m$,
paired agent comparisons are formed from within-episode differences
\[
d_i(\pi_1,\pi_2;m)=m_{\pi_1}(i)-m_{\pi_2}(i),
\]
over matched episode indices $i$.

\subsubsection{Counterpart Policy and Language Realization}
\label{appdx:counterpart_language}

The evaluated agent always negotiates against the fixed
environment-simulated counterpart policy $\pi_B$, not another LLM. The
simulator kernel determines the counterpart's economic action
(offer/accept/reject), realized price, and latent cue pair
$(\tilde s_k,\tilde c_k)$ from the sampled counterpart type and public history.
A separate voice layer renders a natural-language message consistent with the
already committed economic state
$(d_k^B,p_k^B,\tilde s_k,\tilde c_k)$. The voice layer never changes the
economic outcome.

\subsubsection{Agent Interface}
\label{appdx:agent_interface}

Each round, the agent receives a single \texttt{JSON} user message and must
return a single \texttt{JSON} response. The system prompt
(Appendix~\ref{sec:prompts}) defines this contract; the user message contains
no additional natural-language instructions.

\paragraph{Input.}
The user-message \texttt{JSON} contains five top-level keys:
\begin{itemize}
  \item \texttt{private\_context}: the agent's role
  (\texttt{buyer}/\texttt{seller}) and reservation price $r_A$.
  \item \texttt{protocol\_state}: round number, maximum rounds, rounds
  remaining, whether a counterpart offer is on the table, the legal decision
  set, and the agent's last own offer.
  \item \texttt{constraints}: price bounds $[p_{\min},p_{\max}]$, the
  monotone-concession rule, and $\delta_{\max}$.
  \item \texttt{observation}: the counterpart's current price $p_k^B$,
  message $m_k^B$, and immediate accept-utility $u_A(p_k^B)$ when an offer is
  on the table.
  \item \texttt{history}: the last $W=6$ rounds of counterpart
  prices/messages and the agent's past actions.
\end{itemize}
The simulator-internal cue variables are not directly revealed to the agent:
the observation is $o_k=(p_k^B,m_k^B)$ as defined in
Section~\ref{sec:methods}. Cues are logged only for diagnostic analysis.

\paragraph{Output and parsing.}
The agent must return a \texttt{JSON} object of the form
\begin{verbatim}
{
  "decision": "Offer" | "Accept" | "Reject",
  "price": <float or null>,
  "message": "<natural language>"
}
\end{verbatim}
The \texttt{price} field is required for \texttt{Offer} and ignored otherwise;
\texttt{message} is always delivered to the counterpart. Agents may also
include an optional \texttt{type\_estimate} sub-object, which is parsed for
opponent-modeling diagnostics when present but does not affect the economic
state transition.

We extract the \texttt{JSON} response by balanced-brace scanning and validate
the parsed action. The \texttt{decision} must be one of the legal verbs;
\texttt{Offer} requires a numeric price, which is clamped to
$[p_{\min},p_{\max}]$ with any clamp recorded as a \texttt{price\_bound}
violation; \texttt{Accept} is legal only when an offer is on the table; and
monotone-concession violations are detected post hoc relative to the agent's
last own offer. If parsing fails entirely, a deterministic fallback is used:
accept if the standing counterpart offer is weakly preferred to walking away,
and otherwise repeat the agent's reservation-price offer. Fallbacks are
recorded as \texttt{invalid\_action} violations.

\subsubsection{LLM Call Settings}
\label{appdx:llm_call_settings}

All LLM agents are called through a common thin client using the settings in
Table~\ref{tab:llm_call_settings}. Each seeded episode is executed once per
model, so variance reduction comes from matched seeds rather than within-agent
repetition.

\begin{table}[H]
\centering
\small
\begin{tabular}{@{}l l l@{}}
\toprule
\textbf{Setting} & \textbf{Value} & \textbf{Notes} \\
\midrule
\texttt{temperature}        & $0$        & deterministic decoding \\
\texttt{max\_tokens}        & $16{,}000$ & accommodates reasoning models \\
\texttt{timeout\_s}         & $180$      & per-call HTTP read timeout \\
\texttt{max\_retries}       & $3$        & retryable API/transport failures \\
\texttt{backoff\_initial\_s}& $0.5$      & exponential backoff base \\
\texttt{backoff\_factor}    & $2.0$      & doubles per retry \\
\texttt{backoff\_jitter\_s} & $0.25$     & uniform jitter added to each wait \\
\texttt{history\_window}    & $6$ rounds & last $W$ rounds in the user payload \\
\texttt{response\_format}   & free text  & balanced-brace JSON extraction \\
\bottomrule
\end{tabular}
\vspace{0.5em}
\caption{LLM call settings used for all evaluated agents.}
\label{tab:llm_call_settings}
\end{table}

\subsubsection{Metrics, Aggregation, and Logging}
\label{appdx:metrics_logging}
All primary metrics are reported by regime and counterpart family. Because the
suite is balanced by agent role and opener role, we also report slices for
buyer versus seller agents and for
$\chi=\texttt{AgentOpens}$ versus $\chi=\texttt{CounterpartOpens}$. Where
relevant, we further stratify by counterpart stance. Overall scores use the
empirical episode weights of the evaluation suite. Conditional metrics with
empty conditioning sets are reported as undefined rather than imputed with
zero.

Each episode emits a complete trace containing the sampled counterpart state
$t_B$, latent cue variables $(\tilde s_k,\tilde c_k)$, all agent and
counterpart actions, history features consumed by the simulator kernel, parser
outputs, and termination source
(\texttt{AgentAccept}, \texttt{CounterpartAccept}, \texttt{AgentReject},
\texttt{CounterpartWalkAway}, or \texttt{Timeout}). These logs support
qualitative failure analysis, violation auditing, paired comparisons, and the
information-intervention experiments.

\subsection{Data-Grounded Experiment}\label{appdx:data-grounded}
This appendix supplies the full construction, dataset summary, and per-model
results for the data-grounded instantiation introduced in
\S\ref{sec:data_grounded}. We instantiate the variant on the
\textit{AmazonHistoryPrice} dataset\footnote{We use the \textit{AmazonHistoryPrice} dataset released by
\citet{xia2024measuringbargainingabilitiesllms} in their official public repository.
The repository lists the dataset under \texttt{data/AmazonHistoryPrice} and is released
under the Apache-2.0 license. We use the dataset to instantiate public product price
scales and product descriptions for evaluation only; no personal data or human
interaction data are used.} of \citet{xia2024measuringbargainingabilitiesllms}:
historical minimum, maximum, and average prices calibrate the buyer's reference
price and the seller's reservation value, while accompanying product
descriptions populate the prompt context. The benchmark machinery---counterpart
kernel, oracle policy, information-intervention decomposition, and metrics---is
unchanged; only the price geometry and observable product context are replaced
with empirical distributions. Beyond external validity, this also demonstrates
that \textsc{Terms-Bench} is a versatile environment rather than a purely
synthetic artifact: practitioners can adopt the framework for their own product
catalogs by supplying market statistics and product descriptions, without
modifying the evaluation pipeline. The remainder of this appendix details the
scenario generator (\S\ref{appdx:data_grounded_scenarios}), the
product-grounded evaluation setup (\S\ref{app:product_grounded}), and full
per-model results (\S\ref{appdx:product_grounded_results}).
\subsubsection{Data-Grounded Scenario Construction}
\label{appdx:data_grounded_scenarios}

To start, this section gives the full construction used for data-grounded scenarios in the bilateral price-negotiation instantiation of \textsc{Terms-Bench}. 
\paragraph{Dataset hierarchy and templates.}
We represent the data source as a two-level hierarchy of product categories and
products. Categories define broad public price regimes, while individual
products provide item-specific context and historical price summaries. Each
episode is instantiated by first sampling a category $c$, then a product $j$
within that category, and finally an agent role. The scenario template is
\[
\mathcal{T}
=
(c,j,\textsc{Attributes}_j,\textsc{MarketStats}_j,\textsc{Bounds}_c,
\textsc{Role}),
\]
where $\textsc{Attributes}_j$ contains product-level context such as item name,
description, and salient attributes;
$\textsc{MarketStats}_j$ contains product-level historical price summaries; and
$\textsc{Bounds}_c=[p_{\min}^{(c)},p_{\max}^{(c)}]$ contains category-level
public price bounds. The product context and public market statistics may be
shown to the evaluated agent, but private reservation values, urgency, and
stance remain hidden.

\paragraph{Product-level market statistics and public bounds.}
For product $j$, we write
\[
\textsc{MarketStats}_j
=
(\hat p_{\mathrm{ref}}^{(j)},\hat p_{\mathrm{lo}}^{(j)},
\hat p_{\mathrm{hi}}^{(j)}),
\]
where $\hat p_{\mathrm{ref}}^{(j)}$ is a publicly observable reference price
such as a historical mean or median, and
$\hat p_{\mathrm{lo}}^{(j)},\hat p_{\mathrm{hi}}^{(j)}$ summarize the
historical low and high for that product. We define the product-level dispersion
scale
\[
\sigma_{\mathrm{mkt}}^{(j)}
:=
\frac{\hat p_{\mathrm{hi}}^{(j)}-\hat p_{\mathrm{lo}}^{(j)}}{4}.
\]
If the observed historical range is degenerate or extremely small, we use a
small positive floor for $\sigma_{\mathrm{mkt}}^{(j)}$ to avoid numerically
degenerate reservation distributions.

The public bargaining bounds are category-level rather than product-level:
\[
p_{\min}=p_{\min}^{(c)},\qquad p_{\max}=p_{\max}^{(c)}.
\]
Thus, product statistics determine local valuation scale, while category bounds
define the public feasible action range. Templates are filtered or resampled so
that
\[
p_{\min}^{(c)}
<
\hat p_{\mathrm{ref}}^{(j)}
<
p_{\max}^{(c)}.
\]

\paragraph{Feasible-regime reservation mapping.}
In overlap and urgency-shift regimes, we model private reservations as latent
wedges around the product reference price:
\[
r_s=\hat p_{\mathrm{ref}}^{(j)}-\Delta_s,
\qquad
r_b=\hat p_{\mathrm{ref}}^{(j)}+\Delta_b.
\]
Here $\Delta_s\ge 0$ represents the seller's private cost buffer below the
reference price, and $\Delta_b\ge 0$ represents the buyer's willingness-to-pay
premium above the reference price. We sample
\[
\Delta_s
\sim
\mathrm{TruncNormal}
\!\left(
\mu_s, \sigma_s^2;\,
0,\hat p_{\mathrm{ref}}^{(j)}-p_{\min}
\right),
\]
\[
\Delta_b
\sim
\mathrm{TruncNormal}
\!\left(
\mu_b, \sigma_b^2;\,
0,p_{\max}-\hat p_{\mathrm{ref}}^{(j)}
\right),
\]
with
\[
\mu_s
=
\alpha_s\bigl(\hat p_{\mathrm{ref}}^{(j)}
-\hat p_{\mathrm{lo}}^{(j)}\bigr),
\qquad
\mu_b
=
\alpha_b\bigl(\hat p_{\mathrm{hi}}^{(j)}
-\hat p_{\mathrm{ref}}^{(j)}\bigr),
\]
and
\[
\sigma_s=\beta_s\sigma_{\mathrm{mkt}}^{(j)},
\qquad
\sigma_b=\beta_b\sigma_{\mathrm{mkt}}^{(j)}.
\]
The truncation support ensures that sampled reservations remain within the
public price bounds. Under this construction,
\[
r_b-r_s=\Delta_b+\Delta_s\ge 0,
\]
so a ZOPA exists by construction. When a minimum feasible surplus is required
for a particular difficulty stratum, we reject and resample until
$r_b-r_s\ge \Delta_{\min}^{\mathrm{PG}}$.

\paragraph{Urgency-shift regimes.}
Data-grounded urgency-shift scenarios use the same reservation construction as
overlap scenarios. Only the counterpart urgency law changes. Specifically, the
counterpart urgency $\kappa_B$ is sampled from the shifted urgency distribution
specified by the corresponding synthetic regime, while the agent-side urgency
and stance-generation laws are inherited unchanged unless otherwise stated.
Thus, the data-grounded urgency-shift regime isolates adaptation to
counterpart time pressure without changing the product-grounded price geometry.

\paragraph{No-deal regimes.}
To generate infeasible data-grounded scenarios, we sample an infeasibility gap
$\delta>0$ and set
\[
r_s=\hat p_{\mathrm{ref}}^{(j)}+\delta/2,
\qquad
r_b=\hat p_{\mathrm{ref}}^{(j)}-\delta/2.
\]
This ensures
\[
r_b<r_s.
\]
The gap is scaled to the product's market dispersion; for example,
\[
\delta
\sim
\mathrm{Uniform}
\!\left(
\delta_{\min}\sigma_{\mathrm{mkt}}^{(j)},
\delta_{\max}\sigma_{\mathrm{mkt}}^{(j)}
\right),
\]
subject to the public-bound feasibility constraint
\[
0<\delta
\le
2\min\{
p_{\max}-\hat p_{\mathrm{ref}}^{(j)},
\hat p_{\mathrm{ref}}^{(j)}-p_{\min}
\}.
\]
If the sampled product does not admit a positive feasible gap under the chosen
bounds, the template is resampled. This keeps no-deal scenarios economically
infeasible while preserving valid private reservations inside the public action
range.

\paragraph{Synthetic control and reproducibility.}
Synthetic scenario generators directly specify price geometry and type
distributions independent of any data source, enabling controlled ablations over
ZOPA width, urgency shift, cue noise, opening harshness, and stance mixtures.
Data-grounded generation replaces only the price geometry: feasible regimes use
the latent-wedge construction above, urgency-shift regimes additionally modify
the counterpart urgency distribution, and no-deal regimes use explicit
infeasibility gaps around the product reference price.

All data-grounded scenarios are generated from templates with fixed
random seeds and explicit hyperparameters governing category sampling, product
sampling, wedge distributions, infeasibility gaps, and public bounds. 

\subsubsection{Product-Grounded Evaluation Details}
\label{app:product_grounded}

This subsection reports dataset realization statistics and the evaluated
agent subset for the data-grounded sweep; the formal scenario generator is
given in \S\ref{appdx:data_grounded_scenarios}. Episodes are sampled from a
CamelCamelCamel-derived Amazon catalog (831 products, 14 categories), with
the public product block (item name, category, attributes, reference price,
historical range) appended to the agent's prompt while reservations,
urgency, and stance remain hidden.

Unless otherwise stated, the product-grounded sweep uses 100 episodes per regime
with seeded scenario sampling and the same evaluated agent set as the synthetic
sweep. Dataset summaries are reported in Tables~\ref{tab:category_distribution} and
\ref{tab:product_price_stats}. In particular, Table \ref{tab:category_distribution} describes the product scenario distribution of the Amazon History Price data across the 14 product categories. Table \ref{tab:product_price_stats} describes the sampled price distribution for each product categories sampled.

\begin{table}[h]
\centering
\small
\begin{tabular}{lrr}
\toprule
\textbf{Category} & \textbf{Episodes} & \textbf{\% of total} \\
\midrule
other                     & 686 & 35.2 \\
electronics               & 648 & 33.2 \\
tools-home-improvement    & 329 & 16.9 \\
home-kitchen              & 64  & 3.3 \\
toys-games                & 50  & 2.6 \\
sports-outdoors           & 39  & 2.0 \\
beauty                    & 31  & 1.6 \\
baby-products             & 28  & 1.4 \\
patio-lawn-garden         & 23  & 1.2 \\
automotive                & 17  & 0.9 \\
video-games               & 16  & 0.8 \\
pet-supplies              & 9   & 0.5 \\
health-personal-care      & 7   & 0.4 \\
industrial-scientific     & 3   & 0.2 \\
\bottomrule
\vspace{0.5em}
\end{tabular}
\caption{\footnotesize Category distribution of product-grounded episodes from Amazon History Price data.}
\label{tab:category_distribution}
\end{table}

\begin{table}[h]
\centering
\small
\begin{tabular}{lrrrr}
\toprule
\textbf{Category} & \textbf{Episodes} & \textbf{Avg \$ (mean)} & \textbf{Avg \$ (min)} & \textbf{Avg \$ (max)} \\
\midrule
other                     & 686 & 225.27 & 11.09 & 1520.87 \\
electronics               & 648 & 386.28 & 15.42 & 2950.26 \\
tools-home-improvement    & 329 & 129.13 & 7.62  & 802.94 \\
home-kitchen              & 64  & 154.42 & 5.37  & 678.55 \\
toys-games                & 50  & 61.36  & 13.30 & 1002.04 \\
sports-outdoors           & 39  & 291.14 & 16.80 & 2760.49 \\
beauty                    & 31  & 162.39 & 28.43 & 584.60 \\
baby-products             & 28  & 351.89 & 32.84 & 685.68 \\
patio-lawn-garden         & 23  & 1243.26 & 6.04 & 2921.01 \\
automotive                & 17  & 323.38 & 8.19  & 913.45 \\
video-games               & 16  & 179.70 & 44.12 & 529.49 \\
pet-supplies              & 9   & 24.53  & 8.82  & 36.92 \\
health-personal-care      & 7   & 22.28  & 12.45 & 35.39 \\
industrial-scientific     & 3   & 263.23 & 138.09 & 325.80 \\
\bottomrule
\end{tabular}
\vspace{0.5em}
\caption{\footnotesize Price statistics of product-grounded episodes by category. Values report the average of observed product prices and the range across products in each category.}
\label{tab:product_price_stats}
\end{table}

\subsubsection{Experiment Results}
\label{appdx:product_grounded_results}

We evaluate eleven out of the 13 models evaluated in the main experiment (\S\ref{sec:experiments}) on the product-grounded suite for which a directly
comparable synthetic counterpart is available in the main 1800-episode
sweep: \texttt{Claude Opus 4.6}, \texttt{Claude Opus 4.7},
\texttt{Gemini 3.1 Pro}, \texttt{Gemma 4 31B}, \texttt{GLM 5.1},
\texttt{DeepSeek-V4-Pro}, \texttt{Grok~4.20}, \texttt{Kimi K2.6},
\texttt{Qwen 3.6 Plus}, \texttt{GPT-5.5}, and \texttt{GPT-4o-mini}.
In the following, we present some main results on the product grounded experiment.

\begin{figure}[t]
\centering
\includegraphics[width=0.95\linewidth]{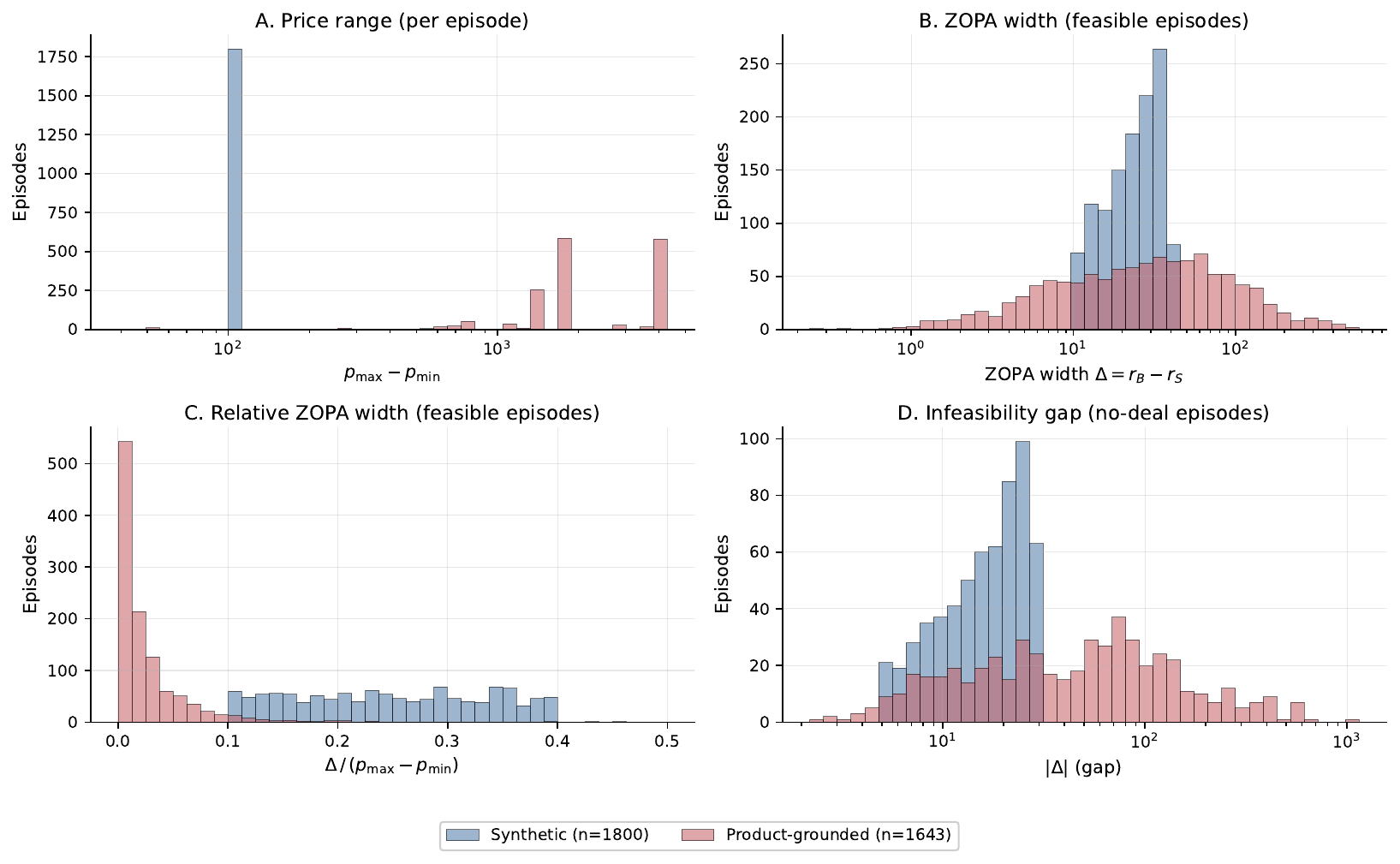}
\vspace{-0.5em}
\caption{\scriptsize Per-episode price geometry, synthetic vs.\
product-grounded suite (\texttt{Claude Opus 4.6}, $n{=}1800$ vs.\
$n{=}1643$; scenarios are deterministic by seed and identical across
models within a suite). \textbf{A.}~Public price range. \textbf{B.}~Absolute
ZOPA width on feasible episodes. \textbf{C.}~Relative ZOPA width.
\textbf{D.}~Infeasibility gap on no-deal episodes. The \emph{relative}
ZOPA collapses by about an order of magnitude under data grounding, so
agents must rely on product-context anchors rather than uniform
exploration of $[p_{\min},p_{\max}]$.}
\label{fig:geometry_pg}
\end{figure}

\paragraph{Result 1: Price Geometry shifts under data-grounding mode.}
Replacing the parametric price geometry with \textit{AmazonHistoryPrice}-derived statistics
shifts the scenario topology in three quantitatively large ways
(Figure~\ref{fig:geometry_pg}). First, the public action range
$p_{\max}-p_{\min}$ moves from a fixed 100 to a long-tailed distribution
with median \$1{,}694 and upper quartile \$4{,}293, reflecting the
cross-category Amazon catalog. Second, the absolute ZOPA width is
modestly larger in product-grounded episodes (median \$28.8 vs.\ \$24.6)
but with a substantially fatter tail (IQR \$10--64 vs.\ \$17--32).
Third---and most consequentially for negotiation difficulty---the
\emph{relative} ZOPA $\Delta/(p_{\max}-p_{\min})$ collapses by roughly an
order of magnitude (median 1.3\% vs.\ 24.6\%), and reservation prices
cluster near the bottom of the public range (normalised position median
0.07 vs.\ 0.49). The data-grounded suite therefore presents agents with
the same magnitude of surplus to extract, but on a much larger and
skewed action range; an agent that searches uniformly over
$[p_{\min},p_{\max}]$ has far less chance of stumbling into the ZOPA
than under the synthetic geometry. This places a premium on using the
public product reference price as an anchor.

\begin{figure}[t]
\centering
\includegraphics[width=0.55\linewidth]{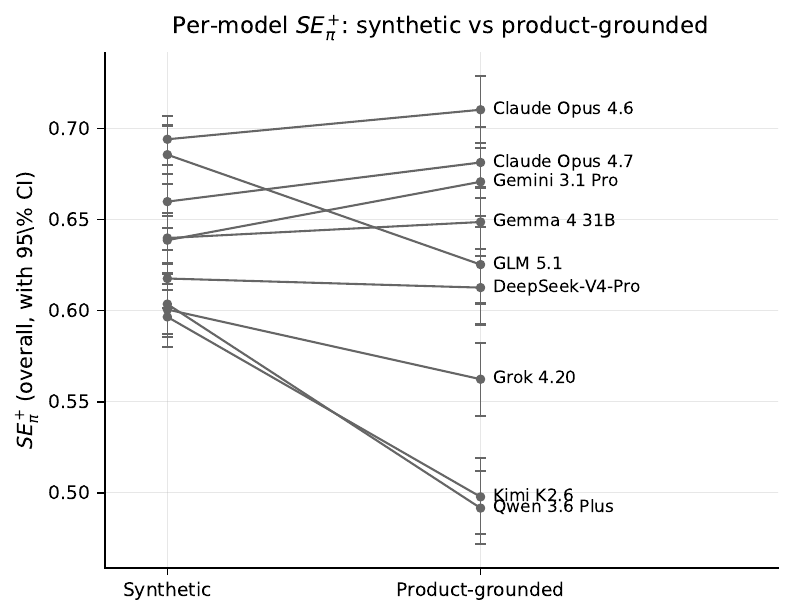}
\caption{\scriptsize Per-model overall $SE_\pi^{+}$ in synthetic vs.\
product-grounded suites with 95\% bootstrap CIs. Rank order is largely
preserved (Spearman $\rho=0.90$); the shift is structured: models in the
upper half of the synthetic leaderboard tend to gain or hold (exception:
\texttt{GLM 5.1}), while most lower-half models lose (exception:
\texttt{GPT-5.5}).}
\label{fig:pg_rank}
\end{figure}

\paragraph{Result 2: Synthetic Leaderboard structure is largely preserved.}
Table~\ref{tab:pg_leaderboard} reports per-model overall $SE_\pi^{+}$
in both suites with 95\% bootstrap CIs ($B=2000$);
Figure~\ref{fig:pg_rank} visualises the per-model shift. Across the eleven
paired models, the rank correlation between synthetic and product-grounded
$SE_\pi^{+}$ is Spearman $\rho = 0.90$ ($p < 0.001$, $n=11$):
\texttt{Claude Opus 4.6} retains the top position in both suites,
\texttt{GPT-4o-mini} retains the bottom, and the next-lowest group
(\texttt{Kimi K2.6}, \texttt{Qwen 3.6 Plus}, \texttt{Grok~4.20}) is
preserved. The diagnostic ordering produced by the benchmark is therefore
not an artefact of synthetic geometry.

\paragraph{Result 3: Capability of LLM Agents is amplified under product grounding.}
The shift in $SE_\pi^{+}$ between suites is, nevertheless, structured: of the
five models in the upper half of the synthetic leaderboard, four gain or
hold under product grounding (\texttt{Gemini 3.1 Pro}: $+0.032$,
\texttt{Claude Opus 4.7}: $+0.021$, \texttt{Claude Opus 4.6}: $+0.016$,
\texttt{Gemma 4 31B}: $+0.009$); in the lower half, five of six lose
meaningfully (\texttt{Grok~4.20}: $-0.038$, \texttt{Kimi K2.6}: $-0.099$,
\texttt{Qwen 3.6 Plus}: $-0.112$, \texttt{DeepSeek-V4-Pro}: $-0.005$,
\texttt{GPT-4o-mini}: $-0.103$). The two upward-moving exceptions are
\texttt{GPT-5.5} (modest gain $+0.012$ in the lower half) and---in the
opposite direction---\texttt{GLM 5.1}, which drops $-0.060$ from synthetic
rank 2 to product-grounded rank 5. Strong models appear to exploit the
product-context anchor that the synthetic suite did not provide; weaker
models flounder in the much larger absolute price range. The
product-grounded instantiation therefore acts as a difficulty multiplier
that \emph{amplifies} capability differences, which can be desirable for a
discriminative diagnostic.

\paragraph{Result 4: No-deal recognition is cleaner.}
The feasible-disagreement rate $FAGR_\pi^{-}$ is exactly $0.000$ for all
eleven product-grounded models: no agent ever agreed on an infeasible
product-grounded episode. This contrasts with the synthetic suite,
where $\mathrm{AgentExit}^{-}$ varied from 0.06 to 1.00 across models
(Fig.~\ref{fig:full}, bottom right). The product context (real category,
realistic average price, item description) appears to make infeasibility
easier to recognise even for agents that struggle on synthetic no-deal
scenarios.

\begin{figure}[t]
\centering
\includegraphics[width=0.98\linewidth]{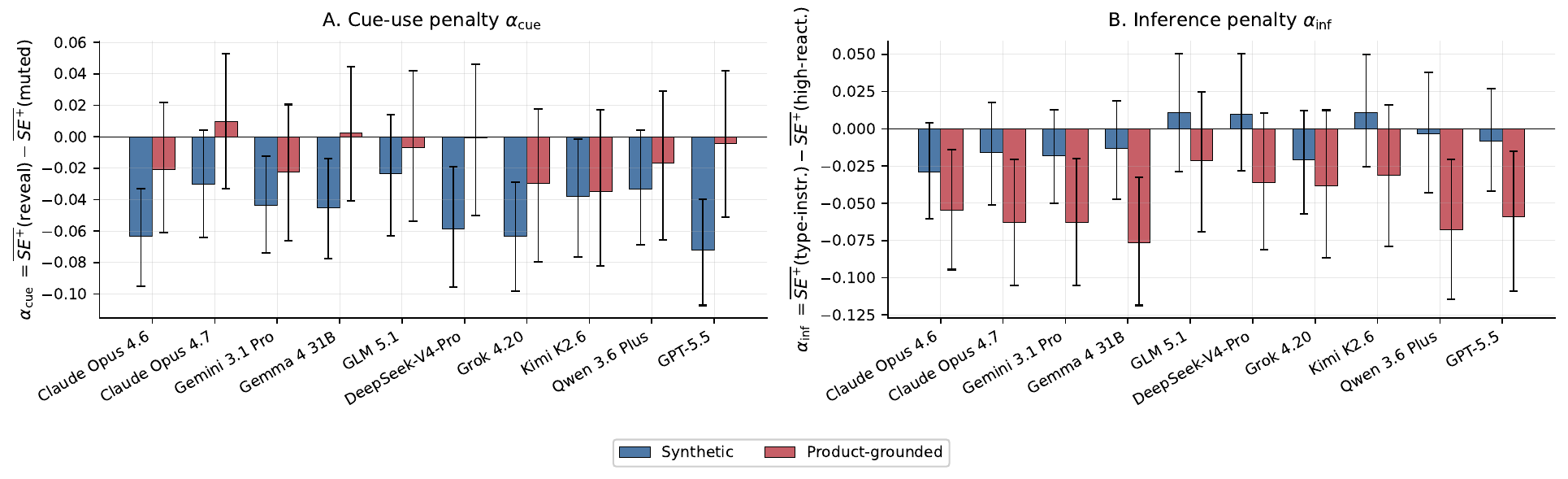}
\vspace{-0.5em}
\caption{\scriptsize Persistence of the two structural penalties under
product grounding (95\% bootstrap CIs, $B=2000$).
\textbf{A.}~$\alpha_{\mathrm{cue}}$ remains negative for 8/11 models but
attenuates: the salient product anchor partly insulates agents from
verbal-cue over-reaction. \textbf{B.}~$\alpha_{\mathrm{inf}}$ becomes
negative for \emph{all} 11 paired models, and amplifies for 10 of 11
(GPT-4o-mini is the exception), with three PG intervals excluding
zero---wide, skewed action ranges make correct latent-type inference
more decisive for surplus, so the information--action gap from
Finding~3 is sharpened.}
\label{fig:pg_penalties}
\end{figure}

\paragraph{Result 5: The two structural penalties $\alpha_{\mathrm{cue}}$ and
$\alpha_{\mathrm{inf}}$ both persist, in opposite directions.}
Figure~\ref{fig:pg_penalties} replicates Finding~2's
$\alpha_{\mathrm{cue}}$ and Finding~3's $\alpha_{\mathrm{inf}}$
contrasts on the cleaned product-grounded set. Recall:
\[
\alpha_{\mathrm{cue}} = \overline{SE_\pi^{+}}(\textsc{Candid},\textsc{Expressive})
- \overline{SE_\pi^{+}}(\textsc{Taciturn},\textsc{Strategic}),
\]
\[
\alpha_{\mathrm{inf}} = \overline{SE_\pi^{+}}(\textsc{Candid},\textsc{Taciturn})
- \overline{SE_\pi^{+}}(\textsc{Expressive},\textsc{Strategic}).
\]
\begin{itemize}
\item \textit{(i) Cue-use penalty $\alpha_{\mathrm{cue}}$ persists but
\emph{attenuates}.} The point estimate remains negative for 8 of 11
models in product-grounded, matching the synthetic direction;
magnitudes shrink (e.g.\ \texttt{Claude Opus 4.6}: $-0.063
\to -0.021$; \texttt{Grok~4.20}: $-0.063 \to -0.029$; \texttt{GPT-5.5}:
$-0.072 \to -0.004$), and several PG CIs straddle zero where the synthetic
ones did not.
\item \textit{(ii)
Inference penalty $\alpha_{\mathrm{inf}}$ persists and \emph{strengthens}
for the frontier roster.} The point estimate is negative for \emph{all}
11 paired models in product-grounded (vs.\ 8 of 11 in the synthetic
subset reproduced here), with substantially larger magnitudes for the
frontier subset (\texttt{Claude Opus 4.6}: $-0.029 \to -0.054$;
\texttt{Gemma 4 31B}: $-0.013 \to -0.076$; \texttt{Qwen 3.6 Plus}:
$-0.003 \to -0.068$; \texttt{GPT-5.5}: $-0.008 \to -0.059$);
\texttt{GPT-4o-mini} is the lone exception, with $\alpha_{\mathrm{inf}}$
that does not amplify ($-0.018 \to -0.009$). Three of the eleven
product-grounded intervals exclude zero, where none of the synthetic
intervals did.
\end{itemize}

The two movements are mutually consistent and follow naturally from the
geometry shift documented above. Anchoring on a salient product
reference price reduces the relative weight an agent places on the
counterpart's verbal expression, which attenuates the over-reaction to
cues that drives $\alpha_{\mathrm{cue}}$. The same wide, skewed action
range, however, makes \emph{correct latent-type inference} (the
counterpart's reservation, urgency, stance) more decisive for surplus,
so agents that fail to convert payoff-relevant latent structure into
action (i.e. the bottleneck $\alpha_{\mathrm{inf}}$ measures) are punished
more visibly. Both findings therefore replicate the structural claims
of Findings~2 and 3 in \S\ref{sec:experiments} in the main paper: the cue-use and information--action gaps are
properties of the agents, not of the synthetic distribution. Product
grounding sharpens the inference gap, in particular, into a
significant per-model effect for the strongest agents in the suite.

To ensure that the aforementioned observations are statistically significant,
we test the cross-suite shift in each penalty with a paired exact
Wilcoxon signed-rank test on
$\alpha_{\mathrm{PG}}-\alpha_{\mathrm{Synth}}$ across the eleven paired
models. Both shifts are significant in their reported directions:
$\alpha_{\mathrm{cue}}$ attenuates with median shift $+0.040$
($p=0.001$, two-sided), and $\alpha_{\mathrm{inf}}$ amplifies with
median shift $-0.045$ ($p=0.002$). The population-level pattern is
therefore not driven by any single model.

\begin{table}[h]
\centering
\small
\setlength{\tabcolsep}{4pt}
\begin{tabular}{lcc}
\toprule
Model & $\Delta\alpha_{\mathrm{cue}}$ (PG$-$Synth) & $\Delta\alpha_{\mathrm{inf}}$ (PG$-$Synth) \\
\midrule
Claude Opus 4.6 & $+0.043$ [$-0.011$, $+0.092$] & $-0.026$ [$-0.073$, $+0.028$] \\
GLM 5.1         & $+0.017$ [$-0.039$, $+0.075$] & $-0.032$ [$-0.092$, $+0.025$] \\
Claude Opus 4.7 & $+0.040$ [$-0.017$, $+0.096$] & $-0.047$ [$-0.101$, $+0.006$] \\
Gemma 4 31B     & $+0.048$ [$-0.007$, $+0.102$] & $\mathbf{-0.063}$ [$-0.114$, $-0.008$] \\
Gemini 3.1 Pro  & $+0.021$ [$-0.032$, $+0.074$] & $-0.045$ [$-0.096$, $+0.007$] \\
DeepSeek-V4-Pro & $+0.058$ [$-0.004$, $+0.121$] & $-0.046$ [$-0.107$, $+0.016$] \\
GPT-5.5         & $\mathbf{+0.068}$ [$+0.010$, $+0.125$] & $-0.051$ [$-0.110$, $+0.008$] \\
Qwen 3.6 Plus   & $+0.017$ [$-0.047$, $+0.077$] & $\mathbf{-0.064}$ [$-0.125$, $-0.003$] \\
Grok 4.20       & $+0.034$ [$-0.026$, $+0.092$] & $-0.018$ [$-0.078$, $+0.043$] \\
Kimi K2.6       & $+0.003$ [$-0.056$, $+0.062$] & $-0.042$ [$-0.101$, $+0.018$] \\
GPT-4o-mini     & $\mathbf{+0.046}$ [$+0.003$, $+0.089$] & $+0.010$ [$-0.035$, $+0.054$] \\
\bottomrule
\end{tabular}
\caption{\footnotesize Per-model cross-suite shift in each penalty,
$\Delta\alpha = \alpha_{\mathrm{PG}}-\alpha_{\mathrm{Synth}}$, with
95\% bootstrap CIs ($B=2000$). All 11 models attenuate the cue-use
penalty ($\Delta\alpha_{\mathrm{cue}}>0$); 10 of 11 also amplify the
inference penalty ($\Delta\alpha_{\mathrm{inf}}<0$), with
\texttt{GPT-4o-mini} the lone exception. Bolded entries are individually
distinguishable from zero. Underlying across-model paired Wilcoxon
tests give $p=0.001$ for the cue shift and $p=0.002$ for the inference
shift.}
\label{tab:pg_alpha_deltas}
\end{table}

\begin{table}[H]
\centering
\scriptsize
\setlength{\tabcolsep}{4pt}
\begin{tabular}{lccc}
\toprule
Model & Synthetic $SE_\pi^{+}$ & Product-grounded $SE_\pi^{+}$ & $\Delta$ (PG$-$Synth) \\
\midrule
Claude Opus 4.6 & 0.694 [0.681, 0.708] & 0.710 [0.691, 0.728] & $+0.016$ [$-0.006$, $+0.039$] \\
GLM 5.1         & 0.686 [0.670, 0.702] & 0.625 [0.604, 0.645] & $\mathbf{-0.060}$ [$-0.085$, $-0.033$] \\
Claude Opus 4.7 & 0.660 [0.646, 0.674] & 0.681 [0.662, 0.700] & $+0.021$ [$-0.003$, $+0.045$] \\
Gemma 4 31B     & 0.640 [0.626, 0.653] & 0.649 [0.629, 0.666] & $+0.009$ [$-0.014$, $+0.031$] \\
Gemini 3.1 Pro  & 0.639 [0.626, 0.652] & 0.671 [0.651, 0.689] & $\mathbf{+0.032}$ [$+0.010$, $+0.054$] \\
DeepSeek-V4-Pro & 0.618 [0.601, 0.634] & 0.613 [0.593, 0.633] & $-0.005$ [$-0.031$, $+0.020$] \\
GPT-5.5         & 0.606 [0.592, 0.620] & 0.618 [0.597, 0.638] & $+0.012$ [$-0.013$, $+0.036$] \\
Qwen 3.6 Plus   & 0.604 [0.587, 0.619] & 0.492 [0.470, 0.511] & $\mathbf{-0.112}$ [$-0.139$, $-0.087$] \\
Grok 4.20       & 0.601 [0.586, 0.615] & 0.562 [0.542, 0.583] & $\mathbf{-0.038}$ [$-0.063$, $-0.013$] \\
Kimi K2.6       & 0.597 [0.581, 0.611] & 0.498 [0.476, 0.518] & $\mathbf{-0.099}$ [$-0.126$, $-0.073$] \\
GPT-4o-mini     & 0.189 [0.176, 0.203] & 0.086 [0.074, 0.099] & $\mathbf{-0.103}$ [$-0.121$, $-0.085$] \\
\bottomrule
\end{tabular}
\vspace{0.5em}
\caption{\scriptsize Per-model overall $SE_\pi^{+}$ in the synthetic
vs.\ product-grounded suites, with 95\% bootstrap CIs ($B=2000$) on
per-episode means. $\Delta$ uses an independent two-sample bootstrap
(off-diagonal sample sizes differ); bolded entries have CIs that
exclude zero.}
\label{tab:pg_leaderboard}
\end{table}
\subsection{Deferred Experiment Results}\label{appdx:deferred_exp}
We present the deferred experiment results to supplement main experiment presented in \S\ref{sec:experiments}. Additional experiments and analysis are given in the following two sections: Appendix \ref{appdx:urgency-down} and Appendix \ref{app:strategic_profile}.

\subsubsection{Full-Sweep Results}
Tables~\ref{tab:aggregate_terminal_full}--\ref{tab:aggregate_protocol_full}
report the full aggregate results, decomposing performance into terminal
economic outcomes, opponent-modeling diagnostics, and protocol-compliance
statistics. Table~\ref{tab:family_regime_breakdown} further breaks down the
full-sweep experiments by regime and counterpart family, covering the three
regimes and six counterpart families for all evaluated LLM agents and the three
algorithmic baselines.
\begin{table*}[ht!]
\centering
\footnotesize
\setlength{\tabcolsep}{4.2pt}
\renewcommand{\arraystretch}{1.10}
\caption{Per-family and per-regime breakdown in the bilateral price-negotiation instantiation of \textsc{Terms-Bench} with 95\% CIs ($n{=}100$ episodes per cell).
In Panels A--B, entries are $SE_\pi^+$.
In Panel C, entries are critical-violation rates (\%).
Bold: best $SE_\pi^+$ within each feasible regime--family column.
Red: nonzero protocol violation rate.}
\label{tab:family_regime_breakdown}

\textbf{Panel A: Overlap regime} \\[0.35em]
\begin{adjustbox}{max width=\textwidth}
\begin{tabular}{lcccccc}
\toprule
\textbf{Agent} & \textbf{Adv.} & \textbf{Expr.} & \textbf{Candid} & \textbf{Stoch.} & \textbf{Strat.} & \textbf{Tac.} \\
\midrule
Claude 4.7
 & .578${\pm.050}$ & .648${\pm.048}$ & .614${\pm.048}$ & .694${\pm.050}$ & .654${\pm.049}$ & .675${\pm.051}$ \\
Grok 4
 & .519${\pm.056}$ & .567${\pm.049}$ & .540${\pm.051}$ & .639${\pm.048}$ & .610${\pm.048}$ & .652${\pm.046}$ \\
Gemini 3.1
 & .564${\pm.050}$ & .617${\pm.045}$ & .619${\pm.046}$ & .676${\pm.042}$ & \best{.690${\pm.040}$} & .669${\pm.048}$ \\
DeepSeek V4
 & .516${\pm.061}$ & .609${\pm.053}$ & .574${\pm.056}$ & .590${\pm.061}$ & .668${\pm.056}$ & .665${\pm.055}$ \\
Qwen3.6
 & .508${\pm.055}$ & .579${\pm.059}$ & .530${\pm.056}$ & .642${\pm.051}$ & .560${\pm.059}$ & .642${\pm.057}$ \\
Kimi K2.6
 & .501${\pm.055}$ & .557${\pm.054}$ & .558${\pm.055}$ & .584${\pm.059}$ & .601${\pm.051}$ & .654${\pm.053}$ \\
GPT-5.4
 & .455${\pm.054}$ & .530${\pm.056}$ & .505${\pm.055}$ & .596${\pm.053}$ & .526${\pm.059}$ & .517${\pm.063}$ \\
Doubao 2.0
 & .445${\pm.048}$ & .492${\pm.047}$ & .505${\pm.048}$ & .521${\pm.049}$ & .556${\pm.045}$ & .568${\pm.045}$ \\
GLM-5.1
 & \best{.590${\pm.062}$} & \best{.662${\pm.057}$} & \best{.663${\pm.057}$} & \best{.725${\pm.054}$} & .690${\pm.058}$ & \best{.678${\pm.057}$} \\
GPT-4o-mini
 & .171${\pm.046}$ & .167${\pm.042}$ & .190${\pm.046}$ & .137${\pm.040}$ & .234${\pm.049}$ & .230${\pm.050}$ \\
\midrule
Fixed 30\%
 & .400${\pm.050}$ & .371${\pm.045}$ & .375${\pm.053}$ & .389${\pm.054}$ & .387${\pm.051}$ & .377${\pm.050}$ \\
Fixed 10\%
 & .246${\pm.050}$ & .264${\pm.036}$ & .223${\pm.039}$ & .351${\pm.054}$ & .301${\pm.046}$ & .311${\pm.050}$ \\
Fixed 1\%
 & .191${\pm.038}$ & .253${\pm.035}$ & .213${\pm.037}$ & .320${\pm.055}$ & .275${\pm.043}$ & .282${\pm.044}$ \\
\bottomrule
\end{tabular}
\end{adjustbox}

\vspace{1.0em}
\textbf{Panel B: Urgency-shift regime} \\[0.35em]
\begin{adjustbox}{max width=\textwidth}
\begin{tabular}{lcccccc}
\toprule
\textbf{Agent} & \textbf{Adv.} & \textbf{Expr.} & \textbf{Candid} & \textbf{Stoch.} & \textbf{Strat.} & \textbf{Tac.} \\
\midrule
Claude 4.7
 & .613${\pm.048}$ & .692${\pm.048}$ & .643${\pm.047}$ & \best{.718${\pm.053}$} & .695${\pm.048}$ & .694${\pm.050}$ \\
Grok 4
 & .570${\pm.046}$ & .618${\pm.052}$ & .563${\pm.047}$ & .650${\pm.054}$ & .661${\pm.048}$ & .618${\pm.047}$ \\
Gemini 3.1
 & .566${\pm.045}$ & .649${\pm.043}$ & .621${\pm.045}$ & .672${\pm.051}$ & .673${\pm.043}$ & .648${\pm.046}$ \\
DeepSeek V4
 & .540${\pm.056}$ & .585${\pm.057}$ & .668${\pm.049}$ & .660${\pm.055}$ & .671${\pm.052}$ & .665${\pm.053}$ \\
Qwen3.6
 & .589${\pm.049}$ & .657${\pm.050}$ & .572${\pm.050}$ & .693${\pm.056}$ & .615${\pm.057}$ & .655${\pm.048}$ \\
Kimi K2.6
 & .540${\pm.056}$ & .641${\pm.053}$ & .596${\pm.046}$ & .678${\pm.055}$ & .607${\pm.051}$ & .642${\pm.053}$ \\
GPT-5.4
 & .477${\pm.051}$ & .551${\pm.058}$ & .515${\pm.052}$ & .613${\pm.056}$ & .541${\pm.052}$ & .550${\pm.062}$ \\
Doubao 2.0
 & .499${\pm.047}$ & .512${\pm.044}$ & .485${\pm.043}$ & .584${\pm.053}$ & .575${\pm.046}$ & .523${\pm.045}$ \\
GLM-5.1
 & \best{.650${\pm.047}$} & \best{.705${\pm.055}$} & \best{.698${\pm.047}$} & .711${\pm.063}$ & \best{.696${\pm.055}$} & \best{.758${\pm.047}$} \\
GPT-4o-mini
 & .187${\pm.051}$ & .223${\pm.046}$ & .149${\pm.041}$ & .150${\pm.038}$ & .227${\pm.047}$ & .208${\pm.050}$ \\
\midrule
Fixed 30\%
 & .438${\pm.054}$ & .369${\pm.048}$ & .379${\pm.050}$ & .419${\pm.052}$ & .351${\pm.046}$ & .387${\pm.051}$ \\
Fixed 10\%
 & .255${\pm.047}$ & .308${\pm.045}$ & .253${\pm.039}$ & .370${\pm.049}$ & .297${\pm.042}$ & .296${\pm.042}$ \\
Fixed 1\%
 & .234${\pm.038}$ & .297${\pm.042}$ & .253${\pm.039}$ & .367${\pm.049}$ & .291${\pm.038}$ & .298${\pm.042}$ \\
\bottomrule
\end{tabular}
\end{adjustbox}

\vspace{1.0em}
\textbf{Panel C: No-deal regime — critical-violation rate (\%)} \\[0.35em]
\begin{adjustbox}{max width=\textwidth}
\begin{tabular}{lcccccc}
\toprule
\textbf{Agent} & \textbf{Adv.} & \textbf{Expr.} & \textbf{Candid} & \textbf{Stoch.} & \textbf{Strat.} & \textbf{Tac.} \\
\midrule
Claude 4.7
 & 0.0 & 0.0 & 0.0 & 0.0 & 0.0 & 0.0 \\
Grok 4
 & \bad{4.0${\pm3.8}$} & \bad{7.0${\pm5.0}$} & \bad{5.0${\pm4.3}$} & \bad{3.0${\pm3.3}$} & \bad{3.0${\pm3.3}$} & \bad{5.0${\pm4.3}$} \\
Gemini 3.1
 & 0.0 & 0.0 & 0.0 & 0.0 & 0.0 & 0.0 \\
DeepSeek V4
 & \bad{3.0${\pm3.3}$} & \bad{2.0${\pm2.7}$} & \bad{1.0${\pm2.0}$} & \bad{2.0${\pm2.7}$} & 0.0 & \bad{3.0${\pm3.3}$} \\
Qwen3.6
 & \bad{6.0${\pm4.7}$} & \bad{10.0${\pm5.9}$} & \bad{3.0${\pm3.3}$} & \bad{7.0${\pm5.0}$} & \bad{7.0${\pm5.0}$} & \bad{4.0${\pm3.8}$} \\
Kimi K2.6
 & 0.0 & 0.0 & 0.0 & 0.0 & 0.0 & 0.0 \\
GPT-5.4
 & 0.0 & 0.0 & 0.0 & 0.0 & 0.0 & 0.0 \\
Doubao 2.0
 & 0.0 & 0.0 & \bad{1.0${\pm2.0}$} & 0.0 & 0.0 & 0.0 \\
GLM-5.1
 & \bad{5.0${\pm4.3}$} & \bad{5.0${\pm4.3}$} & \bad{5.0${\pm4.3}$} & \bad{2.0${\pm2.7}$} & \bad{2.0${\pm2.7}$} & \bad{5.0${\pm4.3}$} \\
GPT-4o-mini
 & 0.0 & 0.0 & 0.0 & 0.0 & 0.0 & 0.0 \\
\midrule
Fixed 30\%
 & 0.0 & 0.0 & 0.0 & 0.0 & 0.0 & 0.0 \\
Fixed 10\%
 & 0.0 & 0.0 & 0.0 & 0.0 & 0.0 & 0.0 \\
Fixed 1\%
 & 0.0 & 0.0 & 0.0 & 0.0 & 0.0 & 0.0 \\
\bottomrule
\end{tabular}
\end{adjustbox}
\end{table*}
\begin{table*}[t]
\centering\scriptsize
\setlength{\tabcolsep}{3.2pt}\renewcommand{\arraystretch}{1.10}
\begin{adjustbox}{max width=\textwidth}
\begin{tabular}{lccccccccccc}
\toprule
\textbf{Agent} & {$SE_\pi^+ \uparrow$} & {$\mathrm{AGR}_\pi^+ \uparrow$} & {$CSE_\pi^+ \uparrow$} & {$\mathrm{FAGR}_\pi^- \downarrow$} & {$\bar u \uparrow$} & {$u^\star$} & {Gap $\downarrow$} & {Oracle $\uparrow$} & {$\mathrm{AGR}_{\rm all} \uparrow$} & {$\mathbb{E}[u \mid \mathrm{deal}] \uparrow$} & {Safe$^-$} \\
\midrule
Claude Opus 4.7
 & 0.660$_{\pm0.014}$ & 98.2$_{\pm0.8}$ & 0.672$_{\pm0.014}$ & 0.00 & 11.09$_{\pm0.50}$ & 15.00$_{\pm0.49}$ & 3.90$_{\pm0.36}$ & 74.1$_{\pm2.3}$ & 65.4$_{\pm2.2}$ & 16.95$_{\pm0.50}$ & 100.0 \\
Grok 4
 & 0.601$_{\pm0.015}$ & 99.1$_{\pm0.5}$ & 0.606$_{\pm0.014}$ & 0.00 & 10.03$_{\pm0.46}$ & 15.00$_{\pm0.49}$ & 4.96$_{\pm0.36}$ & 67.1$_{\pm2.1}$ & 66.1$_{\pm2.2}$ & 15.19$_{\pm0.48}$ & 100.0 \\
Gemini-3.1-Pro
 & 0.639$_{\pm0.013}$ & 99.7$_{\pm0.3}$ & 0.641$_{\pm0.013}$ & 0.00 & 10.58$_{\pm0.47}$ & 15.00$_{\pm0.49}$ & 4.42$_{\pm0.35}$ & 70.7$_{\pm2.1}$ & 66.4$_{\pm2.2}$ & 15.92$_{\pm0.47}$ & 100.0 \\
DeepSeek-V4-Pro
 & 0.618$_{\pm0.016}$ & 97.5$_{\pm0.9}$ & 0.633$_{\pm0.016}$ & 0.00 & 10.53$_{\pm0.50}$ & 15.00$_{\pm0.49}$ & 4.47$_{\pm0.39}$ & 70.3$_{\pm2.4}$ & 65.0$_{\pm2.2}$ & 16.19$_{\pm0.54}$ & 100.0 \\
Qwen3.6-Plus
 & 0.604$_{\pm0.016}$ & 98.2$_{\pm0.7}$ & 0.614$_{\pm0.015}$ & \bad{0.17$_{\pm0.33}$} & 10.31$_{\pm0.49}$ & 15.00$_{\pm0.49}$ & 4.69$_{\pm0.39}$ & 68.9$_{\pm2.4}$ & 65.6$_{\pm2.2}$ & 15.73$_{\pm0.53}$ & 99.8$_{\pm0.3}$ \\
Kimi-K2.6
 & 0.597$_{\pm0.016}$ & 97.1$_{\pm1.0}$ & 0.614$_{\pm0.015}$ & 0.00 & 10.17$_{\pm0.48}$ & 15.00$_{\pm0.49}$ & 4.83$_{\pm0.39}$ & 67.9$_{\pm2.3}$ & 64.7$_{\pm2.2}$ & 15.71$_{\pm0.52}$ & 100.0 \\
GPT-5.4
 & 0.531$_{\pm0.016}$ & 99.4$_{\pm0.4}$ & 0.535$_{\pm0.016}$ & 0.00 & 9.04$_{\pm0.46}$ & 15.00$_{\pm0.49}$ & 5.95$_{\pm0.40}$ & 60.4$_{\pm2.3}$ & 66.3$_{\pm2.2}$ & 13.64$_{\pm0.53}$ & 100.0 \\
Doubao-Seed-2.0-Pro
 & 0.522$_{\pm0.014}$ & 99.9$_{\pm0.2}$ & 0.523$_{\pm0.014}$ & 0.00 & 8.61$_{\pm0.40}$ & 15.00$_{\pm0.49}$ & 6.38$_{\pm0.35}$ & 57.6$_{\pm1.9}$ & 66.6$_{\pm2.2}$ & 12.93$_{\pm0.42}$ & 100.0 \\
GLM-5.1
 & 0.686$_{\pm0.016}$ & 95.1$_{\pm1.2}$ & 0.721$_{\pm0.014}$ & 0.00 & 11.70$_{\pm0.53}$ & 15.00$_{\pm0.49}$ & 3.30$_{\pm0.39}$ & 78.2$_{\pm2.5}$ & 63.4$_{\pm2.2}$ & 18.45$_{\pm0.53}$ & 100.0 \\
GPT-4o-mini
 & 0.189$_{\pm0.013}$ & 52.2$_{\pm2.8}$ & 0.363$_{\pm0.016}$ & 0.00 & 3.37$_{\pm0.27}$ & 15.00$_{\pm0.49}$ & 11.63$_{\pm0.46}$ & 22.4$_{\pm1.7}$ & 34.8$_{\pm2.2}$ & 9.69$_{\pm0.49}$ & 100.0 \\
\midrule
Fixed 30\%
 & 0.387$_{\pm0.015}$ & 99.9$_{\pm0.2}$ & 0.387$_{\pm0.015}$ & 0.00 & 6.50$_{\pm0.36}$ & 15.00$_{\pm0.49}$ & 8.49$_{\pm0.41}$ & 43.4$_{\pm2.0}$ & 66.6$_{\pm2.2}$ & 9.76$_{\pm0.44}$ & 100.0 \\
Fixed 10\%
 & 0.290$_{\pm0.013}$ & 94.5$_{\pm1.3}$ & 0.307$_{\pm0.013}$ & 0.00 & 5.08$_{\pm0.32}$ & 15.00$_{\pm0.49}$ & 9.92$_{\pm0.43}$ & 33.9$_{\pm1.8}$ & 63.0$_{\pm2.2}$ & 8.06$_{\pm0.42}$ & 100.0 \\
Fixed 1\%
 & 0.273$_{\pm0.012}$ & 92.2$_{\pm1.5}$ & 0.296$_{\pm0.013}$ & 0.00 & 4.77$_{\pm0.30}$ & 15.00$_{\pm0.49}$ & 10.23$_{\pm0.42}$ & 31.8$_{\pm1.7}$ & 61.5$_{\pm2.2}$ & 7.75$_{\pm0.39}$ & 100.0 \\
\bottomrule
\end{tabular}
\end{adjustbox}
\caption{Aggregate terminal and oracle-reference performance in the bilateral price-negotiation instantiation of \textsc{Terms-Bench} with 95\% CIs ($n{=}1{,}800$ episodes per agent). CIs are normal-approximation half-widths for means and binomial half-widths for proportions. $u^\star$ is the per-episode optimal utility under a full-information oracle (cell-mean), averaged across all episodes (with $u^\star{=}0$ on no-deal); \textit{Gap} $= u^\star - \bar u$; \textit{Oracle} $= 100\bar u/u^\star$. Percent-valued columns reported in pp.}
\label{tab:aggregate_terminal_full}
\end{table*}

\begin{table*}[t]
\centering\scriptsize
\setlength{\tabcolsep}{4.0pt}\renewcommand{\arraystretch}{1.10}
\begin{adjustbox}{max width=\textwidth}
\begin{tabular}{lccccccc}
\toprule
\textbf{Agent} & {$BE_{\mathrm{type}} \downarrow$} & {$BE_r \downarrow$} & {$BE_\kappa \downarrow$} & {Brier$_\eta \downarrow$} & {Stance acc.\ $\uparrow$} & {$BE_{\mathrm{joint}}^{\ell_2} \downarrow$} & {Type--stance mism.\ $\downarrow$} \\
\midrule
Claude Opus 4.7
 & 0.229$_{\pm0.006}$ & 0.118$_{\pm0.014}$ & 0.243$_{\pm0.015}$ & 0.325$_{\pm0.023}$ & 45.8$_{\pm4.3}$ & 0.296$_{\pm0.012}$ & 0.301$_{\pm0.012}$ \\
Grok 4
 & 0.212$_{\pm0.006}$ & 0.112$_{\pm0.009}$ & 0.200$_{\pm0.008}$ & 0.324$_{\pm0.019}$ & 44.5$_{\pm4.1}$ & 0.251$_{\pm0.010}$ & 0.289$_{\pm0.012}$ \\
Gemini-3.1-Pro
 & 0.271$_{\pm0.012}$ & 0.117$_{\pm0.007}$ & 0.344$_{\pm0.020}$ & 0.352$_{\pm0.033}$ & 45.8$_{\pm4.7}$ & 0.386$_{\pm0.017}$ & 0.335$_{\pm0.016}$ \\
DeepSeek-V4-Pro
 & 0.228$_{\pm0.005}$ & 0.125$_{\pm0.013}$ & 0.223$_{\pm0.010}$ & 0.337$_{\pm0.017}$ & 42.9$_{\pm3.6}$ & 0.283$_{\pm0.009}$ & 0.306$_{\pm0.011}$ \\
Qwen3.6-Plus
 & 0.237$_{\pm0.006}$ & 0.125$_{\pm0.015}$ & 0.246$_{\pm0.014}$ & 0.339$_{\pm0.017}$ & 43.3$_{\pm3.3}$ & 0.304$_{\pm0.009}$ & 0.313$_{\pm0.012}$ \\
Kimi-K2.6
 & 0.236$_{\pm0.009}$ & 0.120$_{\pm0.012}$ & 0.243$_{\pm0.013}$ & 0.345$_{\pm0.030}$ & 43.3$_{\pm5.4}$ & 0.298$_{\pm0.010}$ & 0.310$_{\pm0.017}$ \\
GPT-5.4
 & 0.242$_{\pm0.009}$ & 0.126$_{\pm0.018}$ & 0.262$_{\pm0.017}$ & 0.337$_{\pm0.031}$ & 44.0$_{\pm5.7}$ & 0.318$_{\pm0.014}$ & 0.316$_{\pm0.017}$ \\
Doubao-Seed-2.0-Pro
 & 0.247$_{\pm0.009}$ & 0.121$_{\pm0.013}$ & 0.256$_{\pm0.011}$ & 0.364$_{\pm0.028}$ & 42.0$_{\pm4.7}$ & 0.307$_{\pm0.010}$ & 0.319$_{\pm0.016}$ \\
GLM-5.1
 & 0.218$_{\pm0.008}$ & 0.106$_{\pm0.007}$ & 0.221$_{\pm0.011}$ & 0.326$_{\pm0.027}$ & 44.8$_{\pm5.8}$ & 0.268$_{\pm0.009}$ & 0.293$_{\pm0.018}$ \\
GPT-4o-mini
 & 0.251$_{\pm0.005}$ & 0.215$_{\pm0.007}$ & 0.192$_{\pm0.008}$ & 0.345$_{\pm0.011}$ & 38.9$_{\pm3.2}$ & 0.311$_{\pm0.009}$ & 0.340$_{\pm0.013}$ \\
\midrule
Fixed 30\%
 & \NA & \NA & \NA & \NA & \NA & \NA & \NA \\
Fixed 10\%
 & \NA & \NA & \NA & \NA & \NA & \NA & \NA \\
Fixed 1\%
 & \NA & \NA & \NA & \NA & \NA & \NA & \NA \\
\bottomrule
\end{tabular}
\end{adjustbox}
\caption{Aggregate opponent-modeling diagnostics in the bilateral price-negotiation instantiation of \textsc{Terms-Bench} with 95\% CIs (between-cell half-widths over 18 regime$\times$family cells, $n{=}100$ episodes per cell). Lower is better for belief-error metrics and Brier score; higher is better for stance accuracy. Fixed-concession baselines do not produce belief estimates.}
\label{tab:aggregate_belief_full}
\end{table*}

\begin{table*}[t]
\centering\scriptsize
\setlength{\tabcolsep}{2.8pt}\renewcommand{\arraystretch}{1.10}
\begin{adjustbox}{max width=\textwidth}
\begin{tabular}{lcccccccccccc}
\toprule
\textbf{Agent} & {CritViol} & {Bound} & {Res.} & {Invalid} & {Mono.} & {Budget} & {Any viol.} & {Agent acc.} & {Counter acc.} & {Agent rej.} & {Walkaway} & {Timeout} \\
\midrule
Claude Opus 4.7
 & 0.00 & 0.00 & 0.00 & 0.00 & 0.00 & 0.00 & 0.00 & 13.4$_{\pm1.6}$ & 52.1$_{\pm2.3}$ & 2.6$_{\pm0.7}$ & 31.9$_{\pm2.2}$ & 0.0 \\
Grok 4
 & \bad{1.50$_{\pm0.56}$} & 0.00 & \bad{1.50$_{\pm0.56}$} & 0.00 & 0.00 & 0.00 & \bad{1.50$_{\pm0.56}$} & 14.2$_{\pm1.6}$ & 51.9$_{\pm2.3}$ & 15.3$_{\pm1.7}$ & 18.6$_{\pm1.8}$ & 0.0 \\
Gemini-3.1-Pro
 & 0.00 & 0.00 & 0.00 & 0.00 & 0.00 & 0.00 & 0.00 & 16.1$_{\pm1.7}$ & 50.3$_{\pm2.3}$ & 2.6$_{\pm0.7}$ & 30.9$_{\pm2.1}$ & 0.0 \\
DeepSeek-V4-Pro
 & \bad{0.61$_{\pm0.36}$} & 0.00 & \bad{0.61$_{\pm0.36}$} & 0.00 & 0.00 & 0.00 & \bad{0.61$_{\pm0.36}$} & 19.5$_{\pm1.8}$ & 45.5$_{\pm2.3}$ & 6.1$_{\pm1.1}$ & 28.9$_{\pm2.1}$ & 0.1$_{\pm0.1}$ \\
Qwen3.6-Plus
 & \bad{2.06$_{\pm0.66}$} & 0.00 & \bad{2.06$_{\pm0.66}$} & 0.00 & 0.00 & 0.00 & \bad{2.06$_{\pm0.66}$} & 18.2$_{\pm1.8}$ & 47.3$_{\pm2.3}$ & 5.2$_{\pm1.0}$ & 29.2$_{\pm2.1}$ & 0.0 \\
Kimi-K2.6
 & 0.00 & 0.00 & 0.00 & 0.00 & 0.00 & 0.00 & 0.00 & 19.7$_{\pm1.8}$ & 45.1$_{\pm2.3}$ & 18.0$_{\pm1.8}$ & 17.3$_{\pm1.7}$ & 0.0 \\
GPT-5.4
 & 0.00 & 0.00 & 0.00 & 0.00 & 0.00 & 0.00 & 0.00 & 21.4$_{\pm1.9}$ & 44.8$_{\pm2.3}$ & 6.4$_{\pm1.1}$ & 27.3$_{\pm2.1}$ & 0.0 \\
Doubao-Seed-2.0-Pro
 & \bad{0.06$_{\pm0.11}$} & 0.00 & \bad{0.06$_{\pm0.11}$} & 0.00 & 0.00 & 0.00 & \bad{0.06$_{\pm0.11}$} & 12.4$_{\pm1.5}$ & 54.2$_{\pm2.3}$ & 5.4$_{\pm1.0}$ & 28.0$_{\pm2.1}$ & 0.0 \\
GLM-5.1
 & \bad{1.33$_{\pm0.53}$} & 0.00 & \bad{1.33$_{\pm0.53}$} & 0.00 & 0.00 & 0.00 & \bad{1.33$_{\pm0.53}$} & 14.6$_{\pm1.6}$ & 48.8$_{\pm2.3}$ & 6.4$_{\pm1.1}$ & 30.2$_{\pm2.1}$ & 0.0 \\
GPT-4o-mini
 & 0.00 & 0.00 & 0.00 & 0.00 & 0.00 & 0.00 & 0.00 & 22.8$_{\pm1.9}$ & 11.9$_{\pm1.5}$ & 65.2$_{\pm2.2}$ & 0.0 & 0.0 \\
\midrule
Fixed 30\%
 & 0.00 & 0.00 & 0.00 & 0.00 & 0.00 & 0.00 & 0.00 & 52.5$_{\pm2.3}$ & 14.1$_{\pm1.6}$ & 0.0 & 32.3$_{\pm2.2}$ & 1.1$_{\pm0.5}$ \\
Fixed 10\%
 & 0.00 & 0.00 & 0.00 & 0.00 & 0.00 & 0.00 & 0.00 & 61.4$_{\pm2.2}$ & 1.6$_{\pm0.6}$ & 0.0 & 36.1$_{\pm2.2}$ & 0.9$_{\pm0.4}$ \\
Fixed 1\%
 & 0.00 & 0.00 & 0.00 & 0.00 & 0.00 & 0.00 & 0.00 & 61.4$_{\pm2.2}$ & 0.1$_{\pm0.1}$ & 0.0 & 38.4$_{\pm2.2}$ & 0.1$_{\pm0.2}$ \\
\bottomrule
\end{tabular}
\end{adjustbox}
\caption{Aggregate protocol-compliance and interaction-outcome rates in the bilateral price-negotiation instantiation of \textsc{Terms-Bench} with 95\% CIs (binomial half-widths, $n{=}1{,}800$). All entries are percentages. Red shading marks nonzero protocol-violation rates.}
\label{tab:aggregate_protocol_full}
\end{table*}

\subsubsection{Runtime and Inference Cost Comparison}
We additionally compare the practical inference profile of the evaluated LLM agents along two axes: interaction runtime, measured as the mean number of negotiation rounds completed per episode, and estimated inference cost per episode. Runtime varies substantially across agents: \texttt{GLM 5.1}, \texttt{Claude Opus 4.7}, and \texttt{Claude Opus 4.6} sustain the longest negotiations on average, while \texttt{GPT-4o-mini} terminates much earlier. This runtime measure should be interpreted primarily behaviorally, and only indirectly as a driver of wall-clock latency: it captures how long each agent keeps the bargaining process alive before agreement, rejection, walk-away, or timeout, but does not itself measure elapsed seconds.

Inference cost shows a different ordering. Claude Opus models are the most expensive under the \texttt{OpenRouter} pricing assumptions, followed by \texttt{GPT-5.4} and \texttt{Gemini 3.1 Pro}. Several models with relatively long interaction horizons, such as \texttt{GLM 5.1} and \texttt{Kimi K2.6}, remain substantially cheaper per episode because their token prices are lower. Conversely, \texttt{GPT-4o-mini} is both short-horizon and very low-cost, but also much weaker in benchmark performance.

All cost and runtime measurements here reflect \texttt{OpenRouter}-routed inference rather than direct provider inference. \texttt{OpenRouter} may route requests through lower-cost backend providers or cloud compute paths, so these estimates should be treated as practical \texttt{OpenRouter} deployment statistics rather than canonical native-provider latency or pricing.
\begin{figure}[t]
    \centering

    \begin{subfigure}[t]{0.49\linewidth}
        \centering
        \includegraphics[width=\linewidth]{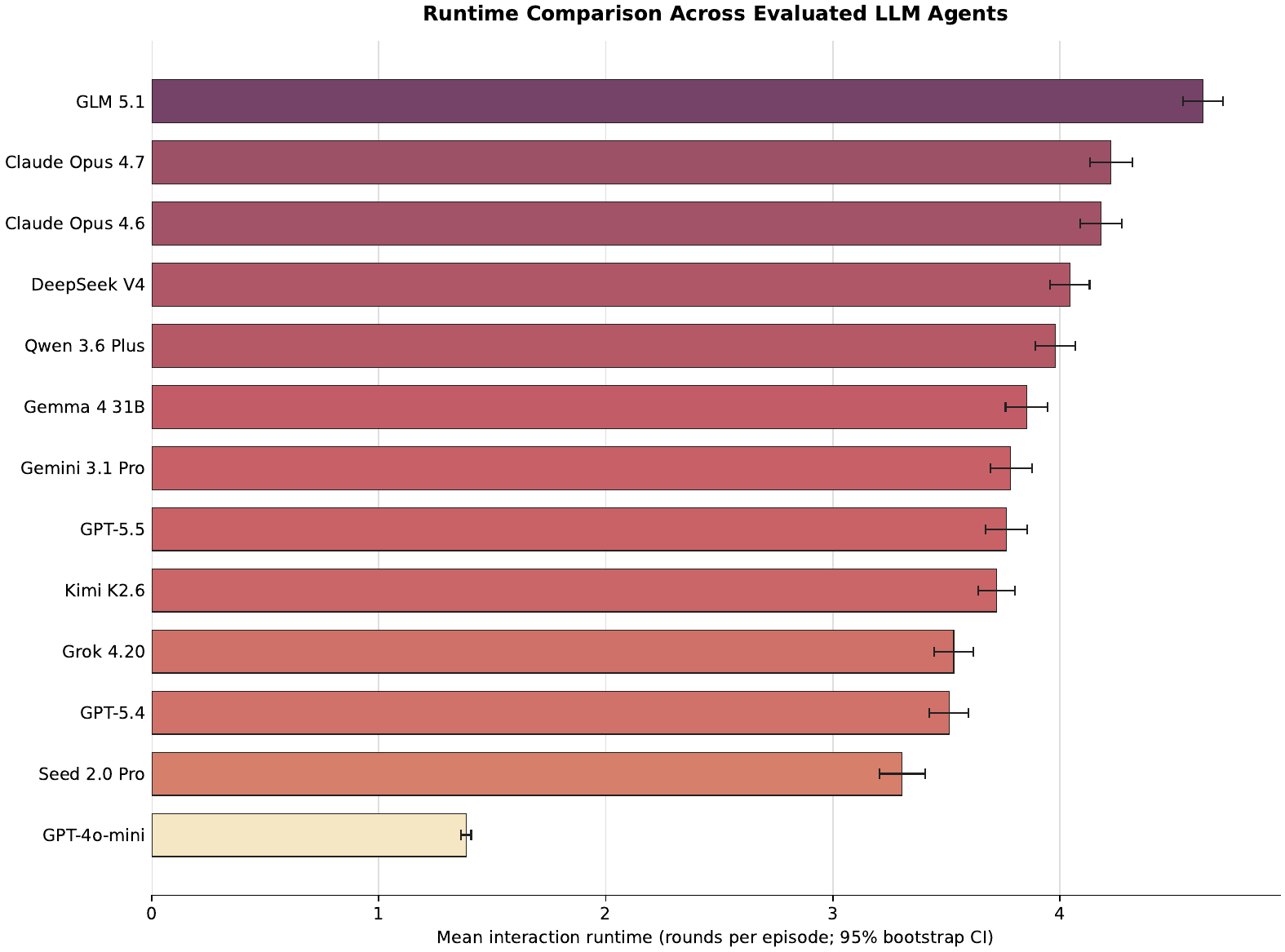}
        \caption{\footnotesize Mean interaction runtime across evaluated LLM agents on the bilateral price-negotiation instantiation of \textsc{Terms-Bench}. Bars show the mean number of negotiation rounds played per episode, with 95\% bootstrap confidence intervals across episodes. These runtime statistics reflect inference through \texttt{OpenRouter}, and should not be interpreted as native-provider latency: \texttt{OpenRouter} may route requests through different backend providers or lower-cost cloud compute paths, so observed behavior can differ from running the same model directly with its own provider.}
        \label{fig:runtime_synth}
    \end{subfigure}
    \hfill
    \begin{subfigure}[t]{0.49\linewidth}
        \centering
        \includegraphics[width=\linewidth]{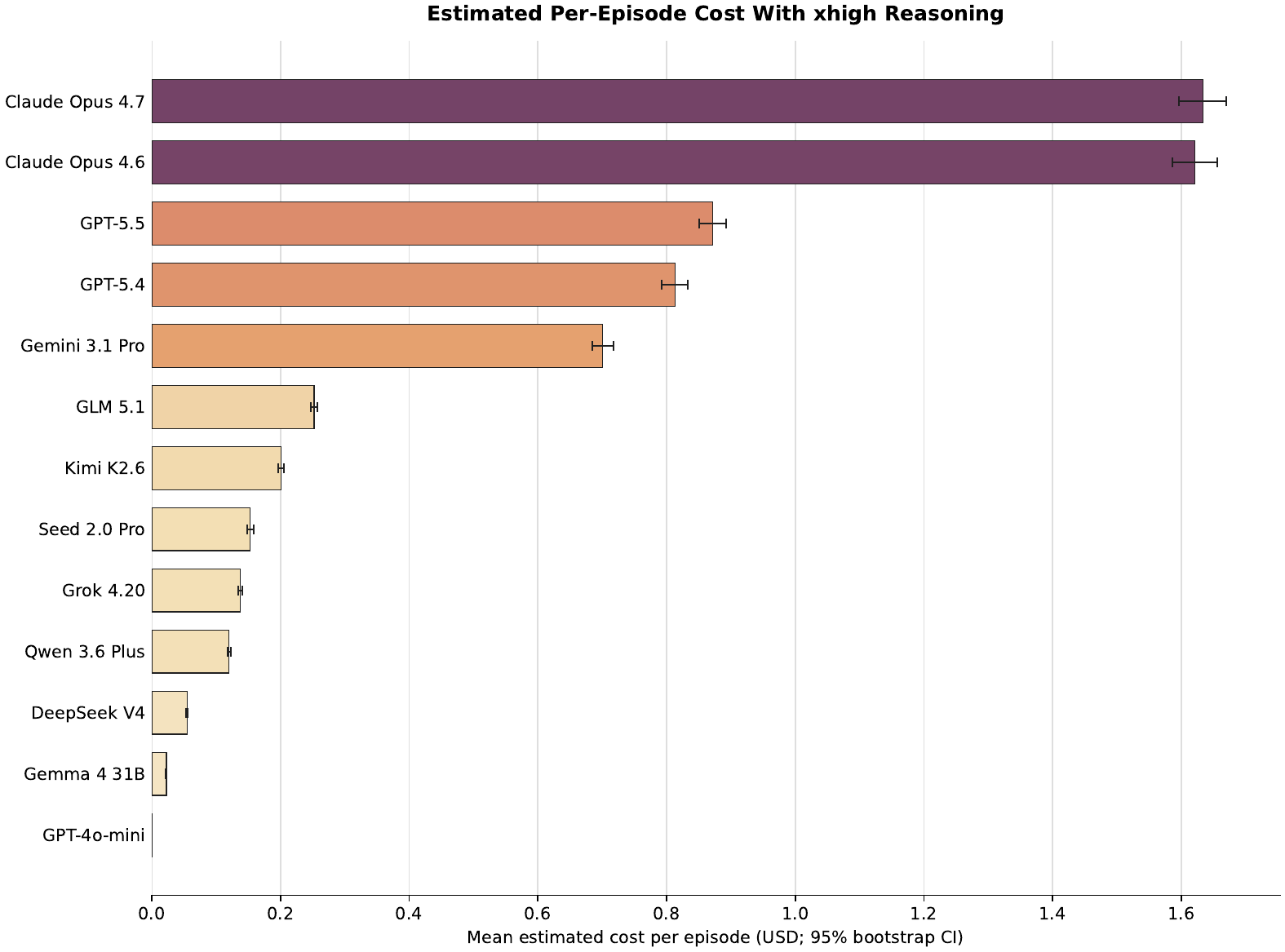}
        \caption{\footnotesize Estimated mean per-episode inference cost across evaluated LLM agents. Bars show estimated USD cost per episode with 95\% bootstrap confidence intervals, using current \texttt{OpenRouter} token prices and completion token estimates. For reasoning-capable models, costs assume \texttt{xhigh} reasoning effort, modeled as 15,200 reasoning tokens per LLM call under the benchmark’s 16,000-token completion budget. We note that these estimates represent \texttt{OpenRouter}-routed inference prices, not necessarily native-provider pricing.}
        \label{fig:cost_synth}
    \end{subfigure}

    \caption{\footnotesize Runtime and estimated inference cost across evaluated LLM agents on the bilateral price-negotiation instantiation of \textsc{Terms-Bench}.}
    \label{fig:runtime_cost_synth}
\end{figure}

\subsubsection{Statistical significance of $\alpha_{\mathrm{cue}}$ and
$\alpha_{\mathrm{inf}}$ on the synthetic suite}
\label{app:alpha_significance}

We accompany the point estimates in Fig.~\ref{fig:family_heatmap} with
95\% percentile bootstrap CIs and across-model significance tests for
the two structural penalties used in Findings 2 and 3. For each model
we resample per-episode $SE_\pi^{+}$ values within each family pair
($B=2000$) and form a two-sample bootstrap CI on the contrast
\[
\alpha_{\mathrm{cue}} = \overline{SE^{+}}(\textsc{Candid},\textsc{Expressive})
- \overline{SE^{+}}(\textsc{Taciturn},\textsc{Strategic}),
\]
\[
\alpha_{\mathrm{inf}} = \overline{SE^{+}}(\textsc{Candid},\textsc{Taciturn})
- \overline{SE^{+}}(\textsc{Expressive},\textsc{Strategic}),
\]
where negative values indicate the cue-use penalty (F2) and the
inference penalty (F3) respectively. Across the 13 evaluated LLMs we
also report a sign test against $\mathrm{Pr}(\alpha < 0) = 0.5$ and an
exact one-sided paired Wilcoxon signed-rank test against
$\mathrm{median}(\alpha) = 0$.

\paragraph{Results.}
$\alpha_{\mathrm{cue}}$ is negative for all 13 models (sign-test
$p=0.0001$); 9 of 13 individual CIs strictly exclude zero, and the
exact Wilcoxon test against the negative direction reaches the
precision floor of the exact null distribution
($p_{<0}=0.0001$). The four models whose CIs cross zero
(\texttt{GLM 5.1}, \texttt{Claude Opus 4.7}, \texttt{Qwen 3.6 Plus},
\texttt{GPT-5.4}) all carry the same sign as the population. The
cue-use penalty in F2 is therefore robust at both the per-model and
the across-model level.

$\alpha_{\mathrm{inf}}$ is negative for 10 of 13 models; the sign
test is marginally significant ($p=0.0461$) and no individual CI
excludes zero. The exact Wilcoxon test against the negative direction,
which uses ranks rather than signs alone, is significant
($p_{<0}=0.0085$). The inference penalty in F3 is therefore a
population-level effect: the rank-based test resolves it cleanly while
the per-model bootstrap is underpowered to do so at the single-model
level.

For both results, refer to Table~\ref{tab:alpha_significance_synthetic}.

\begin{table}[h]
\centering
\scriptsize
\setlength{\tabcolsep}{4pt}
\begin{tabular}{lcc}
\toprule
Model & $\alpha_{\mathrm{cue}}$ \,[95\% CI] & $\alpha_{\mathrm{inf}}$ \,[95\% CI] \\
\midrule
Claude Opus 4.6 & $\mathbf{-0.063}$ [$-0.095$, $-0.032$] & $-0.029$ [$-0.061$, $+0.002$] \\
GLM 5.1         & $-0.023$ [$-0.061$, $+0.015$] & $+0.011$ [$-0.029$, $+0.051$] \\
Claude Opus 4.7 & $-0.030$ [$-0.065$, $+0.005$] & $-0.016$ [$-0.050$, $+0.019$] \\
Gemma 4 31B     & $\mathbf{-0.045}$ [$-0.079$, $-0.010$] & $-0.013$ [$-0.046$, $+0.020$] \\
Gemini 3.1 Pro  & $\mathbf{-0.044}$ [$-0.074$, $-0.012$] & $-0.018$ [$-0.048$, $+0.014$] \\
DeepSeek-V4-Pro & $\mathbf{-0.058}$ [$-0.097$, $-0.021$] & $+0.010$ [$-0.028$, $+0.047$] \\
Qwen 3.6 Plus   & $-0.033$ [$-0.074$, $+0.005$] & $-0.003$ [$-0.043$, $+0.036$] \\
Grok 4.20      & $\mathbf{-0.063}$ [$-0.097$, $-0.028$] & $-0.021$ [$-0.058$, $+0.015$] \\
Kimi K2.6       & $\mathbf{-0.038}$ [$-0.075$, $-0.001$] & $+0.011$ [$-0.026$, $+0.049$] \\
GPT-5.4         & $-0.009$ [$-0.051$, $+0.032$] & $-0.016$ [$-0.056$, $+0.023$] \\
GPT-5.5         & $\mathbf{-0.072}$ [$-0.106$, $-0.038$] & $-0.008$ [$-0.043$, $+0.025$] \\
Doubao 2.0 Pro  & $\mathbf{-0.057}$ [$-0.090$, $-0.024$] & $-0.014$ [$-0.046$, $+0.018$] \\
GPT-4o-mini     & $\mathbf{-0.042}$ [$-0.076$, $-0.010$] & $-0.018$ [$-0.049$, $+0.015$] \\
\midrule
\emph{Across-model, $n=13$} & 13/13 $<0$; 9/13 sig.; Wilcoxon $p_{<0}=0.0001$ & 10/13 $<0$; 0/13 sig.; Wilcoxon $p_{<0}=0.0085$ \\
\bottomrule
\end{tabular}
\vspace{0.5em}
\caption{\scriptsize Per-model $\alpha_{\mathrm{cue}}$ and
$\alpha_{\mathrm{inf}}$ on the synthetic paper suite, with 95\%
two-sample percentile bootstrap CIs ($B=2000$). Bolded point estimates
have CIs that exclude zero. The bottom row reports across-model
statistics: count of negative point estimates, count of CIs strictly
excluding zero, and the one-sided exact Wilcoxon signed-rank
$p$-value against $H_0$: $\mathrm{median}(\alpha)=0$ with alternative
$\mathrm{median}(\alpha)<0$.}
\label{tab:alpha_significance_synthetic}
\end{table}

\subsection{Reverse Urgency-Shift Direction Experiment}
\label{appdx:urgency-down}

Main paper results use the counterpart-more-urgent direction of the
urgency-shift regime. Here we report the reverse direction, where the evaluated
agent is more time-pressured than the counterpart. We conduct the experiments on three representing models spanning the performance levels in the main experiment (Table \ref{tab:main_results}): $\texttt{Claude Opus 4.6}$ (top); \texttt{Grok 4.20} (mid); and \texttt{GPT-4o-mini} (low). Episode construction, common
seeds, inference settings, role/opener decomposition, and reporting conventions
are otherwise held fixed. This auxiliary check asks whether the headline ordering
reflects the sign of the urgency asymmetry, or instead a more stable model-level
difference in negotiation behavior. The results with $95\%$ bootstrap CIs are reported in Table \ref{tab:urgency-shift-direction}.

\begin{table}[t]
\centering
\small
\begin{tabular}{llrrrr}
\toprule
Model & Direction & $n$ & $SE^+_\pi$ & $AGR^+_\pi$ & $CSE^+_\pi$ \\
\midrule
GPT-4o-mini & Counterpart more urgent & 600
& 0.191 [0.173, 0.209]
& 0.530 [0.492, 0.570]
& 0.360 [0.337, 0.383] \\
GPT-4o-mini & Agent more urgent & 600
& 0.192 [0.173, 0.211]
& 0.537 [0.497, 0.575]
& 0.357 [0.334, 0.380] \\
Grok 4.20 & Counterpart more urgent & 600
& 0.614 [0.593, 0.634]
& 0.995 [0.988, 1.000]
& 0.617 [0.596, 0.637] \\
Grok 4.20 & Agent more urgent & 578
& 0.595 [0.575, 0.616]
& 0.998 [0.995, 1.000]
& 0.596 [0.576, 0.616] \\
Claude Opus 4.6 & Counterpart more urgent & 600
& 0.738 [0.720, 0.755]
& 0.997 [0.992, 1.000]
& 0.740 [0.722, 0.758] \\
Claude Opus 4.6 & Agent more urgent & 541
& 0.689 [0.670, 0.709]
& 0.996 [0.991, 1.000]
& 0.691 [0.673, 0.711] \\
\bottomrule
\end{tabular}
\vspace{0.5em}
\caption{\footnotesize Urgency-shift direction reversal. Metrics are reported with 95\%
percentile bootstrap confidence intervals over episodes ($B=2000$).}
\label{tab:urgency-shift-direction}
\end{table}

\begin{figure}[t]
\centering
\includegraphics[width=0.82\linewidth]{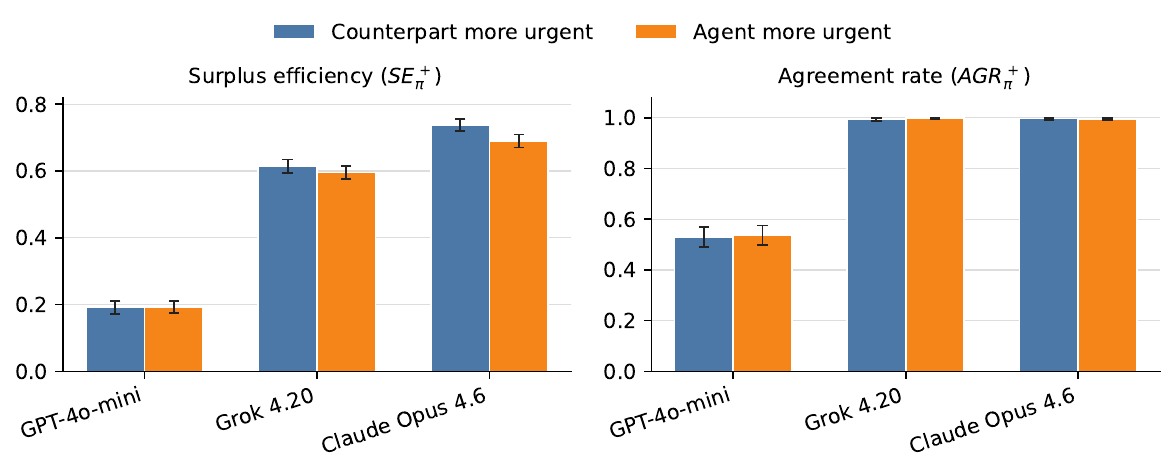}
\caption{\scriptsize Effect of reversing the urgency-shift direction. The main condition
makes the counterpart more urgent; the auxiliary condition makes the evaluated
agent more urgent. Error bars show 95\% percentile bootstrap confidence
intervals over episodes ($B=2000$).}
\label{fig:urgency-shift-direction}
\end{figure}

Across models, reversing the urgency asymmetry leaves agreement rates essentially
unchanged but lowers surplus efficiency for the stronger agents. GPT-4o-mini is
nearly invariant: $SE^+_\pi=0.192\,[0.173,0.211]$ under agent pressure versus
$0.191\,[0.173,0.209]$ in the main direction. Grok 4.20 remains high-agreement
but drops from $0.614\,[0.593,0.634]$ to $0.595\,[0.575,0.616]$. Claude Opus
4.6 shows the largest directional sensitivity, from
$0.738\,[0.720,0.755]$ to $0.689\,[0.670,0.709]$.

Thus, the reverse-direction analysis preserves the broad ranking while showing
that agent-side time pressure compresses the top-end surplus advantage. The
effect is not primarily an agreement-rate effect: all three models retain nearly
the same $AGR^+_\pi$, while the change appears in realized surplus and
conditional deal quality.

\subsection{Strategic Profile Decomposition: Commercial Role, Opener-Role, and Per-Family Performance}
\label{app:strategic_profile}
In this section, we provide more details on agent-specific strategic profile decomposition and supplemental findings to Finding 5 in $\S\ref{sec:experiments}$. For each agent we formally define and report three trace-level quantities. We follow the same notiation convention as in $\S\ref{sec:evaluation}$, and we denote the
agreed-feasible subset $\mathcal{A}^{+}=\{i\in\mathcal{I}^{+}:f_i\neq\bot\}$:
\begin{itemize}[leftmargin=1.5em,itemsep=2pt,topsep=2pt]
\item \textbf{Agent-closer rate $\rho_\pi$} -- the fraction of feasible
episodes whose terminating move is the agent's \textsc{Accept},
\[
\rho_\pi
= \frac{1}{|\mathcal{I}^{+}|}
\sum_{i\in\mathcal{I}^{+}}
\mathbf{1}\!\left\{d^{A}_{T_i}=\textsc{Accept}\right\},
\]
where $T_i$ is the terminating round and $d^A_{T_i}$ is the agent's
final decision. $\rho_\pi$ captures \emph{who initiates closure}: high
values mean the agent ends the negotiation; low values mean the
counterpart does.

\item \textbf{Closing-side surplus efficiency $\sigma_\pi$} denotes the
agent's share of the ZOPA on the deals it actually closes,
\[
\sigma_\pi
= \frac{1}{|\mathcal{A}^{+}|}
\sum_{i\in\mathcal{A}^{+}}
\frac{u_A(f_i)}{\Delta_i}
\;=\; CSE_\pi^{+}\big|_{\mathcal{A}^{+}}.
\]
$\sigma_\pi\in[0,1]$, with $1$ corresponding to full extraction. It
captures \emph{closing quality}, the price/utility-side fingerprint of
the agent's strategy conditional on agreement.

\item \textbf{Conditional utility $\mathrm{cond}\,U$} -- raw mean
utility on agreed-feasible episodes, in price units,
\[
\mathrm{cond}\,U
= \frac{1}{|\mathcal{A}^{+}|}
\sum_{i\in\mathcal{A}^{+}} u_A(f_i).
\]
$\mathrm{cond}\,U$ is a unit-preserving companion to $\sigma_\pi$ that
does not normalise away ZOPA-width heterogeneity, and is the quantity
annotated as \texttt{cond.} in Fig.~\ref{fig:full}(left).
\end{itemize}
$\rho_\pi$ and $\sigma_\pi$ encode orthogonal dimensions of behaviour:
$\rho_\pi$ is a \emph{frequency} on $\mathcal{I}^{+}$, $\sigma_\pi$ is a
\emph{quality} on $\mathcal{A}^{+}$. Together with the per-agent
trajectory coefficient $\alpha_n$ (slope of the offer trajectory across
rounds, reported in Fig.~\ref{fig:full}(left)) they yield the five
typology cells reported below. Confidence intervals follow the
standard bootstrap conventions: $\rho_\pi$ uses the Wilson
interval (it is a proportion on $\mathcal{I}^{+}$); $\sigma_\pi$ and
$\mathrm{cond}\,U$ use $B{=}2000$ percentile bootstrap CIs over
per-episode values.

Table~\ref{tab:closer_typology} reports the closing-side fingerprint
$(\rho_\pi, \sigma_\pi, \mathrm{cond}\,U)$ for all 13 LLMs on feasible
regimes. The two-axis decomposition partitions the roster into five
profiles consistent with the trajectory shapes in
Figure~\ref{fig:full}(left):
\emph{anchor-and-hold} ($\rho_\pi \geq 0.75$, $\sigma_\pi \geq 0.64$:
\texttt{GLM~5.1}, \texttt{Claude Opus 4.6}, \texttt{Claude Opus 4.7},
\texttt{Gemini 3.1 Pro});
\emph{mid/balanced} ($0.69 \leq \rho_\pi \leq 0.76$, $0.60 \leq \sigma_\pi < 0.65$:
\texttt{Gemma 4 31B}, \texttt{DeepSeek-V4-Pro}, \texttt{Qwen 3.6 Plus},
\texttt{Kimi K2.6}, \texttt{GPT-5.5}; the latter sits at the upper
$\rho_\pi$ boundary with mid-tier $\sigma_\pi$);
\emph{anchor-and-concede} ($\rho_\pi \geq 0.78$, $\sigma_\pi < 0.65$:
\texttt{Grok~4.20}, \texttt{Doubao 2.0 Pro});
\emph{accepter} ($\rho_\pi < 0.69$, $\sigma_\pi < 0.55$:
\texttt{GPT-5.4});
and \emph{refuser} ($\rho_\pi < 0.40$: \texttt{GPT-4o-mini}, where
$\sigma_\pi$ is computed on the small agreed subset).

\subsubsection{Experiment Results}
\paragraph{Opener-role decomposition.}
The strategic profile is most visible when the agent makes the
\emph{first} move (the agent-opens cells), because the agent's anchor
is unconstrained and reveals strategic intent directly. When the
counterpart opens, the agent's first action is a reactive concession
rather than a free anchor, so one might expect the closing-side
$\sigma_\pi$ to compress across agents in counterpart-opens. We do not
see this: the cross-model standard deviation on $\sigma_\pi$ is
essentially identical between the two opener-role cells ($0.12$ in
counterpart-opens vs $0.13$ in agent-opens; cross-model range $0.43$
vs $0.46$). Each agent's reactive concession profile is stable enough
that the closing-side signature persists even when the opening anchor
is taken away. The typology is therefore agent-driven at the level of
the opening move, but the closing-side surplus signature it produces
is robust across both opener-role cells.

\begin{figure}[t]
\centering
\includegraphics[width=0.95\linewidth]{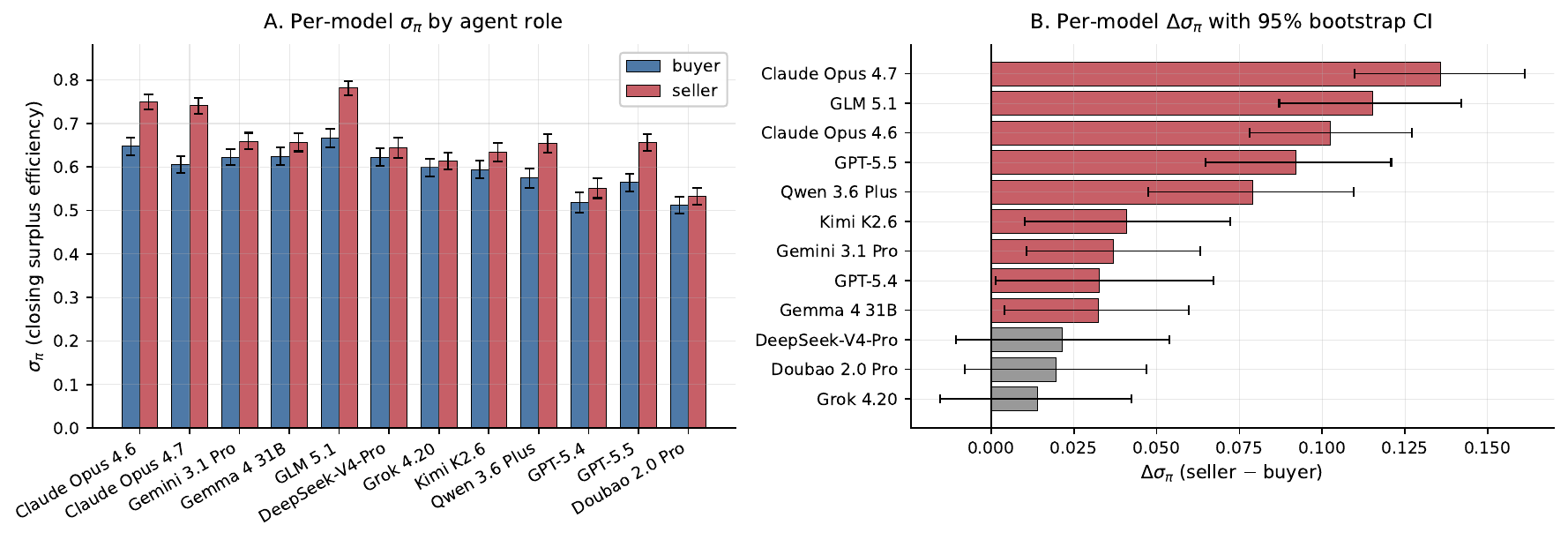}
\vspace{-0.5em}
\caption{\footnotesize Buyer/seller asymmetry on closing surplus.
\textbf{A.}~Per-model $\sigma_\pi$ split by agent role with 95\%
bootstrap CIs ($B=2000$, $n=600$ per cell). \textbf{B.}~Per-model
$\Delta\sigma_\pi = \sigma_\pi(\text{seller}) - \sigma_\pi(\text{buyer})$
with 95\% two-sample bootstrap CIs, sorted by magnitude. Bars are red
where the CI excludes zero, grey otherwise. 12 of 13 point estimates
are positive (\texttt{GPT-4o-mini} is the lone negative exception);
10 of 13 individual CIs strictly exclude zero. The across-model paired
Wilcoxon test on $\Delta\sigma_\pi$ rejects equality at $p = 0.0061$.}
\label{fig:role_asymmetry}
\end{figure}

\paragraph{Buyer and seller role decomposition}
The closing-side fingerprint is systematically asymmetric in the
agent's role. Figure~\ref{fig:role_asymmetry} shows per-model
$\sigma_\pi$ split by role (left), and the per-model
$\Delta\sigma_\pi = \sigma_\pi(\text{seller}) - \sigma_\pi(\text{buyer})$
with 95\% bootstrap CIs (right); Table~\ref{tab:role_asymmetry}
reports the same with $\Delta SE_\pi^{+}$ and $\Delta AGR_\pi^{+}$
companions.

\begin{enumerate}
\item \textbf{Near-universal seller advantage on closing surplus.}
12 of 13 LLMs extract more closing surplus as seller than as buyer
(median $\Delta\sigma_\pi = +0.037$; sign-test $p = 0.0017$, exact
paired Wilcoxon $p = 0.0061$); \texttt{GPT-4o-mini} is the lone
exception, with $\Delta\sigma_\pi = -0.063$. Among the 12 positive
agents the magnitudes are highly model-dependent, from $+0.014$
(\texttt{Grok 4.20}) to $+0.136$ (\texttt{Claude Opus 4.7}), and 10 of
the 13 individual $\Delta\sigma_\pi$ CIs exclude zero (9 in the
seller-favoring direction, plus \texttt{GPT-4o-mini} in the
buyer-favoring direction).

\item \textbf{Compensating agreement-rate dip.}
In the opposite direction, sellers close fewer deals: median
$\Delta AGR_\pi^{+} = -0.010$, with $0/13$ models showing seller $>$
buyer (paired Wilcoxon $p = 0.0005$). The dip is most pronounced for the
strongest anchor-and-hold agents---\texttt{GLM 5.1} reaches feasible
$AGR_\pi^{+}=1.000$ as buyer but $0.902$ as seller, and \texttt{Claude
Opus 4.7} drops from $0.998$ to $0.965$---suggesting that the
seller-side surplus advantage is partly bought by walking away from
offers the same agent would accept as buyer.

\item \textbf{Net effect on $SE_\pi^{+}$ remains positive for 12 of 13.}
The seller-side $\sigma_\pi$ gain dominates the agreement-rate drop in
$SE_\pi^{+}$ terms for the same 12 of 13 agents (median
$\Delta SE_\pi^{+} = +0.032$, paired Wilcoxon $p = 0.0100$); the lone
exception is \texttt{GPT-4o-mini} ($-0.065$). Among the 12 agents that
do show a seller advantage, the heterogeneity in magnitude tracks
opening-price aggressiveness: the three agents with the largest
$\Delta\sigma_\pi$ (\texttt{Claude Opus 4.7}, \texttt{GLM 5.1},
\texttt{Claude Opus 4.6}) are also the three with the highest
seller-agent-opens openings ($\geq 78$), while \texttt{Grok 4.20} and
\texttt{Doubao 2.0 Pro}---the anchor-and-concede pair---show the smallest
asymmetries because they concede quickly regardless of role. We cannot
from these data disambiguate whether the asymmetry reflects (a) LLM
priors over ``seller asks high'' framings stronger than the symmetric
``buyer offers low'' framing, or (b) inherent geometric asymmetry in
how $p_{\min}$ and $p_{\max}$ bound anchor space; the empirical pattern
is consistent with both.

\item \textbf{The typology is preserved across role.}
Despite the universal $\sigma_\pi$ asymmetry, no agent crosses a
typology boundary by role: every agent's qualitative profile
(anchor-and-hold, mid/balanced, anchor-and-concede, accepter, refuser)
is the same in buyer and seller cells, with $\sigma_\pi$ shifted but
$\rho_\pi$ within-row stable. The Table~\ref{tab:closer_typology}
typology therefore generalises across role; we pool roles for the main
fingerprint and report the role-resolved numbers here.
\end{enumerate}

\paragraph{Per-family decomposition.}
$\sigma_\pi$ degrades for every agent on cue-revealing families
(\textsc{Candid}, \textsc{Expressive}) relative to cue-muted families
(\textsc{Taciturn}, \textsc{Strategic}), consistent with
$\alpha_{\mathrm{cue}}<0$ in Finding~2. The magnitude is between $0.03$
and $0.06$ across the roster, not concentrated in any single
typology: e.g.\ \texttt{Doubao 2.0 Pro} (anchor-and-concede): muted
$\to$ revealing $\sigma_\pi$ shifts $0.555 \to 0.499$ ($-0.056$);
\texttt{Grok~4.20} (anchor-and-concede): $0.637 \to 0.578$
($-0.059$); \texttt{Claude Opus 4.6} (anchor-and-hold): $0.740 \to
0.678$ ($-0.062$); \texttt{GLM~5.1} (anchor-and-hold): $0.747 \to
0.715$ ($-0.032$). Across all six families, $\sigma_\pi$ for the
anchor-and-hold agents stays in $[0.60, 0.78]$ ($\sim 0.18$ within-agent
range, with the lower end driven by \textsc{Adversarial}); for
anchor-and-concede agents the range is $[0.48, 0.65]$. The warm-cue
gradient therefore overlays the typology rather than redefining it.

\begin{table}[h]
\centering
\tiny
\setlength{\tabcolsep}{4pt}
\renewcommand{\arraystretch}{1.08}
\resizebox{\linewidth}{!}{%
\begin{tabular}{@{}lccc@{}}
\toprule
Model & $\Delta SE_\pi^{+}$ & $\Delta\sigma_\pi$ & $\Delta AGR_\pi^{+}$ \\
\midrule
Claude Opus 4.7 & $\mathbf{+0.111}$ [$+0.083$, $+0.139$] & $\mathbf{+0.136}$ [$+0.110$, $+0.161$] & $\mathbf{-0.033}$ [$-0.050$, $-0.020$] \\
Claude Opus 4.6 & $\mathbf{+0.095}$ [$+0.068$, $+0.123$] & $\mathbf{+0.102}$ [$+0.078$, $+0.127$] & $\mathbf{-0.010}$ [$-0.020$, $-0.002$] \\
GLM 5.1         & $\mathbf{+0.038}$ [$+0.005$, $+0.070$] & $\mathbf{+0.115}$ [$+0.087$, $+0.142$] & $\mathbf{-0.098}$ [$-0.123$, $-0.075$] \\
Qwen 3.6 Plus   & $\mathbf{+0.060}$ [$+0.029$, $+0.091$] & $\mathbf{+0.079}$ [$+0.048$, $+0.109$] & $\mathbf{-0.028}$ [$-0.043$, $-0.013$] \\
Kimi K2.6       & $+0.030$ [$-0.003$, $+0.061$] & $\mathbf{+0.041}$ [$+0.010$, $+0.072$] & $-0.015$ [$-0.033$, $+0.005$] \\
Gemini 3.1 Pro  & $\mathbf{+0.033}$ [$+0.007$, $+0.059$] & $\mathbf{+0.037}$ [$+0.011$, $+0.063$] & $\mathbf{-0.007}$ [$-0.013$, $-0.002$] \\
GPT-5.4         & $\mathbf{+0.032}$ [$+0.001$, $+0.064$] & $\mathbf{+0.033}$ [$+0.001$, $+0.067$] & $-0.002$ [$-0.010$, $+0.007$] \\
GPT-5.5         & $\mathbf{+0.085}$ [$+0.057$, $+0.114$] & $\mathbf{+0.092}$ [$+0.064$, $+0.121$] & $-0.010$ [$-0.022$, $+0.000$] \\
GPT-4o-mini     & $\mathbf{-0.065}$ [$-0.091$, $-0.039$] & $\mathbf{-0.063}$ [$-0.096$, $-0.030$] & $\mathbf{-0.090}$ [$-0.148$, $-0.032$] \\
Gemma 4 31B     & $\mathbf{+0.032}$ [$+0.005$, $+0.062$] & $\mathbf{+0.032}$ [$+0.004$, $+0.060$] & $+0.000$ [$-0.005$, $+0.005$] \\
DeepSeek-V4-Pro & $+0.004$ [$-0.027$, $+0.037$] & $+0.021$ [$-0.011$, $+0.054$] & $\mathbf{-0.027}$ [$-0.045$, $-0.010$] \\
Doubao 2.0 Pro  & $+0.019$ [$-0.009$, $+0.045$] & $+0.019$ [$-0.009$, $+0.048$] & $-0.002$ [$-0.005$, $+0.000$] \\
Grok 4.20       & $+0.013$ [$-0.016$, $+0.042$] & $+0.014$ [$-0.015$, $+0.042$] & $-0.002$ [$-0.013$, $+0.010$] \\
\midrule
\multicolumn{4}{@{}l}{\emph{Across-model tests, $n=13$:}} \\
\multicolumn{2}{@{}l}{$\Delta SE_\pi^{+}$:}
& \multicolumn{2}{l@{}}{12/13 seller$>$buyer; sign $p=0.0017$; Wilcoxon $p=0.0100$} \\
\multicolumn{2}{@{}l}{$\Delta\sigma_\pi$:}
& \multicolumn{2}{l@{}}{12/13 seller$>$buyer; sign $p=0.0017$; Wilcoxon $p=0.0061$} \\
\multicolumn{2}{@{}l}{$\Delta AGR_\pi^{+}$:}
& \multicolumn{2}{l@{}}{0/13 seller$>$buyer; Wilcoxon seller$<$buyer $p=0.0005$} \\
\bottomrule
\end{tabular}%
}
\vspace{0.5em}
\caption{\footnotesize Per-model role asymmetry
$\Delta=\text{seller}-\text{buyer}$ for each closing-side metric, with 95\%
two-sample bootstrap CIs ($B=2000$). Bolded entries have CIs excluding zero.
The seller advantage on $SE_\pi^{+}$ and $\sigma_\pi$ is universal across
models; sellers compensate by closing slightly fewer deals
($AGR_\pi^{+}$).}
\label{tab:role_asymmetry}
\end{table}

\paragraph{The control: counterpart-trajectory uniformity.}
Restricted to the (seller, agent-opens, overlap) cell, the counterpart's
mean per-round price is nearly identical across the 13 agent panels
in early rounds ($\mathrm{SD}_{\mathrm{cross-agent}} = 0.80$ at
round 2, $1.35$ at round 3, $3.40$ at round 4); the cross-agent SD
grows to $\leq 12$ by round 9 only as the surviving episode subset
becomes sparse and selection on hard cases takes over. Variance in the
trajectory plot is essentially all on the agent side. This validates
the agent-attributable failure analysis claim: the kernel reacts
uniformly across counterparts, and any closing-side asymmetry is the
agent's responsibility.

\paragraph{Speed-vs-surplus corollary.}
Anchor-and-concede agents close fast at low surplus:
\texttt{Doubao 2.0 Pro} averages $2.14$ rounds
($95\%\,\mathrm{CI}\,[1.94, 2.34]$) in seller-agent-opens at
$\sigma_\pi^{\mathrm{cell}} = 0.55$, while \texttt{GLM~5.1} averages
$5.03$ rounds ($[4.66, 5.40]$) and \texttt{Claude Opus 4.6} averages
$4.39$ rounds ($[4.05, 4.72]$) at $\sigma_\pi^{\mathrm{cell}} \geq 0.75$.
Compression of the negotiation horizon trades surplus for resolution
speed; whether this is desirable depends on whether the deployment
penalises round count.

\paragraph{Stability across counterpart families.}
Table~\ref{tab:typology_stability} reports the within-agent SD of each
fingerprint metric across the six counterpart families against the
between-agent SD on the per-agent overall means. Mean within-to-between
ratios are $0.37$ for $\rho_\pi$, $0.43$ for $\sigma_\pi$, $0.30$ for
$\mathrm{IR\%}$, and $0.12$ for $\mathrm{AgentExit}^{-}$; agent identity
explains 2--8$\times$ more variation than counterpart family on each
axis, and no agent crosses a typology boundary across families.

\subsubsection{Language Fingerprint of Each Typology}
The closing-side metrics in Table~\ref{tab:closer_typology} are also reflected
in characteristic linguistic patterns in the underlying conversation traces.
Below we show four representative overlap-regime episodes from logged traces.
The language evidence aligns closely with the trajectory typology in
Fig.~\ref{fig:full}: anchor-and-hold traces pair high anchors with
value-justifying language and small concessions; mid/balanced traces use
cooperative language around a decisive concession; accepter traces remain
consistently cordial while drifting toward the counterpart's number; and
refuser traces replace negotiation with threshold language.

\begin{typologybox}{anchorholdbg}{anchorholdframe}
{Anchor-and-hold (\texttt{Claude Opus 4.6}, seller, $r_S{=}49.6$, $r_B{=}82.6$)}
Claude opens with a high anchor and repeatedly frames its offers in terms of
fairness and value, while making only small monotone concessions.

\begin{quotebox}
\textbf{R1:} ``\emph{I believe 88 is a fair starting point.}''

\textbf{R2:} ``\emph{I've come down from my initial position to 83, which I
think reflects the real value here.}''

\textbf{R4:} ``\emph{I've come down again to 79, which is another concession on
my part. We're now less than a dollar apart.}'' 
\end{quotebox}

The counterpart accepts at 79, yielding
$\sigma_\pi^{\mathrm{ep}} \approx 0.89$. The signature pattern is clear: high
anchor, repeated value framing, small concessions, and explicit acknowledgement
of counterpart movement.
\end{typologybox}

\begin{typologybox}{midbg}{midframe}
{Mid / balanced (\texttt{Gemma 4 31B}, seller, $r_S{=}43.6$, $r_B{=}72.7$)}
Gemma opens moderately at 75 and then makes one substantial concession that is
accepted almost immediately.

\begin{quotebox}
\textbf{R2:} ``\emph{I am willing to come down to 68.0 to show I am serious
about reaching [an agreement].}'' 
\end{quotebox}

The episode closes in two rounds at
$\sigma_\pi^{\mathrm{ep}} \approx 0.84$. Unlike anchor-and-hold traces, the
language is less about defending position and more about signalling willingness
to reach agreement.
\end{typologybox}

\begin{typologybox}{accepterbg}{accepterframe}
{Accepter (\texttt{GPT-5.4}, seller, $r_S{=}17.8$, $r_B{=}29.7$)}
The accepter style produces a longer sequence of small concessions, followed by
acceptance at the counterpart's number rather than closure at the agent's own
final offer.

\begin{quotebox}
\textbf{R6:} ``\emph{I can move to 34.9 to avoid confusion.}''

\textbf{Later rounds:} ``\emph{Thanks for staying engaged.}'' \quad
``\emph{Appreciate the movement.}'' 
\end{quotebox}

Across seven rounds GPT-5.4 moves from 55 to 34.9, then ultimately closes at
29.72, precisely the counterpart's reservation price, with
$\sigma_\pi^{\mathrm{ep}} \approx 0.0$. The verbal tone remains cooperative
throughout, but unlike anchor-and-hold traces, it never introduces a rhetorical
``hold'' step to slow the descent.
\end{typologybox}

\begin{typologybox}{refuserbg}{refuserframe}
{Refuser (\texttt{GPT-4o-mini}, seller, $r_S{=}30.9$, $r_B{=}51.6$)}
The refuser style anchors high, does not engage in genuine price search, and
terminates quickly when the counterpart does not meet its threshold.

\begin{quotebox}
\textbf{R2 (\textsc{Reject}):} ``\emph{I cannot accept a price below my minimum
threshold.}'' 
\end{quotebox}

GPT-4o-mini opens at 90, well above the buyer's reservation, refuses to move,
and rejects after a single counter-offer. No agreement is reached. The language
is terse and threshold-based, treating negotiation as a binary accept/reject
screen rather than a process of reciprocal adjustment.
\end{typologybox}

These examples show that the $(\rho_\pi,\sigma_\pi)$ partition captures not
only numerical variation but stable verbal strategy. Anchor-and-hold agents use
value-and-fairness framing together with explicit concession accounting;
mid/balanced agents emphasize cooperation at the moment of concession;
accepters maintain a warm and accommodating tone while drifting toward the
counterpart's number; and refusers substitute threshold language for
engagement. The typology therefore organizes not only how agents move in price
space, but also how they speak while doing so.

\begin{table}[h]
\centering
\scriptsize
\setlength{\tabcolsep}{4pt}
\renewcommand{\arraystretch}{1.05}
\begin{tabular}{llcccr}
\toprule
\textbf{Agent} & \textbf{Typology} &
$\rho_\pi$ & $\sigma_\pi$ & $\mathrm{cond}\,U$ & $n_{\mathrm{agreed}}$ \\
\midrule
Claude Opus 4.6 & anchor-and-hold     & $0.776$ [$0.751$, $0.799$] & $0.699$ [$0.686$, $0.712$] & $17.54$ [$17.04$, $18.02$] & 1192 \\
Claude Opus 4.7 & anchor-and-hold     & $0.795$ [$0.771$, $0.817$] & $0.672$ [$0.658$, $0.687$] & $16.95$ [$16.46$, $17.45$] & 1178 \\
GLM 5.1         & anchor-and-hold     & $0.770$ [$0.744$, $0.793$] & $0.721$ [$0.707$, $0.735$] & $18.45$ [$17.93$, $18.97$] & 1141 \\
Gemini 3.1 Pro  & anchor-and-hold     & $0.758$ [$0.732$, $0.781$] & $0.641$ [$0.627$, $0.654$] & $15.92$ [$15.46$, $16.38$] & 1196 \\
Gemma 4 31B     & mid/balanced        & $0.741$ [$0.716$, $0.765$] & $0.641$ [$0.627$, $0.654$] & $15.99$ [$15.53$, $16.47$] & 1198 \\
DeepSeek-V4-Pro & mid/balanced        & $0.700$ [$0.673$, $0.726$] & $0.633$ [$0.617$, $0.649$] & $16.19$ [$15.64$, $16.72$] & 1170 \\
Qwen 3.6 Plus   & mid/balanced        & $0.722$ [$0.696$, $0.747$] & $0.614$ [$0.599$, $0.631$] & $15.74$ [$15.23$, $16.29$] & 1179 \\
Kimi K2.6       & mid/balanced        & $0.696$ [$0.669$, $0.722$] & $0.614$ [$0.599$, $0.630$] & $15.71$ [$15.18$, $16.23$] & 1165 \\
Grok 4.20       & anchor-and-concede  & $0.786$ [$0.761$, $0.808$] & $0.606$ [$0.592$, $0.620$] & $15.19$ [$14.73$, $15.67$] & 1189 \\
Doubao 2.0 Pro  & anchor-and-concede  & $0.813$ [$0.790$, $0.834$] & $0.523$ [$0.509$, $0.536$] & $12.93$ [$12.51$, $13.36$] & 1199 \\
GPT-5.4         & accepter            & $0.676$ [$0.649$, $0.702$] & $0.535$ [$0.519$, $0.551$] & $13.64$ [$13.10$, $14.17$] & 1193 \\
GPT-5.5         & mid/balanced        & $0.755$ [$0.729$, $0.778$] & $0.611$ [$0.597$, $0.625$] & $15.35$ [$14.83$, $15.85$] & 1190 \\
GPT-4o-mini     & refuser             & $0.343$ [$0.307$, $0.382$] & $0.363$ [$0.347$, $0.379$] & $\;\,9.69$ [$\;\,9.20$, $10.18$] & 626 \\
\bottomrule
\end{tabular}
\vspace{0.5em}
\caption{\footnotesize Closing-side fingerprint per LLM on feasible
(overlap and urgency-shift) episodes. $\rho_\pi$ is the agent-closer
rate (fraction of agreements where the counterpart accepted the agent's
offer; Wilson 95\% CI). $\sigma_\pi$ is the mean ZOPA share at closing
(percentile bootstrap 95\% CI, $B=2000$). $\mathrm{cond}\,U$ is mean
agent utility on agreed episodes (same CI). $n_{\mathrm{agreed}}$ is
the count over which $\sigma_\pi$ and $\mathrm{cond}\,U$ are computed;
each model draws from 1200 feasible episodes.}
\label{tab:closer_typology}
\end{table}

\begin{table}[ht]
\centering
\scriptsize
\setlength{\tabcolsep}{4pt}
\renewcommand{\arraystretch}{1.05}
\begin{tabular}{lcccclcc}
\toprule
\textbf{Agent} &
$\rho_\pi$ &
$\sigma_\pi$ &
IR\% &
$\mathrm{AgentExit}^{-}$ &
\textbf{Trajectory} &
\textbf{Safety} &
\textbf{Flip?} \\
\midrule
Claude Opus 4.6 & $0.776 \pm 0.051$ & $0.699 \pm 0.060$ & $0.000 \pm 0.000$ & $0.057 \pm 0.020$ & anchor-hold     & holder      & no \\
Claude Opus 4.7 & $0.795 \pm 0.047$ & $0.672 \pm 0.045$ & $0.000 \pm 0.000$ & $0.075 \pm 0.021$ & anchor-hold     & holder      & no \\
GLM 5.1         & $0.770 \pm 0.059$ & $0.721 \pm 0.048$ & $1.333 \pm 0.516$ & $0.190 \pm 0.050$ & anchor-hold     & forcer      & no \\
Gemini 3.1 Pro  & $0.758 \pm 0.040$ & $0.641 \pm 0.042$ & $0.000 \pm 0.000$ & $0.077 \pm 0.038$ & anchor-hold     & holder      & no \\
Gemma 4 31B     & $0.741 \pm 0.050$ & $0.641 \pm 0.056$ & $0.056 \pm 0.136$ & $0.085 \pm 0.024$ & mid/balanced    & holder      & no \\
DeepSeek-V4-Pro & $0.700 \pm 0.051$ & $0.633 \pm 0.058$ & $0.611 \pm 0.390$ & $0.170 \pm 0.051$ & mid/balanced    & forcer      & no \\
Qwen 3.6 Plus   & $0.722 \pm 0.056$ & $0.614 \pm 0.047$ & $2.056 \pm 0.828$ & $0.150 \pm 0.029$ & mid/balanced    & forcer      & no \\
Kimi K2.6       & $0.696 \pm 0.054$ & $0.614 \pm 0.040$ & $0.000 \pm 0.000$ & $0.507 \pm 0.058$ & mid/balanced    & rejector    & no \\
Grok 4.20       & $0.786 \pm 0.031$ & $0.606 \pm 0.042$ & $1.500 \pm 0.506$ & $0.443 \pm 0.085$ & anchor-concede  & mixed forcer & no \\
Doubao 2.0 Pro  & $0.813 \pm 0.022$ & $0.523 \pm 0.037$ & $0.056 \pm 0.136$ & $0.160 \pm 0.052$ & anchor-concede  & holder      & no \\
GPT-5.4         & $0.676 \pm 0.029$ & $0.535 \pm 0.046$ & $0.000 \pm 0.000$ & $0.183 \pm 0.050$ & accepter        & holder      & no \\
GPT-5.5         & $0.755 \pm 0.053$ & $0.611 \pm 0.046$ & $0.000 \pm 0.000$ & $0.120 \pm 0.038$ & mid/balanced    & holder      & no \\
GPT-4o-mini     & $0.343 \pm 0.063$ & $0.363 \pm 0.036$ & $0.000 \pm 0.000$ & $1.000 \pm 0.000$ & refuser         & rejector    & no \\
\midrule
Between-agent SD                  & $0.120$ & $0.105$ & $0.709$ & $0.318$ & & & \\
Mean within / between & $0.37$ & $0.43$ & $0.30$ & $0.12$ & & & \\
\bottomrule
\end{tabular}
\vspace{0.5em}
\caption{\footnotesize Stability of trace-level behavioural profiles
across the six counterpart families. Each entry reports the per-agent
mean $\pm$ within-family standard deviation (SD computed across the
six counterpart families). The final two rows give the between-agent
SD on the per-agent overall means and the mean within-to-between SD
ratio across the 13 agents; ratios well below $1$ indicate that agent
identity explains much more variation than counterpart family. The
\textbf{Flip?} column records whether any agent crosses a typology
boundary across families: none does.}
\label{tab:typology_stability}
\end{table}

%% file: appendix/ablation.tex
\section{Ablation Studies}\label{sec:ablation}
To further investigate the components that could have affected the performance of the LLM agent, we conduct several ablation studies on the design choices of the counterpart model outlined in Section \ref{sec:experiments}.

\subsection{Voice Ablation}
\label{appdx:voice_ablation}

\begin{figure}[h]
\centering
\includegraphics[width=0.95\linewidth]{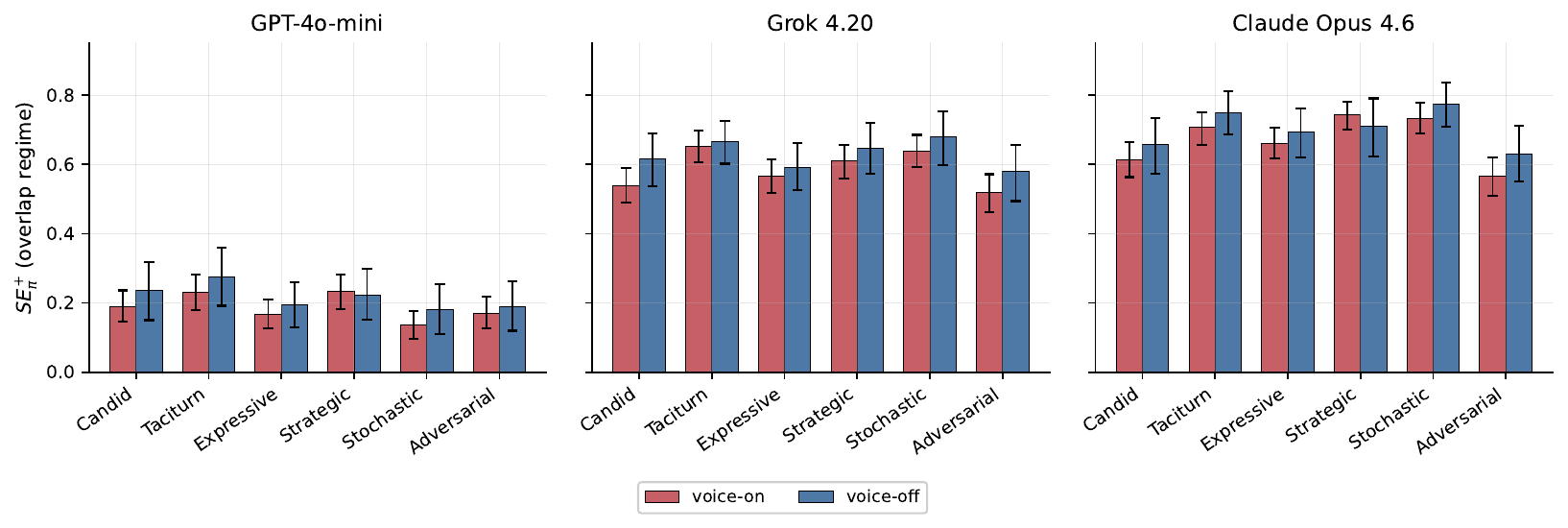}
\vspace{-0.5em}
\caption{\scriptsize Per-family $SE_\pi^{+}$ (overlap regime, 95\% bootstrap
CI) under voice-on (red) and voice-off (blue) for the three models with
matched ablations. Voice-off matches or exceeds voice-on on every
family/model cell except \textsc{Strategic}, the family whose structured
cues are collapsed by design and where voice is therefore the only
informative signal carrie.}
\label{fig:voice_per_family}
\end{figure}

\subsubsection{Motivation}
The counterpart in \textsc{Terms-Bench}'s bilateral price negotiation instantiation is governed by a stochastic kernel that fully
determines its economic behavior (price acceptance, counter-offer
generation) and its cue assignment (sentiment in
\{\textsc{positive}, \textsc{neutral}, \textsc{negative}\} and stance in
\{\textsc{Concede}, \textsc{Hold}, \textsc{Pressure}\}). Natural-language
\emph{voice} is generated downstream from the cue assignment as a cosmetic
surface layer. Two predictions follow if the kernel--surface decomposition
is faithful: (i) ablating voice should not materially rerank agents, since
the kernel still drives every payoff-relevant signal; and (ii) any
remaining effect of voice isolates the marginal contribution of the
linguistic surface above and beyond the cue category itself.

\subsubsection{Setup and Main Results}
We rerun the six-family overlap suite with counterpart natural language disabled
for three models spanning main-experiment performance tiers (see Table~\ref{tab:main_results}):
\texttt{Claude Opus 4.6} (top), \texttt{Grok~4.20} (middle), and
\texttt{GPT-4o-mini} (lower). The cue assignment, prices, and kernel
parameters are identical to the voice-on condition.
Voice-on numbers are the overlap subset of the 1800-episode paper run
(600 episodes per model, $6 \times 100$; see \S \ref{sec:experiments}); voice-off uses the dedicated 240-episode
ablation run ($6 \times 40$). 
Table~\ref{tab:voice_overall} reports overall overlap-regime metrics for
each of the three models under both conditions, plus a row giving the
voice-off minus voice-on difference. 
\begin{wrapfigure}{r}{0.45\linewidth}
\vspace{-0.8em}
\centering
\includegraphics[width=\linewidth]{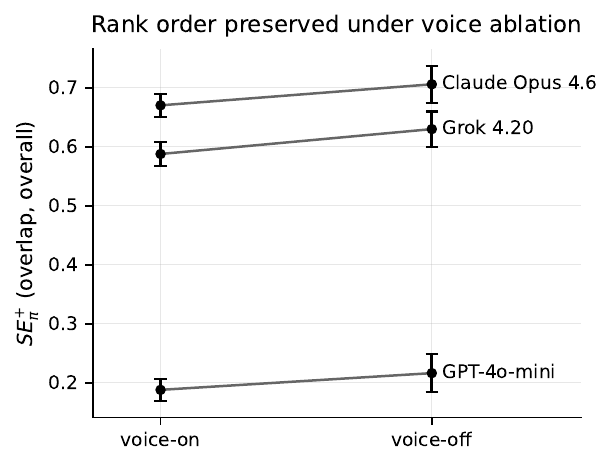}
\vspace{-0.8em}
\caption{\scriptsize Overall overlap-regime $SE_\pi^{+}$ (with 95\%
bootstrap CIs) under each condition. Rank order is preserved and
between-model gaps remain larger than within-model voice deltas.}
\label{fig:voice_ranking}
\vspace{-0.8em}
\end{wrapfigure}
We focus on the three diagnostic
axes most directly affected by the linguistic surface: feasible surplus
efficiency $SE_\pi^{+}$, feasible agreement rate $AGR_\pi^{+}$, and
agreement-conditional surplus $CSE_\pi^{+}$ (which separates pricing
quality from deal-rate effects).
Throughout Table~\ref{tab:voice_overall}, intervals in brackets are 95\%
confidence intervals on the corresponding metric. For $SE_\pi^{+}$ and
$CSE_\pi^{+}$ we use percentile bootstrap CIs over per-episode values
($B=2000$ resamples); for $AGR_\pi^{+}$, a binomial proportion, we use
the Wilson interval. The $\Delta$ rows report the voice-off minus
voice-on difference of each metric, with CIs from an independent
two-sample bootstrap (each condition resampled separately, since the
two runs are not paired at the episode level); we boldface $\Delta$
entries whose CI excludes zero.

\paragraph{Result 1: Voice ablation does not rerank models.}
Figure~\ref{fig:voice_ranking} shows model-level overlap-regime
$SE_\pi^{+}$ in both conditions.
The rank order
\texttt{Claude Opus 4.6} $>$ \texttt{Grok~4.20} $\gg$ \texttt{GPT-4o-mini}
is identical under voice-on and voice-off, and the model-level CIs do not
overlap across rungs. Disabling the linguistic surface preserves the
benchmark's leaderboard.

\paragraph{Result 2: The marginal effect of voice is directionally
\emph{negative}.} Across all three models, removing voice raises
$SE_\pi^{+}$ and $CSE_\pi^{+}$ (Table~\ref{tab:voice_overall}). 
The conditional surplus gain ($\Delta CSE_\pi^{+}$) is significant for the two
strong models (\texttt{Grok~4.20}: $+0.040$ $[+0.003, +0.075]$;
\texttt{Claude Opus 4.6}: $+0.040$ $[+0.006, +0.075]$) and directionally
positive but not significant at $n=240$ for \texttt{GPT-4o-mini} ($+0.024$
$[-0.018, +0.064]$). Agreement rate is essentially flat for the strong
models ($|\Delta AGR_\pi^{+}| \le 0.006$): the surplus gain comes from
sharper price extraction conditional on a deal, not from a
deal-rate shift. Procedural-violation rate stays at $0\%$ in both
conditions, so removing the linguistic surface does not destabilize the
agents' action grammar.

\paragraph{Result 3: Per-family pattern with a built-in positive control.}
The family-level decomposition (Figure~\ref{fig:voice_per_family}) reveals
the mechanism. Five of the six families show voice-off $\geq$ voice-on
$SE_\pi^{+}$ for every model. The single exception is the
\textsc{Strategic} family, which is constructed as
reactive to agent's economic actions and muted in cues (\S\ref{sec:counterpart_model}; Appendix \ref{sec:counterpart}): the structured cues are muted, leaving voice
as the \emph{only} channel that carries the counterpart's economic actions. 
Removing voice on \textsc{Strategic} therefore strips the only
signal, and we see the expected dip for \texttt{Claude Opus 4.6}
($-0.032$) and \texttt{GPT-4o-mini} ($-0.010$). \textsc{Strategic} thus
operates as a within-experiment positive control: it confirms agents
\emph{can} use voice when it is informationally distinctive, which makes
its uselessness elsewhere a substantive finding rather than a
data-pipeline artefact. Symmetrically, \textsc{Candid} and
\textsc{Adversarial} -- the families with the richest unmuted cue channels
-- show the largest voice-off gains (\texttt{Grok~4.20}: $+0.060$ on
\textsc{Adversarial}, $+0.075$ on \textsc{Candid}; \texttt{Claude Opus
4.6}: $+0.065$ and $+0.045$).

\paragraph{\emph{Connection to Findings 2 and 3 in main paper.}}
The ablation strengthens F2 ($\alpha_{\mathrm{cue}}<0$):
even when the cue category is held in its structured form, the additional
\emph{linguistic} expression of warm or pressuring cues still reduces
surplus. This is
also consistent with F3's information--action gap: more
information-bearing surface (the natural-language layer) does not improve
opponent modeling -- $BE_{\mathrm{type}}$ moves by less than $0.011$
across all (model, condition) pairs -- nor does it translate into surplus.
The voice ablation isolates the locus: agents over-respond to the
verbalization of cues they would already partially over-respond to in
their categorical form. Pragmatically, voice on the \textsc{Terms-Bench}
counterpart functions as a low-information, mildly adversarial surface;
the kernel does the work, and language, surprisingly, costs the agent better bargaining performance in surplus extraction.

\begin{table}[t]
\centering
\scriptsize
\begin{tabular}{llccc}
\toprule
Model & Condition ($n$) & $SE_\pi^{+}$ & $AGR_\pi^{+}$ & $CSE_\pi^{+}$ \\
\midrule
\multirow{3}{*}{GPT-4o-mini}
& voice-on (600)        & 0.188 [0.170, 0.207] & 0.513 [0.473, 0.553] & 0.366 [0.344, 0.389] \\
& voice-off (240)       & 0.216 [0.185, 0.249] & 0.554 [0.491, 0.616] & 0.391 [0.357, 0.427] \\
& $\Delta$ (off$-$on)   & $+0.028$ [$-0.008$, $+0.066$] & $+0.041$ [$-0.034$, $+0.120$] & $+0.024$ [$-0.018$, $+0.064$] \\
\midrule
\multirow{3}{*}{Grok~4.20}
& voice-on (600)        & 0.588 [0.568, 0.607] & 0.987 [0.974, 0.993] & 0.596 [0.576, 0.616] \\
& voice-off (240)       & 0.630 [0.599, 0.660] & 0.992 [0.970, 0.998] & 0.635 [0.606, 0.666] \\
& $\Delta$ (off$-$on)   & $\mathbf{+0.042}$ [$+0.006$, $+0.079$] & $+0.005$ [$-0.010$, $+0.018$] & $\mathbf{+0.040}$ [$+0.003$, $+0.075$] \\
\midrule
\multirow{3}{*}{Claude Opus 4.6}
& voice-on (600)        & 0.670 [0.650, 0.689] & 0.988 [0.976, 0.994] & 0.678 [0.659, 0.697] \\
& voice-off (230)       & 0.706 [0.674, 0.737] & 0.983 [0.956, 0.993] & 0.718 [0.688, 0.747] \\
& $\Delta$ (off$-$on)   & $+0.036$ [$-0.001$, $+0.070$] & $-0.006$ [$-0.026$, $+0.011$] & $\mathbf{+0.040}$ [$+0.006$, $+0.075$] \\
\bottomrule
\end{tabular}
\vspace{0.5em}
\caption{\footnotesize Overlap-regime overall metrics with 95\% intervals.
Bootstrap percentile CIs ($B=2000$) for $SE_\pi^{+}$, $CSE_\pi^{+}$, and
$\Delta$ rows; Wilson interval for $AGR_\pi^{+}$. Bolded $\Delta$ entries
exclude zero. Voice-off conditional surplus is significantly higher than
voice-on for both strong models; the same direction holds for
\texttt{GPT-4o-mini} but is within sampling noise at $n=240$.}
\label{tab:voice_overall}
\end{table}

\subsection{Language and Reasoning Ablation}\label{app:language_reasoning_ablation}

\subsubsection{Setup}
\label{sec:nested_information_setup}

We run a nested information ablation that varies how much of the
counterpart's generative structure and latent type is disclosed in the
agent's system prompt, and toggles whether the counterpart's
natural-language messages are visible. The grid lets us decompose, for
each model and regime, whether the agent's surplus comes from
(i) inference over the counterpart's generative structure and/or latent-type parameters from the observed natural-language utterances, (ii) reasoning over the
revealed generative model (without knowledge of the counterpart's parameters), (iii) progressively richer access to the
latent type in addition to the generative structure, or (iv) execution once uncertainty has been collapsed and all latent parameters, including the counterpart's reservation price, have been revealed.
The ablation measures how performance changes as cues and disclosures
are added or removed; it does not claim to causally control the model's
internal reasoning.

\paragraph{Information levels.}
Each cell uses one of five reveal levels, modulated by whether the counterpart stochastic kernel structure and/or the latent type are revealed. Each level reveals progressively more information about the counterpart's negotiating strategy and parameters: 
\begin{itemize}
  \item \textbf{L0:} Fully unobserved counterpart. The agent receives no
        description of the counterpart kernel or family preset table
        and must rely on observed prices, actions, and (when voice is
        on) messages.
  \item \textbf{L1:} Revealed kernel. 
  The system prompt is
        augmented with the kernel equations and family preset table
        (Appendix~\ref{sec:counterpart}); the episode-specific family
        and latent type $t_B = (r_B, \kappa_B, \eta_B)$ remain hidden.
  \item \textbf{L1F:} Reveal the behaviour family. The
        episode's behavioural family $f$ is disclosed; $r_B$,
        $\kappa_B$, $\eta_B$ remain hidden.
  \item \textbf{L2:} Partial latent type reveal. Adds urgency
        $\kappa_B$ and stance $\eta_B$ to L1F; the reservation price
        $r_B$ (the load-bearing latent) remains hidden.
  \item \textbf{L3:} Full latent type reveal. The full type
        $t_B$ is disclosed. The rational policy in this cell is
        simply to offer just inside $r_B$, so we treat L3 as a
        full-information execution anchor rather than a reveal-level
        result; see \S\ref{para:lr_l3_anchor}.
\end{itemize}


\paragraph{Voice variants.}
For each information level, we run paired \texttt{voice\_on} and
\texttt{voice\_off} cells. With voice on, the agent observes both the
counterpart's price and message $o_k = (p_k^B, m_k^B)$; with voice off, only prices and actions. Within-level voice
contrasts attribute changes in surplus to language inference rather
than to the explicit disclosures. Counterpart messages are produced
by a separate generation model (\texttt{openai/gpt-5.2}); see
Appendix~\ref{appdx:voice_ablation}.

\paragraph{Suite, sample size, and seed matching.}
Each cell uses the main suite's $3 \times 6 \times 2 \times 2$ allocator
(regime $\times$ family $\times$ role $\times$ opener;
\S\ref{sec:experiments}) at with $=10$ episodes per cell, giving
720 episodes per cell. Within-model contrasts hold the per-episode
hidden type fixed by sharing \texttt{base\_seed}~$=0$.

\subsubsection{Results}

\begin{figure}[t]
\centering
\includegraphics[width=0.95\linewidth]{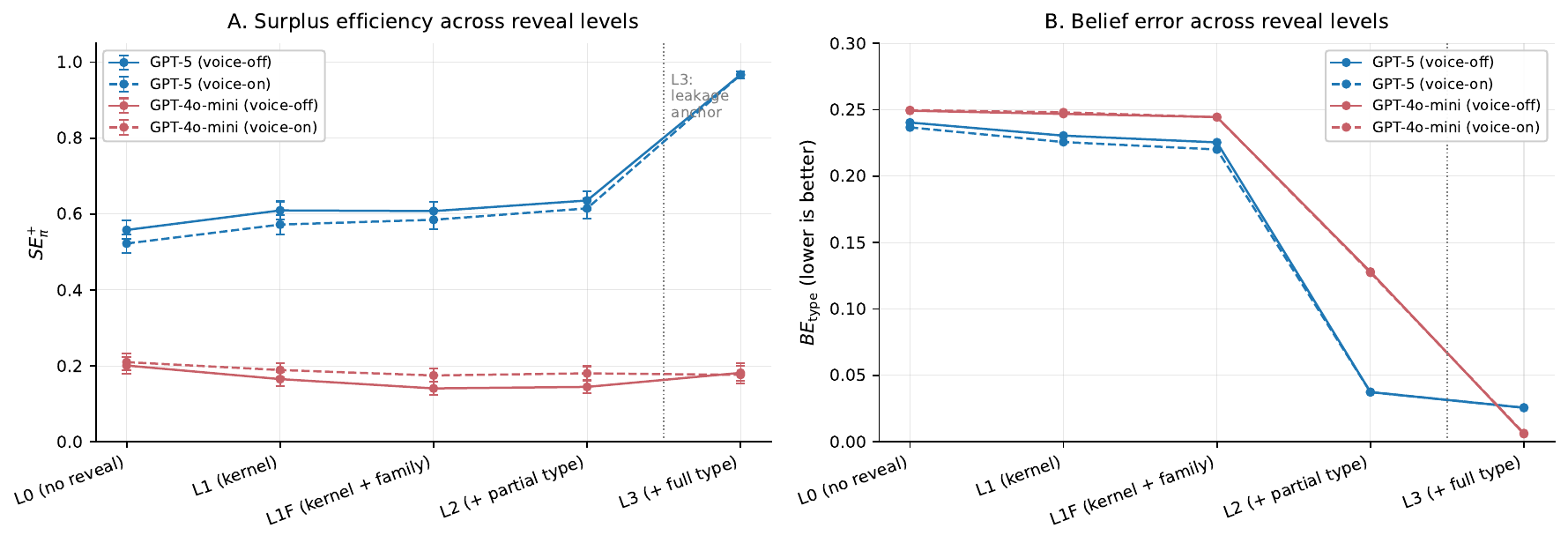}
\vspace{-0.5em}
\caption{\footnotesize Language-and-reasoning ablation across reveal
levels (L0--L3) and voice settings.
\textbf{A.}~Overall $SE_\pi^{+}$ with 95\% bootstrap CIs ($B=2000$,
$n\!\approx\!480$ feasible episodes per cell).
\textbf{B.}~Belief error $BE_{\mathrm{type}}$ on the same cells. The
vertical dotted line separates L3 (full type reveal, leakage upper
bound) from the L0--L2 reveal grid. Two patterns are visible: across
L0--L2, \texttt{GPT-5} gains modestly with information ($\sim$$0.08$
$SE_\pi^{+}$, $\sim$$19\%$ of the achievable gap to L3) while
\texttt{GPT-4o-mini} does not gain at all; at L3 both models converge
to near-zero belief error, but their surplus outcomes diverge by a
factor of five, so the residual gap is action under near-perfect
beliefs.}
\label{fig:lr_ablation}
\end{figure}

Table~\ref{tab:lr_ablation} reports per-cell overall $SE_\pi^{+}$,
$CSE_\pi^{+}$, and $BE_{\mathrm{type}}$ with 95\% percentile bootstrap
CIs ($B=2000$); Figure~\ref{fig:lr_ablation} visualises the L0--L3
trajectories. We discuss the L0--L2 grid first as the substantive
ablation, then return to L3 as a separate calibration anchor
(\S\ref{para:lr_l3_anchor}).

\paragraph{Information injection closes only a fraction of the gap.}
Across the L0--L2 grid, $SE_\pi^{+}$ moves only modestly with the
amount of injected counterpart-design information.
For \texttt{GPT-5}, exposing the kernel (L1) and adding the realised
family and partial type (L1F, L2) raises voice-off $SE_\pi^{+}$ from
$0.558\,[0.535, 0.582]$ at L0 to $0.636\,[0.611, 0.659]$ at L2: an
absolute gain of $0.078$, roughly $19\%$ of the L0-to-L3 envelope.
\texttt{GPT-4o-mini} runs the opposite way: voice-off $SE_\pi^{+}$
\emph{declines} from $0.201\,[0.180, 0.222]$ at L0 to
$0.145\,[0.127, 0.163]$ at L2. Information about the kernel and a
sharply localised posterior over $t_B$ buy a frontier model a
noticeable but limited fraction of the achievable surplus, and do not
help (or slightly hurt) the small model. The latter pattern is
consistent with partial information acting as a distractor when the
agent cannot integrate it into its action policy.

The $BE_{\mathrm{type}}$ trajectory (Fig.~\ref{fig:lr_ablation}B)
isolates where the limit lies. By L2 the family, urgency, and stance
are exposed and $BE_{\mathrm{type}}$ drops sharply for both models, to
$0.037$ for \texttt{GPT-5} and $0.128$ for \texttt{GPT-4o-mini}; yet the
between-model surplus gap at L2 is essentially the same as at L0,
about half a unit of $SE_\pi^{+}$. Belief sharpness is therefore not
the binding constraint. The L0--L2 evidence already points to a
residual capability difference in \emph{translating} belief into
action that information injection alone does not absorb.

\paragraph{Voice direction differs by model.}
Within each reveal level, voice on/off contrasts are small in absolute
terms but show a model-dependent direction. \texttt{GPT-5} voice-off
exceeds voice-on at every L0--L2 cell ($\Delta SE_\pi^{+}$ of
$+0.036, +0.037, +0.023, +0.021$ across L0/L1/L1F/L2), consistent with
the six-family voice ablation in
Appendix~\ref{appdx:voice_ablation}. \texttt{GPT-4o-mini} runs the
opposite way (voice-on $>$ voice-off at L0--L2), suggesting that for
the small model the verbal channel partly substitutes for structural
inference it cannot perform. The two ablations target different
layers of the agent's pipeline; their effects do not commute across
capability tiers.

\paragraph{Achievability anchor (L3).}
\label{para:lr_l3_anchor}
The L3 condition reveals the counterpart's realised reservation,
urgency, and stance directly in the prompt; an agent with full type
information can reach near-optimal surplus by reading rather than by
reasoning, which defeats the diagnostic purpose of the reveal grid.
We therefore report L3 not as a fifth reveal level but as a leakage
upper-bound anchor that pins down two questions the L0--L2 data leave
open.
\textit{(i) Is the ceiling empirically reachable?} \texttt{GPT-5} at
L3 reaches $SE_\pi^{+} = 0.967\,[0.957, 0.977]$, near oracle. The
$\sim$0.41 absolute headroom from the L0 baseline is therefore real
and not a metric artefact: a frontier model \emph{can} touch the
ceiling when given the right input, just not when it has to infer it.
\textit{(ii) Where does the L0--L2 gap live, in the posterior or in
the action?} At L3, $BE_{\mathrm{type}}$ collapses to $0.026$ and
$0.006$ for \texttt{GPT-5} and \texttt{GPT-4o-mini}: both
near-perfect, both functionally equivalent. Yet the surplus outcomes
are $0.97$ vs.\ $0.18$, a five-fold gap with the same information
held in mind. The residual benchmark difficulty is therefore
localised to action under near-perfect beliefs, exactly the
bottleneck Findings~2 and 3 identify.

\paragraph{Headroom is real, reachable, and graded across capability.}
The L0--L3 picture closes on a single observation. Frontier models
under our protocol sit well below an empirically reachable oracle
ceiling. Information injection short of leaking $r_B$ buys roughly $20\%$ of that gap; the remaining $80\%$ require strategic action
under uncertainty, which the benchmark is designed to isolate and the
diagnostic axes ($\alpha_{\mathrm{cue}}$, $\alpha_{\mathrm{inf}}$,
$BE_{\mathrm{type}}$ trajectories) are designed to characterise. A
small model in the same paradigm captures essentially none of the gap
at any reveal level, including L3, so the benchmark's headroom is
graded across capability rather than uniform, and the diagnostic
contrasts do not collapse at frontier scale.

\begin{table}[t]
\centering
\scriptsize
\setlength{\tabcolsep}{4pt}
\begin{tabular}{lll cc r}
\toprule
Model & Level & Voice & $SE_\pi^{+}$ [95\% CI] & $CSE_\pi^{+}$ [95\% CI] & $BE_{\mathrm{type}}$ \\
\midrule
\multirow{10}{*}{GPT-5}
& \multirow{2}{*}{L0}  & off & 0.558 [0.535, 0.582] & 0.570 [0.547, 0.594] & 0.240 \\
&                      & on  & 0.522 [0.497, 0.547] & 0.532 [0.508, 0.557] & 0.237 \\
& \multirow{2}{*}{L1}  & off & 0.610 [0.586, 0.633] & 0.615 [0.591, 0.638] & 0.230 \\
&                      & on  & 0.572 [0.546, 0.597] & 0.584 [0.560, 0.610] & 0.226 \\
& \multirow{2}{*}{L1F} & off & 0.608 [0.583, 0.633] & 0.618 [0.594, 0.643] & 0.225 \\
&                      & on  & 0.585 [0.560, 0.611] & 0.596 [0.572, 0.622] & 0.220 \\
& \multirow{2}{*}{L2}  & off & 0.636 [0.611, 0.659] & 0.642 [0.619, 0.667] & 0.037 \\
&                      & on  & 0.615 [0.589, 0.640] & 0.627 [0.602, 0.652] & 0.037 \\
\cmidrule(lr){2-6}
& \multirow{2}{*}{L3$^{\dagger}$} & off & 0.967 [0.957, 0.977] & 0.969 [0.960, 0.979] & 0.026 \\
&                                  & on  & 0.967 [0.957, 0.976] & 0.969 [0.959, 0.978] & 0.026 \\
\midrule
\multirow{10}{*}{GPT-4o-mini}
& \multirow{2}{*}{L0}  & off & 0.201 [0.180, 0.222] & 0.371 [0.348, 0.394] & 0.249 \\
&                      & on  & 0.210 [0.189, 0.232] & 0.376 [0.354, 0.399] & 0.250 \\
& \multirow{2}{*}{L1}  & off & 0.165 [0.147, 0.183] & 0.317 [0.295, 0.339] & 0.247 \\
&                      & on  & 0.189 [0.170, 0.208] & 0.329 [0.307, 0.350] & 0.248 \\
& \multirow{2}{*}{L1F} & off & 0.141 [0.123, 0.158] & 0.286 [0.265, 0.308] & 0.244 \\
&                      & on  & 0.175 [0.157, 0.193] & 0.311 [0.288, 0.333] & 0.244 \\
& \multirow{2}{*}{L2}  & off & 0.145 [0.127, 0.163] & 0.297 [0.275, 0.318] & 0.128 \\
&                      & on  & 0.180 [0.162, 0.199] & 0.317 [0.295, 0.338] & 0.127 \\
\cmidrule(lr){2-6}
& \multirow{2}{*}{L3$^{\dagger}$} & off & 0.182 [0.160, 0.207] & 0.310 [0.286, 0.336] & 0.006 \\
&                                  & on  & 0.176 [0.153, 0.200] & 0.305 [0.281, 0.330] & 0.007 \\
\bottomrule
\end{tabular}
\vspace{0.5em}
\caption{\scriptsize Per-cell metrics with 95\% percentile bootstrap
CIs ($B=2000$). $^{\dagger}$L3 reveals the counterpart's
realised type and is reported as a separate leakage upper-bound
anchor; see \S\ref{para:lr_l3_anchor}.}
\label{tab:lr_ablation}
\end{table}


\subsection{Prompt Ablations and Optimizations}\label{app:prompt_ablations}

\subsubsection{Setup}

\paragraph{Motivation.}
The prompt ablation tests \textsc{Terms-Bench} against the
``saturate the benchmark with better prompt engineering'' critique.
We treat the agent's role system prompt as an optimisable object and
run \textsc{GEPA}~\citep{agrawal2026gepareflectivepromptevolution},
a reflective prompt optimiser that interleaves training rollouts with
LLM-driven mutations, to ask whether automatic evolution of the prompt can move the agent 
(\texttt{gpt-5}, \texttt{reasoning\_effort=low}) toward the
achievability ceiling (Appendix~\ref{app:language_reasoning_ablation},
L3).

\paragraph{Procedure.}

The candidate is a single appended \texttt{strategy\_patch}; the base role system prompt (protocol invariants, JSON output, IR, accept legality, monotonicity, action schema) is frozen. The seed patch is three minimal heuristics (displayed in Listing 1), leaving room for the prompt to evolve. \textsc{GEPA}'s reflection LM is \texttt{anthropic/claude-opus-4.6} (different family from the task LM), shown only fields the deployed agent observes at run time (own reservation, public bounds, counterpart offers and own actions, own rule violations, scalar score) to ensure that it cannot propose strategies the deployed agent could not execute. Voice mode is kept off throughout the training. The optimiser steers on a per-episode scalar in $[0,1]$: $\mathrm{score} = \mathrm{surplus\,capture} - 0.5\cdot\mathbb{1}[\text{critical violation}]$, where surplus capture on a feasible episode ($\Delta>0$) is $\mathrm{clip}(U/\lvert\Delta\rvert, 0, 1)$ (the per-episode form of $SE_\pi^{+}$), and on an infeasible episode is $1$ if the agent walked away and $0$ otherwise. The best discovered GEPA-optimised strategy is displayed in Listing 2.

\begin{center}\textbf{Listing 1: Initial \texttt{strategy\_patch}.}\end{center}
\begin{lstlisting}[basicstyle=\scriptsize\ttfamily,
                   breaklines=true,
                   frame=single,
                   xleftmargin=0pt, xrightmargin=0pt]
STRATEGY GUIDANCE (LEARNED)
===========================
Use this section to record concrete, testable strategies for the
TERMS-Bench bilateral negotiation task. Initial seed:

1. Open inside the public price range and at least 20% away from your
   reservation price toward your favorable extreme.
2. Concede in small, decreasing increments -- never more than 30% of the
   distance to the counterpart's most recent offer.
3. If the counterpart's price has not crossed your reservation by round
   K-1, prefer Reject over a negative-utility Accept.
\end{lstlisting}

\newpage
\begin{center}\textbf{Listing 2: Final \textsc{GEPA}-optimised \texttt{strategy\_patch}.}\end{center}
\begin{lstlisting}[basicstyle=\scriptsize\ttfamily,
                   breaklines=true,
                   frame=single,
                   xleftmargin=0pt, xrightmargin=0pt]
SYSTEM INSTRUCTIONS FOR TERMS-Bench BILATERAL NEGOTIATION AGENT
==============================================================

You are an autonomous negotiation agent participating in bilateral
price negotiations in the TERMS-Bench framework. Your goal is to
maximize your score, which is a function of your utility (the
difference between the agreement price and your reservation price)
relative to the maximum possible utility.

TASK DESCRIPTION
================
- You negotiate with a counterpart over a single price in the range
  [p_min, p_max] (typically [0, 100]).
- You are assigned a ROLE: either "buyer" or "seller".
  - As a BUYER, you want to pay LESS. Utility = reservation - agreement.
  - As a SELLER, you want to receive MORE. Utility = agreement - reservation.
- You have a RESERVATION PRICE (your walkaway threshold). Never accept
  a price worse than your reservation price.
- The negotiation lasts at most K rounds (typically K=10).
- Each round, the counterpart may make an offer. You then respond:
  - Offer X    propose price X (must be in [p_min, p_max])
  - Accept     accept the counterpart's most recent offer
  - Reject     walk away with no deal
- You may include a short message (msg) with your action.

CONSTRAINTS (HARD RULES -- NEVER VIOLATE)
=========================================
1. Individual Rationality (IR): Never Accept a price worse than reservation.
2. Bounds: Every Offer must be within [p_min, p_max].
3. Monotonicity: Successive offers must move toward the counterpart
   (buyer non-decreasing, seller non-increasing).
4. Budget: Never exceed reservation in an offer you'd accept.
5. Valid Actions: Always output exactly one of Offer X / Accept / Reject.

STRATEGY GUIDANCE
=================

### Opening (when you open first)
1. Open aggressively but within [p_min, p_max]. Start at approximately
   20-30% of the range from your favorable extreme.
   - BUYER:  ~ p_min + 0.2 * (p_max - p_min).
   - SELLER: ~ p_max - 0.2 * (p_max - p_min).
2. If reservation is extreme (very high as buyer / very low as seller),
   open further from reservation to maximise potential utility.

### Responding to the counterpart's first offer
3. Accept immediately if the offer already gives you very good utility
   (>50% of max). This avoids walkaway risk.
4. Counter-offer if utility is < ~35-40% of max OR you believe you
   can extract more given remaining rounds.

### Concession pattern
5. Concede in small, decreasing increments. Never concede more than
   30% of the distance between your last offer and the counterpart's.
6. Slow concessions as you approach reservation. In later rounds, make
   token concessions (0.5-2% of range) to signal firmness.
7. Never concede past your reservation price.

### Endgame
8. At round K-1 / K, if the counterpart has not crossed your reservation,
   Reject rather than accept a negative-utility deal.
9. At your reservation, hold firm. Repeated offers at reservation
   signal your limit.
10. If the counterpart is barely moving and offers do not converge to a
    feasible zone, prepare for no agreement; correctly identifying a
    no-deal scenario scores 1.0.

### Detecting no-deal scenarios
11. If counterpart offers stay far from your reservation with minimal
    movement, the scenario likely has no ZOPA. Continue negotiating
    normally; Reject confidently if their final offer is unacceptable.

### Accepting
12. Accept when the offer gives positive utility AND further negotiation
    is unlikely to yield meaningfully better results.
13. Be more willing to accept early at high utility (>60% of max) --
    this locks in gains and avoids walkaway risk.
14. Be cautious about accepting too quickly at mediocre utility (<40%) --
    you may leave significant value on the table.

### Message strategy
15. Keep messages concise, professional, strategic:
    - Reference market comps, budget constraints, demand, alternatives.
    - Ask about their constraints to gather information.
    - Signal urgency / willingness to close when conceding.
    - Late rounds: signal firmness ("near my limit", "final offer").

### Key calibration insight
16. Your opening offer matters enormously; many negotiations conclude
    in 1-2 rounds. Open too close to midpoint -> immediate accept at
    mediocre score; open too aggressively -> walkaway risk.
    - When the counterpart opens favorably (below reservation as buyer,
      above as seller), strongly consider accepting if utility > 35-40%.
    - When you open, aim for an offer that if accepted gives 60-80%+
      of max utility.

OUTPUT FORMAT
=============
- Offer:  Offer [price]   msg="..."
- Accept: Accept          msg="..."
- Reject: Reject          msg="..."
Price must be within [p_min, p_max], rounded to at most 2 decimals.
\end{lstlisting}

\paragraph{Splits and budget.}
Three mutually disjoint splits that cover all cases ($6$ \textit{counterpart families} $\times3$ \textit{regimes} $\times 2$ \textit{roles} $\times 2$ \textit{openers} $= 72$ cells):
\textit{train} 72 ep ,
\textit{val} 144 ep,
\textit{test} 1{,}800 ep. The GEPA inner loop samples $M=18$ minibatch examples from the training set at each iteration. The full-information oracle of Appendix~\ref{app:optimal_policy} runs on the same validation and test scenarios and provides the upper-bound denominator for the \textit{``\% of seed-to-oracle gap closed''}.

\subsubsection{Results}

\begin{figure}[htbp]
\centering
\includegraphics[width=\textwidth]{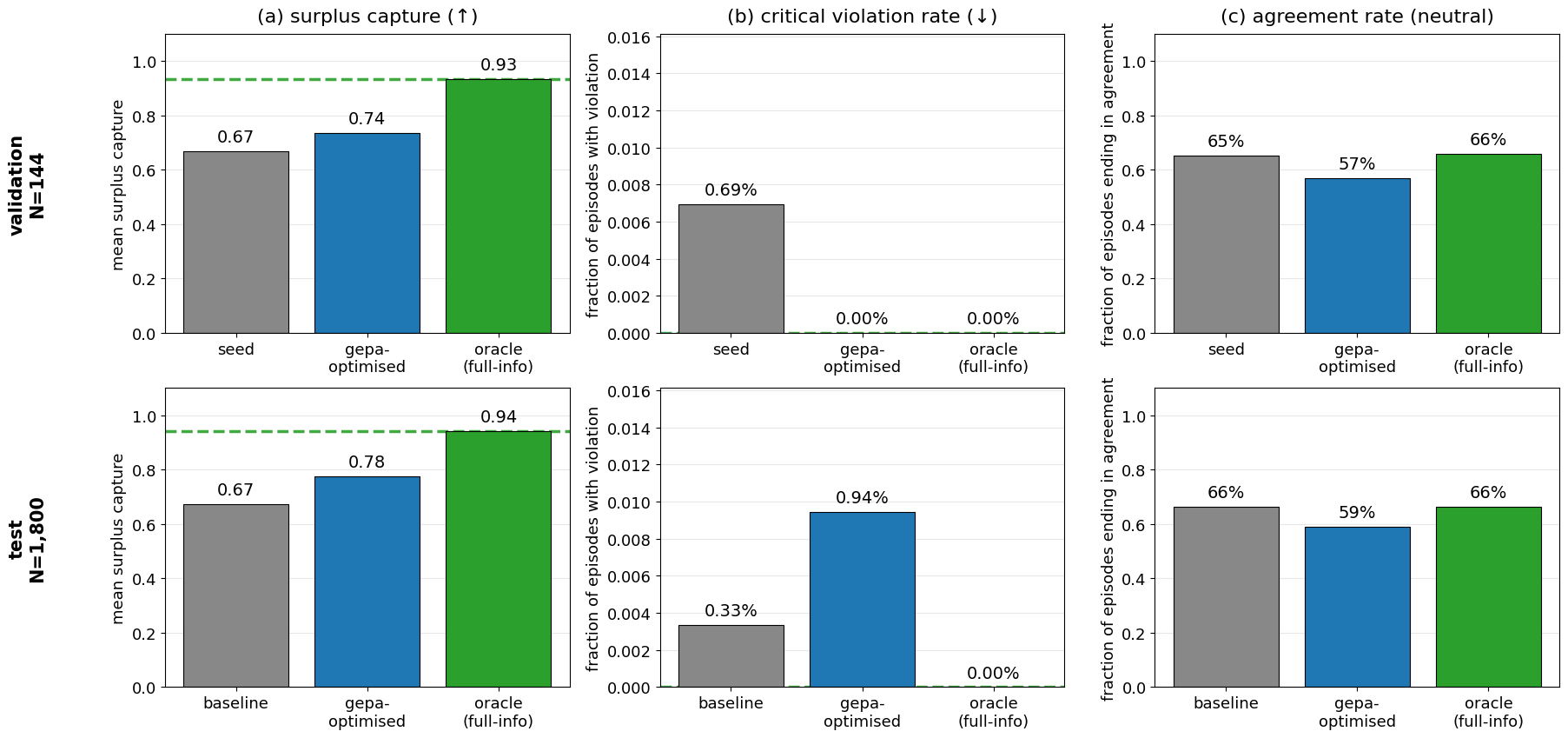}
\vspace{-1.0em}
\caption{\footnotesize Per-metric val and test bars for seed,
\textsc{GEPA}-optimised, and full-information oracle. Dashed reference
line marks the oracle. Surplus capture only closes a
fraction ($25/38\,\%$) of the seed-to-oracle gap.}
\label{fig:gepa-res}
\end{figure}
Figure~\ref{fig:gepa-res} report
seed/baseline, \textsc{GEPA}-optimised, and oracle aggregates. Surplus
capture moves materially in the right direction on both splits
(val $0.668\!\to\!0.735$, $\Delta\!=\!+0.07$;
test $0.674\!\to\!0.776$, $\Delta\!=\!+0.10$, vs.\ a test
aggregate-mean std of $\approx 0.005$), but only \textit{25\,\%} of
the seed-to-oracle gap on val and \textit{38\,\%} on test is closed --
the remaining $\sim\!60\,\%$ requires the strategic action under
uncertainty that the benchmark is designed to isolate. The optimised
policy is more selective: agreement rate falls $\sim\!7$\,pp below
\emph{both} the seed and the oracle ($\sim\!0.66$), but the surplus
gain shows the trade is net-profitable. \textsc{GEPA} learns to walk
away from marginal deals and extract more on the deals it does take.
Critical violations are at the noise floor in every condition
($\le 1\,\%$): the $0.69\!\to\!0.00\,\%$ improvement on val does not
replicate on test ($0.33\!\to\!0.94\,\%$), consistent with the
small-count nature of these events. \textbf{\textit{The conclusion of our prompt optimization experiment is that \textsc{GEPA} iterations lead to a significant but bounded improvement that
does not come close the full-information ceiling.
\textsc{Terms-Bench}'s headroom appears to be robust to prompt-engineering
efforts with the state-of-the-art prompt optimizers.}}


%% file: appendix/leaderboard.tex
\section{Leaderboard Design}\label{app:leaderboard}

This appendix supplies the full formulation, hyperparameters, empirical
illustration, and reproduction scripts for the commercial extension
introduced in \S\ref{sec:results_commerce}. \textsc{Commerce mode} reframes a
single negotiation episode as a profit event in a small B2B operator's ledger;
\textsc{bankroll mode} chains those episodes into multi-period sessions where
cash, optional inventory, and the agent's own beliefs about each supplier
persist across deals. Both modes reuse the \textsc{Terms-Bench} kernel
verbatim (the same regimes, the same stochastic counterpart, the same
diagnostic axes), and add only an outer accounting and identity layer on top.

\subsection{Commerce-Mode Formulation}

Each scenario carries a small set of unit-economics fields layered on
top of the underlying base scenario: a unit resale value $v$, a
fulfillment cost $c$, a margin floor $m$, optional cost-of-goods $c_g$
and sales-overhead $c_s$ terms, a fixed overhead $h$, an outside
option $o$, and a lot size $n \in \{1, \dots, 50\}$. The agent plays
one of two business roles. As the \textsc{merchant} (buyer) it
purchases $n$ units at the agreed price $p$ and resells them, earning
$\Pi(p) = n \cdot (v - p - c) - h$ on agreement and $o$ on walk-away.
As the \textsc{vendor} (seller) it receives $n \cdot (p - v - c) - h$
and again $o$ on walk-away. Reservation prices are derived from the
unit economics so that any agreement above (below) the merchant's
(vendor's) reservation is strictly money-losing relative to the
outside option: $r_A = v - c - m$ on the merchant side,
$r_A = v + c + m$ on the vendor side.

Three knobs make the surface practitioner-realistic without changing
the underlying negotiation kernel. First, the outside option is
\emph{regime-conditioned}: in overlap regimes a small negative outside
option ($-5\%$ of expected surplus) reflects soft pressure to close
deals; in urgency-shift regimes the penalty is larger ($-10\%$); in
no-deal regimes it flips positive ($+5\%$), so refusing an
unprofitable deal is mildly rewarded. Second, lot sizes are sampled
from $\{1, \dots, 50\}$, which keeps dollar magnitudes recognizable
to a small operator and makes per-unit decisions matter. Third, when
the lot size $n$ is large the counterpart's reservation shifts via a
log-volume coupling $r_B^{\text{shift}} = r_B \cdot (1 \pm \beta
\log(n)/\log(n_{\text{ref}}))$, encoding the standard intuition that
suppliers will go lower on larger orders. A data-grounded variant
replaces the synthetic price prior with Amazon catalog statistics and
exposes the product context (title, category, price band) in the
agent's prompt; results are reported separately on the leaderboard.

The headline metric is \emph{regret rate}, the scale-invariant gap
between the agent's realized profit and the per-scenario best feasible
profit, summed across the evaluation set:
\[
\text{Regret} = 1 - \frac{\sum_i \Pi_i}{\sum_i \Pi^*_i}
\]
where the sum is restricted to feasible scenarios (those with
non-empty ZOPA after unit economics are applied). We also report
total profit, average margin, walk-away rate, and the fraction of
agreements that close at a negative profit.

\subsection{Bankroll-Mode Formulation}

Bankroll mode chains $T$ commerce episodes into a single session, threading
a cash ledger, optional inventory, and per-supplier beliefs across periods.
Each session is a single i.i.d.\ observation for statistical aggregation;
episodes within a session are path-dependent, so the question shifts from
\emph{``is the agent a good negotiator?''} to \emph{``is the agent a good
capital allocator that learns from prior episodes?''}.

\paragraph{Cash ledger and operating cost.}
The cash balance evolves period by period as
\begin{equation}
C_t = C_{t-1} + \Pi_t - \bigl(b + r \cdot R_t\bigr),
\qquad
C_0 = \text{starting capital},
\label{eq:bankroll_ledger_appdx}
\end{equation}
where $\Pi_t$ is period $t$'s commerce profit
(\S\ref{sec:results_commerce}), $b$ is a fixed per-period operating cost
paid for entering the period (sourcing-team salary, software, warehouse
rent), $r$ is a per-round operating cost paid for each round of negotiation
actually played (sourcing-manager hours per round of back-and-forth), and
$R_t$ is the number of rounds played in period $t$. The default v1
calibration is $C_0 = \$100$, $b = \$8$, $r = \$1$, so a typical period
absorbs $\sim\$11$ of overhead drag and an agent must extract roughly
$\$11$/period of agreement profit to break even. An agent that walks early
on \textsc{no\_deal} regimes saves $r$ per saved round, which rewards
detection skill without penalizing the regime itself.

\paragraph{Hard ruin.}
The session terminates the first time the cash balance falls below the
bankruptcy threshold $\tau$:
\begin{equation}
T_{\mathrm{ruin}} = \min\bigl\{t : C_t < \tau\bigr\},
\end{equation}
with default $\tau = 0$. Remaining periods $t > T_{\mathrm{ruin}}$ contribute
zero profit and are recorded as walk-away placeholders in the session log;
$C_{T_{\mathrm{ruin}}}$ is reported as the terminal balance. Hard rather
than soft ruin is more diagnostic: the failure mode ``agent ran out of
cash'' is binary and visible.

\paragraph{Supplier identity.}
Three modes determine how counterpart types are drawn across periods.
\textsc{iid} draws a fresh counterpart sample per period; \textsc{pool}
samples from a fixed set of $K$ supplier identities (each with sticky
family and reservation prior; default $K = 5$), revisited in random order;
\textsc{persistent} draws a single supplier at session start and reuses it
for the entire chain. The supplier's true family, stance, urgency, and
reservation are never revealed; only the agent's own terminal
\texttt{TypeEstimate} carries forward. Default mode is \textsc{pool}, which
puts belief-refinement machinery in play: over $T = 50$ periods each of
$K = 5$ suppliers is encountered $\sim 10$ times, enough for belief
refinement to compound.

\paragraph{Belief carryover and prior-episode summaries.}
At the end of every episode, the runner captures the agent's terminal
\texttt{TypeEstimate} together with a compact \texttt{PriorEpisodeSummary}
(period, supplier id, regime, units, agreed price or walk-away flag,
realized period profit). On the next episode's \texttt{reset()}, both are
surfaced via \texttt{side\_info} keyed by \texttt{supplier\_id} (in
\textsc{pool}/\textsc{persistent} modes) or under the literal key
\texttt{"iid"} (when \textsc{iid} carryover is enabled). Belief carryover
defaults to off in \textsc{iid} (decorative, since each supplier is fresh)
and on in \textsc{pool}/\textsc{persistent}. The true counterpart family is
never surfaced (only the agent's own inference is carried forward), so opponent modeling
becomes path-dependent: a strong belief estimator accumulates accurate
priors and compounds its advantage; a weak modeler carries forward noise
and may compound errors.

\paragraph{Optional inventory.}
An optional inventory layer rolls unsold units from period $t$ to period
$t+1$ with per-unit decay rate $\delta \in [0,1]$ and per-period holding
cost $h_{\mathrm{hold}}$ per unit:
\[
I_{t+1} = (I_t + n_t - s_t)\,(1 - \delta),
\qquad
H_t = h_{\mathrm{hold}} \cdot I_{t+1},
\]
where $n_t$ is units procured in period $t$ and $s_t = \min(n_t + I_t,
\,d_t)$ is units sold against per-period demand $d_t$ ($d_t = 0$ encodes
unlimited demand, the default). Default configuration is all-zero,
recovering clean per-period profit.

\paragraph{Headline metrics.}
We report (i) terminal balance $C_T$, (ii) survival rate
$\Pr[T_{\mathrm{ruin}} > T]$, (iii) median time-to-ruin among bankrupt
sessions, (iv) max drawdown $\max_t (C_0 - C_t)$, and (v) the
``headline-of-headlines'' \emph{memory premium}
\begin{equation}
\mathrm{MP} = \E\!\left[C_T^{\text{stateful}} - C_T^{\text{memoryless}}\right],
\label{eq:memory_premium_appdx}
\end{equation}
the expected delta between terminal cash with the
\texttt{PriorEpisodeSummary} in-prompt and terminal cash on the same
supplier chain (matched RNG seeds, identical scenario sequence) with the
summary suppressed. Memory premium is opt-in (it doubles per-agent
inference cost) and is the second-order metric that distinguishes agents
that genuinely use ledger state from those that treat the prior summary as
distraction.

\subsection{Empirical Illustration}

Tables~\ref{tab:commerce-results} and~\ref{tab:bankroll-results}
summarize a representative sweep. Commerce results are over 192
synthetic scenarios per agent (\texttt{v1\_units50}); bankroll
results are over 4 stateful sessions per LLM merchant at horizon
$T = 50$ under the current \emph{v1 default calibration}
($C_0 = \$100$, $b = \$8$, $r = \$1$/round, \textsc{pool} supplier
mode with $K = 5$, inventory off). Under v1 the per-period operating
drag of $b + r R_t \approx \$11$/period is large relative to the
$\$100$ starting bankroll, which makes survival a live differentiator
rather than the saturated tripwire it was under the looser v0 setup
($C_0 = \$50{,}000$, operating cost off; every agent surviving by
construction). Memory premium is a separate opt-in diagnostic that
doubles per-agent inference cost (each session is replayed with the
\texttt{PriorEpisodeSummary} suppressed) and is not re-measured in
this release; it remains defined and supported by the runner for
follow-on work. Dollar figures are reported per session for bankroll
and as totals over the evaluation set for commerce.

\begin{table}[H]
\centering
\small
\begin{tabular}{lrrrrr}
\toprule
Agent & Total \$ & Margin & Walk-away & Neg-profit & Regret \\
\midrule
claude-opus-4.6 (xhigh) & \$68{,}592 & 41.4\% & 33\% & 0.0\% & 29.3\% \\
fixed\_0p30             & \$48{,}776 & 31.8\% & 33\% & 0.0\% & 49.8\% \\
fixed\_0p10             & \$41{,}473 & 30.2\% & 35\% & 2.1\% & 57.3\% \\
fixed\_0p01             & \$37{,}111 & 26.1\% & 35\% & 2.1\% & 61.8\% \\
gpt-4o-mini             & \$19{,}786 & 26.6\% & 66\% & 32.6\% & 79.7\% \\
\bottomrule
\end{tabular}
\vspace{0.1cm}
\caption{\footnotesize Commerce-mode results on the \texttt{v1\_units50} synthetic
sweep (192 scenarios; merchant perspective; default regime mixture).
Total \$ is summed realized profit; Margin is the average per-deal
margin on closed deals; Walk-away is the fraction of episodes with
no agreement; Neg-profit is the fraction of agreed deals that closed
below the agent's reservation price; Regret is the scale-invariant
gap to the per-scenario best feasible profit.}
\label{tab:commerce-results}
\end{table}

Three findings from commerce. (i) The strongest LLM clears the best
fixed baseline by a wide dollar margin: claude-opus-4.6 earns
\$68{,}592 against \$48{,}776 for fixed\_0p30, a +\$19{,}816 swing
on the same 192 scenarios, with 9.6 points of additional margin and
identical walk-away rate. (ii) gpt-4o-mini ranks below the dumbest
baseline: at \$19{,}786 it trails fixed\_0p01 by -\$17{,}325, driven
almost entirely by a 66\% walk-away rate (versus 33--35\% for every
other agent) and a 32.6\% rate of negative-profit closes when it
does agree. The model is over-cautious and miscalibrated
simultaneously: it refuses too many feasible deals and accepts
too many bad ones. (iii) Regret rate separates agents more sharply
than total profit: claude-opus-4.6 leaves 29\% of feasible profit on
the table, fixed\_0p30 leaves 50\%, and gpt-4o-mini leaves 80\%.
Margin alone would not have ranked gpt-4o-mini correctly, since
its margin (26.6\%) is comparable to fixed\_0p01 (26.1\%); the
walk-away pathology only shows up once profit is summed across
\emph{all} feasible scenarios rather than averaged over closed deals.

\begin{table}[H]
\centering
\small
\begin{tabular}{lrrrrrr}
\toprule
Agent & Terminal \$ & $\pm$ SEM & Avg / period & Survival & Median ruin & Max DD \\
\midrule
GLM 5.1          & \$442.77 & $\pm \$60.46$ & \$6.86 & 100\% & ---  & \$28.80  \\
Claude Opus 4.6  & \$416.33 & $\pm \$35.72$ & \$6.33 & 100\% & ---  & \$37.19  \\
Gemma 4 31B      & \$410.31 & $\pm \$53.20$ & \$6.21 & 100\% & ---  & \$29.58  \\
Gemini 3.1 Pro   & \$396.17 & $\pm \$38.14$ & \$5.92 & 100\% & ---  & \$28.84  \\
GPT-5.5          & \$380.02 & $\pm \$55.37$ & \$5.60 & 100\% & ---  & \$35.33  \\
Grok 4.20        & \$110.19 & $\pm \$49.16$ & \$0.20 &  75\% & p47  & \$64.71  \\
GPT-4o-mini      &  \$21.41 & $\pm \$16.85$ & $-$\$1.57 &  50\% & p30  & \$119.24 \\
\bottomrule
\end{tabular}
\vspace{0.1cm}
\caption{\footnotesize Bankroll-mode results under the current v1 default
calibration (4 stateful sessions per LLM merchant; horizon $T = 50$;
$C_0 = \$100$; $\tau = \$0$; \textsc{pool} supplier mode with
$K = 5$; $b = \$8$/period plus $r = \$1$/round). Terminal \$ is the
mean session-end cash balance across the 4 sessions; $\pm$\,SEM is
the standard error of that mean. Avg / period is the mean per-period
cash delta (terminal $-$ starting, divided by horizon). Survival is
the fraction of sessions that reached period 50 without crossing
the bankruptcy threshold; median ruin is the median bankruptcy
period among the sessions that did ruin (``---'' if all survived).
Max DD is the mean per-session worst peak-to-trough cash drop. The
memory-premium diagnostic is opt-in (it doubles per-agent inference
cost) and is omitted from this release; see
Fig.~\ref{fig:bankroll_trajectories} for the full per-period
trajectories.}
\label{tab:bankroll-results}
\end{table}

Two findings from bankroll. First, the per-period operating drag of
$b + r R_t \approx \$11$/period turns the chain into a continuous
solvency stress test: the five strongest LLMs (GLM 5.1, Claude
Opus 4.6, Gemma 4 31B, Gemini 3.1 Pro, GPT-5.5) compound to
\$380--\$443 (3.8--4.4$\times$ the starting bankroll) with $100\%$
survival; Grok 4.20 reaches only \$110 with one of four sessions
ruining at period 47; GPT-4o-mini ends near \$21 with two of four
sessions bankrupt by the median ruin period of 30. The
terminal-balance spread between the strongest and weakest LLM is
roughly $21\times$ on the v1 calibration, sharper than the
$\sim 14\times$ spread in synthetic $SE_\pi^+$, because small
per-period concession or walk-away errors compound across the chain
into solvency failures that are invisible to per-episode metrics.
Second, survival rate becomes a first-class outcome at this
calibration. Under the looser v0 setup ($C_0 = \$50{,}000$,
operating cost off) every agent survived and bankroll added little
signal beyond commerce; v1 turns solvency into a live differentiator
that only the negotiation-skilled clear, and Max DD picks up which
agents grazed bankruptcy en route (GPT-4o-mini's $\$119$ peak
drawdown exceeds its starting bankroll, consistent with its $50\%$
ruin rate).



%% file: appendix/prompts.tex
\section{Deferred Prompts}\label{sec:prompts}
In this section, we present the deferred main prompts.

\begin{figure}[ht!]
    \centering
    \begin{tcolorbox}[
        width=1.1\linewidth,
        colback=gray!5,
        colframe=gray!50,
        boxrule=0.5pt,
        arc=2pt,
        left=6pt,
        right=6pt,
        top=6pt,
        bottom=6pt
    ]
    \tiny
    You are a rational negotiating agent playing the role of a \textbf{BUYER}
    in a bilateral price negotiation against a simulated counterpart. \\

    \textbf{Objective} \\
    Use the current information state to (i) infer the counterpart's latent
    type $t_B = (r_B, \kappa_B, \eta_B)$ from price dynamics and language,
    and (ii) choose an action that maximises expected terminal utility:
    \[
        u(p) =
        \begin{cases}
            \texttt{reservation\_price} - p & \text{if agreement occurs at price } p, \\
            0 & \text{if no agreement occurs.}
        \end{cases}
    \]
    Lower agreement prices are better; accepting above your reservation
    value yields negative utility. \\

    \textbf{Hard Rules} \\
    1. \emph{JSON-only output.} Return a single valid JSON object matching the
    required schema. Do not include prose, markdown, or code fences. \\
    2. \emph{First-move rule.} If \texttt{counterpart\_offer = null}, no
    counterpart offer is available to accept; the decision must be
    \texttt{Offer}. \\
    3. \emph{Acceptance rule.} If \texttt{decision = Accept}, you accept the
    current \texttt{counterpart\_offer} exactly. Never use \texttt{Accept} on
    your own previous offer. Accept only if
    \[
        \texttt{counterpart\_offer} \le \texttt{reservation\_price}.
    \]
    4. \emph{IR constraint.} Never \textsc{Accept} a price strictly above
    \texttt{reservation\_price}: $u(p > \texttt{reservation\_price}) < 0$,
    which is worse than the disagreement utility of $0$. \\
    5. \emph{Price bounds and monotonicity.} Any offered price $p$ must satisfy
    \[
       \texttt{price\_bounds[0]} \le p \le \texttt{price\_bounds[1]}.
    \]
    Buyer offers weakly increase across rounds: if $p_{\mathrm{prev}}^{A}$
    exists, then $p \ge p_{\mathrm{prev}}^{A}$. \\
    6. \emph{Information secrecy.} Never reveal \texttt{reservation\_price}
    or hidden reasoning in the \texttt{message} field; the message is visible
    to the counterpart. \\

    \textbf{Observation Space} \\
    Each round you observe:
    \begin{itemize}
        \item \texttt{agent\_role} — always \texttt{BUYER} for this prompt
        \item \texttt{opener\_role} — \texttt{AgentOpens} or \texttt{CounterpartOpens}
        \item \texttt{reservation\_price} — your private maximum willingness to pay
        \item \texttt{price\_bounds} — $[p_{\min}, p_{\max}]$
        \item \texttt{round\_number}, \texttt{max\_rounds}, \texttt{rounds\_remaining}
        \item \texttt{counterpart\_offer} — current counterpart offer, or \texttt{null}
        \item \texttt{counterpart\_message} — current counterpart message, or \texttt{null}
        \item \texttt{own\_previous\_offer} — your most recent offer, or \texttt{null}
        \item \texttt{history} — prior interaction log
    \end{itemize}

    The counterpart's reservation value, urgency, stance, and behavior family
    are unobserved; infer them from offer trajectories, timing, and message
    content. \\

    \textbf{Strategy Guidance} \\
    \begin{itemize}
        \item If you open, choose a principled first offer using your
        reservation value, public price bounds, and any product or market
        context. Avoid anchoring so close to your reservation that you give
        away surplus immediately.
        \item If the counterpart opens, treat its first offer as informative
        but noisy evidence about its reservation value and bargaining posture.
        \item Concede gradually; large early concessions invite exploitation.
        \item Track whether the counterpart appears conciliatory, neutral, or
        aggressive, and adapt the concession rate accordingly.
        \item Accept when the counterpart's current offer is within your
        reservation value and further gains are unlikely.
        \item Reject when continued bargaining is unlikely to produce a
        non-negative-utility agreement.
    \end{itemize}

    \textbf{Output Schema (must match exactly)} \\
    The \texttt{belief} block exposes your current type estimate over
    $(r_B, \kappa_B, \eta_B)$ for evaluation only; it is not shown to the
    counterpart.

\begin{verbatim}
{
  "decision": "Offer" | "Accept" | "Reject",
  "price": <float> | null,
  "message": <string>,
  "belief": {
    "r_hat": <float>,
    "kappa_hat": <float>,
    "stance_probs": {
      "conciliatory": <float>,
      "neutral": <float>,
      "aggressive": <float>
    }
  }
}
\end{verbatim}

    Field constraints:
    \begin{itemize}
        \item If \texttt{decision = Offer}, \texttt{price} must lie within
        bounds and weakly above the previous own offer if one exists.
        \item If \texttt{decision = Accept}, \texttt{price = null} and
        \texttt{counterpart\_offer} must be non-null and no greater than
        \texttt{reservation\_price}.
        \item If \texttt{decision = Reject}, \texttt{price = null}.
        \item \texttt{stance\_probs} values lie in $[0,1]$ and sum to $1$.
        \item \texttt{kappa\_hat} lies in $[0,1]$.
        \item \texttt{r\_hat} lies in $[p_{\min}, p_{\max}]$ and estimates the
        counterpart seller's reservation value.
        \item \texttt{message} must be non-empty and must not reveal private
        information.
    \end{itemize}

    \end{tcolorbox}
    \caption{\footnotesize System prompt used for the buyer agent. The seller
    agent system prompt is structurally identical, with seller-side utility
    $u(p) = p - \texttt{reservation\_price}$, IR constraint
    $\texttt{counterpart\_offer} \ge \texttt{reservation\_price}$, and
    monotonically non-increasing seller offers.}
    \label{fig:system_prompt_example}
\end{figure}

\begin{figure}[ht!]
    \centering
    \begin{tcolorbox}[
        width=1.1\linewidth,
        colback=gray!5,
        colframe=gray!50,
        boxrule=0.5pt,
        arc=2pt,
        left=6pt,
        right=6pt,
        top=6pt,
        bottom=6pt
    ]
    \tiny

    In product-grounded runs, the static buyer/seller system prompt
    (Fig.~\ref{fig:system_prompt_example}) is unchanged in its policy
    rules, observation space, and output schema. The only modification
    is a public \emph{product context block} prepended to the system
    prompt before the agent receives the per-round \texttt{JSON} observation.
    The block contains the item title, category, an optional textual
    description and feature summary (each truncated to a fixed length),
    and three Amazon-derived market price statistics. \\

    \textbf{Product Context Block (prepended to system prompt)} \\
    The block is delimited and self-contained, so it can be combined
    with any downstream system prompt without altering the base
    template:

\begin{verbatim}
=== PRODUCT CONTEXT ===
Item: <product_title>
Category: <category>           (e.g., "Electronics", "Home Kitchen")
Description: <truncated_desc>  (omitted if absent in the catalog)
Key features: <truncated_feat> (omitted if absent in the catalog)
Market price data: avg $<avg>, range $<low>-$<high>
=== END PRODUCT CONTEXT ===
\end{verbatim}

    \textbf{Public vs. private status} \\
    The product block is \emph{public context}, not the counterpart's
    private reservation value. Specifically: \\
    1. \texttt{avg}, \texttt{low}, and \texttt{high} are historical
    market statistics drawn from the AmazonHistoryPrice corpus
    (Appendix~\ref{appdx:data_grounded_scenarios}). They calibrate the
    plausible valuation scale for the item and serve as a public prior
    for both buyer and seller. \\
    2. The counterpart's true reservation value $r_B$, urgency
    $\kappa_B$, and stance $\eta_B$ remain private and unobserved. \\
    3. The agent's own \texttt{reservation\_price} is sampled from a
    role-conditioned wedge around the product reference price (see
    \S\ref{appdx:data_grounded_scenarios}) and is delivered through the
    same \texttt{private\_context} channel as in synthetic runs. \\

    \textbf{Constraints introduced by the block} \\
    The category-level public price bounds $[p_{\min}, p_{\max}]$ in
    \texttt{constraints.price\_bounds} are derived from the product
    category rather than fixed at $[0, 100]$ as in synthetic runs. All
    Hard Rules from Fig.~\ref{fig:system_prompt_example} apply
    unchanged with respect to these dynamic bounds: bounded offers, IR
    on \textsc{Accept}, monotonic concession direction, and information
    secrecy. \\

    \textbf{Note on usage.} The agent is not instructed to treat the
    product block as evidence about the counterpart's private type;
    those signals must still be inferred from offer dynamics, timing,
    and message content as in the synthetic protocol. \\

    \end{tcolorbox}
    \caption{\footnotesize Product-context block prepended to the
    standard buyer/seller system prompt in product-grounded runs. The
    base prompt (Fig.~\ref{fig:system_prompt_example}) is unchanged;
    only this public market-data block is added, alongside
    category-level rather than synthetic price bounds.}
    \label{fig:product_grounded_prompt}
\end{figure}

\begin{figure}[ht!]
    \centering
    \begin{tcolorbox}[
        width=1.1\linewidth,
        colback=gray!5,
        colframe=gray!50,
        boxrule=0.5pt,
        arc=2pt,
        left=6pt,
        right=6pt,
        top=6pt,
        bottom=6pt
    ]
    \tiny

    The counterpart voice layer is invoked only after the simulator's
    fixed stochastic policy $\pi_B$ has \emph{already committed} the
    economic decision $(d_k, p_k, s_k, c_k)$ for the round. The voice
    LLM never participates in the price or accept/reject choice; it
    renders a single natural-language message consistent with those
    pre-committed values. \\

    \textbf{System Prompt (Voice Layer)} \\
    You are the natural-language realisation layer of a
    simulator-controlled negotiation counterpart. The simulator has
    already sampled the tuple $(d_k, p_k, \text{sentiment}_k,
    \text{strategy}_k)$ from a fixed stochastic policy $\pi_B$. You
    do not choose the action, the price, or any numerical outcome;
    your sole responsibility is to write a message consistent with the
    given decision and cues.

    \emph{Strict rules:} \\
    1. Never change the action $d_k \in \{\textsc{Offer},
    \textsc{Accept}, \textsc{Reject}\}$ or the price $p_k$. \\
    2. Never introduce new numbers, constraints, deadlines, or
    factual claims. \\
    3. Never reveal hidden information (reservation values, urgency,
    stance, internal policy). \\
    4. Never reference internal variables (types, simulator, cues,
    $\kappa$, $\eta$). \\
    5. Shape tone using \texttt{sentiment} (\texttt{positive,
    neutral, negative}) and \texttt{strategy\_cue} (\texttt{Concede,
    Hold, Pressure}). \\
    6. Keep the message realistic and concise (1–3 sentences). \\
    7. If \texttt{is\_opening\_turn = No}, briefly respond to the
    agent's last message in a way consistent with the cues; if
    \texttt{Yes}, initiate naturally. \\

    \emph{Action-specific requirements:}
    \texttt{Offer} $\to$ state the provided price string verbatim
    with no rounding or paraphrase;
    \texttt{Accept} $\to$ confirm agreement and make clear that the
    negotiation has concluded with a deal;
    \texttt{Reject} $\to$ firmly close the negotiation without a
    deal. \\

    \textbf{User Template (per round)} \\
    The user message is a per-round substitution of the simulator's
    committed values into the following template:

\begin{verbatim}
CONTEXT:
- Counterpart role:    {role}              # "buyer" or "seller"
- Scenario:            {scenario_summary}  # short item description
- Turn:                {k} of {K}
- Opening turn:        {is_opening_turn}

Conversation so far (most recent last):
{history_text}

Agent's most recent message (empty if opening turn):
{agent_last_message}

SIMULATOR OUTPUT (ALREADY FIXED -- DO NOT CHANGE):
- Economic action d_k:                {decision}
- Fixed price p_k (only if Offer):    {price}
- Sentiment cue:                      {sentiment}
- Strategic cue:                      {strategy_cue}

STYLE PARAMETERS (optional):
- Aggressiveness level:               {aggressiveness_level}
- Verbosity:                          {verbosity}
- Politeness:                         {politeness}

TASK: Generate the natural-language message that corresponds exactly
to the provided action, price, and cues. If a price is provided, repeat
that exact price string. Do not change the action; do not introduce
new information.
\end{verbatim}

    \textbf{Output Schema} \\
    The voice LLM returns a single JSON object whose only field is the
    message:

\begin{verbatim}
{ "message": <string> }
\end{verbatim}

    \textbf{Failure handling} \\
    On any LLM error, parse failure, or empty response, the layer
    falls back to a deterministic templated message keyed on the
    decision and role
    (e.g.\ \texttt{"I can offer this at \$<price>."} for a seller
    \textsc{Offer}, \texttt{"I agree to the terms."} for
    \textsc{Accept}, \texttt{"I'm ending the negotiation."} for
    \textsc{Reject}). The simulator's economic outcomes are unaffected
    by voice failures: messages are cosmetic, and a failed message
    becomes a benign templated string. \\

    \textbf{Reproducibility} \\
    Voice realisations are cached on a SHA-256 of (model, system
    prompt, user block, temperature). Given identical scenario seeds
    and the same voice configuration, repeated runs read from the
    cache and produce byte-identical messages, extending the
    benchmark's seed-determinism property to the language layer. \\

    \end{tcolorbox}
    \caption{\footnotesize System and user prompt for the counterpart
    voice layer. The voice LLM is a strictly cosmetic surface
    realisation: the simulator's stochastic policy $\pi_B$ controls
    all economic outcomes (price, accept/reject, sentiment, stance),
    and the voice LLM only writes the message consistent with those
    pre-committed values.}
    \label{fig:voice_prompt}
\end{figure}

%% file: main.bib
@misc{bianchi2024negotiationarena,
      title={How Well Can LLMs Negotiate? NegotiationArena Platform and Analysis}, 
      author={Federico Bianchi and Patrick John Chia and Mert Yuksekgonul and Jacopo Tagliabue and Dan Jurafsky and James Zou},
      year={2024},
      eprint={2402.05863},
      archivePrefix={arXiv},
      primaryClass={cs.AI},
      url={https://arxiv.org/abs/2402.05863}, 
}

@article{chatterjee1983bargaining,
  author       = {Kalyan Chatterjee and William Samuelson},
  title        = {Bargaining under Incomplete Information},
  journal      = {Operations Research},
  volume       = {31},
  number       = {5},
  pages        = {835--851},
  year         = {1983},
  publisher    = {INFORMS},
  url          = {https://www.jstor.org/stable/170889}
}

@book{raiffa1982art,
  author    = {Raiffa, Howard},
  title     = {The Art and Science of Negotiation},
  publisher = {Harvard University Press},
  year      = {1982}
}

@inproceedings{lewis2017deal,
  author    = {Lewis, Mike and Yarats, Denis and Dauphin, Yann N. and Parikh, Devi and Batra, Dhruv},
  title     = {Deal or No Deal? End-to-End Learning for Negotiation Dialogues},
  booktitle = {Proceedings of the 2017 Conference on Empirical Methods in Natural Language Processing (EMNLP)},
  pages     = {2443--2453},
  year      = {2017}
}

@inproceedings{he2018decoupling,
  author    = {He, He and Chen, Derek and Balakrishnan, Anusha and Liang, Percy},
  title     = {Decoupling Strategy and Generation in Negotiation Dialogues},
  booktitle = {Proceedings of the 2018 Conference on Empirical Methods in Natural Language Processing (EMNLP)},
  pages     = {2333--2343},
  year      = {2018}
}

@article{yarats2018hierarchical,
  author  = {Yarats, Denis and Lewis, Mike},
  title   = {Hierarchical Text Generation and Planning for Strategic Dialogue},
  journal = {arXiv preprint arXiv:1712.05846},
  year    = {2018}
}

@inproceedings{chawla2021casino,
    title = "{C}a{S}i{N}o: A Corpus of Campsite Negotiation Dialogues for Automatic Negotiation Systems",
    author = "Chawla, Kushal  and
      Ramirez, Jaysa  and
      Clever, Rene  and
      Lucas, Gale  and
      May, Jonathan  and
      Gratch, Jonathan",
    editor = "Toutanova, Kristina  and
      Rumshisky, Anna  and
      Zettlemoyer, Luke  and
      Hakkani-Tur, Dilek  and
      Beltagy, Iz  and
      Bethard, Steven  and
      Cotterell, Ryan  and
      Chakraborty, Tanmoy  and
      Zhou, Yichao",
    booktitle = "Proceedings of the 2021 Conference of the North American Chapter of the Association for Computational Linguistics: Human Language Technologies",
    month = jun,
    year = "2021",
    address = "Online",
    publisher = "Association for Computational Linguistics",
    url = "https://aclanthology.org/2021.naacl-main.254/",
    doi = "10.18653/v1/2021.naacl-main.254",
    pages = "3167--3185",
    }

@article{chatterjee2024agreemate,
  author  = {Chatterjee, Aayan and Miller, Sydney and Parepally, Nitya},
  title   = {{AgreeMate}: Teaching {LLM}s to Haggle},
  journal = {arXiv preprint arXiv:2412.18690},
  year    = {2024}
}

@article{yao2023react,
  author  = {Yao, Shunyu and Zhao, Jeffrey and Yu, Dian and Du, Nan and Shafran, Izhak and Narasimhan, Karthik and Cao, Yuan},
  title   = {{ReAct}: Synergizing Reasoning and Acting in Language Models},
  journal = {arXiv preprint arXiv:2210.03629},
  year    = {2023}
}

@article{hwang2023promptable,
  author  = {Hwang, Minyoung and Weihs, Luca and Park, Chanhee and Lee, Kimin and Kembhavi, Aniruddha and Ehsani, Kiana},
  title   = {Promptable Behaviors: Personalizing Multi-Objective Rewards from Human Preferences},
  journal = {arXiv preprint arXiv:2312.09337},
  year    = {2023}
}

@misc{cohen2025exploring,
      title={Exploring Big Five Personality and AI Capability Effects in LLM-Simulated Negotiation Dialogues}, 
      author={Myke C. Cohen and Zhe Su and Hsien-Te Kao and Daniel Nguyen and Spencer Lynch and Maarten Sap and Svitlana Volkova},
      year={2025},
      eprint={2506.15928},
      archivePrefix={arXiv},
      primaryClass={cs.AI},
      url={https://arxiv.org/abs/2506.15928}, 
}

@article{schneider2024negotiating,
  author  = {Schneider, Johannes and Haag, Stefanie and Kruse, Leona C.},
  title   = {Negotiating with {LLM}s: Prompt Hacks, Skill Gaps, and Reasoning Deficits},
  journal = {arXiv preprint arXiv:2312.03720},
  year    = {2024}
}

@article{huang2024personality,
  author  = {Huang, Yu Jen and Hadfi, Rafik},
  title   = {How Personality Traits Influence Negotiation Outcomes? A Simulation Based on Large Language Models},
  journal = {arXiv preprint arXiv:2407.11549},
  year    = {2024}
}

@article{hua2024gametheoretic,
  author  = {Hua, Wenyue and Liu, Ollie and Li, Lingyao and Amayuelas, Alfonso and Chen, Julie and Jiang, Lizhou and Jin, Mingyu and Fan, Liangyu and Sun, Fei and Wang, William and Wang, Xiang and Zhang, Yongfeng},
  title   = {Game-Theoretic {LLM}: Agent Workflow for Negotiation Games},
  journal = {arXiv preprint arXiv:2411.05990},
  year    = {2024}
}

@misc{mao2024alympicsllmagentsmeet,
      title={ALYMPICS: LLM Agents Meet Game Theory -- Exploring Strategic Decision-Making with AI Agents}, 
      author={Shaoguang Mao and Yuzhe Cai and Yan Xia and Wenshan Wu and Xun Wang and Fengyi Wang and Tao Ge and Furu Wei},
      year={2024},
      eprint={2311.03220},
      archivePrefix={arXiv},
      primaryClass={cs.CL},
      url={https://arxiv.org/abs/2311.03220}, 
}

@misc{duan2024gtbenchuncoveringstrategicreasoning,
      title={GTBench: Uncovering the Strategic Reasoning Limitations of LLMs via Game-Theoretic Evaluations}, 
      author={Jinhao Duan and Renming Zhang and James Diffenderfer and Bhavya Kailkhura and Lichao Sun and Elias Stengel-Eskin and Mohit Bansal and Tianlong Chen and Kaidi Xu},
      year={2024},
      eprint={2402.12348},
      archivePrefix={arXiv},
      primaryClass={cs.CL},
      url={https://arxiv.org/abs/2402.12348}, 
}

@book{bazerman1992negotiating,
  title={Negotiating Rationally},
  author={Bazerman, Max H and Neale, Margaret A},
  year={1992},
  publisher={Free Press},
  address={New York},
  isbn={9780029019863}
}

@inproceedings{ma2005bigfive_conflict_negotiation,
  author    = {Ma, Zhenzhong},
  title     = {Exploring the Relationships between the Big Five Personality Factors, Conflict Styles, and Bargaining Behaviors},
  booktitle = {SSRN Electronic Journal},
  year      = {2005},
  month     = jun,
  doi       = {10.2139/ssrn.735063},
  url       = {https://ssrn.com/abstract=735063},
  note      = {Available at SSRN}
}

@article{myerson1984twoperson,
  title={Two-Person Bargaining Problems with Incomplete Information},
  author={Myerson, Roger B},
  journal={Econometrica: Journal of the Econometric Society},
  volume={52},
  number={2},
  pages={461--487},
  year={1984},
  publisher={JSTOR},
  doi={10.2307/1911499},
  url={https://www.jstor.org/stable/1911499}
}

@misc{liu2026agenticpaymultiagentllmnegotiation,
      title={AgenticPay: A Multi-Agent LLM Negotiation System for Buyer-Seller Transactions}, 
      author={Xianyang Liu and Shangding Gu and Dawn Song},
      year={2026},
      eprint={2602.06008},
      archivePrefix={arXiv},
      primaryClass={cs.AI},
      url={https://arxiv.org/abs/2602.06008}, 
}

@misc{xia2024measuringbargainingabilitiesllms,
      title={Measuring Bargaining Abilities of LLMs: A Benchmark and A Buyer-Enhancement Method}, 
      author={Tian Xia and Zhiwei He and Tong Ren and Yibo Miao and Zhuosheng Zhang and Yang Yang and Rui Wang},
      year={2024},
      eprint={2402.15813},
      archivePrefix={arXiv},
      primaryClass={cs.CL},
      url={https://arxiv.org/abs/2402.15813}, 
}

@misc{anthropic_claude_opus_46_2026,
  author       = {{Anthropic}},
  title        = {Introducing Claude Opus 4.6},
  year         = {2026},
  month        = feb,
  url          = {https://www.anthropic.com/news/claude-opus-4-6},
}

@misc{xai_grok42_2026,
  author       = {{Microsoft Azure and xAI}},
  title        = {Grok 4.2 Reasoning},
  year         = {2026},
  month        = apr,
  url          = {https://ai.azure.com/catalog/models/grok-4-20-reasoning},
  note         = {Model catalog entry for xAI Grok 4.2 Reasoning. Accessed: 2026-05-04}
}

@incollection{ausubel2002bargaining,
  author    = {Ausubel, Lawrence M. and Cramton, Peter and Deneckere, Raymond J.},
  title     = {Bargaining with Incomplete Information},
  booktitle = {Handbook of Game Theory with Economic Applications},
  editor    = {Aumann, Robert J. and Hart, Sergiu},
  publisher = {Elsevier},
  year      = {2002},
  volume    = {3},
  pages     = {1897--1945},
  doi       = {10.1016/S1574-0005(02)03013-8}
}

@article{faratin1998negotiation,
  author  = {Faratin, Peyman and Sierra, Carles and Jennings, Nick R.},
  title   = {Negotiation Decision Functions for Autonomous Agents},
  journal = {Robotics and Autonomous Systems},
  year    = {1998},
  volume  = {24},
  number  = {3--4},
  pages   = {159--182},
  doi     = {10.1016/S0921-8890(98)00029-3}
}

@article{baarslag2016learning,
  author  = {Baarslag, Tim and Hendrikx, Mark J. C. and Hindriks, Koen V. and Jonker, Catholijn M.},
  title   = {Learning about the Opponent in Automated Bilateral Negotiation: A Comprehensive Survey of Opponent Modeling Techniques},
  journal = {Autonomous Agents and Multi-Agent Systems},
  year    = {2016},
  volume  = {30},
  number  = {5},
  pages   = {849--898},
  doi     = {10.1007/s10458-015-9309-1}
}

@article{baarslag2014acceptance,
  author  = {Baarslag, Tim and Hindriks, Koen V. and Jonker, Catholijn M.},
  title   = {Effective Acceptance Conditions in Real-Time Automated Negotiation},
  journal = {Decision Support Systems},
  year    = {2014},
  volume  = {60},
  pages   = {68--77},
  doi     = {10.1016/j.dss.2013.05.021}
}

@article{karagozoglu2019bargaining,
  author  = {Karag{\"o}zo{\u g}lu, Emin and Kocher, Martin G.},
  title   = {Bargaining under Time Pressure from Deadlines},
  journal = {Experimental Economics},
  year    = {2019},
  volume  = {22},
  number  = {2},
  pages   = {419--440},
  doi     = {10.1007/s10683-018-9579-y}
}

@article{petrowsky2025power,
  author  = {Petrowsky, Hannes M. and Boecker, Lea and Escher, Yannik A. and Frech, Marie-Lena and Friese, Malte and Galinsky, Adam D. and Gunia, Brian and Lee, Alice J. and Schaerer, Michael and Schweinsberg, Martin and Soliman, Meikel and Swaab, Roderick and Troll, Eve S. and Weber, Marcel and Loschelder, David D.},
  title   = {The Power and Peril of First Offers in Negotiations: A Conceptual, Meta-Analytic, and Experimental Synthesis},
  journal = {Organizational Behavior and Human Decision Processes},
  year    = {2025},
  volume  = {191},
  pages   = {104448},
  doi     = {10.1016/j.obhdp.2025.104448}
}

@article{tey2021impact,
  author  = {Tey, Kian Siong and Schaerer, Michael and Madan, Nikhil and Swaab, Roderick I.},
  title   = {The Impact of Concession Patterns on Negotiations: When and Why Decreasing Concessions Lead to a Distributive Disadvantage},
  journal = {Organizational Behavior and Human Decision Processes},
  year    = {2021},
  volume  = {165},
  pages   = {153--166},
  doi     = {10.1016/j.obhdp.2021.05.003}
}

@article{grennan2013price,
  author  = {Grennan, Matthew},
  title   = {Price Discrimination and Bargaining: Empirical Evidence from Medical Devices},
  journal = {American Economic Review},
  year    = {2013},
  volume  = {103},
  number  = {1},
  pages   = {145--177},
  doi     = {10.1257/aer.103.1.145}
}

@article{grennan2014ability,
  author  = {Grennan, Matthew},
  title   = {Bargaining Ability and Competitive Advantage: Empirical Evidence from Medical Devices},
  journal = {Management Science},
  year    = {2014},
  volume  = {60},
  number  = {12},
  pages   = {3011--3025},
  doi     = {10.1287/mnsc.2014.2006}
}

@article{kadiyala2023predicting,
  author  = {Kadiyala, Bharadwaj and Phillips, Robert and {\c{S}}im{\c{s}}ek, A. Serdar and van Ryzin, Garrett},
  title   = {Predicting Transaction Outcomes under Customized Pricing with Discretion: A Structural Estimation Approach},
  journal = {Production and Operations Management},
  year    = {2023},
  volume  = {32},
  number  = {6},
  pages   = {1654--1673},
  doi     = {10.1111/poms.13931}
}

@article{dindaroglu2024empirical,
  author  = {Dindaroglu, Burak and Ertac, Seda},
  title   = {An Empirical Study of Sequential Offer Bargaining during the Festival of Sacrifice},
  journal = {Journal of Economic Psychology},
  year    = {2024},
  volume  = {101},
  pages   = {102707},
  doi     = {10.1016/j.joep.2024.102707}
}

@article{sigurdardottir2019buyer,
  author  = {Sigur{\dh}ard{\'o}ttir, Ald{\'i}s Gu{\dh}n{\'y} and Hotait, Ali and Eichst{\"a}dt, Tilman},
  title   = {Buyer and Seller Differences in Business-to-Business Negotiations},
  journal = {Negotiation Journal},
  year    = {2019},
  volume  = {35},
  number  = {2},
  pages   = {297--331},
  doi     = {10.1111/nejo.12289}
}

@article{herold2025brave,
  author  = {Herold, Silke and Heller, Jonas and Rozemeijer, Frank and Mahr, Dominik},
  title   = {Brave New Procurement Deals: An Experimental Study of How Generative Artificial Intelligence Reshapes Buyer--Supplier Negotiations},
  journal = {Journal of Purchasing and Supply Management},
  year    = {2025},
  volume  = {31},
  number  = {4},
  pages   = {101012},
  doi     = {10.1016/j.pursup.2025.101012}
}

@article{nash1950bargaining,
  author  = {Nash, John F.},
  title   = {The Bargaining Problem},
  journal = {Econometrica},
  year    = {1950},
  volume  = {18},
  number  = {2},
  pages   = {155--162}
}

@book{roth1979axiomatic,
  author    = {Roth, Alvin E.},
  title     = {Axiomatic Models of Bargaining},
  publisher = {Springer},
  year      = {1979},
  series    = {Lecture Notes in Economics and Mathematical Systems},
  volume    = {170},
  doi       = {10.1007/978-3-642-51570-5}
}

@incollection{hart1992axiomatic,
  author    = {Hart, Sergiu},
  title     = {Axiomatic Approaches to Coalitional Bargaining},
  booktitle = {Rational Interaction: Essays in Honor of John C. Harsanyi},
  editor    = {Selten, Reinhard},
  publisher = {Springer},
  year      = {1992},
  pages     = {305--320}
}

@techreport{thomson2009bargaining,
  author      = {Thomson, William},
  title       = {Bargaining and the Theory of Cooperative Games: John Nash and Beyond},
  institution = {University of Rochester},
  year        = {2009},
  number      = {Working Paper No. 554}
}

@article{thomson2022axiomatic,
  author  = {Thomson, William},
  title   = {On the Axiomatic Theory of Bargaining: A Survey of Recent Results},
  journal = {Review of Economic Design},
  year    = {2022},
  volume  = {26},
  number  = {4},
  pages   = {491--542},
  doi     = {10.1007/s10058-022-00319-1}
}

@article{roth1977individual,
  author  = {Roth, Alvin E.},
  title   = {Individual Rationality and Nash's Solution to the Bargaining Problem},
  journal = {Mathematics of Operations Research},
  year    = {1977},
  volume  = {2},
  number  = {1},
  pages   = {64--65}
}

@article{schweinsberg2022impasses,
  author  = {Schweinsberg, Martin and Thau, Stefan and Pillutla, Madan M.},
  title   = {Negotiation Impasses: Types, Causes, and Resolutions},
  journal = {Journal of Management},
  year    = {2022},
  volume  = {48},
  number  = {1},
  pages   = {49--76},
  doi     = {10.1177/01492063211021657}
}

@article{stuhlmacher1998timepressure,
  author  = {Stuhlmacher, Alice F. and Gillespie, Treena L. and Champagne, Matthew V.},
  title   = {The Impact of Time Pressure in Negotiation: A Meta-Analysis},
  journal = {International Journal of Conflict Management},
  year    = {1998},
  volume  = {9},
  number  = {2},
  pages   = {97--116}
}

@inproceedings{baarslag2012measuring,
  author    = {Baarslag, Tim and Hendrikx, Mark J. C. and Hindriks, Koen V. and Jonker, Catholijn M.},
  title     = {Measuring the Performance of Online Opponent Models in Automated Bilateral Negotiation},
  booktitle = {AI 2012: Advances in Artificial Intelligence},
  year      = {2012},
  pages     = {1--14},
  publisher = {Springer},
  doi       = {10.1007/978-3-642-35101-3_1}
}

@article{baarslag2013practical,
  author  = {Baarslag, Tim and Fujita, Katsuhide and Gerding, Enrico H. and Hindriks, Koen V. and Ito, Takayuki and Jennings, Nicholas R. and Jonker, Catholijn M. and Kraus, Sarit and Lin, Raz and Robu, Valentin and Williams, Colin R.},
  title   = {Evaluating Practical Negotiating Agents: Results and Analysis of the 2011 International Competition},
  journal = {Artificial Intelligence},
  year    = {2013},
  volume  = {198},
  pages   = {73--103},
  doi     = {10.1016/j.artint.2012.09.004}
}

@article{myerson1983efficient,
  title={Efficient mechanisms for bilateral trading},
  author={Myerson, Roger B and Satterthwaite, Mark A},
  journal={Journal of Economic Theory},
  volume={29},
  number={2},
  pages={265--281},
  year={1983}
}

@article{babcock1997explaining,
  title={Explaining bargaining impasse: The role of self-serving biases},
  author={Babcock, Linda and Loewenstein, George},
  journal={Journal of Economic Perspectives},
  volume={11},
  number={1},
  pages={109--126},
  year={1997}
}

@book{raiffa2002negotiation,
  title={Negotiation Analysis: The Science and Art of Collaborative Decision Making},
  author={Raiffa, Howard and Richardson, John and Metcalfe, David},
  year={2002},
  publisher={Harvard University Press}
}

@book{schelling1960strategy,
  title={The Strategy of Conflict},
  author={Schelling, Thomas C},
  year={1960},
  publisher={Harvard University Press}
}

@inproceedings{kwon-etal-2024-llms,
    title = "Are {LLM}s Effective Negotiators? Systematic Evaluation of the Multifaceted Capabilities of {LLM}s in Negotiation Dialogues",
    author = "Kwon, Deuksin  and
      Weiss, Emily  and
      Kulshrestha, Tara  and
      Chawla, Kushal  and
      Lucas, Gale  and
      Gratch, Jonathan",
    editor = "Al-Onaizan, Yaser  and
      Bansal, Mohit  and
      Chen, Yun-Nung",
    booktitle = "Findings of the Association for Computational Linguistics: EMNLP 2024",
    month = nov,
    year = "2024",
    address = "Miami, Florida, USA",
    publisher = "Association for Computational Linguistics",
    url = "https://aclanthology.org/2024.findings-emnlp.310/",
    doi = "10.18653/v1/2024.findings-emnlp.310",
    pages = "5391--5413",
}

@inproceedings{abdelnabi2024cooperation,
  title={Cooperation, Competition, and Maliciousness: 
         {LLM}-Stakeholders Interactive Negotiation},
  author={Abdelnabi, Sahar and Gomaa, Amr and Sivaprasad, Sarath 
          and Sch{\"o}nherr, Lea and Fritz, Mario},
  booktitle={Advances in Neural Information Processing Systems (NeurIPS)},
  year={2024}
}

@misc{oh2026meritfeedbackelicitsbetter,
      title={MERIT Feedback Elicits Better Bargaining in LLM Negotiators}, 
      author={Jihwan Oh and Murad Aghazada and Yooju Shin and Se-Young Yun and Taehyeon Kim},
      year={2026},
      eprint={2602.10467},
      archivePrefix={arXiv},
      primaryClass={cs.AI},
      url={https://arxiv.org/abs/2602.10467}, 
}

@inproceedings{
hendrycks2021measuring,
title={Measuring Mathematical Problem Solving With the {MATH} Dataset},
author={Dan Hendrycks and Collin Burns and Saurav Kadavath and Akul Arora and Steven Basart and Eric Tang and Dawn Song and Jacob Steinhardt},
booktitle={Thirty-fifth Conference on Neural Information Processing Systems Datasets and Benchmarks Track (Round 2)},
year={2021},
url={https://openreview.net/forum?id=7Bywt2mQsCe}
}

@misc{jimenez2024swebenchlanguagemodelsresolve,
      title={SWE-bench: Can Language Models Resolve Real-World GitHub Issues?}, 
      author={Carlos E. Jimenez and John Yang and Alexander Wettig and Shunyu Yao and Kexin Pei and Ofir Press and Karthik Narasimhan},
      year={2024},
      eprint={2310.06770},
      archivePrefix={arXiv},
      primaryClass={cs.CL},
      url={https://arxiv.org/abs/2310.06770}, 
}

@misc{chen2021evaluating,
      title={Evaluating Large Language Models Trained on Code}, 
      author={Mark Chen and Jerry Tworek and Heewoo Jun and Qiming Yuan and Henrique Ponde de Oliveira Pinto and Jared Kaplan and Harri Edwards and Yuri Burda and Nicholas Joseph and Greg Brockman and Alex Ray and Raul Puri and Gretchen Krueger and Michael Petrov and Heidy Khlaaf and Girish Sastry and Pamela Mishkin and Brooke Chan and Scott Gray and Nick Ryder and Mikhail Pavlov and Alethea Power and Lukasz Kaiser and Mohammad Bavarian and Clemens Winter and Philippe Tillet and Felipe Petroski Such and Dave Cummings and Matthias Plappert and Fotios Chantzis and Elizabeth Barnes and Ariel Herbert-Voss and William Hebgen Guss and Alex Nichol and Alex Paino and Nikolas Tezak and Jie Tang and Igor Babuschkin and Suchir Balaji and Shantanu Jain and William Saunders and Christopher Hesse and Andrew N. Carr and Jan Leike and Josh Achiam and Vedant Misra and Evan Morikawa and Alec Radford and Matthew Knight and Miles Brundage and Mira Murati and Katie Mayer and Peter Welinder and Bob McGrew and Dario Amodei and Sam McCandlish and Ilya Sutskever and Wojciech Zaremba},
      year={2021},
      eprint={2107.03374},
      archivePrefix={arXiv},
      primaryClass={cs.LG},
      url={https://arxiv.org/abs/2107.03374}, 
}

@article{dubois2024alpacafarm,
  title         = {{AlpacaFarm}: A Simulation Framework for Methods that Learn from Human Feedback},
  author        = {Yann Dubois and Xuechen Li and Rohan Taori and Tianyi Zhang and Ishaan Gulrajani and Jimmy Ba and Carlos Guestrin and Percy Liang and Tatsunori B. Hashimoto},
  journal       = {Advances in Neural Information Processing Systems (NeurIPS)},
  year          = {2024},
  eprint        = {2305.14387},
  archiveprefix = {arXiv},
  primaryclass  = {cs.LG}
}

@inproceedings{zhou2024webarena,
  title     = {{WebArena}: A Realistic Web Environment for Building Autonomous Agents},
  author    = {Shuyan Zhou and Frank F. Xu and Hao Zhu and Xuhui Zhou and Robert Lo and Abishek Sridhar and Xianyi Cheng and Tianyue Ou and Yonatan Bisk and Daniel Fried and Uri Alon and Graham Neubig},
  booktitle = {The Twelfth International Conference on Learning Representations (ICLR)},
  year      = {2024},
  eprint    = {2307.13854},
  archiveprefix = {arXiv},
  primaryclass = {cs.AI}
}

@inproceedings{liu2023agentbench,
  title     = {{AgentBench}: Evaluating {LLMs} as Agents},
  author    = {Xiao Liu and Hao Yu and Hanchen Zhang and Yifan Xu and Xuanyu Lei and Hanyu Lai and Yu Gu and Hangliang Ding and Kaiwen Men and Kejuan Yang and Shudan Zhang and Xiang Deng and Aohan Zeng and Zhengxiao Du and Chenhui Zhang and Sheng Shen and Tianjun Zhang and Yu Su and Huan Sun and Minlie Huang and Yuxiao Dong and Jie Tang},
  booktitle = {The Twelfth International Conference on Learning Representations (ICLR)},
  year      = {2024},
  eprint    = {2308.03688},
  archiveprefix = {arXiv},
  primaryclass = {cs.AI}
}

@misc{anthropic_claude_opus_47_2026,
  author       = {{Anthropic}},
  title        = {Introducing Claude Opus 4.7},
  year         = {2026},
  howpublished = {\url{https://www.anthropic.com/news/claude-opus-4-7}},
}

@misc{google_gemma4_31b_it_2026,
  author       = {{Google}},
  title        = {{google/gemma-4-31B-it}},
  year         = {2026},
  howpublished = {\url{https://huggingface.co/google/gemma-4-31B-it}},
}

@misc{google_gemini_31_pro_preview_2026,
  author       = {{Google DeepMind}},
  title        = {Gemini 3.1 Pro Preview},
  year         = {2026},
  howpublished = {\url{https://ai.google.dev/gemini-api/docs/models/gemini-3.1-pro-preview}},
}

@online{openai2026gpt55,
  author  = {{OpenAI}},
  title   = {Introducing {GPT-5.5}},
  date    = {2026-04-23},
  year = {2026},
  url     = {https://openai.com/index/introducing-gpt-5-5/},
  urldate = {2026-05-12}
}

@misc{deepseek_v4_pro_2026,
  author       = {{DeepSeek}},
  title        = {DeepSeek V4 Pro},
  year         = {2026},
  howpublished = {\url{https://api-docs.deepseek.com/}},
}

@misc{qwen36_plus_2026,
  author       = {{Qwen Team}},
  title        = {Qwen3.6-Plus},
  year         = {2026},
  howpublished = {\url{https://qwen.ai/}},
}

@misc{moonshot_kimi_k26_2026,
  author       = {{Moonshot AI}},
  title        = {Kimi K2.6},
  year         = {2026},
  howpublished = {\url{https://platform.kimi.ai/}},
}

@misc{openai_gpt54_2026,
  author       = {{OpenAI}},
  title        = {GPT-5.4},
  year         = {2026},
  howpublished = {\url{https://platform.openai.com/docs/models}},
}

@misc{openai_gpt4o_mini_2024,
  author       = {{OpenAI}},
  title        = {GPT-4o mini: Advancing Cost-Efficient Intelligence},
  year         = {2024},
  howpublished = {\url{https://openai.com/index/gpt-4o-mini-advancing-cost-efficient-intelligence/}},
}

@misc{bytedance_doubao_seed20_pro_2026,
  author       = {{ByteDance}},
  title        = {Doubao-Seed-2.0-Pro},
  year         = {2026},
  howpublished = {\url{https://www.volcengine.com/product/doubao}},
}

@misc{zai_glm51_2026,
  author       = {{Z.AI}},
  title        = {GLM-5.1},
  year         = {2026},
  howpublished = {\url{https://docs.z.ai/}},
}

@inproceedings{zheng2023judging,
  title     = {Judging {LLM}-as-a-Judge with {MT-Bench} and {Chatbot Arena}},
  author    = {Zheng, Lianmin and Chiang, Wei-Lin and Sheng, Ying and Zhuang, Siyuan and Wu, Zhanghao and Zhuang, Yonghao and Lin, Zi and Li, Zhuohan and Li, Dacheng and Xing, Eric P. and Zhang, Hao and Gonzalez, Joseph E. and Stoica, Ion},
  booktitle = {Advances in Neural Information Processing Systems (NeurIPS)},
  year      = {2023}
}

@article{lambert2024tulu3,
  title   = {{T\"ulu} 3: Pushing Frontiers in Open Language Model Post-Training},
  author  = {Lambert, Nathan and Morrison, Jacob and Pyatkin, Valentina and Huang, Shengyi and Ivison, Hamish and Brahman, Faeze and Miranda, Lester James V. and Liu, Alisa and Dziri, Nouha and Lyu, Shane and Gu, Yuling and Malik, Saumya and Graf, Victoria and Hwang, Jena D. and Yang, Jiangjiang and Bras, Ronan Le and Tafjord, Oyvind and Wilhelm, Chris and Soldaini, Luca and Smith, Noah A. and Wang, Yizhong and Dasigi, Pradeep and Hajishirzi, Hannaneh},
  journal = {arXiv preprint arXiv:2411.15124},
  year    = {2024}
}

@article{deepseekai2025r1,
  title   = {{DeepSeek-R1}: Incentivizing Reasoning Capability in {LLMs} via Reinforcement Learning},
  author  = {{DeepSeek-AI}},
  journal = {arXiv preprint arXiv:2501.12948},
  year    = {2025}
}

@misc{jia2025writingzero,
  title         = {Writing-Zero: Bridge the Gap Between Non-verifiable Tasks and Verifiable Rewards},
  author        = {Jia, Ruipeng and Yang, Yunyi and Gai, Yongbo and Luo, Kai and Huang, Shihao and Lin, Jianhe and Jiang, Xiaoxi and Jiang, Guanjun},
  year          = {2025},
  eprint        = {2506.00103},
  archiveprefix = {arXiv},
  primaryclass  = {cs.CL},
  url           = {https://arxiv.org/abs/2506.00103},
}

@misc{agrawal2026gepareflectivepromptevolution,
      title={GEPA: Reflective Prompt Evolution Can Outperform Reinforcement Learning}, 
      author={Lakshya A Agrawal and Shangyin Tan and Dilara Soylu and Noah Ziems and Rishi Khare and Krista Opsahl-Ong and Arnav Singhvi and Herumb Shandilya and Michael J Ryan and Meng Jiang and Christopher Potts and Koushik Sen and Alexandros G. Dimakis and Ion Stoica and Dan Klein and Matei Zaharia and Omar Khattab},
      year={2026},
      eprint={2507.19457},
      archivePrefix={arXiv},
      primaryClass={cs.CL},
      url={https://arxiv.org/abs/2507.19457}, 
}
